\documentclass[10pt,epsfig]{article}

\usepackage{enumerate}
\usepackage{float}
\usepackage{subfig}
\usepackage{color}
\usepackage{slashed}
\usepackage{multirow}
\usepackage{cite}

\let\counterwithin\relax
\usepackage{chngcntr}
\usepackage{amssymb, amsmath,mathrsfs}

\usepackage{graphics}
\usepackage{graphicx}
\usepackage{epsf}
\usepackage{epsfig}
\usepackage{float}
\usepackage{makecell}
\usepackage{multirow}
\usepackage{color}
\usepackage{xcolor}

\usepackage[utf8]{inputenc}
\usepackage{amsmath}
\usepackage{amssymb}
\usepackage{subfig}
\usepackage[normalem]{ulem}

\usepackage{mathtools}
\usepackage{amsmath}

\newcommand\undermat[2]{
  \makebox[0.5pt][l]{$\smash{\underbrace{\phantom{%
    \begin{matrix}#2\end{matrix}}}_{ \let\scriptstyle\textstyle\text{\large $#1$}}}$}#2}
\newcommand\overmat[2]{
  \makebox[-1pt][l]{$\smash{\overbrace{\phantom{%
    \begin{matrix}#2\end{matrix}}}^{ \let\scriptstyle\textstyle\text{\large $#1$}}}$}#2}    
\usepackage{tikz}
\usepackage[vcentermath]{youngtab}
\usepackage{slashed} 
\usepackage[font=small]{caption} 
\usepackage{times} 

\usepackage{bm}       
\usepackage{bbm} 

\usepackage{relsize}  

\usepackage{autobreak}  

\usepackage{makeidx} 
\usepackage{bbding} 
\usepackage{listings} 
\usepackage{ytableau}

\usepackage[colorlinks=true,
            linkcolor=blue,
            urlcolor=red,
            citecolor=red]{hyperref}

\long\def\rpl#1!!#2!!{\textcolor{red}{#1} \textcolor{blue}{#2}}

\usepackage[top=1in, left=0.95in, bottom=1.1in, right=0.95in]{geometry}
\def\baselinestretch{1.27}
\usepackage[toc]{appendix}

\newcommand{\beq}{\begin {equation}}  
\newcommand{\eeq}{\end   {equation}} 
\newcommand{\bea}{\begin {eqnarray}} 
\newcommand{\eea}{\end   {eqnarray}}  
\newcommand{\baa}{\begin {array}   } 
\newcommand{\eaa}{\end   {array}   }     
\newcommand{\bit}{\begin {itemize} }
\newcommand{\eit}{\end   {itemize} }
\newcommand{\be }{\begin {equation}} 
\newcommand{\ee }{\end   {equation}}

\newcommand{\mc}[1]{\mathcal{#1}}

\newcommand{\vev}[1]{ \left\langle {#1}  \right\rangle }

\newcommand{\hc}{{\text{h.c.}}~}

\newcommand{\eq}[1]{\begin{equation}\begin{split} #1 \end{split}\end{equation}}
\newcommand{\eqs}[1]{\begin{align} #1 \end{align}}

\newcommand{\4}{\!\!\!\!/\,}
\newcommand{\comments}[1]{}

\newcommand*\circled[1]{
	\tikz[baseline= -2.5pt]{\node[shape=circle,draw,inner sep=0.3pt] (char) {#1};}
}

\newcolumntype{M}[1]{>{\centering\arraybackslash}m{#1}}
\newcolumntype{N}{@{}m{0pt}@{}}
\DeclareMathOperator*{\Motimes}{\text{\raisebox{0.25ex}{\scalebox{0.5}{$\bigotimes$}}}}

\allowdisplaybreaks
\begin{document}

\begin{center}

{\Large \textbf  {Complete Set of Dimension-8 Operators in the Standard Model Effective Field Theory}}\\[10mm]

Hao-Lin Li$^{a}$\footnote{lihaolin@itp.ac.cn}, Zhe Ren$^{a, b}$, Jing Shu$^{a, b, c, d, e}$\footnote{jshu@itp.ac.cn}, Ming-Lei Xiao$^{a}$\footnote{mingleix@itp.ac.cn}, Jiang-Hao Yu$^{a, b}$\footnote{jhyu@itp.ac.cn}, Yu-Hui Zheng$^{a, b}$\\[10mm]


\noindent 
$^a${\em \small CAS Key Laboratory of Theoretical Physics, Institute of Theoretical Physics, Chinese Academy of Sciences,    \\ Beijing 100190, P. R. China}  \\
$^b${\em \small School of Physical Sciences, University of Chinese Academy of Sciences,   Beijing 100049, P.R. China}   \\
$^c${\em \small CAS Center for Excellence in Particle Physics, Beijing 100049, China} \\
$^d${\em \small Center for High Energy Physics, Peking University, Beijing 100871, China} \\
$^e${\em \small School of Fundamental Physics and Mathematical Sciences, Hangzhou Institute for Advanced Study, University of Chinese Academy of Sciences, Hangzhou 310024, China} \\[10mm]

\date{\today}   
          
\end{center}

\begin{abstract} 

We present a complete list of the dimension 8 operator basis in the standard model effective field theory using group theoretic techniques in a systematic and automated way. We adopt a new form of operators in terms of the irreducible representations of the Lorentz group, and identify the Lorentz structures as states in a $SU(N)$ group. In this way, redundancy from equations of motion is absent and that from integration-by-part is treated using the fact that the independent Lorentz basis forms an invariant subspace of the $SU(N)$ group. We also decompose operators into the ones with definite permutation symmetries among flavor indices to deal with subtlety from repeated fields. For the first time, we provide the explicit form of independent flavor-specified operators in a systematic way. Our algorithm can be easily applied to higher dimensional standard model effective field theory and other effective field theories, making these studies more approachable.

\end{abstract}

\newpage

\tableofcontents

\setcounter{footnote}{0}

\def\baselinestretch{1.5}
\counterwithin{equation}{section}

\newpage
%
%

\section{Introduction}
\label{sec:intro}

The standard model (SM) of particle physics is a great triumph of modern physics. It has successfully explained almost all experimental results and predicted a wide variety of phenomena with unprecedented accuracy. 
Despite its great success, however, the SM fails to account for some basic properties of our universe, e.g., neutrino masses, matter-antimatter asymmetry, the existence of dark matter, etc. 
This has motivated both the theorists and experimentalists to make a dedicated effort to search for pieces of evidence of new physics (NP) beyond the SM.
Until now, direct searches have not yielded anything of significance, which already pushed the NP scale to be above the TeV scale. 
Therefore, it is highly motivated to study NP phenomena involving only the SM particles within the framework of effective theories.

Effective field theory (EFT) provides a systematical framework for parametrizing various NP only based on the field content, the Lorentz invariance, and the gauge symmetries in the SM.
The Lagrangian of such an EFT contains not only the renormalized SM Lagrangian but also all the higher-dimensional invariant operators, which parametrize all the possible deviations from the SM. 
Assuming that NP appears at the scale $\Lambda$ above the electroweak scale~\footnote{New physics could also exist below electroweak scale, but such scenario is not considered here.}, 
the general Lagrangian can be parametrized as
\bea
{\mathcal L}_{\rm SMEFT} = {\mathcal L}_{\rm SM}^{(4)} + \sum_{d > 4 }\left( \frac{1}{\Lambda} \right)^{d-4} \sum_i C_i^{(d)} {\mathcal O}^{(d)}_{i},
\eea
which describes the standard model effective field theory (SMEFT), with $C_i$ identified as the Wilson coefficients. 
The only possible dimension 5 (d = 5) operator is the famous Weinberg operator~\cite{Weinberg:1979sa}, with lepton number violation encoded.
The dimension 6 operators were first listed in Ref.~\cite{Buchmuller:1985jz}, and a subtle problem arises due to redundancies among the operators.
It is often convenient to obtain a complete set of independent operators, namely the non-redundant operator basis. 
This task is highly non-trivial because different structured operators may be related by the equation of motion (EOM), integration by parts (IBP), and Fierz identities. 
These redundancies could be avoided by imposing the EOMs and IBPs explicitly, and the independent dimension 6 operators in the Warsaw basis~\cite{Grzadkowski:2010es} was constructed based on this principle and the complete renormalization group equations are written in Ref.~\cite{Jenkins:2013zja,Jenkins:2013wua,Alonso:2013hga}. 
In Ref.~\cite{Lehman:2014jma, Liao:2016hru} the complete set of dimension 7 operators has been obtained. 
Recently the Hilbert series method~\cite{Feng:2007ur, Jenkins:2009dy, Hanany:2010vu} has been applied to enumerate the SMEFT operators up to dimension 15~\cite{Lehman:2015via, Lehman:2015coa, Henning:2015daa, Henning:2015alf}, but it is only designed to count the number of independent operators in each dimension. Besides, a few other papers \cite{Gripaios:2018zrz,Criado:2019ugp,Fonseca:2019yya,Banerjee:2020bym} also developed programs to count the number of operators in alternative ways.
Although partial lists of the dim-8 and higher dimensional operators have been obtained~\cite{Babu:2001ex, deGouvea:2007qla, Degrande:2013kka, Graesser:2016bpz, Hays:2018zze}, writing down a complete set of the non-redundant operators explicitly at dimension 8 and higher is still a challenging task.

Our goal in this paper is to find a complete set of dimension 8 operators in the SMEFT framework. 
For a physical process, if the leading NP contribution directly comes from the dimension 8 operators, or 
if the contribution from the dimension 6 operators is sub-dominant or highly constrained, the dimension 8 operators should be seriously considered, even though their Wilson coefficients are suppressed by higher inverse power of the NP scale. 
The first example is the neutral triple gauge boson couplings (nTGC) $ZZV$ and $Z\gamma V$, for which no dimension 6 operator contributes and thus dimension 8 operators dominate~\cite{Degrande:2013kka, Ellis:2019zex}. 
Furthermore, in the dimension 6 operator basis, the Wilson coefficient for the quartic gauge boson coupling (QGC) is related to the one in the triple gauge boson coupling (TGC), while at the dimension 8 there is no such correlation in the Wilson coefficients~\cite{Eboli:2016kko}. 
Similar is true for various Higgs gauge boson couplings.  
For the four fermion interactions, let us take the non-standard neutrino interaction (NSI) as an example. 
At the dimension 6 level, new physics which induces the neutral current NSI also induces an operator involving the charged current NSI, which has been tightly constrained by experiments. 
Thus we expect the dimension 8 operators could dominate the neutral current NSI processes~\cite{Davidson:2003ha}. 
The presence of the electric dipole moment (EDM) which can also be generated by the dimension 8 operators directly indicates the existence of the CP violation in the UV theory, and in some cases dimension 8 operators can be the leading order contribution as its counterpart at dimension 6 operator vanishes.
The dimension 8 operators also generate new kinds of four fermion interactions with quite different Lorentz structures. 
Overall, the dimension 8 operator basis deserves a detailed study with all the non-redundant operators written explicitly.

The main difficulty of listing the independent operators arises from how to effectively eliminate the redundancies among operators with derivatives.
Operators with derivatives often involve in two types of redundancies: 
(1) operators differing by the classical EOM are related to each other through field redefinitions; 
(2) operators differing by a total derivative are equivalent in perturbative calculations, the so-called IBP. 
At lower dimensions where limited operators with derivatives are present, the EOM and IBP relations could be imposed explicitly to eliminate all the redundancies, as was done when the dim 6 and 7 operators were written down~\cite{Grzadkowski:2010es, Lehman:2014jma}.
The on-shell amplitude method~\cite{Shadmi:2018xan, Ma:2019gtx, Aoude:2019tzn, Durieux:2019eor, Falkowski:2019zdo} has been applied to the dimension 6 SMEFT~\cite{Ma:2019gtx, Aoude:2019tzn, Durieux:2019eor} which solves the EOM redundancy but still needs to impose conditions to treat the IBP redundancy. 
Nevertheless, at dimension 8 or higher, the number of such operators increases tremendously, which makes the task very tedious and prone to error. 
The Hilbert series method, applied to the SMEFT, deals with these redundancies via decomposing the field derivatives into irreducible representations of the Lorentz group and removing all the descendants while keeping the primaries in each irreducible representation of the conform group. In spite of its efficiency at counting, this method does not help us write down the operator basis explicitly. 
One step forward along this line is Ref.~\cite{Henning:2019enq}, in which independent Lorentz structures were constructed as ``harmonics'' on the sphere of momentum conservation that exempt from the IBP redundancy, but the issue of identical particles, namely the repeated field problem for operators, was not taken into account.
Ref.~\cite{Gripaios:2018zrz} generates an over-complete list of operators at first, and then reduce it to an independent basis by putting all the redundant relations into a matrix,  
which has also been applied~\cite{Hays:2018zze} to write down the partial list of the dimension 8 operators involving only the bosonic fields.

Another difficulty is how to obtain independent flavor structures when repeated fields are present. 
In the literature~\cite{Grzadkowski:2010es,Fonseca:2019yya,Murphy:2020rsh}, the concept of Lagrangian terms is ambiguous, and it is usually subtle to talk about their flavor structures.
In particular, the $Q^3L$ type operators were pointed out to have only one independent term~\cite{Abbott:1980zj,Grzadkowski:2010es} instead of two terms shown in the older literature~\cite{Weinberg:1979sa}, while for both of them, extra efforts are needed to provide independent flavor-specified operators. It is especially confusing when more than one term has to be written down, when the dependence among their flavor-specified operators is even more obscure.
Later when the dimension 7 operators were listed, this issue of flavor structure was completely ignored in Ref.~\cite{Lehman:2014jma} but later addressed in Ref.~\cite{Liao:2019tep} by imposing several flavor relations explicitly with a tedious procedure.
Ref.~\cite{Fonseca:2019yya} provides a systematic way to deal with flavor structures with repeated fields, in which permutation symmetries of all the factor structures (Lorentz, $SU(3)$, $SU(2)$) in the operator are combined via inner product decomposition into irreducible representations of the flavor tensors. Again, this was only for counting, and thus did not work out any of the symmetrized factor structures explicitly.

In this work, we provide a new and systematic method to list all the independent operators by using group theoretic techniques, which solves the two main difficulties mentioned above. 
Inspired by the correspondence between operators and on-shell amplitudes~\cite{Shadmi:2018xan, Ma:2019gtx, Aoude:2019tzn, Durieux:2019eor, Falkowski:2019zdo} and the $SU(N)$ transformation of on-shell amplitudes~\cite{Henning:2019enq, Henning:2019mcv}, we start by adopting a new form of operators constructed from building blocks, fields with or without derivatives, in the irreducible representations (irreps) of the Lorentz group, for which the EOM redundancy is absent. 
The Lorentz structure of operators is then identified as states which transform linearly under the $SU(N)$ group and form a large inter-class space, in which total derivatives form an invariant subspace of the $SU(N)$. 
Group theory indicates that the non-redundant Lorentz structures with respect to the IBP should also form an invariant space consisting of irrep spaces, and a basis for them is easily found by translating the semi-standard Young tableau (SSYT) of these irrep's. 
In addition, we develop a procedure to list all classes of Lorentz structures at a given dimension, and for each of them we can directly obtain the corresponding irrep of the $SU(N)$ and the labels to be filled in, which is sufficient for enumerating all the SSYT's and Lorentz structures. Since we directly obtain an independent basis of Lorentz structures in the process, we never need to actually use the EOM and IBP relations as in other literature, and the correctness of our result is theoretically guaranteed.

For the sake of generality, we treat the gauge group structures systematically which, as far as we know, was not presented yet before. Gauge symmetry demands singlet combinations of fields with various representations, described by tensor product decomposition. To construct these singlets explicitly, we turn all the constituting fields into forms with only fundamental indices, and adopt the Littlewood-Richardson rules to merge their Young diagrams into a singlet Young diagram, during which we keep their fundamental indices in the diagram as labels. What we finally get is a singlet Young tableau, each column representing a Levi-Civita tensor in the group structure that contracts with the indices inside the column. In this way, we get an independent basis of group structures consisting of Levi-Civita's.

Having obtained the complete basis of Lorentz and gauge group structures, it is easy to combine them into a basis of operators with only free flavor indices.
However, as mentioned above, permutation symmetries among repeated fields induce symmetries of the flavor indices, which we shall resolve by constructing the factor structures with definite permutation symmetries and combining them via inner product decomposition. 
According to the plethysm technique, operators with definite permutation symmetry of flavor indices can be systematically addressed by obtaining definite Lorentz and gauge group permutation symmetries of the same set of repeated fields. 
To obtain the symmetrized Lorentz and gauge group structures, we introduce the basis symmetrizer in the minimal left ideal of the symmetric group algebra, which, applied to the factor structures, and generates a basis transforming as irrep of the symmetric group. 
Using these bases, we obtain the flavor-independent operators at the dimension 8 which constitute our main result.

The paper is organized as follows. In section~\ref{sec:eft}, we introduce notations for fields and operators used in our paper and define the terminologies for operators at different levels. In section~\ref{sec:basis}, we first discuss the problem of repeated fields and show in \ref{sec:motiv} that solving this problem leads the demand of finding symmetrized Lorentz and gauge group bases. Then we explain our algorithm to obtain these symmetrized bases in details in \ref{sec:lorentz} and \ref{sec:gauge} for Lorentz and gauge group respectively. In \ref{sec:flavor} we show how to obtain the operators with definite permutation symmetries of flavor indices from ingredients discussed above via inner product decomposition of the symmetric group.  In section~\ref{sec:list}, we exhibit a table showing numbers of operators for each sub-class in terms of fermion flavor number, and list the complete set of Dim-8 SMEFT operators organized by the number of fermions. Our conclusion is presented in section~\ref{sec:conclusion}. In appendix~\ref{app:A}, we list useful identities and examples of format conversions between Lorentz representations, both for fermions and gauge bosons, which are used in presenting our results. In appendix~\ref{app:B}, we introduce some basics of symmetric groups $S_m$ and a few group theory tools we used in the paper, including the basis symmetrizer $b^\lambda$ and the projection operator involved in the inner product decomposition.

\section{Standard Model Effective Field Theory}
\label{sec:eft}

\subsection{SM Fields: Building Block, and Notation}
\label{sec:field}
The Lagrangian of SMEFT should be invariant under the Lorentz group and the SM gauge group. 
We start by defining the building blocks of the effective operators: fields and their covariant derivatives. The building blocks are characterized by their representations under the Lorentz group $SL(2,\mathbb{C}) = SU(2)_{l}\times SU(2)_{r}$ and the SM gauge groups $SU(3)_C \times SU(2)_W \times U(1)_Y$. The representation under Lorentz symmetry is given by $(j_l,j_r)$, the quantum numbers of the $SU(2)_l$ and $SU(2)_r$ components of the Lorentz group $SL(2,\mathbb{C})$.
We adopt the following notations on the field constituents: 
\begin{itemize}
	\item Since all the SM fermions are chiral-like, we use the two-component Weyl spinor notation, which transforms as Irreps of the Lorentz group
	\bea
	\psi_{\alpha }  \in (1/2,0), \quad \psi^{\dagger}_{\dot\alpha } \in (0,1/2), 
	\eea
	where the indices $\alpha, \dot{\alpha}$ denote the fundamental representation of $SU(2)_l$, and $SU(2)_r$, respectively. 
	We further adopt the all-left chirality convention for the fermions: $Q$ and $L$ are the left-handed components of the quark and lepton doublet fields, and $u_{_\mathbb{C}}$, $d_{_\mathbb{C}}$ and $e_{_\mathbb{C}}$ are the left-handed components of the anti-up, anti-down and anti-electron fields. 
	The transformation to the 4-component Dirac Spinor notation is given in the Appendix.~\ref{app:a1}.
	
	\item We use the following notation for the SM Higgs doublet 
	\bea
	H_i \in \left(0,0\right), \quad H^{\dagger i} \in \left(0,0\right),
	\eea
	where the index $i$ denotes the (anti)fundamental representation of $SU(2)_W$. 
	We avoid to use the notation $\tilde{H}=\epsilon H^\dagger$, as it is essentially the same as $H^\dagger$ but with the original $SU(2)_W$ antifundamental indices lowered to the fundamental one by the $\epsilon$ tensor. 
	In our final results all the gauge group indices are left explicit, consequently whenever there is a $\tilde{H}$ present in other literature, it translates into $\epsilon_{ij}H^{\dagger j}$ in our notation.
	
	\item We use the chiral basis of the gauge boson $F_{\rm L/R} = \frac12(F \pm i\tilde{F})$ because they transform under irreps of the Lorentz group, which is important for us to study the constraints on the Lorentz structures. They transform to the normal gauge field strength as
	\eq{
		F_{{\rm L}\alpha\beta} = \frac{i}{2}F_{\mu\nu}\sigma^{\mu\nu}_{\alpha\beta}\in (1,0),  \quad F_{{\rm R} \dot\alpha\dot\beta} = -\frac{i}{2}F_{\mu\nu}\bar\sigma^{\mu\nu}_{\dot\alpha\dot\beta}\in(0,1).
	}
	The spinor indices of $F_{\rm L/R}$ are symmetric in order to form $(1,0)$ or $(0,1)$ representations, as can be proved by property of $\sigma^{\mu\nu}$, defined in the Appendix.~\ref{app:a1}.

\end{itemize}

\begin{table}[h]
	\begin{center}
		\begin{tabular}{|c|cc|ccc|c|}
			\hline
			\text{Fields} & $SU(2)_{l}\times SU(2)_{r}$	& $h$ & $SU(3)_{C}$ & $SU(2)_{W}$ & $U(1)_{Y}$ &  Flavor\tabularnewline
			\hline
			$G_{\rm L\alpha\beta}^A$   & $\left(1,0\right)$  & $-1$    & $\boldsymbol{8}$ & $\boldsymbol{1}$ & 0  & $1$\tabularnewline
			$W_{\rm L\alpha\beta}^I$   & $\left(1,0\right)$  & $-1$           & $\boldsymbol{1}$ & $\boldsymbol{3}$ & 0  & $1$\tabularnewline
			$B_{\rm L\alpha\beta}$   & $\left(1,0\right)$    & $-1$        & $\boldsymbol{1}$ & $\boldsymbol{1}$ & 0  & $1$\tabularnewline
			\hline
			$L_{\alpha i}$     & $\left(\frac{1}{2},0\right)$  & $-1/2$  & $\boldsymbol{1}$ & $\boldsymbol{2}$ & $-1/2$  & $n_f$\tabularnewline
			$e_{_\mathbb{C}\alpha}$ & $\left(\frac{1}{2},0\right)$ & $-1/2$   & $\boldsymbol{1}$ & $\boldsymbol{1}$ & $1$  & $n_f$\tabularnewline
			$Q_{\alpha ai}$     & $\left(\frac{1}{2},0\right)$ & $-1/2$   & $\boldsymbol{3}$ & $\boldsymbol{2}$ & $1/6$  & $n_f$\tabularnewline
			$u_{_\mathbb{C}\alpha}^a$ & $\left(\frac{1}{2},0\right)$ & $-1/2$   & $\overline{\boldsymbol{3}}$ & $\boldsymbol{1}$ & $-2/3$  & $n_f$\tabularnewline
			$d_{_\mathbb{C}\alpha}^a$ & $\left(\frac{1}{2},0\right)$ & $-1/2$   & $\overline{\boldsymbol{3}}$ & $\boldsymbol{1}$ & $1/3$  & $n_f$\tabularnewline
			\hline
			$H_i$     & $\left(0,0\right)$&  0     & $\boldsymbol{1}$ & $\boldsymbol{2}$ & $1/2$  & $1$\tabularnewline
			\hline
		\end{tabular}
		\caption{\label{tab:SMEFT-field-content}
			The field content of the standard model, along with their representations under the Lorentz and gauge symmetries. The representation under Lorentz group is denoted by $(j_l,j_r)$, while the helicity of the field is given by $h = j_r-j_l$ .
			The number of fermion flavors is denoted as $n_f$, which is 3 in the standard model. All of the fields are accompanied with their Hermitian conjugates that are omitted, $(F_{\rm L \alpha\beta})^\dagger = F_{\rm R \dot\alpha\dot\beta}$ for gauge bosons, $(\psi_\alpha)^\dagger = (\psi^\dagger)_{\dot\alpha}$ for fermions, and $H^\dagger$ for the Higgs, which are under the conjugate representations of all the groups. }
	\end{center}
\end{table}

The field constituents without derivatives are given in table~\ref{tab:SMEFT-field-content}. As shown in table~\ref{tab:SMEFT-field-content}, the indices for the (anti)fundamental representation of $SU(2)_l$, $SU(2)_r$, $SU(3)_C$ and $SU(2)_W$ are denoted by $\{\alpha, \beta, \gamma, \delta\}$, $\{\dot{\alpha}, \dot{\beta}, \dot{\gamma}, \dot{\delta}\}$, $\{a, b, c, d\}$ and $\{i, j, k, l\}$, respectively. We use subscripts to indicate the fundamental representation and superscripts to indicate the anti-fundamental representation. 
The indices for the adjoint representation of $SU(3)_C$ and $SU(2)_W$ are denoted by $\{A, B, C, D\}$ and $\{I, J, K, L\}$, respectively. In case flavor indices are needed, we use $\{p, r, s, t\}$. In the final result, the spinor indices are left implicit.

Not only the SM fields but also fields with covariant derivatives are the building blocks, although the covariant derivative itself is not. In our notation, the covariant derivatives only act on the nearest field on the right, and the gauge group indices on that field should be understood as the indices of the whole, for example, $D Q_{pai} = \left(D Q_{p}\right)_{ai}$. Regarding to the Lorentz index on $D$, we also adopt the $SL(2,\mathbb{C})$ notation for convenience
\eq{
	D_{\alpha\dot\alpha} = D_{\mu}\sigma^{\mu}_{\alpha\dot\alpha} \in (1/2,1/2),
}
Thus covariant derivatives of a field $D^{n_D}\Phi$, in which $\Phi$ denotes the SM field, could be expressed in general as 
\eq{\label{eq:field_w_D}
	(D^{r-|h|}\Phi)_{\alpha^{(1)}\cdots\alpha^{(r-h)}\dot\alpha^{(1)}\cdots\dot\alpha^{(r+h)}} \equiv \left\{\begin{array}{ll} D_{\alpha^{(1)}\dot\alpha^{(1)}} \cdots D_{\alpha^{(r+h)}\dot\alpha^{(r+h)}}\Phi_{\alpha^{(r+h+1)}\cdots\alpha^{(r-h)}}, \quad h<0, \\
		D_{\alpha^{(1)}\dot\alpha^{(1)}} \cdots D_{\alpha^{(r-h)}\dot\alpha^{(r-h)}}\Phi_{\dot\alpha^{(r-h+1)}\cdots\dot\alpha^{(r+h)}}, \quad h>0.	\end{array}\right.
}
where $h$ is the helicity of the massless particle annihilated by the field, and $r=|h|+n_D$ is half the total number of spinor indices in this building block. 
One could verify the number of Lorentz indices is correct for any $\Phi$, for example, the $r=5/2$ building block of field $Q$, which has helicity $h=-1/2$, is given by
\eq{
	(D^2Q_{ai})_{\alpha^{(1)}\alpha^{(2)}\alpha^{(3)}\dot\alpha^{(1)}\dot\alpha^{(2)}} = D_{\alpha^{(1)}\dot\alpha^{(1)}}D_{\alpha^{(2)}\dot\alpha^{(2)}}Q_{\alpha^{(3)}ai}.
}

With these building blocks, operators are simply constructed as combinations of Building Blocks that form the singlet representation, under all of the Lorentz group and gauge groups. 
In general, when the constituents of the operator are fixed, the indices of Lorentz and gauge groups are also fixed, then the simplest way to assemble these constituents into a singlet is to contract all the indices with invariant tensors for each group. In this way we obtain a basis of operators that are direct products of three-factor structures:
\begin{eqnarray}
	\Theta^{\{f\}} &=& T_{\rm SU3}^{\{g\}}T_{\rm SU2}^{\{h\}}\left[T_{\rm Lorentz}^{\{\alpha\},\{\dot\alpha\}}\prod_{i=1}^N(D^{r_i-|h_i|}\Phi_i)_{\alpha_i^{(1)}\cdots\alpha_i^{(r-h)}\dot\alpha_i^{(1)}\cdots\dot\alpha_i^{(r+h)}}\right]^{\{f\}}_{\{g\},\{h\}}\nonumber\\
	& = & T_{\rm SU3}^{\{g\}}T_{\rm SU2}^{\{h\}}{\cal M}^{\{f\}}_{\{g\},\{h\}} ,
	\label{eq:term_format}
\end{eqnarray}
where the invariant tensors $T_{\rm SU3}$, $T_{\rm SU2}$ and $T_{\rm Lorentz}$ form polynomial rings generated by the following corresponding ingredients
\eq{
	& SU(3) : \quad f^{ABC},d^{ABC},\delta^{AB},(\lambda^A)_a^b,\epsilon_{abc},\epsilon^{abc}, \\
	& SU(2) : \quad \epsilon^{IJK},\delta^{IJ},(\tau^I)_i^j,\epsilon_{ij},\epsilon^{ij}, \\
	& {\rm Lorentz} : \ \sigma^{\mu\nu}_{\alpha\beta}, \bar\sigma^{\mu\nu}_{\dot\alpha\dot\beta}, \sigma^\mu_{\alpha\dot\alpha}, \bar\sigma^{\mu\dot\alpha\alpha}, \epsilon^{\alpha\beta}, \tilde\epsilon^{\dot\alpha\dot\beta}.
}
In the second line, we collapse $T_{\rm Lorentz}^{\{\alpha\},\{\dot\alpha\}}$ with the building blocks to a formal Lorentz singlet ${\cal M}^{\{f\}}_{\{g\},\{h\}}$, which we will often refer to as a Lorentz structure, with free flavor and gauge group indices that are specified after fixing the constituting fields\footnote{We would like to point out here that formally each field in our notation has a flavor index, even for gauge bosons and the Higgs which can take only one possible value. The reason will be clear in section~\ref{sec:motiv}.}.

The dimension of $\Theta$ can be derived as
\eq{\label{eq:operator_dim}
	\dim(\Theta) = \sum_{i=1}^N (r_i-|h_i|+\dim(\Phi_i)) = N+r ,
}
where $N$ is the particle number, $r=\sum_ir_i$ turns out to be the mass dimension of the on-shell amplitude generated by $\Theta$.

In section~\ref{sec:basis} we will put additional constraints on the form of building blocks eq.~\eqref{eq:term_format}, and our {\it master formula} of operator basis eq.~\eqref{eq:master} will be constructed as particular non-redundant combinations of them.

\subsection{Invariants at Different Levels: Class, Type, Term, and Operator}
\label{sec:inv}
The above subsection defines the building blocks that we use in constructing invariants, i.e., operators, for the sake of clarity, we specify the terminologies used in the rest of the paper that describe the invariants at different levels.  
For practical purpose, we group the effective operators into several levels of clusters defined as below: 
\begin{itemize}
	\item The biggest cluster is called a \textbf{class}, which involves operators with the same kind of fields in terms of spin, and the same number of derivatives, denoted as $F^{n_F}\psi^{n_{\psi}}\phi^{n_\phi}D^{n_D}$. One could be more accurate by setting the definite number of left/right-handed fermions and gauge bosons, so that we get \textbf{sub-classes} such as 
	\begin{equation}
		F_{\rm L}^{n_{-1}}\psi^{n_{-1/2}}\phi^{n_0}\psi^{\dagger n_{1/2}}F_{\rm R}^{n_1}D^{n_D}.\nonumber 
	\end{equation}
	One could come up with any combinations of $n_i$ at this level, and rule out the ones that are not able to form Lorentz invariants later, but we propose a criteria for selecting the Lorentz invariant sub-classes, which makes the program more effective at higher dimensions.
	
	\item In a (sub-)class, we further group together operators with the same constituting fields, selected by the requirement of conservation laws: the combination of fields should be able to form a singlet of any symmetric groups (besides the Lorentz group) the theory has. This level of cluster is called a \textbf{type}, denoted by a sequence of fields and derivatives. An example of the type is:
	\begin{eqnarray}
		Q^2u^\dagger_{_\mathbb{C}}LH^\dagger D,
		\label{eq:typeexmp}
	\end{eqnarray}	
	which corresponds to $n_{-1/2}=3,n_0=1,n_{1/2}=1, n_D =1$.
	Note that Lorentz structures may not be fixed at this level, especially at higher dimensions, when a type could contain quite a number of independent ones. At this level, though, we could identify the groups of repeated fields in a type, which put constraints on the form of independent operators within a type.
	
	\item At this level, we define the (Lagrangian) \textbf{terms} that have different interpretations from the other literature e.g. \cite{Fonseca:2019yya}. 
	We define a term as an operator with free flavor indices that transform as an irreducible tensor of the auxiliary flavor symmetry group $SU(n_f)$ for each set of repeated fields with $n_f$ number of flavors. We occasionally refer the irreducible tensor nature of terms to as flavor symmetry of the operator.  An example of term for the type in eq.~(\ref{eq:typeexmp}) is:	
	
	\begin{eqnarray}
		\Theta^{prst} = i\epsilon^{abc}\epsilon^{jk}\left((L_{pi}Q_{sbk}(Q_{raj}\sigma^\mu u^\dagger_{_\mathbb{C}tc})+(L_{pi}Q_{raj}(Q_{sbk}\sigma^\mu u^\dagger_{_\mathbb{C}tc})\right)D_\mu H^{\dagger k}.
	\end{eqnarray}
	One can verify that $\Theta^{prst}$ is symmetric under exchange of $r,s$ flavor indices of two $Q$s, which indicates $\Theta^{prst}$ is an irreducible tensor represented by {\scriptsize \yng(2)} under $SU(n_f)$ group for $Q$s. This definition of the term renders enumerating independent flavor-specified operators defined below trivial by finding all the SSYTs of the corresponding Young diagram of irrep of $SU(n_f)$. We shall explicitly illustrate this point and describe the algorithm to obtain a complete set of terms up to any dimensions in detail in Sec.~\ref{sec:motiv}.

	\item Finally, the \textbf{flavor-specified operators} are defined as independent flavor assignments in a term. The corresponding example gives
	\begin{eqnarray}
		\Theta^{1111} = i\epsilon^{abc}\epsilon^{jk}\left((L_{1i}Q_{1bk}(Q_{1aj}\sigma^\mu u^\dagger_{_\mathbb{C}1c})+(L_{1i}Q_{1aj}(Q_{1bk}\sigma^\mu u^\dagger_{_\mathbb{C}1c})\right)D_\mu H^{\dagger k}.
	\end{eqnarray}  
	
\end{itemize}

\section{Operator Basis for Lorentz and Gauge Symmetries}
\label{sec:basis}

\subsection{Motivation and Mathematical Preparation}
\label{sec:motiv}
In this subsection, we first explain why we need definite permutation symmetries of the Lorentz and gauge group structures for the term we defined in section~\ref{sec:inv} and how they are related to the permutation symmetry of the flavor indices, 
then we give a gentle introduction to mathematical tools used in generating symmetrized Lorentz and gauge group structures. 

\subsubsection{Why Permutation Symmetry}

Given a type of the operator, one can enumerate all the independent ways to construct a singlet under both Lorentz and gauge symmetries with flavor indices unspecified.
Fixing the flavor indices of such a Lorentz and gauge singlet completely determines the form of a flavor-specified operator.  
If there are no repeated fields, then different choices of flavor for each field correspond to different operators. 
However, the presence of repeated fields complicates the game. 
To demonstrate the problem let us take a look at the dim-5 Weinberg operator in the SMEFT:
\begin{eqnarray}
	\Theta^{\{f_1f_2,11\}}\equiv\epsilon^{i_1j_1}\epsilon^{i_2j_2}\epsilon^{\alpha_1\alpha_2}L_{\alpha_1,i_1}^{f_1} L_{\alpha_2,i_2}^{f_2}H_{j_1} H_{j_2},
	\label{eq:theta12}
\end{eqnarray}
where $\alpha_{1,2}$ are spinor indices, $i_{1,2}$ and $j_{1,2}$ are SU$(2)_W$ indices, $\Theta^{\{f_1f_2,11\}}$ has the same notation defined in eq.~\eqref{eq:term_format} with $\Phi_i$s specified to $LLHH$. 
We have also grouped the flavor indices of each set of repeated fields together where the flavor indices for $H$ have already set to 1 as we only have one Higgs in the SM. In the following, we shall drop the flavor indices of $H$ when their absences do not obscure our explanation.  
One can verify that $\Theta^{f_1f_2}=\Theta^{f_2f_1}$, which we will prove later, is a result of the antisymmetric nature of the Lorentz and gauge structures, therefore the number of independent operators are $n_f(n_f+1)/2$ if we have $n_f$ flavors of $L$.
Generally, to count and enumerate independent operators with flavor indices specified, one can view $\Theta^{\{f_k\}}$ as a tensor of the $SU(n_f)$ group and decompose it into different irreps of the $SU(n_f)$ group, then the independent operators are given by setting the flavor indices according to all the SSYTs of the corresponding irrep with numbers in the tableaux weakly increased in each row and strictly increased in each column.  
Following the example above, $\Theta^{f_1f_2}$ in eq.~\eqref{eq:theta12} is symmetric under exchange of $f_1$ and $f_2$, hence it is represented by the Young tableau: {\scriptsize \young({{f_1}}{{f_2}})}. 
If $n_f=2$, then there are three semi-standard Young tableaux: {\scriptsize \young(11)}, {\scriptsize \young(12)} and {\scriptsize \young(22)}, which correspond to independent operators $\Theta^{11}$, $\Theta^{12}$, $\Theta^{22}$. All the other choices of $f_1$ and $f_2$ can be expressed by the linear combination of these three using Fock's conditions~\cite{ma2007group}, and in this example, we simply have $\Theta^{21}=\Theta^{12}$.

The $LLHH$ example above seems to be too trivial as it is easy to find out the symmetric properties among the flavor indices $f_1,f_2$. However, the situation becomes complicated when the number of repeated fields in a set goes up. The simplest non-trivial example is $Q^3L$ in the dim-6 SMEFT, in the Warsaw basis~\cite{Grzadkowski:2010es} it is expressed as:
\begin{eqnarray}
	\epsilon^{abc}\epsilon_{ji}\epsilon_{km}[(q^{aj}_r)^T C q^{b k}_s][(q^{cm}_t)^TC l^i_p],
	\label{eq:warsawq3l}
\end{eqnarray}
where $q$ and $l$ are four component $SU(2)_W$ quark and lepton doublets of which the relation to our two-component notations are shown in appendix~\ref{app:a1}, $a,b,c$ are $SU(3)_C$ indices, $i,j,k,m$ are $SU(2)_W$ indices, and $p,r,s,t$ are flavor indices of fermions respectively. 
From eq.~\eqref{eq:warsawq3l}, it is very hard to tell the independent components of the flavor-specified operators. 
However, we separate this operator into three terms in our notation, and each term is an irrep of $SU(n_f)$:
\begin{eqnarray}
	\young(r,s,t)\ &:&\ \Theta^{\{p,rst\}}_{[1^3]}=\epsilon^{abc}\left[\epsilon^{ij}\epsilon^{lk}(L_{pi}Q_{sbk})(Q_{raj}Q_{tcl})-\epsilon^{ik}\epsilon^{jl}(L_{pi}Q_{raj})(Q_{sbk}Q_{tcl})\right]\label{eq:q3l1}\\
	\young(rs,t)\ &:&\ \Theta^{\{p,rst\}}_{[2,1]}=\epsilon^{abc}\left[(\epsilon^{il}\epsilon^{jk}-\epsilon^{ij}\epsilon^{kl})(L_{pi}Q_{sbk})(Q_{raj}Q_{tcl})+(\epsilon^{il}\epsilon^{jk}+\epsilon^{ik}\epsilon^{jl})(L_{pi}Q_{raj})(Q_{sbk}Q_{tcl})\right]\nonumber\\
	\\
	\young(rst)\ &:&\ \Theta^{\{p,rst\}}_{[3]}=\epsilon^{abc}\left[(\epsilon^{ik}\epsilon^{jl}+\epsilon^{il}\epsilon^{jk})(L_{pi}Q_{sbk})(Q_{raj}Q_{tcl})+(\epsilon^{il}\epsilon^{jk}-\epsilon^{ij}\epsilon^{kl})(L_{pi}Q_{raj})(Q_{sbk}Q_{tcl})\right],\nonumber\\
	\label{eq:q3l3}
\end{eqnarray}
where the subscripts of $\Theta$ are partitions of the integer 3 that have a one-to-one correspondence with the Young Diagrams (YD) on the left of each equation.
From the above notation, we can immediately write down the independent combinatorial choices of $r,s,t$ by enumerating the SSYTs of each irrep:
\begin{gather*}
	\young(1,2,3),\\
	\young(11,2),\young(11,3),\young(22,3),\young(12,2),\young(13,3),\young(23,3),\young(12,3),\young(13,2),\\
	\young(111),\young(112),\young(113),\young(122),\young(123),\young(133),\young(222),\young(223),\young(233),\young(333),
\end{gather*}
then each Young tableau can pair with three choices of lepton flavor $p=1,2,3$ resulting in totally $3\times(1+8+10)$ independent operators.
In addition to the benefits discussed above, we would like to point out that our definition of term, combined with the algorithm obtaining a complete set of independent terms described below, will automatically avoid the redundancy that needs to be resolved by obscure relations between different terms with flavor indices permuted. 
For example, $Q^3L$ was initially written as two independent terms in ref.~\cite{Grzadkowski:2010es} and was corrected to only one in the form of eq.~\eqref{eq:warsawq3l} later.
We will not proceed with this example further in this subsection as it involves technical details discussed in section~\ref{sec:flavor}, which may blur the big picture we would like to convey. Therefore we shall continue with the $LLHH$ example below and show the roadmap constructing these $SU(n_f)$ irreps. 

The above discussion demonstrates the desire to obtain $\Theta^{\{f_k\}}$ as an irrep of $SU(n_f)$, where $\{f_k\}$ are the flavor indices of a set of repeated fields. 
Due to the Schur-Weyl duality and also pointed out in ref.~\cite{Fonseca:2019yya}, if one can construct a set of tensor $\Theta^{\{f_k\}}_{(\lambda,x)}$ that transform as an irrep $\lambda$ of the symmetric group $S_m$ ($m$ is the number of the repeated fields) in terms of permuting subscripts $k$, 
then any one of the $\Theta^{\{f_k\}}_{(\lambda,x)}$ is the irrep of $SU(n_f)$ with the number of independent components obtained by the Hook Content Formula ${\cal S}(\lambda, n_f)$~\cite{Fonseca:2019yya}.
$\lambda$ is a partition of the integer $m$ that can be written in the form of the subscripts in eq.~(\ref{eq:q3l1}-\ref{eq:q3l3}) serving as a character of irreps of $S_m$.
$x$ goes from 1 to $d_\lambda$ labeling basis vectors in the irrep $\lambda$. 
$\Theta^{\{f_k\}}_{(\lambda,x)}$ by definition satisfies the following relation:
\begin{eqnarray}
	\pi\circ \Theta^{\{f_k\}}_{(\lambda,x)} &\equiv & \Theta^{\{f_{\pi(k)}\}}_{(\lambda,x)}\nonumber \\
	&=&\sum_y\Theta^{\{f_k\}}_{(\lambda,y)}\mc{D}^{\lambda}(\pi)_{yx},
	\label{eq:Thetai}
\end{eqnarray}
for $\pi\in S_m$, where the action $\pi\circ$ on tensors is defined according to the first line, and $\mc{D}^{\lambda}(\pi)_{yx}$ are the matrix representation of the $S_m$ group for irrep $\lambda$. 
In the presence of multiple sets of repeated fields, one can generalize $\Theta^{\{f_k\}}$ into $\Theta^{\{f_k,p_m,...\}}$ as irreps of the groups:
\eq{
	&\overline{SU}=SU(n^1_f)\otimes SU(n^2_f)\otimes... , \\
	&\bar{S} = S_{m_1}\otimes S_{m_2}\otimes...,
} 
where $n^1_f,n^2_f,...$ and $m_1,m_2,...$ are numbers of flavor and fields for different sets of repeated fields respectively. An immediate conclusion that can be drawn from the above discussion is that the operators involving repeated fields with only 1 flavor such as gauge bosons or Higgs must be in a totally symmetric representation , which is simply a result of:
\begin{eqnarray}
	{\cal S}(\lambda,1)=0,\quad \forall \lambda\neq[m],
	\label{eq:oneflavorS}
\end{eqnarray}  
where $[m]$ is the totally symmetric irrep.

Now we are going to show that permuting flavor indices of a given set of repeated fields in a term is equivalent to permute the corresponding indices related to the gauge and Lorentz structures. In general, a term  of a given type can be formally expressed as a linear combination of a set of factorizable $\Theta$s according to eq.~\eqref{eq:term_format} as:
\begin{eqnarray}
	\Theta^{\{f_k,...\}} = T_{{\rm SU3}}^{\{g_k,...\}}T_{{\rm SU2}}^{\{h_k,...\}}{\cal M}^{\{f_k,...\}}_{\{g_k,...\},\{h_k,...\}}, 
\end{eqnarray}
where $T_{{\rm SU3}}^{\{g_k,...\}},T_{{\rm SU2}}^{\{h_k,...\}}$ are the same $T_{{\rm SU3}},T_{{\rm SU2}}$ in eq.~\eqref{eq:term_format} with the indices of each set of repeated fields grouped together, the same argument applies for the correspondence between ${\cal M}^{\{f_k,...\}}_{\{g_k,...\},\{h_k,...\}}$ here and  ${\cal M}^{\{f\}}_{\{g\},\{h\}}$ in eq.~\eqref{eq:term_format}. 
Concretely, in the $LLHH$ example we have:
\begin{eqnarray}
	T_{{\rm SU3}}&=&1\nonumber \\
	T_{{\rm SU2}}^{\{i_1i_2,j_1j_2\}}&=&\epsilon^{i_1j_1}\epsilon^{i_2j_2}\nonumber \\
	{\cal M}^{\{f_1f_2,11\}}_{\{i_1i_2,j_1j_2\}}&=&\epsilon^{\alpha_1\alpha_2}L_{\alpha_1,i_1}^{f_1} L_{\alpha_2,i_2}^{f_2}H_{j_1} H_{j_2}\nonumber
\end{eqnarray}
For a given type, we call in the rest of our paper a complete set of independent $T_{{\rm SU3}}^{\{g_k,...\}},T_{{\rm SU2}}^{\{h_k,...\}}$ and ${\cal M}^{\{f_k,...\}}_{\{g_k,...\},\{h_k,...\}}$ as the Lorentz and gauge basis from which a term is constructed.
It is enough to show the permutation relations between flavor and combined gauge and Lorentz structures hold for these factorizable bases, and the same is true for a general term.   
A permutation of the flavor indices of \textbf{the first set} of repeated fields in $\Theta^{\{f_k,...\}}$ can be expressed as:
\begin{eqnarray}
	\underbrace{\pi\circ \Theta^{\{f_{k},...\}}}_{\rm permute\ flavor} &=&  T_{{\rm SU3}}^{\{g_k,...\}}T_{{\rm SU2}}^{\{h_k,...\}}{\cal M}^{\{f_{\pi(k)},...\}}_{\{g_k,...\},\{h_k,...\}} \nonumber \\
	& = &  T_{{\rm SU3}}^{\{g_{\pi(k)},...\}}T_{{\rm SU2}}^{\{h_{\pi(k)},...\}}{\cal M}^{\{f_{\pi(k)},...\}}_{\{g_{\pi(k)},...\},\{h_{\pi(k)},...\}}\nonumber \\
	& = & \underbrace{\left(\pi\circ T_{{\rm SU3}}^{\{g_k,...\}}\right)\left(\pi\circ T_{{\rm SU2}}^{\{h_k,...\}}\right)}_{\rm permute\ gauge}\underbrace{\left(\pi\circ{\cal M}^{\{f_k,...\}}_{\{g_{k},...\},\{h_{k},...\}}\right)}_{\rm permute\ Lorentz},
	\label{eq:tperm}
\end{eqnarray} 
where $\pi\circ T$ again only permutes the gauge indices of the first set of the repeated fields,
$\left(\pi\circ{\cal M}\right)$ is equal to the permutation of all the subscripts of the Lorentz indices and the associated derivatives of the first set of repeated fields while leaving the gauge and flavor indices unchanged. 
We demonstrate that the second to the third line in eq.~\eqref{eq:tperm} does hold for $\cal M$ and $\left(\pi\circ{\cal M}\right)$ defined above with the $LLHH$ example in eq.~\eqref{eq:theta12} with $\pi=(12)$:
\begin{eqnarray}
	{\cal M}^{\{f_2f_1,11\}}_{\{i_2i_1,j_1j_2\}} &=& \epsilon^{\alpha_1\alpha_2}L_{\alpha_1,i_{\color{red} 2}}^{f_{\color{red} 2}} L_{\alpha_2,i_{\color{red} 1}}^{f_{\color{red} 1}}H_{j_1} H_{j_2}\nonumber \\
	&=&\epsilon^{\alpha_{\color{red} 2}\alpha_{\color{red} 1}} L_{\alpha_{\color{red} 2},i_1}^{f_1}L_{\alpha_{\color{red} 1},i_2}^{f_2}H_{j_1} H_{j_2}\nonumber \\
	& = &\pi\circ{\cal M}^{\{f_1f_2,11\}}_{\{i_1i_2,j_1j_2\}},
\end{eqnarray}
where we have used the Grassmann nature of the lepton field.

So far, it is obvious from eq.~\eqref{eq:tperm} that those of gauge and Lorentz indices determine the permutation property of flavor indices.   
Hence, if a set of ${\cal M}^{\lambda_1}_j$, $T^{\lambda_{2}}_{\rm SU3}$ and $T^{\lambda_{3}}_{\rm SU2}$ transform according to irreps $\lambda_{1,2,3}$ of $S_m$ for the certain set of repeated fields, 
then the direct product space spanned by
\begin{eqnarray}
	\{\Theta_{(\lambda_1,x_1),(\lambda_2,x_2),(\lambda_3,x_3)} = {\cal M}^{\lambda_1}_{x_1} T^{\lambda_2}_{{\rm SU3},x_2} T^{\lambda_3}_{{\rm SU2},x_3}\ \mid \ x_i\in 1,...,d_{\lambda_i}\},
	\label{eq:prodbasis}
\end{eqnarray}
with $d_{\lambda_i}$ the dimension of irreps $\lambda_i$, can be decomposed into invariant subspaces that form different irreps of $\lambda$. 
The inner product decomposition $\lambda_1\odot \lambda_2\odot\lambda_3 = \sum_{\lambda\vdash m} r_\lambda\ \lambda$ tells that the multiplicity of the invariant subspace of irrep $\lambda$ is $r_\lambda$. 
For the $LLHH$ example we discussed above, one can find that ${\cal M}^{\{f_{\pi(k)},11\}}_{\{i_{\pi(k)}\},\{j_m\}}={\cal M}^{\{f_k,11\}}_{\{i_k\},\{j_m\}}={\cal M}^{[2]}_1$ forms a total symmetric representation of $S_2$ of two lepton fields, the same is for the  $SU(3)_C$ gauge group factor $T_{{\rm SU3},1}^{[2]}=1$.
The permuted $SU(2)_W$ gauge group factor $T_{\rm SU2}^{\{i_{\pi(k)},j_m\}}=\epsilon^{i_2j_1}\epsilon^{i_1j_2}$ seems not equal to the unpermuted one. However this is just a result  of the  simplification of the symmetric one $T^{[2]}_{{\rm SU2},1}=(\epsilon^{i_2j_1}\epsilon^{i_1j_2}+\epsilon^{i_1j_1}\epsilon^{i_2j_2})/2$ 
when contracting with $H_{j_1}H_{j_2}$\footnote{This is also an example discussed in eq.~\eqref{eq:oneflavorS} that the repeated fields of one flavor must form a total symmetric representation under permutation. As the Lorentz structure of $H^2$ is trivial, therefore $T_{\rm SU2}^{\{i_{k}\},\{j_m\}}$ must be symmetric under $h_1, h_2$.}, so it also forms a totally symmetric representation of $S_2$. 
One can find from the inner product decomposition $[2]\odot[2]\odot [2] =[2]$ that the only resulting irrep is the symmetric one $[2]$, and indeed the $\Theta^{f_1f_2}$ is totally symmetric under permutation of the flavor indices. 
In general, this decomposition is more complicated and contains the irreps with dimension larger than one. We will present a non-trivial example $Q^3L W_{\rm L}$ in section~\ref{sec:flavor} and refer to appendix~\ref{sec:projection} for more general cases. Although we only consider the permutation symmetry for a single set of repeated fields in the eq.~\eqref{eq:tperm} and eq.~\eqref{eq:prodbasis}, it is straight forward to generalize it to multiple sets of repeated fields under the product group $\bar{S}$ as the permutations acting on different sets of repeated fields simply commute with each other.

Up to now, we have changed the problem of finding a term $\Theta^{\{f_{k},...\}}$ as irreps of $\overline{SU}$ into finding a series of $\Theta_{(\lambda,i),...}^{\{f_{k},...\}}$ as irreps of $\bar{S}$, then further into finding the corresponding $T^{\lambda_2,...}_{{\rm SU3},x_2,...},T^{\lambda_3,...}_{{\rm SU3},x_3,...}$ and ${\cal M}^{\lambda_1,...}_{x_1,...}$ as irreps of $\bar{S}$.
Before we delve into the details of obtaining all the independent symmetrized group factors and Lorentz structures, we first take a digress to introduce the mathematical tools used in the rest of the section. 

\subsubsection{Group algebra and left ideal}\label{sec:groupalgebra}
As mentioned above, in this section, we explain the key mathematical tools to obtain the Lorentz and gauge structure in different irreps $\lambda$ of the symmetric group associated with the repeated fields.
We introduce the idea of group algebra and the method using them to generate a series of symmetrized functions transforming as an irrep of $S_m$ group under permutations defined in \ref{eq:Thetai} from an asymmetrized one.
The first concept is the group algebra space $\tilde{S}_m$ of $S_m$, which is defined as a set consist of formal linear combinations of the group elements in the group $S_m$~\cite{tung1985group}:
\begin{eqnarray}
	\tilde{S}_m:\{r|r=\sum_i r^i \pi_i\ {\rm for}\ r_i\in \mathbb{C},\ \pi_i\in S_m\}.
\end{eqnarray}
The addition and multiplication rules of the elements in the group algebra are:
\begin{eqnarray}
	&&c_1 r+  c_2 q = \sum_i \pi_i(c_1r^i+c_2q^i)\ {\rm for\ } r,q\in \tilde{S}_m,\ c_1,c_2\in \mathbb{C},
	\\
	&&r\cdot q = \sum_{i,j}r^i q^j (\pi_i\cdot \pi_j) = \sum_{i,j,k}\pi_k\Delta^k_{ij}r^i r^j,
\end{eqnarray}
where the matrix $\Delta^k_{ij}$ with only one non-zero element defined by $\pi_i\cdot \pi_j = \sum_k \pi_k\Delta^k_{ij}$ is the regular representation of the group. Obviously, a linear vector space structure is contained in the group algebra.
In this sense, the group algebra elements have a dual role of vectors and linear operators. 

It is well known that the $\tilde{S}_m$~\cite{tung1985group,ma2007group} can be decomposed into invariant subspaces transforming as irrep $\lambda$ of the $S_m$ expanded by a set of group algebra elements $b^\lambda_x = \sum_i c^{\lambda,i}_x \pi_i$ such that 
\begin{eqnarray}
	\pi_i\cdot b^\lambda_x = \sum_y b^\lambda_y \mc{D}^{\lambda}_{yx}(\pi_i), 
	\label{eq:bandD}
\end{eqnarray}
where the indices $x,y$ go from 1 to $d_\lambda$, the dimension of irrep $\lambda$, $\mc{D}^{\lambda}_{yx}(\pi_i)$ is the same one in eq.~\eqref{eq:Thetai}, the matrix representation of $\lambda$ \cite{tung1985group}. The invariant subspaces expanded by $b^\lambda_x$ is actually a minimal left ideal ${\cal L}_\lambda$ of $\tilde{S}_m$ such that 
\begin{eqnarray}
	r\cdot b \in {\cal L}_\lambda, \quad\forall r\in \tilde{S}_m, b\in {\cal L}_\lambda.
\end{eqnarray}

Alternatively, one can view the group algebra elements as symmetrizers that act on a function generating another one by permuting the arguments. It can be shown in appendix~\ref{sec:permoperation} that a series of new functions $F^{\lambda}_{x}(\{p_k\})$ generated by applying $b^\lambda_x$ to a function $F(\{p_k\})$ defined by:
\begin{eqnarray}
	F^\lambda_{x}(\{p_k\}) &=& b^\lambda_x\circ F(\{p_k\})\nonumber \\
	&=& \left(\sum_i c^{\lambda,i}_x \pi_i\right)\circ F(\{p_k\})\nonumber \\
	&\equiv& \sum_i c^{\lambda,i}_x  F(\{p_{\pi_i(k)}\})
\end{eqnarray}
transform as an irrep of $\lambda$  under the permutation:
\begin{eqnarray}
	\pi_i\circ F^\lambda_{x}(\{p_k\}) &=& F_x^\lambda(\{p_{\pi_i(k)}\})\nonumber \\
	&=& \sum_y F^\lambda_{y}(\{p_{k}\}) \mc{D}^{\lambda}_{yx}(\pi_i).
	\label{eq:piF}
\end{eqnarray}
The function here has general meanings, in the $LLHH$ example above, the function can be referred to the gauge group tensor $T_{{\rm SU2}}^{i_1i_2,j_1j_2}$ with arguments $i_{1,2}, j_{1,2}$ or the Lorentz structure $\mc{M} = \mc{M}(\alpha_1,\alpha_2)$ with the arguments $\alpha_{1,2}$.

In addition, we would like to mention that our convention for $b^\lambda_x$ follows the Chapter 6 of the textbook~\cite{ma2007group}, where $b_1^\lambda$ is proportional to the Young symmetrizer of the normal Young tableau of $\lambda$, i.e., the Young tableau with the numbers 1 to $n$ appearing in order from left to right and from the top row to the bottom row. 
For example, $b_1^{[2,1]}$ is proportional to the Young symmetrizer of: {\scriptsize \young(12,3)}, which is equal to the multiplication of $s^\lambda$, the sum over all possible horizontal permutations, and $a^\lambda$, the sum over all possible vertical permutations weighted by their signatures, $\pm 1$ for even and odd permutations respectively. 
For {\scriptsize \young(12,3)}, we have:
\begin{eqnarray}
	s^{[2,1]} &=& E+(12),\\
	a^{[2,1]} &=& E-(13),\\
	b^{[2,1]}_1 \propto \mc{Y}\left[{\scriptsize \young(12,3)}\right]&=&s^{[2,1]}\cdot a^{[2,1]} \\
	&=& E +(12)-(13)-(132).
\end{eqnarray}
When we apply the $b^{[2,1]}_1$ on a tensor $T^{rst}$, we will associate the tensor indices $r,s,t$ to the numbers $1,2,3$ (not to confuse with flavors), then formally we have:
\begin{eqnarray}
	\mc{Y}\left[{\scriptsize \young(rs,t)}\right]T^{rst}
	&=&\mc{Y}\left[{\scriptsize \young(12,3)}\right]T^{123}\nonumber \\
	& = & T^{123}+T^{213}-T^{321}-T^{312}\nonumber\\
	& = & T^{rst}+T^{srt}-T^{tsr}-T^{trs}.
\end{eqnarray}
The resulting symmetrized tensor is symmetric for the permutation of labels $r,s$ as they appear in the same row in the Young tableau, which is a general property of the Young symmetrizers.

\subsection{Lorentz Basis: $SU(N) \times \bar{S} $ Irreps}
\label{sec:lorentz}
To obtain an independent set of Lorentz structures, in literature such as the Warsaw basis~\cite{Grzadkowski:2010es}, one usually writes down all the possible Lorentz invariant combinations of the building blocks, and then removes all the redundancies by imposing following relations among operators repeatedly:
\begin{itemize}
	\item[(a)] Fierz Identity. As explained in the appendix \ref{app:a1}, for Weyl spinors, the Fierz identities can be expressed as
	\eqs{
		& g_{\mu\nu}\sigma^\mu_{\alpha\dot\alpha}\sigma^\nu_{\beta\dot\beta} = 2\epsilon_{\alpha\beta}\epsilon_{\dot\alpha\dot\beta}, \label{eq:fierz_l}\\
		\begin{split}\label{eq:schouten}
			& \epsilon^{\alpha\beta}\delta_{\kappa}^\gamma + \epsilon^{\beta\gamma}\delta_{\kappa}^\alpha + \epsilon^{\gamma\alpha}\delta_{\kappa}^\beta = 0. \\
			& \tilde\epsilon_{\dot\alpha\dot\beta}\delta^{\dot\kappa}_{\dot\gamma} + \tilde\epsilon_{\dot\beta\dot\gamma}\delta^{\dot\kappa}_{\dot\alpha} + \tilde\epsilon_{\dot\gamma\dot\alpha}\delta^{\dot\kappa}_{\dot\beta} = 0
		\end{split}
	}
	For the first identity, we choose to replace the left-hand side whenever it appears in the operator by the right-hand side. As our building blocks do not contain any Lorentz indices $\mu,\nu,$ etc., there would be no chance to use the $\sigma$ matrices in $T_{\rm Lorentz}$, Thus we are left with only the $\epsilon$ tensors for both dotted and undotted spinor indices in the Lorentz invariant tensor. The other two identities, also known as the Schouten identities, will be tackled later.
	
	\item[(b)] $[D_\mu,D_{\nu}]=-iF_{\mu\nu}$. We also choose to replace the left-hand side whenever it appears in the operator by the right-hand side. Note that the replacement changes the type of operator, thus it should not be counted in the original type as an independent operator. Effectively, we treat $[D_\mu,D_\nu]$ as zero while counting operators of a given type. 
	
	\item[(c)] Equation of Motion (EOM). Classically there are the EOM relation for each kind of fields
	\eq{
		D^2 \phi + J_\phi = 0, \quad iD\4 \psi + J_\psi = 0, \quad D_\mu F^{\mu\nu} + J_A^\nu = 0,
	}
	For quantum fields, these are not rigorous operator equations. Nevertheless, operators differing by EOM are related with each other by field redefinitions, and are hence physically equivalent. To remove this redundancy of field redefinition, we choose to replace the first term (the kinetic term) whenever it appears in the operator by the source term $J_\Phi$. Again, because the type is changed during the replacement, we effectively treat the kinetic terms as zero while counting operators of a given type. This choice guarantees that the operator basis we find have non-vanishing on-shell amplitudes, which also form an amplitude basis.
	
	\item[(d)] Integration by Part (IBP). In perturbative QFT, we have
	\eq{
		XD_\mu Y \sim -D_\mu XY.
	}
	In other words, operators are equivalent modulo total derivatives. From the on-shell point of view, it is equivalent to the momentum conservation law.
	This may be the most subtle one, because it is the way people eliminate this redundancy while counting that prevents the listing of the independent operator basis. In this section, we develop a new method to deal with IBP.
\end{itemize}
We aim at a systematic treatment of all of these redundancies before we write down operators, so that we don't need to examine them in an over-complete list. Subsection~\ref{sec:lorentz_inv} tackles with the redundancies (b), (c), and the first half of (a), while subsection~\ref{sec:lorentz_str} deals with the Schouten identities and the IBP. Finally in subsection~\ref{sec:lorentz_perm}, we symmetrize the Lorentz structures over repeated fields for a specific type and obtain the Lorentz basis.

\subsubsection{Lorentz Invariance: Enumerating the Classes}
\label{sec:lorentz_inv}

We start by further analyzing the building block defined in eq.~\eqref{eq:field_w_D}, and reduce them to Irreps of the Lorentz group. By applying the following relations
\eq{
	& D_{[\alpha\dot\alpha}D_{\beta]\dot\beta} = D_{\mu}D_{\nu}\sigma^{\mu}_{[\alpha\dot\alpha}\sigma^{\nu}_{\beta]\dot\beta} = -D^2\epsilon_{\alpha\beta}\epsilon_{\dot\alpha\dot\beta} + \frac{i}{2}[D_{\mu},D_{\nu}]\epsilon_{\alpha\beta}\bar\sigma^{\mu\nu}_{\dot\alpha\dot\beta}, \\
	& D_{[\alpha\dot\alpha}\psi_{\beta]} = D_{\mu}\sigma^{\mu}_{[\alpha\dot\alpha}\psi_{\beta]} = -\epsilon_{\alpha\beta}(D\4\psi)_{\dot\alpha}, \\
	& D_{[\alpha\dot\alpha}F_{{\rm L} \beta]\gamma} = D_{\mu}F_{\nu\rho} \sigma^{\mu}_{[\alpha\dot\alpha}\sigma^{\nu\rho}_{\beta]\gamma} = 2D^{\mu}F_{\mu\nu} \epsilon_{\alpha\beta}\sigma^{\nu}_{\gamma\dot\alpha},
}
we note that any pair of anti-symmetric spinor indices in a building block would lead to factors that vanish according to the redundancies (b) or (c). 
As a consequence, we are left with building blocks in which all spinor indices, dotted or undotted, are totally symmetric respectively. 
After raising the dotted indices, we could express the remaining building blocks as
\eq{\label{eq:field_format}
	(D^{r-|h|}\Phi)_{\left(\alpha^{(1)}\alpha^{(2)}\dots\alpha^{(r-h)}\right)}^{\left(\dot\alpha^{(1)}\dot\alpha^{(2)}\dots\dot\alpha^{(r+h)}\right)} \equiv (D^{r-|h|}\Phi)_{\alpha^{r-h}}^{\dot\alpha^{r+h}} \in \left(\frac{r-h}{2},\frac{r+h}{2}\right)
}
where, without ambiguity, we abbreviate the totally symmetric indices (indicated by the parenthesis) by an index with a power. Now the remaining building block transforms as irreps under the Lorentz group as shown above.

With this notation, together with our treatment of the redundancy (a) using eq.~\eqref{eq:fierz_l}, we arrive at a general form of Lorentz structure modulo (b,c) redundancies as
\eq{\label{eq:lorentz_format}
	\mc{M} = (\epsilon^{\alpha_i\alpha_j})^{\otimes n}(\tilde\epsilon_{\dot\alpha_i\dot\alpha_j})^{\otimes \tilde{n}} \prod_{i=1}^N (D^{r_i-|h_i|}\Phi_i)_{\alpha_i^{r_i-h_i}}^{\dot\alpha_i^{r_i+h_i}} \in [\mc{M}]_{N,n,\tilde{n}},
}
where $N$ is the number of building blocks in the operator, corresponding to the number of particles in the on-shell amplitude it generates. 
Here we recognize the epsilon tensors introduced is the Lorentz invariant tensor $T_{\rm Lorentz}$ in eq.~\eqref{eq:field_w_D}, in which eq.~\eqref{eq:schouten} guarantees that only the epsilon tensors appear. 
The power of the epsilon tensors should be understood as a product of epsilons with possibly different spinor indices, which are only distinguished by the building block eq.~\eqref{eq:field_format} they come from. Such operators with certain $N$ and the numbers of epsilons $(n,\tilde{n})$ form a basis of the linear space $[\mc{M}]_{N,n,\tilde{n}}$, which still have redundancy from the Schouten identity and the IBP. 

To solve the IBP problem, we decompose the space as $[\mc{M}]_{N,n,\tilde{n}} = [\mc{A}]_{N,n,\tilde{n}} \oplus [\mc{B}]_{N,n,\tilde{n}}$,
where the subspace $[\mc{B}]_{N,n,\tilde{n}}$ contains all the Lorentz structures with total derivatives. 
For any Lorentz structure $\mc{M}\in[\mc{M}]$ we have 
\eq{\label{eq:M_decompose}
	\mc{M} = \mc{M}_\mc{A} + \mc{M}_\mc{B}, \quad \mc{M}_\mc{A}\in[\mc{A}],\ \mc{M}_\mc{B}\in[\mc{B}].
}
If this decomposition is possible, the subspace $[\mc{A}]$ would be the space of the non-redundant Lorentz structures, because for any two Lorentz structures in $[\mc{A}]$, their difference is also in $[\mc{A}]$ and cannot be a total derivative. We will achieve this decomposition in the next subsection.

Before that, we would like to show how Lorentz invariance constrains the classes appearing at a certain dimension. We derive some non-trivial constraints among the parameters in eq.~\eqref{eq:lorentz_format}. The contractions of spinor indices lead to the following relations
\eqs{
	& \tilde{n}+n = \sum_ir_i = r, \qquad \tilde{n}-n = \sum_ih_i \equiv h, \label{eq:spinor_index_matching_1}\\
	& n_D = \sum_i(r_i-|h_i|) = 2n + h - \sum_i|h_i| = 2\tilde{n} - h - \sum_i|h_i| \leq \min(2n, 2\tilde{n}). \label{eq:spinor_index_matching_2}
}
Here we find another interpretation of $r$ as the total number of $\epsilon$'s. The second line gives one constraint on the number of derivatives necessary for an operator with given helicity combination $h_i$, that the number must equal $h-\sum_i|h_i|$ mod 2 and is bounded by twice the minimum of $n$ and $\tilde{n}$.
Another constraint is already shown in \cite{Durieux:2019siw} comes from the following fact indicated by $r\geq |h|$,
\eq{
	n \geq r_i-h_i \geq -2h_i, \ \forall i \qquad 			&\Rightarrow \qquad \frac12\sum_i(r_i-h_i) = n \geq -2\min h_i, \\
	\tilde{n} \geq r_i+h_i \geq 2h_i, \ \forall i \qquad 	&\Rightarrow \qquad \frac12\sum_i(r_i+h_i) = \tilde{n} \geq 2\max h_i.
} 
from which we deduce 
\eq{
	n_D = \sum_i(r_i-h_i) - \sum_{h_i<0}2|h_i| \ \geq \ -4\min h_i - \sum_{h_i<0}2|h_i|, \\
	n_D = \sum_i(r_i+h_i) - \sum_{h_i>0}2|h_i| \ \geq \ 4\max h_i - \sum_{h_i>0}2|h_i|.
}
In sum, we arrive at the complete constraint on $n_D$:
\eq{\label{eq:D_constraint}
	\min(2n, 2\tilde{n}) \geq n_D \geq \max\begin{pmatrix} h-\sum_i|h_i|,\ \mod 2 \\ -4\min h_i - \sum_{h_i<0}2|h_i| \\ 4\max h_i - \sum_{h_i>0}2|h_i| \end{pmatrix}.
}
The minimum is a correction to the constraint shown in \cite{Durieux:2019siw}. 

In light of the above relations, we can enumerate the classes of Lorentz structures for a given dimension after the following steps:
\begin{itemize}
	\item From eq.~\eqref{eq:operator_dim} and eq.~\eqref{eq:spinor_index_matching_1}, we get $d = n+\tilde{n}+N$. We start by iterating $N$ from 3\footnote{
		$N=3$ is a special case when there is the so-called special kinematics that renders $n=0$ or $\tilde{n}=0$. Particularly it implies that $n_D=0$ when $N=3$. For example we have $D_\mu\phi_1 D^\mu\phi_2\phi_3 = \frac12(\phi_1\phi_2D^2\phi_3 -\phi_1D^2\phi_2\phi_3 -D^2\phi_1\phi_2\phi_3)$ which is redundant due to EOM of $\phi_i$ in our treatment.}
	to $d$, while for each $N$ we could iterate $n,\tilde{n}$ 
	under the constraint $n+\tilde{n} = d-N$.
	
	\item Given the tuple $(N,n,\tilde{n})$, we iterate $n_D$ from 0 to $\min(2n,2\tilde{n})$ according to eq.~\eqref{eq:D_constraint}. Provided the number of derivatives $n_D$, we have the following relations implied by eq.~\eqref{eq:spinor_index_matching_2}
	\eq{\label{eq:nh_constraint}
		2n_{-1}+n_{-1/2} = \sum_i|h_i|-h = 2n-n_D, \quad 2n_{1}+n_{1/2} = \sum_i|h_i|+h = 2\tilde{n}-n_D.
	}
	\item Then we find all tuples $n_i=(n_{-1},n_{-1/2},n_0,n_{1/2},n_1)$ that satisfy eq.~\eqref{eq:nh_constraint} and $\sum_in_i=N$, making sure that $n_D$ satisfies the minimum given in eq.~\eqref{eq:D_constraint} at the meantime. In this way, we find all the combinations of $(n_i,n_D)$ that could form Lorentz invariant structures, each of which determines a sub-class of operators.
\end{itemize}
At dimension 8, we list all the sub-classes in table~\ref{eq:classes8}.
\begin{table}[h]
	\eq{
		\begin{array}{ll|llll}
			\hline
			N		&	(n,\tilde{n})\footnote{We only list classes with $n\geq\tilde{n}$, while all the classes with $n<\tilde{n}$ are Hermitian conjugate of some classes listed here (denoted as $+\hc$).}
			& 	& \text{Sub-classes}	&	&	\\
			\hline
			4		&	(4,0)	&	F_{\rm L}^4+\hc	&	&	&	\\
			&	(3,1)	&	F_{\rm L}^2\psi\psi^{\dagger}D+\hc		&	\psi^4D^2+\hc				&	F_{\rm L}\psi^2\phi D^2+\hc	&	F_{\rm L}^2\phi^2 D^2+\hc	\\
			&	(2,2)	&	F_{\rm L}^2F_{\rm R}^2	&	F_{\rm L}F_{\rm R}\psi\psi^{\dagger}D	&	\psi^2\psi^{\dagger 2}D^2	&	F_{\rm R}\psi^2\phi D^2+\hc	\\
			&			&	F_{\rm L}F_{\rm R}\phi^2 D^2	&	\psi\psi^{\dagger}\phi^2D^3	&	\phi^4D^4 	& \\
			\hline
			5		&	(3,0)	&	F_{\rm L}\psi^4+\hc	&	F_{\rm L}^2\psi^2\phi+\hc	&	F_{\rm L}^3\phi^2+\hc	&	\\
			&	(2,1)	&	F_{\rm L}\psi^2\psi^{\dagger 2}+\hc	&	F_{\rm L}^2\psi^{\dagger2}\phi+\hc	&	\psi^3\psi^\dagger\phi D+\hc	&	F_{\rm L}\psi\psi^\dagger\phi^2 D+\hc \\
			&			&	\psi^2\phi^3D^2+\hc	&	F_{\rm L}\phi^4D^2+\hc	&	&	\\
			\hline
			6		&	(2,0)	&	\psi^4\phi^2+\hc	&	F_{\rm L}\psi^2\phi^3+\hc	&	F_{\rm L}^2\phi^4+\hc	&	\\
			&	(1,1)	&	\psi^2\psi^{\dagger 2}\phi^2	&	\psi\psi^\dagger\phi^4D	&	\phi^6D^2	& \\
			\hline
			7		&	(1,0)	&	\psi^2\phi^5+\hc	&	&	&	\\
			\hline
			8		&	(0,0)	&	\phi^8	&	&	&	\\
			\hline
		\end{array}\notag
	}
	\caption{All the sub-classes of Lorentz structures at dimension 8.}\label{eq:classes8}
\end{table}

\subsubsection{Lorentz Structures as $SU(N)$ States}
\label{sec:lorentz_str}

Back to the problem of finding the subspace $[\mc{A}]$ from $\mc{M}$ as proposed after eq.~\eqref{eq:M_decompose}, we first claim property of its elements in the format of eq.~\eqref{eq:lorentz_format}, that such Lorentz structure is completely determined by the epsilon tensors, because the numbers of $\alpha_i$ and $\dot\alpha_i$ on these tensors fix all the parameters of the building blocks $D^{r_i-|h_i|}\Phi_i$. For example, given $\epsilon^{\alpha_1\alpha_3}\epsilon^{\alpha_2\alpha_3}\epsilon_{\dot\alpha_3\dot\alpha_4}$, we obtain
\eq{\label{eq:example_operator}
	\epsilon^{\alpha_1\alpha_3}\epsilon^{\alpha_2\alpha_3}\tilde\epsilon_{\dot\alpha_3\dot\alpha_4} \quad \Rightarrow \quad	\mc{M} = \epsilon^{\alpha_1\alpha_3}\epsilon^{\alpha_2\alpha_3}\tilde\epsilon_{\dot\alpha_3\dot\alpha_4}(\psi_1)_{\alpha_1}(\psi_2)_{\alpha_2}(D\psi_3)^{\dot\alpha_3}_{\alpha_3^2}(\psi_4^{\dagger})^{\dot\alpha_4}.
}
Those who are familiar with spinor helicity variables should recognize that these epsilons are nothing but the spinor brackets $\epsilon^{\alpha_i\alpha_j} \sim \vev{ij}$, $\epsilon_{\dot\alpha_i\dot\alpha_j} \sim [ij]$\footnote{Recall the $r$ is the number of $\epsilon$'s, which corresponds to the number of spinor brackets in the on-shell amplitude, or in other words, the mass dimension of the amplitude. It matches with the discovery in eq.~\eqref{eq:operator_dim}.}. Therefore our claim here is exactly the amplitude-operator correspondence \cite{Ma:2019gtx, Jiang:2020sdh}. 
In this subsection, unless stated otherwise, we claim that {\it a product of epsilons refers to the Lorentz structure determined by it, and a linear combination of them refers to the linear combination of the corresponding Lorentz structures.} It gives us a hint on how to identify $[\mc{B}]$, the subspace of Lorentz structures with total derivatives. 
First, a derivative on field $\Phi_i$ has a pair of indices $(\alpha_i,\dot\alpha_i)$, which must also be found in the epsilons. Hence there has to be a factor of $\epsilon^{\alpha_i\alpha_j}\tilde\epsilon_{\dot\alpha_i\dot\alpha_k}$ in the epsilons. Therefore, a total derivative is thus represented by a factor of $\sum_i\epsilon^{\alpha_i\alpha_j}\tilde\epsilon_{\dot\alpha_i\dot\alpha_k}$, which is the character of Lorentz structures in $[\mc{B}]$.

To identify the complement space $[\mc{A}]$, we use a trick: by introducing an $SU(N)$ group for which Lorentz structures $\mc{M}\in[\mc{M}]$ transforms linearly, and both $[\mc{M}]$ and $[\mc{B}]$ are invariant spaces, $[\mc{A}]$ must also be an invariant space that consists of whole representation spaces. This group is defined by the following transformations of the epsilons
\eq{\label{eq:sun_define}
	\epsilon^{\alpha_i\alpha_j}\to\sum_{k,l}\mc{U}_k^i\mc{U}_l^j\epsilon^{\alpha_k\alpha_l}, \quad \tilde\epsilon_{\dot\alpha_i\dot\alpha_j}\to \sum_{k,l}\mc{U}^\dagger{}_i^k\mc{U}^\dagger{}_j^l\tilde\epsilon_{\dot\alpha_k\dot\alpha_l}.
}
In other words, the undotted spinor index $\alpha_i$ with subscript $i$ running from 1 to $N$ transforms as $\mathbf{2}\times\mathbf{N}$ of the $SL(2,\mathbb{C})\times SU(N)$ group, while the dotted index $\dot\alpha_i$ transforms as $\mathbf{\bar{2}}\times\mathbf{\bar{N}}$. Obviously, the transformation does not change the tuple $(N,n,\tilde{n})$, which means that $[\mc{M}]$ is invariant.It is also easy to prove the invariance of $[\mc{B}]$ 
\eq{\label{eq:total_derivative}
	\sum_i \epsilon^{\alpha_i\alpha_j}\tilde\epsilon_{\dot\alpha_i\dot\alpha_k} \to \sum_{m,n}\mc{U}_m^j\mc{U}^\dagger{}_k^n \sum_i\epsilon^{\alpha_i\alpha_m}\tilde\epsilon_{\dot\alpha_i\dot\alpha_n}.
}
Now the task is converted to finding irreducible representation spaces of $SU(N)$ in $[\mc{M}]_{N,n,\tilde{n}}$, and classifying them into $[\mc{A}]$ and $[\mc{B}]$. Specifically, due to eq.~\eqref{eq:sun_define}, it amounts to the decomposition of the tensor representations formed by products of the epsilons.

In terms of $SU(N)$ YD in which a box represents fundamental representation, $\epsilon$ and $\tilde\epsilon$ form irreducible representations ${\scriptsize \yng(1,1)}=[1^2]$ and ${\scriptsize \overline{\yng(1,1)}}=[1^{N-2}]$ respectively, due to the antisymmetry of their indices. Given specifically labels $i,j$ for the epsilons, they are states in these representation spaces, indicated by Young Tableau. For example when $N=5$, we have
\eq{\label{eq:lorentz-YT}
	\epsilon^{\alpha_2\alpha_3} \sim \young(2,3) \quad,\quad \tilde\epsilon_{\dot\alpha_1\dot\alpha_3} =-\mc{E}^{24513}\tilde\epsilon_{\dot\alpha_1\dot\alpha_3} \sim -\,\young(2,4,5) .
}
where $\mc{E}$ is the Levi-Civita tensor of $SU(N)$.

Then we use the Littlewood-Richardson (LR) rules~\cite{ma2007group} to decompose their products. First, we examine the tensor power of each type of the epsilons. Since
\eq{\label{eq:LR-epsilon^n}
	\epsilon^{\otimes2} = \yng(1,1) \otimes \yng(1,1) = \yng(2,2) \oplus \yng(2,1,1) \oplus \yng(1,1,1,1) \quad,\qquad \tilde\epsilon^{\otimes2} = \overline{\yng(1,1)} \otimes \overline{\yng(1,1)} = \overline{\yng(2,2)} \oplus \overline{\yng(2,1,1)} \oplus \overline{\yng(1,1,1,1)}
} 
We can use the Schouten identity to eliminate representations with more than two rows, either dotted or undotted
\eq{
	\young(il,j,k) \sim \epsilon^{\alpha_i\alpha_j}\epsilon^{\alpha_k\alpha_l} + \epsilon^{\alpha_k\alpha_i}\epsilon^{\alpha_j\alpha_l} + \epsilon^{\alpha_j\alpha_k}\epsilon^{\alpha_i\alpha_l} = 0,
}
and hence we are left with only the first term in the decomposition eq.~\eqref{eq:LR-epsilon^n}. Similarly if we multiply more epsilons of the same kind, we should only be left with the following YD's:
\eq{
	\epsilon^{\otimes n} = \underbrace{\yng(1,1)\ ...\ \yng(1,1)}_{\let\scriptstyle\textstyle\text{\large $n$}} \quad , \qquad \tilde\epsilon^{\otimes \tilde{n}} = \rotatebox[]{90}{\text{$N-2$}}  \left\{
	\begin{array}{lll}
		\yng(1,1) &\ldots{}& \yng(1,1) \\
		\quad\vdotswithin{}& & \quad \vdotswithin{}\\
		\undermat{\tilde{n}}{\yng(1,1) &\ldots{}& \yng(1,1)} 
	\end{array}
	\right. .
}

\vspace{0.5cm}

\noindent This reflects the fact that the spinor indices only take two values, forbidding antisymmetry over more than two of them.
The independent basis of the representation space is given by the SSYT's, where labels filled in the YD are increasing down the columns and non-decreasing along the rows. The Fock's conditions~\cite{ma2007group} for such YD's are nothing but the Schouten identities. Therefore, choosing the SSYT basis automatically eliminates the redundancy form the Schouten identity. For example\footnote{
	Note that it seems like we did not perform the row symmetrization for the Young tableau, which was done in ref.~\cite{Henning:2019enq}. It is due to our different treatments of the action of permutations on the $SU(N)$ tensors: in ref.~\cite{Henning:2019enq} the action of permutation, say $(12)$, means permuting the specific labels 1 and 2, while we treat the action of $(12)$ as permuting the 1st and 2nd indices in the tensor. While the two treatments give different sets of Lorentz structures, both of them are the independent basis of the same space. In our treatment, Young tableau would have indicated the column anti-symmetry rather than the row symmetry of the resulting tensor. That is why we only need to translate each column to an $\epsilon$ tensor to generate this desired feature. The same translation of Young tableau is used in the next subsection regarding gauge group tensors.}
\eq{
	\begin{array}{lll}
		\quad \young(12,43) 									&= \quad -\ \young(13,24) 										&+ \quad \young(12,34) \quad , \\[1em]
		\epsilon^{\alpha_1\alpha_4}\epsilon^{\alpha_2\alpha_3} 	&= -\epsilon^{\alpha_1\alpha_2}\epsilon^{\alpha_3\alpha_4} 		&+ \epsilon^{\alpha_1\alpha_3}\epsilon^{\alpha_2\alpha_4}, \\[1em]
		\psi_1^\alpha\psi_2^\beta\psi_{3\beta}\psi_{4\alpha}	&= -\psi_1^\alpha\psi_{2\alpha}\psi_3^\beta\psi_{4\beta}		&+ \psi_1^\alpha\psi_2^\beta\psi_{3\alpha}\psi_{4\beta}.
	\end{array}
}
The two terms on the right side corresponding to the SSYT's are our standard basis, and the left side corresponding to non-SSYT can be expressed by the standard basis via the Schouten identity.

Finally we use the LR rule to obtain the tensor product of these two YD's. 
Using the LR rules we place boxes from $[n^2]$ into the YD $[\tilde{n}^{N-2}]$
\begin{eqnarray}
	\rotatebox[]{90}{\text{$N-2$}} \left\{
	\arraycolsep=0pt\def\arraystretch{1}    \begin{array}{ccc}
		\yng(1,1) &\ \ldots{}\ & \yng(1,1) \\
		\vdotswithin{}& &    \vdotswithin{}\\
		\undermat{\tilde{n}}{\yng(1,1) &\ \ldots{}\ & \yng(1,1)} 
	\end{array}
	\right.\quad \otimes\quad \underbrace{\yng(1,1)\ ...\ \yng(1,1)}_{\let\scriptstyle\textstyle\text{\large $n$}} \quad =\quad \arraycolsep=0pt\def\arraystretch{1}
	\rotatebox[]{90}{\text{$N-2$}} 
	\left\{
	\begin{array}{cccccc}
		\yng(1,1) &\ \ldots{}&\ \yng(1,1)& \overmat{n}{\yng(1,1)&\ \ldots{}\  &\yng(1,1)} \\
		\vdotswithin{}& & \vdotswithin{}&&&\\
		\undermat{\tilde{n}}{\yng(1,1)\ &\ldots{}&\ \yng(1,1)} &&&
	\end{array}
	\right.\quad +\quad \dots\nonumber\\
	\nonumber\\
	\label{eq:LR-epsilon-epsilon}
\end{eqnarray}
where the first term describes the case when all the boxes are put to the right of the original $[\tilde{n}^{N-2}]$ boxes, and the terms in $\dots$ are cases when boxes are put under the original YD. Following the correspondence eq.~\eqref{eq:lorentz-YT}, whenever a box $\young(i)$ from $[n^2]$ is placed under a column in $[\tilde{n}^{N-2}]$, the resulting column with $N-1$ rows represent a tensor
\eq{
	\sum_{l_1,l_2}\mc{E}^{k_1,\dots,k_{N-2},l_1,l_2}\tilde\epsilon_{\alpha_{l_1}\alpha_{l_2}}\epsilon^{\alpha_i \alpha_j} + (\text{anti-sym over }k_1,\dots,k_{N-2},i) = \sum_y \mc{E}^{k_1,\dots,k_{N-2},i,x}\tilde\epsilon_{\alpha_x\alpha_y}\epsilon^{\alpha_y\alpha_j}.
}
which contains a factor of total derivative as discussed previously. According to eq.~\eqref{eq:total_derivative}, the whole representation space is contained in the subspace $\mc{B}$ of $[\mc{M}]_{N,n,\tilde{n}}$. It rules out all but the first term in eq.~\eqref{eq:LR-epsilon-epsilon}, thus we conclude that the remaining YD, which we define as the primary YD, is the only irrep contained in $[\mc{A}]$, namely
\vspace{0.5cm}
\begin{eqnarray}\label{eq:YD_shape}
	[\mc{A}]_{N,n,\tilde{n}} \quad = \quad \arraycolsep=0pt\def\arraystretch{1}
	\rotatebox[]{90}{\text{$N-2$}} \left\{
	\begin{array}{cccccc}
		\yng(1,1) &\ \ldots{}&\ \yng(1,1)& \overmat{n}{\yng(1,1)&\ \ldots{}\  &\yng(1,1)} \\
		\vdotswithin{}& & \vdotswithin{}&&&\\
		\undermat{\tilde{n}}{\yng(1,1)\ &\ldots{}&\ \yng(1,1)} &&&
	\end{array}
	\right.
	\\
	\nonumber 
\end{eqnarray}
The primary YD $[\mc{A}]_{N,n,\tilde{n}}$ is exactly the space of Lorentz structures without any of the redundancies listed at the beginning of this section. Our next task is to obtain a complete basis of this space, which is again the SSYT's. The Fock's conditions between non-SSYT and the SSYT basis are equivalent to the Schouten identity or the IBP. An example of the Fock's condition reflecting the IBP is
\eq{
	\begin{array}{llll}
		\quad \young(21,35,4) 									&= \quad \young(14,25,3) 										&+ \quad \young(12,35,4) 					&- \quad \young(13,25,4) \quad , \\[1em]
		-\epsilon_{\dot{\alpha}_1\dot{\alpha}_5}\epsilon^{\alpha_1\alpha_5} &= \epsilon_{\dot{\alpha}_4\dot{\alpha}_5}\epsilon^{\alpha_4\alpha_5} 		&+ \epsilon_{\dot{\alpha}_2\dot{\alpha}_5}\epsilon^{\alpha_2\alpha_5}	&+ \epsilon_{\dot{\alpha}_3\dot{\alpha}_5}\epsilon^{\alpha_3\alpha_5}, \\[1em]
		-D_\mu\phi_1\phi_2\phi_3\phi_4D^\mu\phi_5				&= \phi_1\phi_2\phi_3D_\mu\phi_4D^\mu\phi_5						&+ \phi_1D_\mu\phi_2\phi_3\phi_4D^\mu\phi_5	&+ \phi_1\phi_2D_\mu\phi_3\phi_4D^\mu\phi_5	
	\end{array}
}
where $D^2\phi_5$ is understood to be eliminated by the EOM.

As shown by Table.\ref{eq:classes8}, $[\mc{M}]$ usually contains more than one class of operators, and so does $[\mc{A}]$. The full set of its SSYT basis includes a lot of non-physical fillings that involve fields with large helicities (gravitino, graviton and even higher). Thus it would be wise to single out a subset of them as the Lorentz structures for a given class. 
By obtaining the tuple $(N,n,\tilde{n})$ from the helicities and $n_D$ according to eq.~\eqref{eq:nh_constraint}, we can easily find the $[\mc{A}]$ that the class is included, in which the labels in it come from either $\epsilon^{\alpha_i\alpha_j}$ or $\mc{E}^{i\dots jk}\tilde{\epsilon}_{\dot\alpha_j\dot\alpha_k}$. The number of the former, $\epsilon$ with index $\alpha_i$, equals the number of $\alpha$ indices in the building block $i$, which is $r_i-h_i$ according to eq.~\eqref{eq:lorentz_format}, while the number of the latter, same as the number of $\tilde\epsilon$ without index $\dot\alpha_i$, equals $\tilde{n}-(r_i+h_i)$.
Together with eq.~\eqref{eq:nh_constraint}, we get
\eq{\label{eq:state2class}
	\#i = \tilde{n} - 2h_i = \frac12n_D + \sum_{h_i>0}|h_i| - 2h_i,
}
which surprisingly do not depend on $r_i$, and are hence completely and uniquely determined by the class information $(\{h_i\},n_D)$. 

Our strategy is now clear: for each subclass, we find the YD of eq.~\eqref{eq:YD_shape} determined by eq.~\eqref{eq:nh_constraint}, and use eq.~\eqref{eq:state2class} to deduce the tuple of labels $\{1^{\#1},\dots,N^{\# N}\}$ to fill in the YD; the SSYT's obtained this way\footnote{Note that in \cite{Henning:2019enq} the complete basis is given by the so-called reduced SSYT's, which eliminates the over-counting of classes while enumerating the SSYT's. But since we start from a certain class, we do not suffer from the over-counting of classes. Thus the condition of SSYT is sufficient. } correspond to the complete and independent basis of Lorentz structures.
As an example, consider the class of operators $\psi\psi\psi\psi^{\dagger}D$ at dimension 7, with $h_i = \{-1/2,-1/2,-1/2,1/2\}$ and $n_D=1$. With eq.~\eqref{eq:nh_constraint} and eq.~\eqref{eq:state2class} we have
$ \#1 = \#2 = \#3 = 2,\ \#4 = 0$ and $n = 2,\tilde{n} = 1$. The only SSYT is given by 
\eq{\label{eq:example}
	\young(112,233) \sim \mc{E}^{1234}\epsilon_{\dot\alpha_3\dot\alpha_4}\epsilon^{\alpha_1\alpha_3}\epsilon^{\alpha_2\alpha_3},
}
which leads to the Lorentz structure eq.~\eqref{eq:example_operator}. It means that eq.~\eqref{eq:example_operator} is the only independent Lorentz structure of this class, which sounds counter-intuitive.
Indeed, in \cite{Liao:2016qyd} the authors pointed out several redundancies of the dim 7 operators listed in \cite{Bhattacharya:2015vja} and found the correct independent operator basis. One of the redundancies was about this particular class of operators, for which they explicitly apply the identity relations (a$\sim$d) shown at the beginning of this section to prove the redundancies. With our strategy, the redundancy relations, like eq.~(35-37) in \cite{Liao:2016qyd}, are nothing but the Fock's conditions between Young tableau, which is automatically tackled by choosing the SSYT. 

Another example where a class contains several independent Lorentz structures is $F_{\rm L}\psi^4$, which has $\#1=2,\#2=\#3=\#4=\#5=1$ and $n=3,\tilde{n}=0$. The YD has the same shape as the above example, but they indicate different representation spaces due to their different $(N,n,\tilde{n})$. The SSYT basis of Lorentz structures for this class is given by
\eq{\label{eq:example2}
	\begin{array}{lll}
		\young(112,345)															&	\young(113,245)															&	\young(114,235)	\\[1em]
		\epsilon^{\alpha_1\alpha_3}\epsilon^{\alpha_1\alpha_4}\epsilon^{\alpha_2\alpha_5}	&	\epsilon^{\alpha_1\alpha_2}\epsilon^{\alpha_1\alpha_4}\epsilon^{\alpha_3\alpha_5}	&	\epsilon^{\alpha_1\alpha_2}\epsilon^{\alpha_1\alpha_3}\epsilon^{\alpha_4\alpha_5}	\\[1em]
		F_1^{\alpha\beta}\psi_2^\gamma\psi_{3\alpha}\psi_{4\beta}\psi_{5\gamma}	&	F_1^{\alpha\beta}\psi_{2\alpha}\psi_3^\gamma\psi_{4\beta}\psi_{5\gamma}	&	F_1^{\alpha\beta}\psi_{2\alpha}\psi_{3\beta}\psi_4^\gamma\psi_{5\gamma}
	\end{array}
}
To count the number of the basis for a given class, we can treat the YD of eq.~\eqref{eq:YD_shape} as a product of YD's with the same labels, since the latter is determined by the class information $\#i$: they have to be totally symmetric one-row YD $[\#i]$. 
For the case eq.~\eqref{eq:example}, we have $[\#1]=[\#2]=[\#3]=\yng(2)$ (label 4 does not contribute), hence we examine the decomposition of their product
\eq{\label{eq:example_count}
	\yng(2)^{\,\otimes 3} &= \yng(6)\times 1 + \yng(5,1)\times 2 + \yng(4,2)\times 3 + {\color{blue} \yng(3,3)\times 1} \\
	& \qquad  + \yng(4,1,1)\times 1 + \yng(3,2,1)\times 2 + \yng(2,2,2)\times 1 .
}
and find only one target YD $\scriptsize\yng(3,3)$ in it, which means only one SSYT with certain filling exists. Similarly, for the case eq.~\eqref{eq:example2} we have
\eq{\label{eq:example2_count}
	&\yng(2) \otimes \yng(1)^{\otimes 4} = \yng(6)\times 1 + \yng(5,1)\times 4 + \yng(4,2)\times 6 + {\color{blue} \yng(3,3)\times 3} \\
	& \qquad + \yng(4,1,1)\times 6 + \yng(3,2,1)\times 8 + \yng(2,2,2)\times 2 + \yng(3,1,1,1)\times 4 + \yng(2,2,1,1)\times 3.
}
where the multiplicity of the target YD $\scriptsize\yng(3,3)$ precisely reproduces the number of SSYT we listed in eq.~\eqref{eq:example2}.

In summary, by identifying Lorentz structures as states in the $SU(N)$ representation space $[\mc{A}]_{N,n,\tilde{n}}$, not only can we quickly count the number of independent basis, but we can also write them down by a translation from the SSYT's. This makes our approach superior to the competitors, and allows us to achieve a systematic way to list the operators in generic effective field theories.

\subsubsection{Permutation: Counting and Listing the Lorentz Basis}
\label{sec:lorentz_perm}
The Lorentz structures we obtained as SSYT's in the above subsection did not take into account the permutation symmetries of possible repeated fields when we specify the type. 
For the purpose of counting, we adopt the technique of plethysm. Since the repeated fields with same helicities must have a equal amount of labels to be filled into the YD, instead of taking a direct product of the $[\#i]$ as in eq.~\eqref{eq:example_count}, we take the plethysm with particular permutation symmetry. In particular, for any YD $\mc{Y}$ we have
\eq{
	\mc{Y}^{\otimes m} = \sum_{\lambda\vdash m} d_{\lambda}\  \mc{Y}\, \circled{p}\,\lambda ,
}
where $d_{\lambda}$ is the dimension of the $S_m$ irrep $\lambda$. In the example of eq.~\eqref{eq:example} and eq.~\eqref{eq:example_count}, suppose the three $\psi$'s are repeated fields like in the type of operator $Q^3He_{_\mathbb{C}}^{\dagger}D$, where the three $Q$'s could have permutation symmetries $[3]$, $[2,1]$ and $[1^3]$, for which we derive the plethysm
\eq{\label{eq:example_plethysm}
	& {\footnotesize\yng(2)} \ \circled{p} \,[3] = \footnotesize\yng(6) + \yng(4,2) + \yng(2,2,2) , \\
	& {\footnotesize\yng(2)} \ \circled{p} \,[2,1] = \footnotesize\yng(5,1) + \yng(3,2,1) + \yng(4,2) , \\
	& {\footnotesize\yng(2)} \ \circled{p} \,[1^3] = \footnotesize\yng(4,1,1) + {\color{blue} \yng(3,3)} .
}
These are nothing but a classification of the result in eq.~\eqref{eq:example_count}. Note that $d_{[2,1]}=2$, so the YD's in the second line should be counted twice while matching with eq.~\eqref{eq:example_count}.
Among the results, we find the target YD, namely $\scriptsize \yng(3,3)\ $, which only appears in $[1^3]$ symmetry. 
The permutation symmetry $\lambda$ obtained here, which in general should include all sets of repeated fields $\lambda = \prod_\Phi\lambda_\Phi$, is slightly different from that of the Lorentz structure $\mc{M}$ itself, which we defined in sec.~\ref{sec:motiv} as $\lambda_1$.
There are two sources of differences: 
\begin{itemize}
	\item $\lambda$ characterizes the permutation symmetry of labels filled in the YD, which are indices of the combination of the Lorentz structure and $\tilde{n}$ factors of $\mc{E}$ from the Hodge duals of the $\tilde\epsilon$'s. As $\mc{E}$ is totally antisymmetric for any subset of labels, each $\mc{E}$ contributes total antisymmetry $[1^{m_i}]$ to the $i$th repeated fields. 
	
	\item The SSYT does not know about spin-statistics, hence the permutation symmetry of fermionic repeated fields has not taken into account their Grassmann feature, which should have contributed an extra $[1^{m_i}]$. 
\end{itemize}
The property of inner product $\lambda \odot [1^m] = \lambda^T,\ \forall\lambda\vdash m$ then suggests that the final permutation symmetry of the Lorentz structure is given by
\eq{
	&\lambda_1 = \prod_{\rm fermion}\lambda_\Phi^T \times \prod_{\rm boson}\lambda_\Phi, \quad \tilde{n} \text{ is even}, \\
	&\lambda_1 = \prod_{\rm fermion}\lambda_\Phi \times \prod_{\rm boson}\lambda_\Phi^T, \quad \tilde{n} \text{ is odd}.
}
Take the example in eq.~\eqref{eq:example} where the only Lorentz structure has the permutation symmetry $\lambda = [1^3]$ as shown in eq.~\eqref{eq:example_plethysm}, the type of operators $Q^3He_{_\mathbb{C}}^{\dagger}D$ has $\tilde{n}=1$ and the repeated field $Q$ is a fermion, which means $\lambda_M = \lambda = [1^3]$. As for the case in eq.~\eqref{eq:example2}, take the type $W_{\rm L}Q^3L$ as an example which has repeated field $Q$, we compute the plethysm
\eq{
	\yng(2) \otimes \left({\footnotesize\yng(1)} \ \circled{p} \,[3]\right) \otimes \yng(1) &= \footnotesize\yng(6) + \yng(5,1)\times2 + \yng(4,2)\times2 + {\color{blue} \yng(3,3)} + \yng(4,1,1) + \yng(3,2,1) , \\
	\yng(2) \otimes \left({\footnotesize\yng(1)} \ \circled{p} \,[2,1]\right) \otimes \yng(1) &= \scriptsize\yng(5,1) + \yng(4,2)\times2 + {\color{blue} \yng(3,3)} + \yng(4,1,1)\times2 + \yng(3,2,1)\times3 + \yng(2,2,2) + \yng(3,1,1,1) + \yng(2,2,1,1) , \\
	\yng(2) \otimes \left({\footnotesize\yng(1)} \ \circled{p} \,[1^3]\right) \otimes \yng(1) &= \footnotesize\yng(4,1,1) + \footnotesize\yng(3,2,1) + \footnotesize\yng(3,1,1,1)\times2 + \footnotesize\yng(2,2,1,1) .
}
It indicates that the 3 SSYT's obtained in eq.~\eqref{eq:example2} are grouped into a $[3]^T = [1^3]$ and a $[2,1]^T = [2,1]$ representation spaces. 

In order to construct the basis of these representation spaces as combinations of the original SSYT states $\mc{M}_\xi$, we apply the projectors $b^\lambda_x$ introduced in section~\ref{sec:groupalgebra} to all of the SSYT's 
\eq{
	\mc{M}^{\lambda_1}_{\xi,x} \equiv b^{\lambda}_x \mc{M}_\xi,\quad x=1,\dots,d_{\lambda},
}
where the difference between the symmetries $\lambda \to \lambda_1$ should be noticed. Each of the projections either forms a representation space of symmetry $[\lambda_1]$ according to eq.~\eqref{eq:piF}
\eq{
	\pi \circ \mc{M}^{\lambda_1}_{\xi,x} = \sum_y\mc{M}^{\lambda_1}_{\xi,y} \mc{D}(\pi)_{yx}, \quad \pi \in \bar{S},
}
or vanishes by the projection. For the example \eqref{eq:example2} where we got three independent Lorentz structures for the type $W_{\rm L}Q^3L$, which we denote as $\mc{M}_{\xi=1,2,3}$ respectively, we obtain
\eq{\label{eq:lorentz_basis_example}
	\mc{M}^{[1^3]}_1 \equiv & \mc{M}^{[1^3]}_{1,1} = \mc{M}^{[1^3]}_{2,1} = \mc{M}^{[1^3]}_{3,1} = \frac13\left(\mc{M}_1 + \mc{M}_2 + \mc{M}_3\right), \\
	\mc{M}^{[2,1]}_x \equiv & \mc{M}^{[2,1]}_{1,x} = -\mc{M}^{[2,1]}_{3,x} = \left\{\ \frac13\left(\mc{M}_1 + \mc{M}_2 - 2\mc{M}_3\right)\ ,\ \frac13\left(\mc{M}_1 - 2\mc{M}_2 + \mc{M}_3\right)\ \right\}_x, \\
	& \mc{M}^{[2,1]}_{2,x} = \{0,0\},
}
hence we get the symmetrized Lorentz structures as $\mc{M}^{[1^3]}_1$ and $\mc{M}^{[2,1]}_x,\ x=1,2$. 
Note that $b^{[2,1]}_x$ acting on $\mc{M}_1$ and $\mc{M}_3$ produce the same representation space. In general, when there are multiple numbers of the same representation space $[\lambda_1]$, 
picking out linearly independent spaces from the non-vanishing projections of $b^\lambda_x$ is non-trivial, which is why we use the Fock's conditions to convert the symmetrized Lorentz structures to combinations of the original basis $\mc{M}_\xi$. Generically we obtain
\eq{
	\mc{M}^{\lambda_1}_{\xi,x} = \sum_{\zeta} \mc{K}^{\lambda_1,x}_{\xi\zeta}\mc{M}_\zeta .
}
where the coefficient matrix $\mc{K}^{\lambda_1,x}$ has rank\footnote{Actually, the linear dependence among the rows of the matrix $\mc{K}^{\lambda_1,x}$ should not depend on $x$, just as the projector $b^\lambda_x$ either projects out a full representation space, non-vanishing for all $x$, or annihilate a Lorentz structure, vanishing for all $x$. Therefore, the rank $\mc{N}^\lambda_1$ is also independent of $x$. } $\mc{N}^{\lambda_1}$. Now we can select $\mc{N}^{\lambda_1}$ number of rows from $\mc{K}^{\lambda_1,x}$ as $\bar\xi=\xi_1,\dots,\xi_{\mc{N}^{\lambda_1}}$ which provide an independent set of $[\lambda_1]$-symmetry Lorentz basis as $\mc{M}^{\lambda_1}_{\bar\xi,x}$. In the above example, we have 
\eq{
	\mc{K}^{[1^3],1} = \begin{pmatrix} \frac13 & \frac13 & \frac13 \\ \frac13 & \frac13 & \frac13 \\ \frac13 & \frac13 & \frac13 \end{pmatrix}\ , \qquad 
	\mc{K}^{[2,1],1} = \begin{pmatrix} \frac13 & \frac13 & -\frac23 \\ 0 & 0 & 0 \\ -\frac13 & -\frac13 & \frac23 \end{pmatrix}\ , \quad
	\mc{K}^{[2,1],2} = \begin{pmatrix} \frac13 & -\frac23 & \frac13 \\ 0 & 0 & 0 \\ -\frac13 & \frac23 & -\frac13 \end{pmatrix}\ ,
}
which have ranks $\mc{N}^{[3]}=\mc{N}^{[2,1]}=1$. In that there are no multiplicities of the representation spaces, we are allowed to omit the subscript $\bar\xi$ as in eq.~\eqref{eq:lorentz_basis_example}.

\subsection{Gauge Basis: Littelwood-Richardson Rule}
\label{sec:gauge}
After obtaining the symmetrized Lorentz structures ${\cal M}^\lambda_x$, we are now ready to find a set of symmetrized gauge group factors $T^\lambda_{{\rm SU3},x}$ and $T^\lambda_{{\rm SU2},x}$ in eq.~\eqref{eq:term_format}, the procedure is similar to finding the symmetrized Lorentz structures discussed above. We shall find all the independent group factors $T_\xi$ first, then symmetrize them by applying $b^\lambda_x$s discussed in section~\ref{sec:groupalgebra} to the gauge group indices of the repeated fields:
\begin{eqnarray}
	T^\lambda_{\xi,x}=b^\lambda_x\circ T_\xi.
\end{eqnarray}

In principle, one can obtain all the independent  $T_\xi$ by recursively using CGCs of the corresponding gauge group, however this method cannot give nice forms of group factors expressed in terms of invariants using Levi-Civita tensors. Here we postulate a way to express all $T_\xi$ in terms of Levi-Civita tensors of $SU(N)$ group provided that each field is expressed in a tensor of fundamental indices only. 
The algorithm is to use the Littlewood-Richardson (LR) rule repeatedly but with indices associated with the corresponding irreps filled in during the construction of a singlet YD. 
From this procedure, one can obtain different singlet Young tableaux with $N$ rows as different ways to construct a $SU(N)$ singlet, each Young tableau then translates into a $T_\xi$ as a product of $\epsilon$ tensors with the indices setting to the corresponding indices in each column in a consistent manner. 
We illustrate the procedure by constructing the $SU(2)_W$ group factor of the operator $Q^3LW_{\rm L}$. 
Suppose the $SU(2)_W$ indices for three $Q$'s and $L$ are ${j,k,l}$ and $i$ respectively,  while that for $W_{\rm L}$ is $I$.
The first step is to convert all the non-fundamental indices into fundamental ones. The only field needs this preprocessing in our case is $W^I_{\rm L}$, and we convert it by contracting with  $(\tau^I)_{m_1}^{\ x}\epsilon_{xm_2}$, which leads to \begin{eqnarray}
	W_{{\rm L},m_1m_2}=W^I_{\rm L}(\tau^I)_{m_1}^{\ x}\epsilon_{xm_2},
\end{eqnarray}
where the summation over the repeated indices is implied. Next, we are going to form the Young tableaux with indices $j,k,l,i,m_1,m_2$ according to the LR rule. 
There are three different $T_{{\rm SU2},\xi}$'s which correspond to three different paths to construct $3\times 2$ YDs. We illustrate them in the following:
\begin{eqnarray}
	&&Q\xrightarrow{\large Q} Q^2\xrightarrow{Q} Q^3 \xrightarrow{L} Q^3L \xrightarrow{W_{\rm L}}Q^3LW_{\rm L},\\
	&&\yng(1)\xrightarrow{\young(1)}\young({{}}{{1}})\xrightarrow{\young(1)}\young({{}}{{}}{{1}})\xrightarrow{\young(1)}\young({{}}{{}}{{}},1)\xrightarrow{\young(11)}\young({{}}{{}}{{}},{{}}11),\\
	&&\yng(1)\xrightarrow{\young(1)}\young({{}}{{1}})\xrightarrow{\young(1)}\young({{}}{{}},1)\xrightarrow{\young(1)}\young({{}}{{}}1,{{}})\xrightarrow{\young(11)}\young({{}}{{}}{{}},{{}}11),\\
	&&\yng(1)\xrightarrow{\young(1)}\young({{}},{{1}})\xrightarrow{\young(1)}\young({{}}{{1}},{{}})\xrightarrow{\young(1)}\young({{}}{{}}1,{{}})\xrightarrow{\young(11)}\young({{}}{{}}{{}},{{}}11),
\end{eqnarray}
where the first line tells the order of the fields in forming the singlet YD's.
We follow the above paths to fill each box with the corresponding indices of the field and translate them into products of $\epsilon$'s:
\begin{eqnarray}
	&&\Yboxdim15pt Q:\ \young(j),\ Q:\ \young(k),\ Q:\ \young(l), \ L:\ \young(i),\ W:\ \young({{m_1}}{{m_2}}),\\
	&&\Yboxdim15pt \young(j)\xrightarrow{\young(k)}\young({{j}}{{k}})\xrightarrow{\young(l)}\young({{j}}{{k}}{{l}})\xrightarrow{\young(i)}\young({{j}}{{k}}{{l}},i)\xrightarrow{\young({{m_1}}{{m_2}})}\young({{j}}{{k}}{{l}},{{i}}{{m_1}}{{m_2}})=\epsilon^{ji}\epsilon^{km_1}\epsilon^{lm_2}=T_{{\rm SU2},1},\\
	&&\Yboxdim15pt\young(j)\xrightarrow{\young(k)}\young({{j}}{{k}})\xrightarrow{\young(l)}\young({{j}}{{k}},{{l}})\xrightarrow{\young(i)}\young({{j}}{{k}}{{i}},{{l}})\xrightarrow{\young({{m_1}}{{m_2}})}\young({{j}}{{k}}{{i}},{{l}}{{m_1}}{{m_2}})=\epsilon^{jl}\epsilon^{km_1}\epsilon^{im_2}=T_{{\rm SU2},2},\\
	&&\Yboxdim15pt\young(j)\xrightarrow{\young(k)}\young({{j}},{{k}})\xrightarrow{\young(l)}\young({{j}}{{l}},{{k}})\xrightarrow{\young(i)}\young({{j}}{{l}}i,{{k}})\xrightarrow{\young({{m_1}}{{m_2}})}\young({{j}}{{l}}{{i}},{{k}}{{m_1}}{{m_2}})=\epsilon^{jk}\epsilon^{lm_1}\epsilon^{im_2}=T_{{\rm SU2},3}.
\end{eqnarray}
With this set of $T_{{\rm SU2},\xi}$, we can project out the corresponding $T^{\lambda}_{{\rm SU2},x}$ by using the symmetrizers $b^\lambda_x$. 
To find out which $\lambda$ the three $Q$'s can take, we first need to enumerate all the $SU(2)_W$ irreps constructed by $Q$'s that can form a singlet with the rest of the fields $L$ and $W$. In this example both the quadruplet and doublet are capable.
Next, one can pick out the $\lambda$ that after taking plethysm with the $SU(2)_W$ irrep of $Q$'s are able to produce the quadruplet and doublet:
\begin{eqnarray}
	&&\yng(1)\ \circled{p}\,[3] = \yng(3)\ \text{and} \\
	&&\yng(1) \ \circled{p} \,[2,1] = \yng(2,1). 
\end{eqnarray}
From the above equation, we find that [3] and [2,1] are the possible choices,
and we have:
\begin{eqnarray}
	T^{[3]}_{{\rm SU2},1}&=&b^{[3]}_1\circ T_{{\rm SU2},1} = \frac{1}{6}[\epsilon^{ji}\epsilon^{km_1}\epsilon^{lm_2} + ({\rm perm\ }i,j,k)]\nonumber \\
	& = & T_{{\rm SU2},1}-\frac{1}{3}(T_{{\rm SU2},2}+T_{{\rm SU2},3})\label{eq:T3}\\
	T^{[2,1]}_{{\rm SU2},1}&=&b^{[2,1]}_1\circ T_{{\rm SU2},1} = \frac{1}{3}[\epsilon^{jm_1}\epsilon^{ki}\epsilon^{lm_2}+\epsilon^{ji}\epsilon^{km_1}\epsilon^{lm_2}-\epsilon^{jm_2}\epsilon^{km_1}\epsilon^{li}-\epsilon^{jm_1}\epsilon^{km_2}\epsilon^{li}]\nonumber \\
	& = &\frac{2}{3}T_{{\rm SU2},2}-\frac{1}{3}T_{{\rm SU2},3}\label{eq:T211}\\
	T^{[2,1]}_{{\rm SU2},2}&=&b^{[2,1]}_2\circ T_{{\rm SU2},1} = \frac{1}{3}[\epsilon^{jm_1}\epsilon^{km_2}\epsilon^{li}+\epsilon^{ji}\epsilon^{km_2}\epsilon^{lm_1}-\epsilon^{jm_2}\epsilon^{ki}\epsilon^{lm_1}-\epsilon^{jm_1}\epsilon^{ki}\epsilon^{lm_2}]\nonumber \\
	& = &-\frac{1}{3}T_{{\rm SU2},2}+\frac{2}{3}T_{{\rm SU2},3}.\label{eq:T212}
\end{eqnarray}
From the first to the second lines in the above equations, we have used the Schouten identity and the fact that any terms proportional to $\epsilon^{m_1m_2}$ can be dropped as $W_{{\rm L},m_1m_2}$ is a symmetric tensor.
In addition, one can verify that the projection of $b^{[3]}_1$ on $T_{{\rm SU2},2}$ or $T_{{\rm SU2},3}$ gives a null space while that of $b^{[2,1]}_x$'s generate the same space as the one we generate above from eq.~\eqref{eq:T211} to eq.~\eqref{eq:T212}. 

Readers can follow this method to derive the $SU(3)_C$ group factor for this type of operator, which is quite trivial yielding $T^{[1^3]}_{{\rm SU3},1}=\epsilon^{abc}$ given that the indices of three $Q$'s are ${a,b,c}$. It is obvious that this group factor is in the $[1^3]$ representation of $S_3$.

The above construction can be generalized to operator types with more than one set of repeated fields. The projection operations for different sets of repeated fields simply commute with each other. Therefore one can obtain a set of symmetrized group factors transforming as irreps of the direct product symmetric group $\bar{S}$ defined in section~\ref{sec:motiv}.

\subsection{Flavor Basis: Inner Product Decomposition}
\label{sec:flavor}
The above two subsections describe the systematic ways to generate the Lorentz structures and the group factors as irreps of $\bar{S}$. Now we are at the stage to show how to use these ingredients to construct operators with certain flavor permutation symmetry. Still, we shall take the $Q^3L W_{\rm L}$ as an example to demonstrate the procedure of the inner product decomposition of a single symmetric group $S_3$, the generalization to arbitrary sets of repeated fields will be manifest.

We use the projection operator defined in Theorem-4.2 in ref.~\cite{tung1985group} to obtain the generalized CGCs $C_{(\lambda,x),j}^{(\lambda_{1},x_1),(\lambda_{2},x_2),(\lambda_{3},x_3)}$ of the symmetric group with the definition:
\begin{eqnarray}\label{eq:master}
	\Theta_{(\lambda,x),j} = \sum_{x_1,x_2,x_3} C_{(\lambda,x),j}^{(\lambda_{1},x_1),(\lambda_{2},x_2),(\lambda_{3},x_3)}\mc{M}^{\lambda_{1}}_{x_1}\otimes T^{\lambda_2}_{{\rm SU3},x_2}\otimes T^{\lambda_3}_{{\rm SU2},x_3}
\end{eqnarray}
where $\Theta_{(\lambda,x),j}$ is the $x$th basis vector in the $j$th (label of multiplicity) irrep $\lambda$ from the decomposition, which is essentially a linear combination of various factorizable terms defined in eq.~\eqref{eq:term_format}. 
The details of using projection operator to extract CGCs are given in the appendix~\ref{sec:projection}, here we directly provide the relevant CGCs of $S_3$ for our example $Q^3L W_{\rm L}$. 
As we have obtained in the above two sections, the permutation symmetries of the Lorentz structure can be $[1^3]$ or $[2,1]$, those of the $SU(2)_W$ group factor can be $[3]$ or $[2,1]$, while the $SU(3)_C$ group factor only takes $[1^3]$. 
Therefore there are four possibilities to form direct product representations, of which the inner product decompositions are:
\begin{eqnarray}\arraycolsep=1.4pt\def\arraystretch{2.2}
	\begin{array}{ccccccc}
		SU(3)_c& &SU(2)_w & &{\rm Lorentz}& &{\rm Flavor} \\
		\yng(1,1,1)&\odot &\yng(3)&\odot & \yng(1,1,1) &=& 1\times\yng(3) \\
		\yng(1,1,1)&\odot &\yng(3)&\odot &  \yng(2,1)  &=& 1\times\yng(2,1) \\
		\yng(1,1,1)&\odot &\yng(2,1)&\odot &  \yng(1,1,1)  &=& 1\times\yng(2,1) \\
		\yng(1,1,1)&\odot &\yng(2,1)&\odot &  \yng(2,1)  &=& 1\times\yng(3)\oplus 1\times\yng(2,1)\oplus 1\times\yng(1,1,1) .\\
	\end{array} 
	\label{eq:decompose}
\end{eqnarray}
One can observe that the first three combinations of the permutation symmetries are trivial as the decomposition only results in a single irreps of $S_3$, so we only show the detail for the last one in the following.
The relevant CGCs for the last decomposition are summarized in table~\ref{tab:decomp}. 
\begin{table}[ht]
	\centering
	\begin{tabular}{ |c| c| } 
		\hline
		Flavor Sym & Relevant CGCs  \\ 
		\hline
		\multirow{4}{2em}{\centering \yng(3)} & $C_{([3],1,1)}^{([2,1],1),([1^3],1),([2,1],1)}=\frac{2}{3}$ \\ 
		& $C_{([3],1),1}^{([2,1],1),([1^3],1),([2,1],2)}=\frac{1}{3}$ \\  
		& $C_{([3],1),1}^{([2,1],2),([1^3],1),([2,1],1)}=\frac{1}{3}$\\ 
		&$C_{([3],1),1}^{([2,1],2),([1^3],1),([2,1],2)}=\frac{2}{3}$ \\
		\hline
		\multirow{4}{2em}{\centering\yng(2,1)} & $C_{([2,1],1)}^{([2,1],1),([1^3],1),([2,1],1)}=\frac{1}{3}$ \\ 
		& $C_{([2,1],1),1}^{([2,1],1),([1^3],1),([2,1],2)}=\frac{1}{3}$ \\  
		& $C_{([2,1],1),1}^{([2,1],2),([1^3],1),([2,1],1)}=\frac{1}{3}$\\ 
		&$C_{([2,1],1),1}^{([2,1],2),([1^3],1),([2,1],2)}=0$  \\ 
		\hline 
		\multirow{4}{2em}{\centering\yng(1,1,1)} & $C_{([1^3],1)}^{([2,1],1),([1^3],1),([2,1],1)}=0$ \\
		& $C_{([1^3],1),1}^{([2,1],1),([1^3],1),([2,1],2)}=\frac{1}{2}$ \\ 
		& $C_{([1^3],1),1}^{([2,1],2),([1^3],1),([2,1],1)}=-\frac{1}{2}$\\ 
		&$C_{([1^3],1),1}^{([2,1],2),([1^3],1),([2,1],2)}=0$  \\
		\hline
	\end{tabular}
	\caption{The relevant CGCs of $S_3$ inner product decomposition}
	\label{tab:decomp}
\end{table}
Since in our case, the multiplicities of each irreps is 1, the last indices of the subscripts of $C$ are all 1.  
Also as discussed in ref.~\cite{Fonseca:2019yya}, for irreps with dimension larger than 1, we only need to choose one of the basis vectors from the decomposed invariant space as others generate the same flavor space. Here we always select the first vector in our basis, this is why the third subscript indices of $C$ is always 1, in principle it is equivalent to select any one of the basis vector, the reason we choose the first one in our convention is that it is equal to the Young symmetrizer of the normal Young tableau of the corresponding YD discussed in section~\ref{sec:groupalgebra}, which helps us simplify our forms of operators in section~\ref{sec:list}. We shall come back to this point later in section~\ref{sec:prev}. 

Therefore we obtain three \textbf{terms} from the last line of eq.~\eqref{eq:decompose}:
\begin{eqnarray}
	\Theta^{prst}_{([3],1),1} &=& \frac{2}{3}T^{[1^3]}_{{\rm SU3},1} \left(T^{[2,1]}_{{\rm SU2},1} {\cal M}^{[2,1]}_1+ T^{[2,1]}_{{\rm SU2},2} {\cal M}^{[2,1]}_2\right)+\frac{1}{3}T^{[1^3]}_{{\rm SU3},1} \left( T^{[2,1]}_{{\rm SU2},1} {\cal M}^{[2,1]}_2+ T^{[2,1]}_{{\rm SU2},2} {\cal M}^{[2,1]}_1\right)\\
	\Theta^{prst}_{([2,1],1),1} &=& \frac{1}{3}T^{[1^3]}_{{\rm SU3},1} \left( T^{[2,1]}_{{\rm SU2},1} {\cal M}^{[2,1]}_1+ T^{[2,1]}_{{\rm SU2},2} {\cal M}^{[2,1]}_2+ T^{[2,1]}_{{\rm SU2},2} {\cal M}^{[2,1]}_1\right)\\
	\Theta^{prst}_{([1^3],1),1} &=& \frac{1}{2}T^{[1^3]}_{{\rm SU3},1} \left( T^{[2,1]}_{{\rm SU2},2} {\cal M}^{[2,1]}_1- T^{[2,1]}_{{\rm SU2},1} {\cal M}^{[2,1]}_2\right),
\end{eqnarray}
where $r,s,t$ and $p$ are the flavor indices of $Q$'s and $L$ respectively. As each factor is rather lengthy, we only show the full expression of $\Theta^{prst}_{([1^3],1),1}$ here:
\begin{eqnarray}
	\Theta^{prst}_{([1^3],1),1}&=& \frac{i}{12}\epsilon^{abc}\left(\tau^I\right)_{m_1}^{i}W_{{\rm L}\mu\nu}^I \left\{(2\epsilon^{jk}\epsilon^{lm_1}-\epsilon^{jl}\epsilon^{km_1})[(L_{pi}\sigma^{\mu\nu}Q_{sbk})(Q_{raj}Q_{tcl})-(L_{pi}\sigma^{\mu\nu}Q_{raj})(Q_{skb}Q_{tcl})]\nonumber\right. \\
	&& \left.-(2\epsilon^{jl}\epsilon^{km_1}-\epsilon^{jk}\epsilon^{lm_1})[(L_{pi}\sigma^{\mu\nu}Q_{sbk})(Q_{raj}Q_{tcl})+2(L_{pi}\sigma^{\mu\nu}Q_{raj})(Q_{skb}Q_{tcl})]\right\}\nonumber \\
	&=& -\frac{i}{4}\epsilon^{abc}\left(\tau^I\right)_{m_1}^{i}W_{{\rm L}\mu\nu}^I \left[\epsilon^{jm_1}\epsilon^{kl}(L_{pi}\sigma^{\mu\nu}Q_{sbk})(Q_{raj}Q_{tcl})+\epsilon^{jl}\epsilon^{km_1}(L_{pi}\sigma^{\mu\nu}Q_{raj})(Q_{skb}Q_{tcl})\right],
	\label{eq:Theta111}
\end{eqnarray}
where the Schouten identity has been used in the last line. One can verify that $\Theta^{prst}_{([1^3],1),1}$ is indeed totally antisymmetric about indices $r,s,t$ as it should be.

\begin{figure}
	\center{\includegraphics[width=0.7\textwidth]{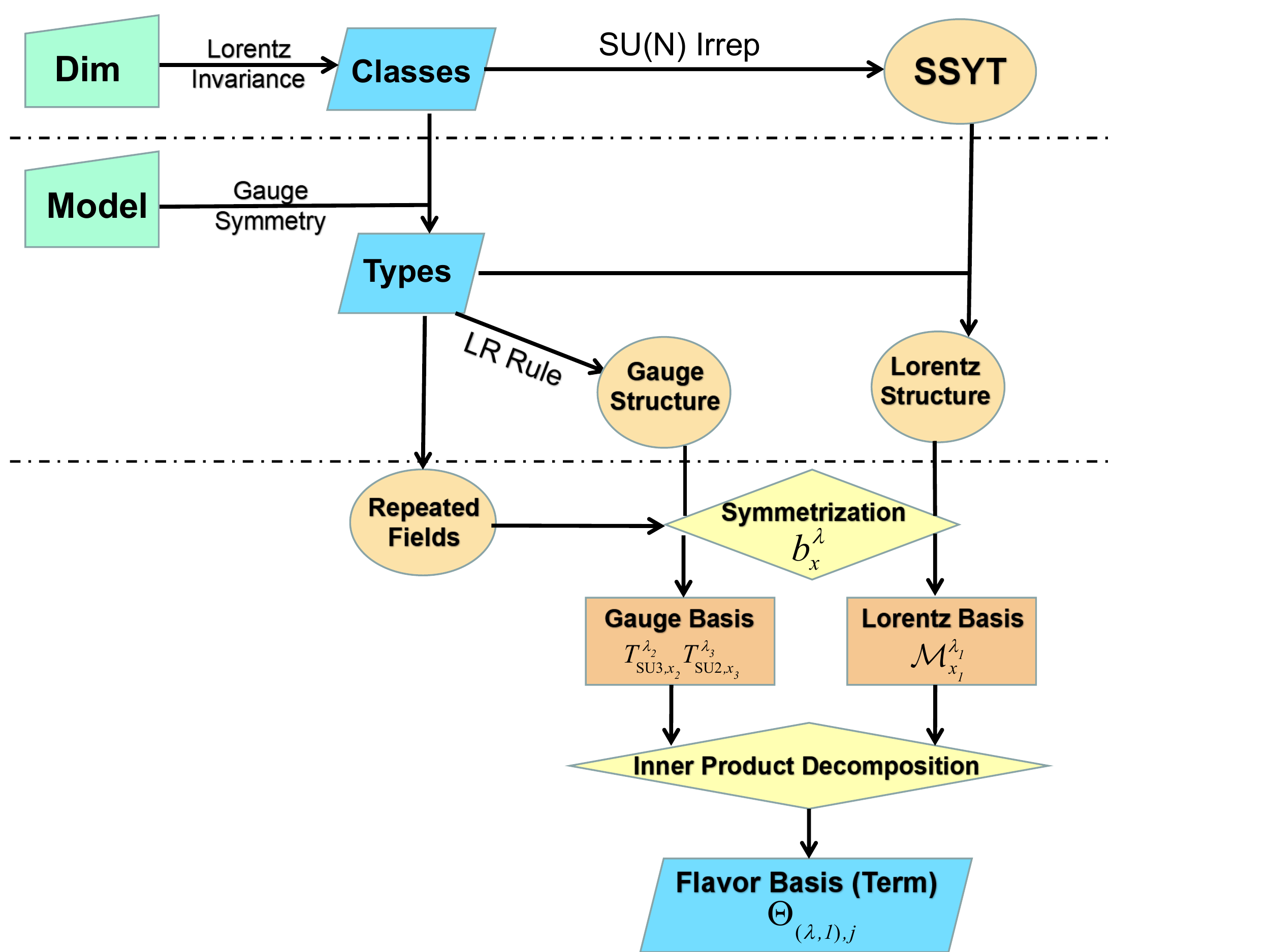}}
	\caption{\label{fig:fchart}Flow chart for finding all the independent terms at a given dimension. The content above the first dash-dotted line is model-independent and can be applied to any EFT. The content below the second dash-dotted line is our main contribution to this work. We automatize the whole procedure in a \textsf{Mathematica} code.}
\end{figure}

So far, we have demonstrated the whole process to obtain a term with a concrete example $Q^3L W_{\rm L}$. We summarize our algorithm to find a complete set of independent terms for a given dimension in a flow chart in figure~\ref{fig:fchart} and realize automated treatment in a \textsf{Mathematica} code.

Given a dimension, one can enumerate the operator classes that determine the number of fields of
each spin and the number of the derivative. Further, by finding the corresponding SSYTs, one can obtain
the Lorentz structure candidates without EOM and IBP redundancy. All of these above the first
dash-dotted line in the figure are model-independent, which can be applied to any Lorentz invariant EFTs.

After specifying the UV model, one can determine the types of operators for each class, and indeed determine the independent Lorentz and gauge structures ${\cal M}_\xi$'s and $T_{\xi}$'s. Afterward, taking into account the information of repeated fields from the specific type, one can symmetrize the  ${\cal M}_\xi$'s and $T_{\xi}$'s to obtain a set of Lorentz and gauge group basis that transform as irreps of $\bar{S}$. Finally, by putting these ingredients together to form the Lorentz and gauge singlets that transform as direct product representations of $\bar{S}$, and using inner product decomposition to decompose them back into the irreps of $\bar{S}$, one obtains several irrep spaces, each corresponding to an independent term with a definite flavor permutation symmetry. The symmetrization and the inner product decomposition below the second dashed-dotted line are our unique contributions that are not present in the literature yet.

\section{Lists of Operator Basis}
\label{sec:list}

\subsection{Preview of the Result}
\label{sec:prev}
In this section, we list all the dimension 8 terms of operators, grouped by the classes of Lorentz structures. 
In Table.~\ref{tab:stat} we show the statistics of all the SMEFT dimension 8 results organized by subclasses, with links referring to the corresponding lists of operators in the following subsections. 
The subclasses with non-trivial polynomials of $n_f$, the fermion flavor number, as the total number of operators are those for which we need to take care of the repeated field issues.
The statistics for B violation ($\Delta B=\pm1$) are listed with underlines, while the lepton number violation is not shown because $B-L$ is conserved at dimension 8. 

\begin{table}[htbp]
	\centering	
	\begin{align}{\small
			\begin{array}{cc|c|c|c|c|c}
				\hline
				N & (n,\tilde{n}) & \text{Subclasses} & \mathcal{N}_{\text{type}} & \mathcal{N}_{\text{term}} & \mathcal{N}_{\text{operator}} & \text{Equations}\\
				\hline\hline
				4 & (4,0) & F^4_{\rm L}+h.c. &  14 & 26 &26& (\ref{cl:F4})\\
				\cline{2-7}
				& (3,1) & F^2_{\rm L}\psi\psi^{\dagger}D+h.c. & 22 & 22 &22n^2_f& (\ref{cl:F2ppdc})\\
				& & \psi^4D^2+h.c. & 4\underline{+4} & 18\underline{+14} &12n^4_f\underline{+n^3_f(5n_f-1)}&(\ref{cl:q4d2c},\ref{cl:q3l1d2},\ref{cl:q2l2d2c}) \\
				& & F_{\rm L}\psi^2\phi D^2+h.c. & 16 & 32 &32n^2_f& (\ref{cl:Fpphd2})\\
				& & F^2_{\rm L}\phi^2D^2+h.c. & 8 & 12 &12& (\ref{cl:F2h2d2})\\
				\cline{2-7}
				& (2,2) & F^2_{\rm L}F^2_{\rm R} & 14 & 17 &17& (\ref{cl:F4})\\
				& & F_{\rm L}F_{\rm R}\psi\psi^{\dagger}D & 27 & 35 &35n^2_f& (\ref{cl:F2ppdr},\ref{cl:F2ppdc})\\
				& & \psi^2\psi^{\dagger 2}D^2 & 17\underline{+4} & 54\underline{+8} & \frac12n^2_f(75n^2_f+11)\underline{+6n^4_f} &(\ref{cl:q4d2r},\ref{cl:q2l2d2r}\text{-}\ref{cl:l4d2r}) \\
				& & F_{\rm R}\psi^2\phi D^2+h.c. & 16 & 16 &16n^2_f& (\ref{cl:Fpphd2})\\
				& & F_{\rm L}F_{\rm R}\phi^2D^2 & 5 & 6 &6& (\ref{cl:F2h2d2})\\
				& & \psi\psi^{\dagger}\phi^2D^3 & 7 & 16 &16n^2_f& (\ref{cl:pph2d3r},\ref{cl:pph2d3c})\\
				& & \phi^4D^4 & 1 & 3 &3& (\ref{cl:h})\\
				\hline
				5 & (3,0) & F_{\rm L}\psi^4+h.c. & 12\underline{+10} & 66\underline{+54} &42n^4_f\underline{+2n^3_f(9n_f+1)} &(\ref{cl:Fq4c},\ref{cl:Fq3l11},\ref{cl:Fq3l12},\ref{cl:Fq2l2c}) \\
				& & F^2_{\rm L}\psi^2\phi+h.c. & 32 & 60 &60n^2_f& (\ref{cl:F2pph1},\ref{cl:F2pph2})\\
				& & F^3_{\rm L}\phi^2+h.c. & 6 & 6 &6& (\ref{cl:F3h})\\
				\cline{2-7}
				& (2,1) & F_{\rm L}\psi^2\psi^{\dagger 2}+h.c. & 84\underline{+24} & 172\underline{+32} &2n^2_f(59n^2_f-2)\underline{+24n^4_f}&(\ref{cl:Fq4r1}\text{-}\ref{cl:Fq4r2}),(\ref{cl:Fq3l11}\text{-}
				\ref{cl:Fl4}) \\
				& & F^2_{\rm R}\psi^2\phi+h.c. & 32 & 36 &36n^2_f& (\ref{cl:F2pph1},\ref{cl:F2pph2})\\
				& & \psi^3\psi^{\dagger}\phi D+h.c. & 32\underline{+14} & 180\underline{+56} & n^3_f(135n_f-1)\underline{+n^3_f(29n_f+3)} &(\ref{cl:q4hd},\ref{cl:q3l1hd}\text{-}\ref{cl:l4hd})\\
				& & F_{\rm L}\psi\psi^{\dagger}\phi^2D+h.c. & 38 & 92 &92n^2_f& (\ref{cl:Fpph2d1},\ref{cl:Fpph2d2})\\
				& & \psi^2\phi^3D^2+h.c. & 6 & 36 &36n^2_f& (\ref{cl:pph3d2})\\
				& & F_{\rm L}\phi^4D^2+h.c. & 4 & 6 &6& (\ref{cl:Fh4d2})\\
				\hline
				6 & (2,0) & \psi^4\phi^2+h.c. & 12\underline{+4} & 48\underline{+18} & 5(5n^4_f+n_f^2)\underline{+\frac23(8n^4_f+n^2_f)}&(\ref{cl:q4h2c},\ref{cl:q3l1h2},\ref{cl:q2l2h2c},\ref{cl:l4h2c})\\
				& & F_{\rm L}\psi^2\phi^3+h.c. & 16 & 22 &22n^2_f& (\ref{cl:Fpp3})\\
				& & F^2_{\rm L}\phi^4+h.c. & 8 & 10 &10& (\ref{cl:F2h})\\
				\cline{2-7}
				& (1,1) & \psi^2\psi^{\dagger 2}\phi^2 & 23\underline{+10} & 57\underline{+14} & n^2_f(42n^2_f+n_f+2)\underline{+3n^3_f(3n_f-1)} &(\ref{cl:q4h2r},\ref{cl:q4h2c},\ref{cl:q3l1h2}\text{-}\ref{cl:l4h2r})\\
				& & \psi\psi^{\dagger}\phi^4D & 7 & 13 & 13n_f^2 & (\ref{cl:pph4dr},\ref{cl:pph4dc})\\
				& & \phi^6D^2 & 1 & 2 &2& (\ref{cl:h})\\
				\hline
				7 & (1,0) & \psi^2\phi^5+h.c. & 6 & 6 & 6n_f^2 & (\ref{cl:pph5})\\
				\hline
				8 & (0,0) & \phi^8 & 1 & 1 &1& (\ref{cl:h})\\
				\hline\hline
				\multicolumn{2}{c|}{\text{Total}} & 48 & 471\underline{+70} & 1070\underline{+196} & 993(n_f=1),\quad 44807(n_f=3)\\
				\hline
			\end{array}\notag}
	\end{align}
	\caption{A complete statistics of dimension 8 operators in the SMEFT, while the numbers with underlines are for the B-violating operators. $N$ in the leftmost column shows the number of particles. $(n,\tilde{n})$ are the numbers of $\epsilon$ and $\tilde\epsilon$ in the Lorentz structure, which determines the primary YD $[\mc{A}]$ the subclasses belong to. Note that our definition of ``term'' is different from the other literature, and the numbers are larger than those in, for instance, \cite{Fonseca:2019yya} because they did an extra step of merging before the counting. However, the number of operators is exactly the same as in \cite{Henning:2015alf,Fonseca:2019yya}. The links in the rightmost column refer to the list(s) of the terms in given subclasses. $\mc{N}_{\text{type}},\mc{N}_{\text{term}}$, and $\mc{N}_{\text{operator}}$ show the number of types, terms and Hermitian operators respectively (independent conjugates are counted). }\label{tab:stat}
\end{table}

For readers' convenience, we further perform several notation changes and simplifications on the basis of the terms directly produced by our algorithm:
\begin{itemize}
	\item In the Lorentz structures, we convert the derivatives and the gauge bosons to the form with Lorentz indices $\mu,\nu,\rho,\dots$ This is done by grouping the spinor contractions into chains that start and end at fermions, and traces that start and end at the same $F$ or $D$. On one hand, we reduce the $\sigma$ products in the chains to the three basic bilinear forms $\psi\chi$, $\psi\sigma^\mu\chi^\dagger$ and $\psi\sigma^{\mu\nu}\chi$ and their conjugates, where all spinor indices are suppressed and $\psi,\chi$ are both left-handed Weyl spinors as in our convention for the fermion fields. On the other hand, all the traces are reduced to products of $g^{\mu\nu}$, $\sigma^{\mu\nu}$ and $\epsilon^{\mu\nu\rho\eta}$. Relevant formulas are listed in appendix~\ref{app:a1}.
	
	\item We are using two-component spinors for all the fermion fields as they are the most natural way to deal with chiral fermions. Conversion rules to four-component spinor notation are provided in appendix~\ref{app:a1}. Due to the way we deal with the Fierz identity eq.\eqref{eq:fierz_l}, the Lorentz structures we exhibit do not contain any vector, axial or tensor couplings for four-fermion interactions. Readers could use Fierz identities also presented in the appendix~\ref{app:a1} to convert the operators to any forms they like. Examples are also provided beside the lists in section~\ref{sec:example3}.
	
	\item We also convert the chiral basis of gauge bosons $F_{\rm L/R}$ to the Hermitian fields $F,\tilde{F}$ by using formula in the appendix~\ref{app:a2}. After this is done, some of the types, even from different subclasses, merge into one. 
	
	\item The following common notations are adopted to reduce some of our terms
	\eq{
		& X D^{\mu}Y - D^{\mu}X Y \equiv X\overleftrightarrow{D}^{\mu}Y, \qquad  D_\mu D^{\mu} \equiv \Box, \\
		& H^{\dagger i}H_i \equiv (H^\dagger H), \qquad H^{\dagger i} (\tau^I)_i^j H_j \equiv (H^\dagger \tau^I H), \\
		& F_{1\mu\nu}F_2^{\mu\nu} \equiv (F_1F_2) \stackrel{F_1=F_2=F}{\equiv}F^2, 
	}
	
	\item The most subtle simplification is trying to superficially reduce the length of ``terms" \footnote{In this paragraph, \emph{term} without quote only indicates a monomial in a polynomial expression rather than a level of operators in our construction.} in order to present them in the paper better. Take eq.~\eqref{eq:Theta111} as an example, where two terms exist after expansion. The two terms together guarantee the total antisymmetry of the $Q$ flavors $r,s,t$ in $\Theta^{prst}_{([1^3],1),1}$. It is fair to guess that by performing total antisymmetrization on one of the terms over $r,s,t$ should reproduce $\Theta^{prst}_{([1^3],1),1}$, such as
	\begin{eqnarray}\label{eq:desym}
		\Theta^{prst}_{([1^3],1),1} \sim \mathcal{Y}\left[\tiny{\young(r,s,t)}\right]i  \epsilon ^{abc} \epsilon ^{jl} \epsilon ^{km} \left(\tau ^I\right)_m^i W_{\rm{L}}^{I}{}_{\mu }{}_{\nu } \left(Q_{ra j} Q_{tc l}\right) \left(L_{pi} \sigma^{\mu }{}^{\nu } Q_{sb k}\right)
	\end{eqnarray}
	where the Young symmetrizer specified by the Young tableaux $\tiny\young(r,s,t)$, equivalent to the projector $b^{[1^3]}_1$ as explained in section~\ref{sec:groupalgebra}, acts on the flavor indices $r,s,t$, so that the permutation symmetry over these indices are guaranteed by the property of the projector eq.~\eqref{eq:bandD} to eq.~\eqref{eq:piF}. An example of non-trivial symmetrizer for the mixed symmetry is
	\begin{eqnarray}
		\mathcal{Y}\left[\tiny{\young(rs,t)}\right] \Theta^{...,rst,...} = \Theta^{...,rst,...}+\Theta^{...,srt,...}-\Theta^{...,trs,...}-\Theta^{...,tsr,...},
	\end{eqnarray}
	where $...$ represents the possible presence of flavor indices of other sets of repeated fields. For ``terms" involving more than one set of repeated fields, the symmetrizer is specified by several Young tableaux, for example
	\begin{align}
		\mathcal{Y}\left[\tiny{\young(pr)},\tiny{\young(st)}\right] \left(L_{pi} L_{rj}\right) \left(L^{\dagger}{}_{s}^{i} L^{\dagger}{}_{t}^{k}\right)  H^{\dagger}{}^{j} H_{k} &= \left(L_{pi} L_{rj} + L_{pj} L_{ri}\right) \left(L^{\dagger}{}_{s}^{i} L^{\dagger}{}_{t}^{k} + L^{\dagger}{}_{s}^{k} L^{\dagger}{}_{t}^{i}\right)  H^{\dagger}{}^{j} H_{k}.
	\end{align}
	
	Back to the example of eq.~\eqref{eq:Theta111}, where only one $[1^3]$ operator exists for the type $W_{\rm L}Q^3L$, there is no doubt that eq.~\eqref{eq:desym} can reproduce it up to an overall constant factor. In this way, we reduce the length of our ``terms" with the definite flavor symmetry and hide the complexity into the corresponding Young symmetrizer leaving rather simple forms exhibited in the following sections. 
	The case becomes more complicated when a certain type containing more than one copys of irrep of the same flavor symmetry $\lambda$. In this cases a dedicated ``de-symmetrization" procedure~\cite{Li:2020xlh} is performed to find out the monomials such that after acting on the corresponding Young symmetrizer they are symmetrized to independent terms. 
	Moreover, if several irreducible flavor tensors can be written as different Young symmetrizers acting on the same term, these tensors can be merged into one tensor with reducible flavor symmetry, which is how ``term'' was used in \cite{Fonseca:2019yya}. The $Q^3L$ operator at dimension 6 is one of such examples, which is why there is only one term for it in the Warsaw basis \cite{Grzadkowski:2010es}. 
	However, the principle for the merging does not exist so far, the number of such ``term" is an ambiguous quantity as discussed in \cite{Fonseca:2019yya}. 
	We emphasize that our ``term'' defined in section~\ref{sec:inv} does not have such ambiguity, and we prefer not to do the merging but instead shorten our notation with the trick of the Young symmetrizer mentioned above.

	\item Finally, instead of listing subclasses sorted by the tuple $(N,n,\tilde{n})$, we list chirality-blind classes sorted by the number of fermions to fit the needs of phenomenologists. Within a class, operators are listed as either ``complex'' types or ``real'' types. We refer to a type of operators whose conjugate are of a different type as a ``complex'' type, and a self-conjugate type as a ``real'' type. Since we do not present conjugates of the ``complex'' types, operators of these types should be counted twice in the sense of Hermitian degrees of freedom. For the ``real'' types, although the operators presented may not be Hermitian on their own, their conjugates must be combinations of operators in the same type and should not be counted separately, so these operators are only counted as one Hermitian degree of freedom. The numbers presented in the Table.~\ref{tab:stat} are all counted in this manner. We have also listed the B violating operators separately in section~\ref{sec:example3}.
	
\end{itemize}

\subsection{Classes involving Bosons only}
\label{sec:example1}
In the following sections, we list our operators in terms of subclasses, ordered by the number of fermions and gauge bosons. The subclasses shown here are summarized in Table~\ref{eq:classes8}, while those not showing up are redundant according to our treatments of various redundancy relations listed at the beginning of section~\ref{sec:lorentz}.

\underline{Class $\phi^{8-n_D}D^{n_D}$}: Operators with only scalars. The subclasses available in Table~\ref{eq:classes8} are $n_D = 0,2,4$. 
The Lorentz structure of the all-scalar subclass $\phi^8$ is trivial, shown as
\begin{align}
	\phi _1 \phi _2 \phi _3 \phi _4 \phi _5 \phi _6 \phi _7 \phi _8.
\end{align}
For the subclass $\phi^6D^2$, all the Lorentz structures are given by the algorithm in section~\ref{sec:lorentz_str} as follows
\eq{
	&\phi _1 \left(D \phi _2\right)^{\alpha}_{\dot\alpha} \phi _3 \phi _5 \left(D \phi _4\right)_{\alpha}^{\dot\alpha} \phi _6,\quad 
	\phi _1 \left(D \phi _2\right)^{\alpha}_{\dot\alpha} \phi _3 \phi _4 \left(D \phi _5\right)_{\alpha}^{\dot\alpha} \phi _6,\quad 
	\phi _1 \left(D \phi _2\right)^{\alpha}_{\dot\alpha} \phi _3 \phi _4 \phi _5 \left(D \phi _6\right)_{\alpha}^{\dot\alpha},\\
	&\phi _1 \phi _2 \left(D \phi _3\right)^{\alpha}_{\dot\alpha} \left(D \phi _4\right)_{\alpha}^{\dot\alpha} \phi _5 \phi _6,\quad 
	\phi _1 \phi _2 \left(D \phi _3\right)^{\alpha}_{\dot\alpha} \phi _4 \left(D \phi _5\right)_{\alpha}^{\dot\alpha} \phi _6,\quad 
	\phi _1 \phi _2 \left(D \phi _3\right)^{\alpha}_{\dot\alpha} \phi _4 \phi _5 \left(D \phi _6\right)_{\alpha}^{\dot\alpha},\\
	&\phi _1 \phi _2 \phi _3 \left(D \phi _4\right)^{\alpha}_{\dot\alpha} \left(D \phi _5\right)_{\alpha}^{\dot\alpha} \phi _6,\quad 
	\phi _1 \phi _2 \phi _3 \left(D \phi _4\right)^{\alpha}_{\dot\alpha} \phi _5 \left(D \phi _6\right)_{\alpha}^{\dot\alpha},\quad 
	\phi _1 \phi _2 \phi _3 \phi _4 \left(D \phi _5\right)^{\alpha}_{\dot\alpha} \left(D \phi _6\right)_{\alpha}^{\dot\alpha}.
}
while those for the subclass $\phi^4D^4$ are given as
\begin{align}
	\phi _1 \left(D^2 \phi _2\right)_{\dot{\alpha }\dot{\beta }}^{\alpha \beta } \phi _3 \left(D^2 \phi _4\right)_{\alpha \beta }^{\dot{\alpha }\dot{\beta }},\quad 
	\phi _1 \left(D \phi _2\right)^{\alpha}_{\dot{\alpha }} \left(D \phi _3\right)^{\beta}_{\dot{\beta }} \left(D^2 \phi _4\right)_{\alpha \beta}^{\dot{\alpha }\dot{\beta }},\quad 
	\phi _1 \phi _2 \left(D^2 \phi _3\right)_{\dot{\alpha }\dot{\beta }}^{\alpha \beta } \left(D^2 \phi _4\right)_{\alpha \beta }^{\dot{\alpha }\dot{\beta }} .
\end{align}

Plugging fields from Table.~\ref{tab:SMEFT-field-content} to these Lorentz structures, making sure that the total hypercharge is zero, we get the following three types of operators, and by going through our algorithm, we obtain 6 terms in all as a non-redundant basis:
\begin{align}\begin{array}{c|l}
		
		\mathcal{O}_{H{}^4 H^{\dagger} {}^4   }
		&\left(H^{\dagger}H\right)^4 \vspace{2ex}\\
		
		\mathcal{O}_{H{}^3 H^{\dagger} {}^3 D^2 }^{(1,2)}
		
		& \left(H^{\dagger}H\right)^2 \Box \left(H^{\dagger}H\right),\quad \left(H^{\dagger}H\right)\left|H^{\dagger}D_{\mu}H\right|^2 \vspace{2ex}\\
		\mathcal{O}_{H{}^2 H^{\dagger} {}^2 D^4}^{(1\sim3)}
		
		& \left(H^{\dagger}H\right)\Box^2\left(H^{\dagger}H\right),\quad \left|H^{\dagger}D_{\mu}D_{\nu}H\right|^2,\quad \left(H^{\dagger}D_{\mu}H\right)^*\Box\left(H^{\dagger}D^{\mu}H\right).
		
	\end{array}\label{cl:h}\end{align}
The superscripts of the $\mc{O}$'s label the terms in the particular type, in the order of left to right and up to bottom.
The first operator modifies the shape of the Higgs potential, and the rest could renormalize the Higgs field and thus modify the Higgs couplings uniformly.

\underline{Class $F\phi^{6-n_D}D^{n_D}$}: Operators with one gauge boson and arbitrary scalars. According to Table~\ref{eq:classes8}, only one subclass $F_{\rm L}\phi^4D^2$ survives our criteria, which contains the following 3 independent Lorentz structures:
\begin{align}
	&F_{\rm{L}1}{}^{\alpha\beta } \phi _2 \left(D \phi _3\right){}_{\alpha }{}_{\dot{\alpha }} \left(D \phi _4\right){}_{\beta }^{\dot{\alpha }} \phi _5,\quad 
	F_{\rm{L}1}{}^{\alpha\beta } \phi _2 \left(D \phi _3\right)_{\alpha\dot{\alpha }} \phi _4 \left(D \phi _5\right)_{\beta }^{\dot{\alpha }},\quad 
	F_{\rm{L}1}{}^{\alpha\beta } \phi _2 \phi _3 \left(D \phi _4\right){}_{\alpha\dot{\alpha }} \left(D \phi _5\right)_{\beta }^{\dot{\alpha }} .
\end{align}
Together with their Hermitian conjugates, they combine into the form with $F,\tilde{F}$, that become real in this notation.
In the SMEFT, we have the following real types:
\begin{align}\begin{array}{c|l}
		\multirow{2}*{$\mathcal{O}_{WH{}^2 H^{\dagger} {}^2D^2}^{(1\sim4)}$}   &W^{I}{}^{\mu }{}^{\nu }\left(H^{\dagger}H\right)\left(D_{\mu}H^{\dagger}\tau^I D_{\nu}H\right),\quad \tilde{W}^{I}{}^{\mu }{}^{\nu }\left(H^{\dagger}H\right) \left(D_{\mu}H^{\dagger}\tau ^ID_{\nu} H\right),\\
		& W^{I}{}^{\mu }{}^{\nu }\left(D_{\mu}H^{\dagger}D_{\nu}H\right)\left(H^{\dagger}\tau^I H\right),\quad \tilde{W}^{I}{}^{\mu }{}^{\nu }\left(D_{\mu}H^{\dagger}D_{\nu}H\right) \left(H^{\dagger}\tau ^I H\right), \vspace{2ex}\\ 
		\mathcal{O}_{BH{}^2 H^{\dagger} {}^2D^2}^{(1,2)} & B^{\mu }{}^{\nu }\left(H^{\dagger}H\right)\left(D_{\mu}H^{\dagger}D_{\nu}H\right)	,\quad \tilde{B}^{\mu }{}^{\nu }\left(H^{\dagger}H\right)\left(D_{\mu}H^{\dagger}D_{\nu}H\right)	.
	\end{array}\label{cl:Fh4d2}\end{align}

\underline{Class $F^2\phi^{4-n_D}D^{n_D}$}: Operators with two gauge bosons and arbitrary scalars. Table~\ref{eq:classes8} contains two subclasses of this form, with $n_D=0,2$.
The only Lorentz structure in the subclass $F^2\phi^4$ is
\begin{align}
	F_{\rm{L}1}{}^{\alpha\beta }F_{\rm{L}2}{}_{\alpha\beta }\phi _3 \phi _4 \phi _5 \phi _6  .
\end{align}
In the SMEFT we get the following types under this subclass:
\begin{align}\begin{array}{c|l}
		
		\mathcal{O}_{G^2 H{}^2 H^{\dagger} {}^2   }^{(1,2)}
		
		&G^2\left(H^{\dagger}H\right)^2,\quad (G^A_{\mu\nu}\tilde{G}^{A\mu\nu})\left(H^{\dagger}H\right)^2 ,\vspace{2ex}\\
		
		\multirow{2}*{$\mathcal{O}_{W^2H{}^2 H^{\dagger} {}^2}^{(1\sim4)}$}  
		
		&W^{I}{}_{\mu }{}_{\nu } W^{J}{}^{\mu }{}^{\nu }\left(H^{\dagger}\tau^IH\right)\left(H^{\dagger}\tau^JH\right),\quad W^2\left(H^{\dagger}H\right)^2,\\
		&W^{I}{}_{\mu }{}_{\nu } \tilde{W}^{J}{}^{\mu }{}^{\nu }\left(H^{\dagger}\tau^IH\right)\left(H^{\dagger}\tau^JH\right),\quad (W^I\tilde{W}^I)\left(H^{\dagger}H\right)^2 ,\vspace{2ex}\\
		
		\mathcal{O}_{B^2 H{}^2 H^{\dagger} {}^2   }^{(1,2)}
		
		& B^2\left(H^{\dagger}H\right)^2\vspace{2ex},\quad (B\tilde{B})\left(H^{\dagger}H\right)^2 ,\vspace{1ex}\\
		
		\mathcal{O}_{B W H{}^2 H^{\dagger} {}^2   }^{(1,2)}
		
		& B_{\mu }{}_{\nu } W^{I}{}^{\mu }{}^{\nu }\left(H^{\dagger}\tau^IH\right)\left(H^{\dagger}H\right),\quad B_{\mu }{}_{\nu } \tilde{W}^{I}{}^{\mu }{}^{\nu }\left(H^{\dagger}\tau^IH\right)\left(H^{\dagger}H\right).
		
	\end{array}\label{cl:F2h}\end{align}
When all the Higgs bosons are put to their vev, the operators normalize the kinetic terms of gauge bosons and thus modify the corresponding gauge couplings uniformly.

On the other hand, there are 3 independent Lorentz structures in the subclass $F^2\phi^2D^2$:
\begin{align}
	&F_{\rm{L}1}{}^{\alpha\beta } F_{\rm{L}2\alpha}{}^{\gamma } \left(D \phi _3\right)_{\beta\dot{\alpha }} \left(D \phi _4\right)_\gamma^{ \dot{\alpha }},\quad
	F_{\rm{L}1}{}^{\alpha\beta } F_{\rm{L}2}{}_{\alpha\beta } \left(D \phi _3\right)^\gamma_{ \dot{\alpha }} \left(D \phi _4\right)_\gamma^{ \dot{\alpha }},\quad F_{\rm{L}1}{}^{\alpha\beta } \phi _2 \left(D^2 \phi _3\right)_{\alpha\beta \dot{\alpha }\dot{\beta }} F_{\rm{R}4}^{\dot{\alpha }\dot{\beta }}\;.
\end{align}
Again, combined with their Hermitian conjugates, we obtain the following real types
\begin{align}\begin{array}{c|l}
		
		\mathcal{O}_{G^2 H H^{\dagger}  D^2 }^{(1\sim3)}
		
		& G^2 \left(D^{\mu } H^{\dagger}D_{\mu } H\right),\quad (G^A\tilde{G}^A) \left(D^{\mu } H^{\dagger}D_{\mu } H\right),\quad G^{A}{}^{\mu }{}{}_{\lambda } G^{A}{}^{\nu }{}^{\lambda } \left(D_{\mu } H^{\dagger}D_{\nu } H\right)\vspace{2ex}\\
		
		\multirow{3}*{$\mathcal{O}_{W^2 H H^{\dagger}  D^2}^{(1\sim6)}$} 
		
		&W^2\left(D^{\mu}H^{\dagger}D_{\mu}H\right),\quad i \epsilon ^{IJK} W^{I}{}^{\mu }{}{}_{\lambda } W^{J}{}^{\nu }{}^{\lambda } \left(D_{\mu } H^{\dagger}\tau^KD_{\nu } H\right),\\
		
		&(W^I\tilde{W}^I)\left(D^{\mu}H^{\dagger}D_{\mu}H\right),\quad i \epsilon ^{IJK} W^{I}{}^{[\mu }{}{}_{\lambda } \tilde{W}^{J}{}^{\nu] }{}^{\lambda } \left(D_{\mu } H^{\dagger}\tau^KD_{\nu } H\right),\\
		
		& W^{I}{}^{\mu }{}{}_{\lambda } W^{I}{}^{\nu }{}^{\lambda } \left(D_{\mu } H^{\dagger}D_{\nu } H\right),\quad i \epsilon ^{IJK} W^{I}{}^{(\mu }{}{}_{\lambda } \tilde{W}^{J}{}^{\nu) }{}^{\lambda } \left(D_{\mu } H^{\dagger}\tau^KD_{\nu } H\right)\vspace{2ex}\\
		
		\mathcal{O}_{B^2 H H^{\dagger}  D^2 }^{(1\sim3)}
		
		& B^2\left(D^{\mu } H^{\dagger}D_{\mu } H\right),\quad (B\tilde{B})\left(D^{\mu } H^{\dagger}D_{\mu } H\right),\quad B^{\mu }{}{}_{\lambda } B^{\nu }{}^{\lambda } \left(D_{\mu } H^{\dagger}D_{\nu } H\right)\vspace{2ex}\\
		
		\multirow{3}*{$\mathcal{O}_{B W H H^{\dagger}  D^2}^{(1\sim6)}$} 
		
		& (BW^I)\left(D^{\mu}H^{\dagger}\tau^ID_{\mu}H\right),\quad (B\tilde{W}^I)\left(D^{\mu}H^{\dagger}\tau^ID_{\mu}H\right),\\
		&iB^{[\mu}{}_{\lambda}W^{I\nu]\lambda}\left(D_{\nu}H^{\dagger}\tau^ID_{\mu}H\right),\quad iB^{[\mu}{}_{\lambda}\tilde{W}^{I\nu]\lambda}\left(D_{\nu}H^{\dagger}\tau^ID_{\mu}H\right),\\
		&B^{(\mu}{}_{\lambda}W^{I\nu)\lambda}\left(D_{\nu}H^{\dagger}\tau^ID_{\mu}H\right),\quad     B^{(\mu}{}_{\lambda}\tilde{W}^{I\nu)\lambda}\left(D_{\nu}H^{\dagger}\tau^ID_{\mu}H\right)
		
	\end{array}\label{cl:F2h2d2}\end{align}
where brackets for the indices are shorthand notations for (anti-)symmetrization $F_1^{[\mu}{}_{\lambda}F_2^{\nu]\lambda} \equiv F_1^{\mu}{}_{\lambda}F_2^{\nu\lambda}-F_1^{\nu}{}_{\lambda}F_2^{\mu\lambda}$ and $F_1^{(\mu}{}_{\lambda}F_2^{\nu)\lambda} \equiv F_1^{\mu}{}_{\lambda}F_2^{\nu\lambda}+F_1^{\nu}{}_{\lambda}F_2^{\mu\lambda}$.
The operators of these types contribute to the neutral triple gauge boson couplings, which do not appear at lower dimensions \cite{Degrande:2013kka}.

\underline{Class $F^3\phi^2$}: Operators with triple gauge bosons. 
Note that the operators of class $F^3 D^2$ are absent due to our treatment about EOM.
The only Lorentz structure in the subclass $F^3\phi^2$ is 
\begin{align}
	F_{\rm{L}1}{}^{\alpha\beta } F_{\rm{L}2}{}_{\alpha }{}^{\gamma }F_{\rm{L}3}{}_{\beta }{}_{\gamma } \phi _4 \phi _5 .
\end{align}
Note that the types $ B^3   H H^{\dagger}   ,\;  
B G^2   H H^{\dagger}   ,\;  
B^2 W   H H^{\dagger} $, and $
G^2 W   H H^{\dagger}  $ cannot exist, even though they are able to form Lorentz invariant gauge singlets. 
The reason is that the only Lorentz structure shown above is totally antisymmetric for the three gauge bosons. In case no antisymmetric structures from the gauge group sectors, like the structure constants, are available, the operators must vanish due to the commuting nature of any repeated gauge bosons in it. The non-vanishing types, which all involve totally antisymmetric structure constants, are shown below
\begin{align}\begin{array}{c|l}
		\mathcal{O}_{G^3HH^{\dagger}}^{(1,2)} & f^{ABC} G^{A}{}_{\mu }{}_{\nu } G^{B}{}^{\mu }{}{}_{\lambda } G^{C}{}^{\nu }{}^{\lambda } H^{\dagger}H,\quad f^{ABC} G^{A}{}_{\mu }{}_{\nu } G^{B}{}^{\mu }{}{}_{\lambda } \tilde{G}^{C}{}^{\nu }{}^{\lambda } H^{\dagger}H\vspace{1.5ex}\\
		\mathcal{O}_{W^3HH^{\dagger}}^{(1,2)} &\epsilon ^{IJK} W^{I}{}_{\mu }{}_{\nu } W^{J}{}^{\mu }{}{}_{\lambda } W^{K}{}^{\nu }{}^{\lambda }H^{\dagger}H,\quad \epsilon ^{IJK} W^{I}{}_{\mu }{}_{\nu } W^{J}{}^{\mu }{}{}_{\lambda } \tilde{W}^{K}{}^{\nu }{}^{\lambda }H^{\dagger}H\vspace{1.5ex}\\
		
		\mathcal{O}_{BW^2HH^{\dagger}}^{(1,2)} & \epsilon ^{IJK} B_{\mu }{}_{\nu } W^{I}{}^{\mu }{}{}_{\lambda } W^{J}{}^{\nu }{}^{\lambda } H^{\dagger}\tau^KH,\quad \epsilon ^{IJK} B_{\mu }{}_{\nu } W^{I}{}^{\mu }{}{}_{\lambda } \tilde{W}^{J}{}^{\nu }{}^{\lambda } H^{\dagger}\tau^KH
	\end{array}\label{cl:F3h}
\end{align}
These operators contribute to the anomalous triple gauge boson couplings. 

\underline{Class $F^4$}:  Operators with four gauge bosons. 
There is one Lorentz structure of subclass $F_{\rm L}^2F_{\rm R}^2$ and three Lorentz structures of subclass $F_{\rm L}^4$
\begin{align}
	&F_{\rm{L}1}{}{}^{\alpha\beta } F_{\rm{L}2}{}_{\alpha }{}_{\beta } F_{\rm{R}3}{}_{\dot\alpha\dot\beta}F_{\rm{R}4}{}^{\dot\alpha\dot\beta } ,\\
	&F_{\rm{L}1}{}{}^{\alpha\beta } F_{\rm{L}2}{}^{\gamma }{}{}^{\delta }F_{\rm{L}3}{}_{\alpha }{}_{\beta } F_{\rm{L}4}{}_{\gamma }{}_{\delta } ,\quad 
	F_{\rm{L}1}{}{}^{\alpha\beta } F_{\rm{L}2}{}_{\alpha }{}{}^{\gamma } F_{\rm{L}3}{}_{\beta }{}{}^{\delta }F_{\rm{L}4}{}_{\gamma }{}_{\delta } ,\quad
	F_{\rm{L}1}{}{}^{\alpha\beta } F_{\rm{L}2}{}_{\alpha }{}_{\beta } F_{\rm{L}3}{}{}^{\gamma }{}{}^{\delta }F_{\rm{L}4}{}_{\gamma }{}_{\delta } ,
\end{align}
After symmetrization described in section~\ref{sec:lorentz_perm}, we find no Lorentz structure that is antisymmetric over the gauge bosons, which implies that the type $B W^3 $ whose $SU(2)_W$ structure has to be totally antisymmetric must vanish. The non-vanishing types are given below
\begin{align}\begin{array}{c|l}
		\multirow{3}*{$\mathcal{O}_{G^4}^{(1\sim9)} $ }
		&(G^AG^B)(G^AG^B),\quad d^{ACE}d^{BDE}(G^AG^B)(G^CG^D),\quad f^{ACE}f^{BDE}(G^AG^B)(G^CG^D),\\
		&(G^AG^B)(G^A\tilde{G}^B),\quad d^{ACE}d^{BDE}(G^AG^B)(G^C\tilde{G}^D),\quad f^{ACE}f^{BDE}(G^AG^B)(G^C\tilde{G}^D),\\
		&(G^A\tilde{G}^B)(G^A\tilde{G}^B),\quad d^{ACE}d^{BDE}(G^A\tilde{G}^B)(G^C\tilde{G}^D),\quad f^{ACE}f^{BDE}(G^A\tilde{G}^B)(G^C\tilde{G}^D)\vspace{2ex}\\
		
		\multirow{2}*{$\mathcal{O}_{W^4}^{(1\sim6)} $ }
		&(W^IW^I)(W^JW^J),\quad (W^IW^I)(W^J\tilde{W}^J),\quad (W^I\tilde{W}^I)(W^J\tilde{W}^J),\\
		&(W^IW^J)(W^IW^J),\quad (W^IW^J)(W^I\tilde{W}^J),\quad (W^I\tilde{W}^J)(W^I\tilde{W}^J)\vspace{2ex}\\
		
		\mathcal{O}_{B^4}^{(1\sim3)}
		&(B^2)(B^2),\quad (B^2)(B\tilde{B}),\quad (B\tilde{B})(B\tilde{B})\vspace{2ex}\\
		
		\multirow{2}*{$\mathcal{O}_{G^2W^2}^{(1\sim7)}$}
		&G^2W^2,\quad G^2(W^I\tilde{W}^I),\quad (G^A\tilde{G}^A)W^2,\quad (G^A\tilde{G}^A)(W^I\tilde{W}^I),\\
		&(G^AW^I)(G^AW^I),\quad (G^AW^I)(G^A\tilde{W}^I),\quad (G^A\tilde{W}^I)(G^A\tilde{W}^I)\vspace{2ex}\\
		
		\multirow{2}*{$\mathcal{O}_{G^2B^2}^{(1\sim7)}$}
		&G^2B^2,\quad G^2(B\tilde{B}),\quad (G^A\tilde{G}^A)B^2,\quad (G^A\tilde{G}^A)(B\tilde{B}),\\
		&(G^AB)(G^AB),\quad (G^AB)(G^A\tilde{B}),\quad (G^A\tilde{B})(G^A\tilde{B})\vspace{2ex}\\
		
		\multirow{2}*{$\mathcal{O}_{W^2B^2}^{(1\sim7)}$}
		&B^2W^2,\quad W^2(B\tilde{B}),\quad (W^I\tilde{W}^I)B^2,\quad (W^I\tilde{W}^I)(B\tilde{B}),\\
		&(W^IB)(W^IB),\quad (W^IB)(W^I\tilde{B}),\quad (W^I\tilde{B})(W^I\tilde{B})\vspace{2ex}\\    
		
		\multirow{2}*{$\mathcal{O}_{BG^3}^{(1\sim4)}$}
		&d^{ABC}(BG^A)(G^BG^C),\quad d^{ABC}(BG^A)(G^B\tilde{G}^C),\\
		&d^{ABC}(B\tilde{G}^A)(G^BG^C),\quad d^{ABC}(B\tilde{G}^A)(G^B\tilde{G}^C)
	\end{array}\label{cl:F4}\end{align}

\subsection{Classes involving Two-fermions}
\label{sec:example2}
\subsubsection{No gauge boson involved}

In this subsection we deal with the classes $\psi^2 \phi^{5-n_D} D^{n_D}$. Note from eq.~\eqref{eq:classes8} that for odd $n_D$ we have fermions of opposite helicities, or chirality conserving, and for even $n_D$ we have them with the same helicities or chirality violating.

\noindent\underline{Class $\psi^2 \phi^5$}: The only Lorentz Structure of this subclass is
\begin{align} 
	\psi _1{}{}^{\alpha }\psi _2{}_{\alpha } \phi _3 \phi _4 \phi _5 \phi _6 \phi _7 .
\end{align}
In the SMEFT, these are Yukawa term with additional Higgses, which are all complex types:
\begin{align}\begin{array}{c|l}
		\mathcal{O}_{Q u_{_\mathbb{C}} H^3H^{\dagger 2} }  
		& \epsilon ^{il}\left(Q_{pa i} u_{_\mathbb{C}}{}_{r}^{a}\right) H_{l} \left(H^{\dagger}H\right)^2\vspace{1.5ex}\\
		\mathcal{O}_{Q d_{_\mathbb{C}} H^2H^{\dagger 3}}
		& \left(d_{_\mathbb{C}}{}_{p}^{a} Q_{ra i}\right)H^{\dagger}{}^{i} \left(H^{\dagger}H\right)^2\vspace{1.5ex}\\
		
		\mathcal{O}_{e_{_\mathbb{C}}L H^2H^{\dagger 3}}
		&\left(e_{_\mathbb{C} p} L_{ri}\right)H^{\dagger}{}^{i} \left(H^{\dagger}H\right)^2
	\end{array}\label{cl:pph5}
\end{align}
After taking the Higgs vev, they give rise to additional contributions to the SM fermion Yukawa couplings. 
According to the appendix~\ref{app:a1}, the relevant bilinear of two-component spinors can be converted to the four-component notation as
\bea\label{eq:pp}
\left(Q_{pa i}\Gamma u_{_\mathbb{C}}{}_{r}^{a} \right)= \left(\bar{u}^a_r \Gamma q_{pai}\right),\quad \left(d_{_\mathbb{C}}{}_{p}^{a}\Gamma Q_{ra i}\right)=\left(\bar{d}^a_p\Gamma q_{rai}\right),\quad \left(e_{_\mathbb{C} p}\Gamma L_{ri}\right)=\left(\bar{e}_p\Gamma l_{ri}\right), \qquad \Gamma=\mathbbm{1},\tau^I,\lambda^A,D^\mu.
\eea

\noindent\underline{Class $\psi^2 \phi^4 D$}: 
The subclass has to be $\psi\psi^\dagger\phi^4D$, which has the following Lorentz structures
\begin{align}
	\psi _1{}{}^{\alpha }\phi _2 \left(D \phi _3\right){}_{\alpha }{}_{\dot{\alpha }} \phi _4 \phi _5 \psi^{\dagger}_6{}{}^{\dot{\alpha }} ,\quad 
	\psi _1{}{}^{\alpha }\phi _2 \phi _3 \left(D \phi _4\right){}_{\alpha }{}_{\dot{\alpha }} \phi _5 \psi^{\dagger}_6{}{}^{\dot{\alpha }} ,\quad 
	\psi _1{}{}^{\alpha }\phi _2 \phi _3 \phi _4 \left(D \phi _5\right){}_{\alpha }{}_{\dot{\alpha }} \psi^{\dagger}_6{}{}^{\dot{\alpha }} 	
\end{align}
In the SMEFT, all but one of the types are real:
\begin{align}\begin{array}{c|l}
		
		\multirow{2}*{$  \mathcal{O}_{Q Q^{\dagger}  H{}^2 H^{\dagger} {}^2 D }^{(1\sim4)}$}
		
		&i\left(Q_{pa i} \sigma^{\mu } Q^{\dagger}{}_{r}^{a i}\right)\left(H^{\dagger}\overleftrightarrow{D}_{\mu}H\right)\left(H^{\dagger}H\right),\quad i\left(Q_{pa i} \sigma^{\mu } Q^{\dagger}{}_{r}^{a j}\right)H^{\dagger i}H_j \left(H^{\dagger}\overleftrightarrow{D}_{\mu}H\right)
		
		\\&\left(Q_{pa i} \sigma^{\mu } Q^{\dagger}{}_{r}^{a j}\right)H^{\dagger i}H_j D_{\mu}\left(H^{\dagger}H\right),\quad i\left(Q_{pa i} \sigma^{\mu } Q^{\dagger}{}_{r}^{a j}\right) H^{\dagger}{}^{i}\overleftrightarrow{D}_{\mu}H_{j} \left(H^{\dagger}H\right)\vspace{2ex}\\
		
		\mathcal{O}_{u_{_\mathbb{C}} u_{_\mathbb{C}}^{\dagger}  H{}^2 H^{\dagger} {}^2 D }
		
		& i\left(u_{_\mathbb{C}}{}_{p}^{a} \sigma^{\mu } u_{_\mathbb{C}}^{\dagger}{}_{ra}\right)\left(H^{\dagger}\overleftrightarrow{D}_{\mu}H\right)\left(H^{\dagger}H\right)\vspace{2ex}\\
		
		\mathcal{O}_{d_{_\mathbb{C}} d_{_\mathbb{C}}^{\dagger}  H{}^2 H^{\dagger} {}^2 D} 
		
		&i\left(d_{_\mathbb{C}}{}_{p}^{a} \sigma^{\mu } d_{_\mathbb{C}}^{\dagger}{}_{ra}\right)\left(H^{\dagger}\overleftrightarrow{D}_{\mu}H\right)\left(H^{\dagger}H\right)\vspace{2ex}\\
		
		\multirow{2}*{$ \mathcal{O}_{ L L^{\dagger}  H{}^2 H^{\dagger} {}^2 D}^{(1\sim4)} $}
		
		&i\left(L_{pi} \sigma^{\mu } L^{\dagger}{}_{r}^{i}\right)\left(H^{\dagger}\overleftrightarrow{D}_{\mu}H\right)\left(H^{\dagger}H\right),\quad i\left(L_{pi} \sigma^{\mu } L^{\dagger}{}_{r}^{j}\right)H^{\dagger i}H_j \left(H^{\dagger}\overleftrightarrow{D}_{\mu}H\right),
		\\& \left(L_{pi} \sigma^{\mu } L^{\dagger}{}_{r}^{j}\right)H^{\dagger i}H_j D_{\mu}\left(H^{\dagger}H\right),\quad i\left(L_{pi} \sigma^{\mu } L^{\dagger}{}_{r}^{j}\right) H^{\dagger}{}^{i}\overleftrightarrow{D}_{\mu}H_{j} \left(H^{\dagger}H\right)\vspace{2ex}\\
		
		\mathcal{O}_{e_{_\mathbb{C}} e_{_\mathbb{C}}^{\dagger}  H{}^2 H^{\dagger} {}^2 D }
		
		&i\left(e_{_\mathbb{C} p} \sigma^{\mu } e_{_\mathbb{C}}^{\dagger}{}_{r}\right)\left(H^{\dagger}\overleftrightarrow{D}_{\mu}H\right)\left(H^{\dagger}H\right)
		
	\end{array}\label{cl:pph4dr}\end{align}
The only complex type is: 
\begin{align}\begin{array}{c|l}
		
		\mathcal{O}_{u_{_\mathbb{C}} d_{_\mathbb{C}}^{\dagger}  H{}^3 H^{\dagger}  D }
		
		&\epsilon^{jk}\left(u_{_\mathbb{C}}{}_{p}^{a} \sigma^{\mu } d_{_\mathbb{C}}^{\dagger}{}_{ra}\right)\left(H^{\dagger}H\right)H_j D_{\mu}H_k
		
	\end{array}\label{cl:pph4dc}\end{align}
After taking vev for two of the Higgses, these are the 5 neutral fermion currents and 1 charged fermion current coupled with the neutral and charged Higgs current, which are already present at dimension 6, but with additional $v^2/\Lambda^2$ suppression. Note that for the left-handed fermions $Q_i,L_i$, new terms exist due to the richness of the $SU(2)_W$ structures.

The conversion of these fermion currents to four-component spinor notation is shown by the following examples
\bea
\left(e_{_\mathbb{C} p} \sigma^{\mu }\Gamma e_{_\mathbb{C}}^{\dagger}{}_{r}\right)=\left(\bar{e}_p\gamma^{\mu}\Gamma e_r \right),\quad \left(L_{pi} \sigma^{\mu } \Gamma L^{\dagger}{}_{r}^{i}\right)=-\left(\bar{l}_r^i\gamma^{\mu}\Gamma l_{pi}\right), \qquad \Gamma=\mathbbm{1},\tau^I,\lambda^A,D^\mu.\label{eq:pgp}
\eea

\noindent\underline{Class $\psi^2\phi^3D^2$}: 
The subclass $\psi^2\phi^3D^2$ contains 6 independent Lorentz structures:
\eq{
	&	\psi _1{}{}^{\alpha } \psi_2{}^{\beta } \left(D \phi _3\right)_{\alpha\dot{\alpha }} \left(D \phi _4\right)_{\beta }^{\dot{\alpha }}\phi _5 ,\quad 
	\psi _1{}{}^{\alpha } \psi_2{}^{\beta } \left(D \phi _3\right)_{\alpha\dot{\alpha }}\phi _4 \left(D \phi _5\right)_{\beta }^{\dot{\alpha }} ,\quad 
	\psi _1{}{}^{\alpha } \psi_2{}^{\beta } \phi _3 \left(D \phi _4\right)_{\alpha\dot{\alpha }} \left(D \phi _5\right)_{\beta }^{\dot{\alpha }},\\
	&	\psi _1{}{}^{\alpha } \psi_2{}_{\alpha } \left(D \phi _3\right)^{\beta}_{\dot{\alpha }} \left(D \phi _4\right)_{\beta}^{\dot{\alpha }}\phi _5 ,\quad 
	\psi _1{}{}^{\alpha } \psi_2{}_{\alpha } \left(D \phi _3\right)^{\beta}_{\dot{\alpha }} \phi _4 \left(D \phi _5\right)_{\beta}^{\dot{\alpha }} ,\quad 
	\psi _1{}{}^{\alpha } \psi_2{}_{\alpha } \phi _3 \left(D \phi _4\right)^{\beta}_{\dot{\alpha }} \left(D \phi _5\right)_{\beta}^{\dot{\alpha }} .
}
Types of this subclass in the SMEFT are similar to the Yukawa terms, which are all complex, with additional Higgs and derivatives:
\begin{align}\begin{array}{c|l}
		\multirow{3}*{$ \mathcal{O}_{Q u_{_\mathbb{C}} H^2H^{\dagger}  D^2}^{(1\sim6)} $}
		& i\epsilon ^{ik}\left(Q_{pa i} u_{_\mathbb{C}}{}_{r}^{a}\right)D_{\mu}H_k\left(H^{\dagger}\overleftrightarrow{D}^{\mu}H\right),\quad 
		i \epsilon ^{ik}\left(Q_{pa i} \sigma^{\mu }{}^{\nu } u_{_\mathbb{C}}{}_{r}^{a}\right)D_{\mu}H_k\left(H^{\dagger}\overleftrightarrow{D}_{\nu}H\right),\\
		& \epsilon ^{ik} \left(Q_{pa i} u_{_\mathbb{C}}{}_{r}^{a}\right) H_{k} \left(D_{\mu } H^{\dagger} D^{\mu } H\right),\quad  
		\epsilon ^{ik}\left(Q_{pa i} \sigma^{\mu }{}^{\nu } u_{_\mathbb{C}}{}_{r}^{a}\right) H_{k} \left(D_{\mu } H^{\dagger} D_{\nu } H\right) \\
		& \epsilon ^{ik}\left(Q_{pa i} u_{_\mathbb{C}}{}_{r}^{a}\right)D_{\mu}H_kD^{\mu}\left(H^{\dagger}H\right),\quad 
		\epsilon ^{ik}\left(Q_{pa i} \sigma^{\mu }{}^{\nu } u_{_\mathbb{C}}{}_{r}^{a}\right) D_{\mu } H_{k} D_{\nu } \left(H^{\dagger}H\right) \vspace{2ex}\\
		\multirow{3}*{$\mathcal{O}_{Q d_{_\mathbb{C}} HH^{\dagger 2} D^2}^{(1\sim6)} $}
		& i\left(d_{_\mathbb{C}}{}_{p}^{a} Q_{ra i}\right) D^{\mu } H^{\dagger}{}^{i} \left(H^{\dagger}\overleftrightarrow{D}_{\mu}H\right),\quad 
		i \left(d_{_\mathbb{C}}{}_{p}^{a} \sigma^{\mu }{}^{\nu } Q_{ra i}\right) D_{\mu } H^{\dagger}{}^{i} \left(H^{\dagger}\overleftrightarrow{D}_{\nu}H\right),\\
		& \left(d_{_\mathbb{C}}{}_{p}^{a} Q_{ra i}\right)H^{\dagger i}\left(D^{\mu}H^{\dagger}D_{\mu}H\right),\quad 
		\left(d_{_\mathbb{C}}{}_{p}^{a} \sigma^{\mu }{}^{\nu } Q_{ra i}\right)H^{\dagger i}\left(D_{\mu}H^{\dagger}D_{\nu}H\right),\\
		& \left(d_{_\mathbb{C}}{}_{p}^{a} Q_{ra i}\right) D^{\mu } H^{\dagger}{}^{i} D_{\mu}\left(H^{\dagger}H\right),\quad 
		\left(d_{_\mathbb{C}}{}_{p}^{a} \sigma^{\mu }{}^{\nu } Q_{ra i}\right) D_{\mu } H^{\dagger}{}^{i} D_{\nu}\left(H^{\dagger}H\right)\vspace{2ex}\\
		\multirow{3}*{$\mathcal{O}_{e_{_\mathbb{C}}L HH^{\dagger 2} D^2}^{(1\sim6)} $}
		& i\left(e_{_\mathbb{C} p} L_{ri}\right) D^{\mu } H^{\dagger}{}^{i} \left(H^{\dagger}\overleftrightarrow{D}_{\mu}H\right),\quad
		i \left(e_{_\mathbb{C} p} \sigma^{\mu }{}^{\nu } L_{ri}\right) D_{\mu } H^{\dagger}{}^{i} \left(H^{\dagger}\overleftrightarrow{D}_{\nu}H\right),\\
		&\left(e_{_\mathbb{C} p} L_{ri}\right)H^{\dagger i}\left(D^{\mu}H^{\dagger}D_{\mu}H\right),\quad 
		\left(e_{_\mathbb{C} p} \sigma^{\mu }{}^{\nu } L_{ri}\right)H^{\dagger i}\left(D_{\mu}H^{\dagger}D_{\nu}H\right),\\
		&\left(e_{_\mathbb{C} p} L_{ri}\right) D^{\mu } H^{\dagger}{}^{i} D_{\mu}\left(H^{\dagger}H\right),\quad 
		\left(e_{_\mathbb{C} p} \sigma^{\mu }{}^{\nu } L_{ri}\right) D_{\mu } H^{\dagger}{}^{i} D_{\nu}\left(H^{\dagger}H\right)
	\end{array}\label{cl:pph3d2}\end{align}
but due to the derivatives, these are new Lorentz structures at dimension 8.
In some of the terms, the dipole moment bilinear appears, which are converted to four-component notation as
\bea
\left(e_{_\mathbb{C} p}\sigma^{\mu\nu}\Gamma L_{ri}\right)= \left(\bar{e}_p\sigma^{\mu\nu}\Gamma l_{ri}\right),\quad \left(Q_{pa i} \sigma^{\mu }{}^{\nu }\Gamma u_{_\mathbb{C}}{}_{r}^{a}\right)=\left(\bar{u}^a_r\sigma^{\mu\nu}\Gamma q_{pai}\right), \qquad \Gamma=\mathbbm{1},\tau^I,\lambda^A,D^\mu.\label{eq:pggp}
\eea

\noindent\underline{Class $\psi^2\phi^2D^3$}: With three derivatives, we only have 2 independent Lorentz structures as follows
\begin{align}
	\psi _1{}^{\alpha } \left(D \phi _2\right)^{\beta}_{\dot{\alpha }}\left(D \phi _3\right)_{\alpha \dot{\beta }}\left(D \psi^{\dagger}_4\right)^{\dot{\alpha }\dot{\beta }}_\beta , \quad 
	\psi _1{}^{\alpha } \phi _2 \left(D^2 \phi _3\right)^\beta_{\alpha \dot{\alpha }\dot{\beta }} \left(D \psi^{\dagger}_4\right)^{\dot{\alpha }\dot{\beta }}_\beta ,	
\end{align}
which can be easily checked by enumerating SSYT of shape $\tiny\yng(4,4)$ and labels $\{1,1,1,2,2,3,3,4\}$, c.f. section~\ref{sec:lorentz_str}. 
The types in the SMEFT are very similar to those of the subclass $\psi^2\phi^4D$, with 5 real types:
\begin{align}\begin{array}{c|l}
		\multirow{2}*{$ \mathcal{O}_{Q Q^{\dagger} HH^{\dagger}  D^3 }^{(1\sim4)}$}
		&i\left(Q_{pa i} \sigma_{\mu } Q^{\dagger}{}_{r}^{a i}\right)\Box \left(H^{\dagger}\overleftrightarrow{D}^{\mu}H\right) ,\quad i\left(Q_{pa i} \sigma_{\mu } \overleftrightarrow{D}_{\nu } Q^{\dagger}{}_{r}^{a i}\right)\left(D^{\mu}H^{\dagger}D^{\nu}H\right),\\
		&i\left(Q_{pa i} \sigma_{\mu } Q^{\dagger}{}_{r}^{a j}\right)\Box \left(H^{\dagger i}\overleftrightarrow{D}^{\mu}H_j\right) ,\quad i\left(Q_{pa i} \sigma_{\mu } \overleftrightarrow{D}_{\nu } Q^{\dagger}{}_{r}^{a j}\right)\left(D^{\mu}H^{\dagger i}D^{\nu}H_j\right)\vspace{2ex}\\
		\mathcal{O}_{u_{_\mathbb{C}} u_{_\mathbb{C}}^{\dagger} HH^{\dagger}  D^3}^{(1,2)}
		&i\left(u_{_\mathbb{C}}{}_{p}^{a} \sigma_{\mu } u_{_\mathbb{C}}^{\dagger}{}_{ra}\right)\Box \left(H^{\dagger}\overleftrightarrow{D}^{\mu}H\right),\quad i\left(u_{_\mathbb{C}}{}_{p}^{a} \sigma_{\mu } \overleftrightarrow{D}_{\nu } u_{_\mathbb{C}}^{\dagger}{}_{ra}\right)\left(D^{\mu}H^{\dagger}D^{\nu}H\right)\vspace{2ex}\\
		\mathcal{O}_{d_{_\mathbb{C}} d_{_\mathbb{C}}^{\dagger}  HH^{\dagger}  D^3}^{(1,2)}
		&i\left(d_{_\mathbb{C}}{}_{p}^{a} \sigma_{\mu } d_{_\mathbb{C}}^{\dagger}{}_{ra}\right)\Box \left(H^{\dagger}\overleftrightarrow{D}^{\mu}H\right),\quad i\left(d_{_\mathbb{C}}{}_{p}^{a} \sigma_{\mu } \overleftrightarrow{D}_{\nu } d_{_\mathbb{C}}^{\dagger}{}_{ra}\right)\left(D^{\mu}H^{\dagger}D^{\nu}H\right)\vspace{2ex}\\
		\multirow{2}*{$ \mathcal{O}_{L L^{\dagger} HH^{\dagger} D^3 }^{(1\sim4)}$}
		&i\left(L_{p i} \sigma_{\mu } L^{\dagger}{}_{r}^{ i}\right)\Box \left(H^{\dagger}\overleftrightarrow{D}^{\mu}H\right) ,\quad i\left(L_{p i} \sigma_{\mu } \overleftrightarrow{D}_{\nu } L^{\dagger}{}_{r}^{ i}\right)\left(D^{\mu}H^{\dagger}D^{\nu}H\right),\\
		&i\left(L_{p i} \sigma_{\mu } L^{\dagger}{}_{r}^{ j}\right)\Box \left(H^{\dagger i}\overleftrightarrow{D}^{\mu}H_j\right) ,\quad i\left(L_{p i} \sigma_{\mu } \overleftrightarrow{D}_{\nu } L^{\dagger}{}_{r}^{ j}\right)\left(D^{\mu}H^{\dagger i}D^{\nu}H_j\right)\vspace{2ex}\\
		\mathcal{O}_{e_{_\mathbb{C}} e_{_\mathbb{C}}^{\dagger}  HH^{\dagger} D^3}^{(1,2)}
		&i\left(e_{_\mathbb{C}}{}_{p} \sigma_{\mu } e_{_\mathbb{C}}^{\dagger}{}_{r}\right)\Box \left(H^{\dagger}\overleftrightarrow{D}^{\mu}H\right),\quad i\left(e_{_\mathbb{C}}{}_{p} \sigma_{\mu } \overleftrightarrow{D}_{\nu } e_{_\mathbb{C}}^{\dagger}{}_{r}\right)\left(D^{\mu}H^{\dagger}D^{\nu}H\right)
	\end{array}\label{cl:pph2d3r}\end{align}
and 1 complex type:
\begin{align}\begin{array}{c|l}
		\mathcal{O}_{u_{_\mathbb{C}}d_{_\mathbb{C}}^{\dagger}  H^2 D^3 }
		& i \epsilon ^{ij}\left(u_{_\mathbb{C}}{}_{p}^{a} \sigma^{\nu } D{}^{\mu } d_{_\mathbb{C}}^{\dagger}{}_{ra}\right) D_{\mu } H_{i} D_{\nu } H_{j} .
	\end{array}\label{cl:pph2d3c}\end{align}
If we use the Fierz identity of $SU(N)$ group eq.~\eqref{eq:Fierz3}, we can perform the following transformation
\begin{align}
	i\left(Q_{pa i} \sigma_{\mu } Q^{\dagger}{}_{r}^{a j}\right)\Box \left(H^{\dagger i}\overleftrightarrow{D}^{\mu}H_j\right)=&\frac12i\left(\bar{q}_r\gamma_{\mu} q_{p}\right)\Box \left(H^{\dagger}\overleftrightarrow{D}^{\mu}H\right)+i\left(\bar{q}_r\gamma_{\mu}\tau^I q_{p}\right)\Box \left(H^{\dagger}\tau^I\overleftrightarrow{D}^{\mu}H\right).\label{eq:Fierzq}
\end{align}
It could help convert our terms in eq.~\eqref{cl:pph2d3r} to more common forms, such as
\begin{align}
	\left\{\begin{array}{c}
		i\left(Q_{pa i} \sigma_{\mu } Q^{\dagger}{}_{r}^{a i}\right)\Box \left(H^{\dagger}\overleftrightarrow{D}^{\mu}H\right)\\
		i\left(Q_{pa i} \sigma_{\mu } Q^{\dagger}{}_{r}^{a j}\right)\Box \left(H^{\dagger i}\overleftrightarrow{D}^{\mu}H_j\right) \\ 
		i\left(Q_{pa i} \sigma_{\mu } \overleftrightarrow{D}_{\nu } Q^{\dagger}{}_{r}^{a i}\right)\left(D^{\mu}H^{\dagger}D^{\nu}H\right)\\
		i\left(Q_{pa i} \sigma_{\mu } \overleftrightarrow{D}_{\nu } Q^{\dagger}{}_{r}^{a j}\right)\left(D^{\mu}H^{\dagger i}D^{\nu}H_j\right)
	\end{array}\right.
	\Longrightarrow\left\{
	\begin{array}{c}
		i\left(\bar{q}_r\gamma_{\mu} q_{p}\right)\Box \left(H^{\dagger}\overleftrightarrow{D}^{\mu}H\right)\\
		i\left(\bar{q}_r\gamma_{\mu}\tau^I q_{p}\right)\Box \left(H^{\dagger}\tau^I\overleftrightarrow{D}^{\mu}H\right)\\
		i\left(\bar{q}_r\gamma_{\mu}\overleftrightarrow{D}_{\nu}q_p\right)\left(D^{\mu}H^{\dagger}D^{\nu}H\right)\\
		i\left(\bar{q}_r\gamma_{\mu}\tau^I\overleftrightarrow{D}_{\nu}q_p\right)\left(D^{\mu}H^{\dagger i}\tau^I D^{\nu}H_j\right)
	\end{array}\right.
\end{align}

\subsubsection{One gauge boson involved}

\noindent\underline{Class $F\psi^2\phi^3$}: The only independent Lorentz structure of this subclass is
\begin{align}
	F_{\rm{L}1}{}{}^{\alpha }{}{}^{\beta }\psi _2{}_{\alpha } \psi _3{}_{\beta } \phi _4 \phi _5 \phi _6 
\end{align}
The operators with this Lorentz structures in the dimension 8 SMEFT are
\begin{align}\begin{array}{c|l}
		\mathcal{O}_{G Q u_{_\mathbb{C}} H^2H^{\dagger}}
		& \epsilon^{ik} G^{A}_{\mu\nu}\left(Q_{pa i} \sigma^{\mu\nu}\left(\lambda ^A\right)_b^a u_{_\mathbb{C}}{}_{r}^{b}\right)H_k\left(H^{\dagger}H\right)\vspace{2ex}\\
		\mathcal{O}_{W Q u_{_\mathbb{C}} H^2H^{\dagger}}^{(1,2)}
		& \epsilon ^{km}\left(\tau ^I\right)_m^j W^{I}_{\mu\nu} \left(Q_{pa i} \sigma^{\mu\nu} u_{_\mathbb{C}}{}_{r}^{a}\right)H_{j} H_{k} H^{\dagger}{}^{i} , \quad \epsilon ^{km} \left(\tau ^I\right)_m^i W^{I}_{\mu\nu}\left(Q_{pa i} \sigma^{\mu\nu} u_{_\mathbb{C}}{}_{r}^{a}\right) H_{k} \left(H^{\dagger}H\right)  \vspace{2ex}\\
		\mathcal{O}_{B Q u_{_\mathbb{C}} H^2H^{\dagger}}
		& \epsilon ^{ik} B_{\mu\nu}\left(Q_{pa i} \sigma^{\mu\nu} u_{_\mathbb{C}}{}_{r}^{a}\right) H_{k}\left(H^{\dagger}H\right) \vspace{2ex}\\	
		\mathcal{O}_{G Q d_{_\mathbb{C}} HH^{\dagger 2}}
		&  G^{A}_{\mu\nu}\left(d_{_\mathbb{C}}{}_{p}^{a} \sigma^{\mu\nu} \left(\lambda ^A\right)_a^b Q_{rb i}\right)H^{\dagger i}\left(H^{\dagger}H\right)\vspace{2ex}\\
		\mathcal{O}_{W Q d_{_\mathbb{C}} HH^{\dagger 2}}^{(1,2)}
		& W^{I}_{\mu\nu}\left(d_{_\mathbb{C}}{}_{p}^{a} \sigma^{\mu\nu} Q_{ra i}\right) H^{\dagger}{}^{i}\left(H^{\dagger}\tau ^I H\right) ,\quad \left(\tau ^I\right)_l^iW^{I}_{\mu\nu} \left(d_{_\mathbb{C}}{}_{p}^{a} \sigma^{\mu\nu} Q_{ra i}\right)H^{\dagger}{}^{l} \left(H^{\dagger}H\right) \vspace{2ex}\\
		\mathcal{O}_{B Q d_{_\mathbb{C}} HH^{\dagger 2}}
		& B_{\mu\nu}\left(d_{_\mathbb{C}}{}_{p}^{a} \sigma^{\mu\nu} Q_{ra i}\right)H^{\dagger i}\left(H^{\dagger}H\right) \vspace{2ex}\\
		\mathcal{O}_{W e_{_\mathbb{C}} L HH^{\dagger 2}}^{(1,2)}
		& W^{I}_{\mu\nu} \left(e_{_\mathbb{C} p} \sigma^{\mu\nu} L_{ri}\right)H^{\dagger}{}^{i}\left(H^{\dagger}\tau ^I H\right) ,\quad  \left(\tau ^I\right)_l^iW^{I}_{\mu\nu} \left(e_{_\mathbb{C} p} \sigma^{\mu\nu} L_{ri}\right)H^{\dagger}{}^{l}\left(H^{\dagger}H\right)  \vspace{2ex}\\
		\mathcal{O}_{B e_{_\mathbb{C}} L HH^{\dagger 2}}
		& B_{\mu\nu} \left(e_{_\mathbb{C} p} \sigma^{\mu\nu} L_{ri}\right)H^{\dagger i}\left(H^{\dagger}H\right) 
	\end{array}\label{cl:Fpp3}\end{align}
Note that these are all complex types, whose real part and imaginary part contribute to the electric and magnetic dipole moments of the fermions, respectively, after the Higgses take their VEV. 
One may refer to eq.~(\ref{eq:pggp}) for the conversion to four-component spinor notation.

\noindent\underline{Class $\mathcal{O}(F\psi^2\phi^2D)$}: In this class, the two spinors have opposite helicities, and form a fermion current, while the gauge boson couples with both of the fermion current and Higgs current. 2 independent Lorentz structures are present:
\begin{align}
	
	F_{\rm{L}1}{}{}^{\alpha }{}{}^{\beta }\psi _2{}_{\alpha } \left(D \phi _3\right){}_{\beta }{}_{\dot{\alpha }}\phi _4 \psi^{\dagger}_5{}^{\dot{\alpha }}  ,\quad F_{\rm{L}1}{}{}^{\alpha }{}{}^{\beta }\psi _2{}_{\alpha } \phi _3 \left(D \phi _4\right){}_{\beta }{}_{\dot{\alpha }} \psi^{\dagger}_5{}^{\dot{\alpha }}.
	
\end{align}
In the SMEFT, real types with neutral fermion currents are as follows
\begin{align}\begin{array}{c|l}
		\multirow{4}*{$\mathcal{O}_{G Q Q^{\dagger}  HH^{\dagger}  D }^{(1\sim8)}$}
		&G^{A}_{\mu \nu} \left(Q_{pa i} \sigma^{\nu }\left(\lambda ^A\right)_b^a Q^{\dagger}{}_{r}^{b i}\right) D^{\mu }\left(H^{\dagger} H\right),\quad iG^{A}_{\mu \nu} \left(Q_{pa i} \sigma^{\nu }\left(\lambda ^A\right)_b^a Q^{\dagger}{}_{r}^{b i}\right)\left(H^{\dagger}\overleftrightarrow{D}^{\mu}H\right),\\
		&\tilde{G}^{A}_{\mu \nu} \left(Q_{pa i} \sigma^{\nu }\left(\lambda ^A\right)_b^a Q^{\dagger}{}_{r}^{b i}\right) D^{\mu }\left(H^{\dagger} H\right) ,\quad i\tilde{G}^{A}_{\mu \nu} \left(Q_{pa i} \sigma^{\nu }\left(\lambda ^A\right)_b^a Q^{\dagger}{}_{r}^{b i}\right)\left(H^{\dagger}\overleftrightarrow{D}^{\mu}H\right),\\
		&G^{A}_{\mu \nu} \left(Q_{pa i} \sigma^{\nu }\left(\lambda ^A\right)_b^a Q^{\dagger}{}_{r}^{b j}\right) D^{\mu }\left(H^{\dagger i} H_j\right),\quad iG^{A}_{\mu \nu} \left(Q_{pa i} \sigma^{\nu }\left(\lambda ^A\right)_b^a Q^{\dagger}{}_{r}^{b j}\right)\left(H^{\dagger i}\overleftrightarrow{D}^{\mu}H_j\right),\\
		&\tilde{G}^{A}_{\mu \nu} \left(Q_{pa i} \sigma^{\nu }\left(\lambda ^A\right)_b^a Q^{\dagger}{}_{r}^{b j}\right) D^{\mu }\left(H^{\dagger i} H_j\right) ,\quad i\tilde{G}^{A}_{\mu \nu} \left(Q_{pa i} \sigma^{\nu }\left(\lambda ^A\right)_b^a Q^{\dagger}{}_{r}^{b j}\right)\left(H^{\dagger i}\overleftrightarrow{D}^{\mu}H_j\right)\vspace{2ex}\\
		
		\multirow{6}*{$\mathcal{O}_{W Q Q^{\dagger}  H H^{\dagger} D }^{(1\sim12)}$}
		& W^{I}_{\mu\nu} \left(Q_{pa i} \sigma^{\nu } Q^{\dagger}{}_{r}^{a i}\right)D_{\mu} \left(H^{\dagger}\tau ^IH\right),\quad iW^{I}_{\mu\nu} \left(Q_{pa i} \sigma^{\nu } Q^{\dagger}{}_{r}^{a i}\right)\left(H^{\dagger}\tau ^I\overleftrightarrow{D}^{\mu}H\right) ,\\
		& \tilde{W}^{I}_{\mu\nu} \left(Q_{pa i} \sigma^{\nu } Q^{\dagger}{}_{r}^{a i}\right)D_{\mu} \left(H^{\dagger}\tau ^IH\right),\quad i\tilde{W}^{I}_{\mu\nu} \left(Q_{pa i} \sigma^{\nu } Q^{\dagger}{}_{r}^{a i}\right)\left(H^{\dagger}\tau ^I\overleftrightarrow{D}^{\mu}H\right), \\
		& \left(\tau^I\right)^k_j W^{I}_{\mu\nu} \left(Q_{pa i} \sigma^{\nu } Q^{\dagger}{}_{r}^{a j}\right)D_{\mu} \left(H^{\dagger i}H_k\right),\quad i\left(\tau^I\right)^k_j W^{I}_{\mu\nu} \left(Q_{pa i} \sigma^{\nu } Q^{\dagger}{}_{r}^{a j}\right)\left(H^{\dagger i}\overleftrightarrow{D}^{\mu}H_k\right) ,\\
		& \left(\tau^I\right)^k_j \tilde{W}^{I}_{\mu\nu} \left(Q_{pa i} \sigma^{\nu } Q^{\dagger}{}_{r}^{a j}\right)D_{\mu} \left(H^{\dagger i}H_k\right),\quad i\left(\tau^I\right)^k_j \tilde{W}^{I}_{\mu\nu} \left(Q_{pa i} \sigma^{\nu } Q^{\dagger}{}_{r}^{a j}\right)\left(H^{\dagger i}\overleftrightarrow{D}^{\mu}H_k\right), \\
		& \left(\tau^I\right)^i_k W^{I}_{\mu\nu} \left(Q_{pa i} \sigma^{\nu } Q^{\dagger}{}_{r}^{a j}\right)D_{\mu} \left(H^{\dagger k}H_j\right),\quad i\left(\tau^I\right)^i_k W^{I}_{\mu\nu} \left(Q_{pa i} \sigma^{\nu } Q^{\dagger}{}_{r}^{a j}\right)\left(H^{\dagger k}\overleftrightarrow{D}^{\mu}H_j\right) ,\\
		& \left(\tau^I\right)^i_k \tilde{W}^{I}_{\mu\nu} \left(Q_{pa i} \sigma^{\nu } Q^{\dagger}{}_{r}^{a j}\right)D_{\mu} \left(H^{\dagger k}H_j\right),\quad i\left(\tau^I\right)^i_k \tilde{W}^{I}_{\mu\nu} \left(Q_{pa i} \sigma^{\nu } Q^{\dagger}{}_{r}^{a j}\right)\left(H^{\dagger k}\overleftrightarrow{D}^{\mu}H_j\right)

	\end{array}\label{cl:Fpph2dr}\end{align}
\begin{align}\begin{array}{c|l}
		\multirow{4}*{$\mathcal{O}_{B Q Q^{\dagger}  HH^{\dagger}  D }^{(1\sim8)}$}
		&B_{\mu \nu} \left(Q_{pa i} \sigma^{\nu } Q^{\dagger}{}_{r}^{a i}\right) D^{\mu }\left(H^{\dagger} H\right),\quad iB_{\mu \nu} \left(Q_{pa i} \sigma^{\nu } Q^{\dagger}{}_{r}^{a i}\right)\left(H^{\dagger}\overleftrightarrow{D}^{\mu}H\right),\\
		&\tilde{B}_{\mu \nu} \left(Q_{pa i} \sigma^{\nu } Q^{\dagger}{}_{r}^{a i}\right) D^{\mu }\left(H^{\dagger} H\right) ,\quad i\tilde{B}_{\mu \nu} \left(Q_{pa i} \sigma^{\nu } Q^{\dagger}{}_{r}^{a i}\right)\left(H^{\dagger}\overleftrightarrow{D}^{\mu}H\right),\\
		&B_{\mu \nu} \left(Q_{pa i} \sigma^{\nu } Q^{\dagger}{}_{r}^{a j}\right) D^{\mu }\left(H^{\dagger i} H_j\right),\quad iB_{\mu \nu} \left(Q_{pa i} \sigma^{\nu } Q^{\dagger}{}_{r}^{a j}\right)\left(H^{\dagger i}\overleftrightarrow{D}^{\mu}H_j\right),\\
		&\tilde{B}_{\mu \nu} \left(Q_{pa i} \sigma^{\nu } Q^{\dagger}{}_{r}^{a j}\right) D^{\mu }\left(H^{\dagger i} H_j\right) ,\quad i\tilde{B}_{\mu \nu} \left(Q_{pa i} \sigma^{\nu } Q^{\dagger}{}_{r}^{a j}\right)\left(H^{\dagger i}\overleftrightarrow{D}^{\mu}H_j\right)\vspace{2ex}\\
		
		\multirow{2}*{$\mathcal{O}_{G u_{_\mathbb{C}} u_{_\mathbb{C}}^{\dagger}  HH^{\dagger}  D}^{(1\sim4)}$}
		&G^{A}_{\mu \nu} \left(u_{_\mathbb{C}}{}_{p}^{a} \sigma^{\nu }\left(\lambda ^A\right)_a^b u_{_\mathbb{C}}^{\dagger}{}_{rb}\right) D^{\mu }\left(H^{\dagger} H\right),\quad iG^{A}_{\mu \nu} \left(u_{_\mathbb{C}}{}_{p}^{a} \sigma^{\nu }\left(\lambda ^A\right)_a^b u_{_\mathbb{C}}^{\dagger}{}_{rb}\right)\left(H^{\dagger}\overleftrightarrow{D}^{\mu}H\right),\\
		&\tilde{G}^{A}_{\mu \nu} \left(u_{_\mathbb{C}}{}_{p}^{a} \sigma^{\nu }\left(\lambda ^A\right)_a^b u_{_\mathbb{C}}^{\dagger}{}_{rb}\right) D^{\mu }\left(H^{\dagger} H\right) ,\quad i\tilde{G}^{A}_{\mu \nu} \left(u_{_\mathbb{C}}{}_{p}^{a} \sigma^{\nu }\left(\lambda ^A\right)_a^b u_{_\mathbb{C}}^{\dagger}{}_{rb}\right)\left(H^{\dagger}\overleftrightarrow{D}^{\mu}H\right) \vspace{2ex}\\
		
		\multirow{2}*{$\mathcal{O}_{W u_{_\mathbb{C}} u_{_\mathbb{C}}^{\dagger}  HH^{\dagger}  D}^{(1\sim4)}$}
		&W^{I}_{\mu \nu} \left(u_{_\mathbb{C}}{}_{p}^{a} \sigma^{\nu } u_{_\mathbb{C}}^{\dagger}{}_{ra}\right) D^{\mu }\left(H^{\dagger}\tau^I H\right),\quad iW^{I}_{\mu \nu} \left(u_{_\mathbb{C}}{}_{p}^{a} \sigma^{\nu } u_{_\mathbb{C}}^{\dagger}{}_{ra}\right)\left(H^{\dagger}\tau^I\overleftrightarrow{D}^{\mu}H\right),\\
		&\tilde{W}^{I}_{\mu \nu} \left(u_{_\mathbb{C}}{}_{p}^{a} \sigma^{\nu } u_{_\mathbb{C}}^{\dagger}{}_{ra}\right) D^{\mu }\left(H^{\dagger}\tau^I H\right) ,\quad i\tilde{W}^{I}_{\mu \nu} \left(u_{_\mathbb{C}}{}_{p}^{a} \sigma^{\nu } u_{_\mathbb{C}}^{\dagger}{}_{ra}\right)\left(H^{\dagger}\tau^I\overleftrightarrow{D}^{\mu}H\right) \vspace{2ex}\\
		
		\multirow{2}*{$\mathcal{O}_{B u_{_\mathbb{C}} u_{_\mathbb{C}}^{\dagger}  HH^{\dagger}  D}^{(1\sim4)}$}
		&B_{\mu \nu} \left(u_{_\mathbb{C}}{}_{p}^{a} \sigma^{\nu } u_{_\mathbb{C}}^{\dagger}{}_{ra}\right) D^{\mu }\left(H^{\dagger} H\right),\quad iB_{\mu \nu} \left(u_{_\mathbb{C}}{}_{p}^{a} \sigma^{\nu } u_{_\mathbb{C}}^{\dagger}{}_{ra}\right)\left(H^{\dagger}\overleftrightarrow{D}^{\mu}H\right),\\
		&\tilde{B}_{\mu \nu} \left(u_{_\mathbb{C}}{}_{p}^{a} \sigma^{\nu } u_{_\mathbb{C}}^{\dagger}{}_{ra}\right) D^{\mu }\left(H^{\dagger} H\right) ,\quad i\tilde{B}_{\mu \nu} \left(u_{_\mathbb{C}}{}_{p}^{a} \sigma^{\nu } u_{_\mathbb{C}}^{\dagger}{}_{ra}\right)\left(H^{\dagger}\overleftrightarrow{D}^{\mu}H\right) \vspace{2ex}\\
		
		\multirow{2}*{$\mathcal{O}_{G d_{_\mathbb{C}} d_{_\mathbb{C}}^{\dagger}  HH^{\dagger}  D}^{(1\sim4)}$}
		&G^{A}_{\mu \nu} \left(d_{_\mathbb{C}}{}_{p}^{a} \sigma^{\nu }\left(\lambda ^A\right)_a^b d_{_\mathbb{C}}^{\dagger}{}_{rb}\right) D^{\mu }\left(H^{\dagger} H\right),\quad iG^{A}_{\mu \nu} \left(d_{_\mathbb{C}}{}_{p}^{a} \sigma^{\nu }\left(\lambda ^A\right)_a^b d_{_\mathbb{C}}^{\dagger}{}_{rb}\right)\left(H^{\dagger}\overleftrightarrow{D}^{\mu}H\right),\\
		&\tilde{G}^{A}_{\mu \nu} \left(d_{_\mathbb{C}}{}_{p}^{a} \sigma^{\nu }\left(\lambda ^A\right)_a^b d_{_\mathbb{C}}^{\dagger}{}_{rb}\right) D^{\mu }\left(H^{\dagger} H\right) ,\quad i\tilde{G}^{A}_{\mu \nu} \left(d_{_\mathbb{C}}{}_{p}^{a} \sigma^{\nu }\left(\lambda ^A\right)_a^b d_{_\mathbb{C}}^{\dagger}{}_{rb}\right)\left(H^{\dagger}\overleftrightarrow{D}^{\mu}H\right) \vspace{2ex}\\
		
		\multirow{2}*{$\mathcal{O}_{W d_{_\mathbb{C}} d_{_\mathbb{C}}^{\dagger}  HH^{\dagger}  D}^{(1\sim4)}$}
		&W^{I}_{\mu \nu} \left(d_{_\mathbb{C}}{}_{p}^{a} \sigma^{\nu } d_{_\mathbb{C}}^{\dagger}{}_{ra}\right) D^{\mu }\left(H^{\dagger}\tau^I H\right),\quad iW^{I}_{\mu \nu} \left(d_{_\mathbb{C}}{}_{p}^{a} \sigma^{\nu } d_{_\mathbb{C}}^{\dagger}{}_{ra}\right)\left(H^{\dagger}\tau^I\overleftrightarrow{D}^{\mu}H\right),\\
		&\tilde{W}^{I}_{\mu \nu} \left(d_{_\mathbb{C}}{}_{p}^{a} \sigma^{\nu } d_{_\mathbb{C}}^{\dagger}{}_{ra}\right) D^{\mu }\left(H^{\dagger}\tau^I H\right) ,\quad i\tilde{W}^{I}_{\mu \nu} \left(d_{_\mathbb{C}}{}_{p}^{a} \sigma^{\nu } d_{_\mathbb{C}}^{\dagger}{}_{ra}\right)\left(H^{\dagger}\tau^I\overleftrightarrow{D}^{\mu}H\right) \vspace{2ex}\\
		
		\multirow{2}*{$\mathcal{O}_{B d_{_\mathbb{C}} d_{_\mathbb{C}}^{\dagger}  HH^{\dagger}  D}^{(1\sim4)}$}
		&B_{\mu \nu} \left(d_{_\mathbb{C}}{}_{p}^{a} \sigma^{\nu } d_{_\mathbb{C}}^{\dagger}{}_{ra}\right) D^{\mu }\left(H^{\dagger} H\right),\quad iB_{\mu \nu} \left(d_{_\mathbb{C}}{}_{p}^{a} \sigma^{\nu } d_{_\mathbb{C}}^{\dagger}{}_{ra}\right)\left(H^{\dagger}\overleftrightarrow{D}^{\mu}H\right),\\
		&\tilde{B}_{\mu \nu} \left(d_{_\mathbb{C}}{}_{p}^{a} \sigma^{\nu } d_{_\mathbb{C}}^{\dagger}{}_{ra}\right) D^{\mu }\left(H^{\dagger} H\right) ,\quad i\tilde{B}_{\mu \nu} \left(d_{_\mathbb{C}}{}_{p}^{a} \sigma^{\nu } d_{_\mathbb{C}}^{\dagger}{}_{ra}\right)\left(H^{\dagger}\overleftrightarrow{D}^{\mu}H\right) \vspace{2ex}\\
		
		\multirow{6}*{$\mathcal{O}_{W L L^{\dagger}  HH^{\dagger}  D }^{(1\sim12)}$}
		& W^{I}_{\mu\nu} \left(L_{p i} \sigma^{\nu } L^{\dagger}{}_{r}^{ i}\right)D_{\mu} \left(H^{\dagger}\tau ^IH\right),\quad iW^{I}_{\mu\nu} \left(L_{p i} \sigma^{\nu } L^{\dagger}{}_{r}^{ i}\right)\left(H^{\dagger}\tau ^I\overleftrightarrow{D}^{\mu}H\right) ,\\
		& \tilde{W}^{I}_{\mu\nu} \left(L_{p i} \sigma^{\nu } L^{\dagger}{}_{r}^{ i}\right)D_{\mu} \left(H^{\dagger}\tau ^IH\right),\quad i\tilde{W}^{I}_{\mu\nu} \left(L_{p i} \sigma^{\nu } L^{\dagger}{}_{r}^{ i}\right)\left(H^{\dagger}\tau ^I\overleftrightarrow{D}^{\mu}H\right), \\
		& \left(\tau^I\right)^k_j W^{I}_{\mu\nu} \left(L_{p i} \sigma^{\nu } L^{\dagger}{}_{r}^{ j}\right)D_{\mu} \left(H^{\dagger i}H_k\right),\quad i\left(\tau^I\right)^k_j W^{I}_{\mu\nu} \left(L_{p i} \sigma^{\nu } L^{\dagger}{}_{r}^{ j}\right)\left(H^{\dagger i}\overleftrightarrow{D}^{\mu}H_k\right) ,\\
		& \left(\tau^I\right)^k_j \tilde{W}^{I}_{\mu\nu} \left(L_{p i} \sigma^{\nu } L^{\dagger}{}_{r}^{ j}\right)D_{\mu} \left(H^{\dagger i}H_k\right),\quad i\left(\tau^I\right)^k_j \tilde{W}^{I}_{\mu\nu} \left(L_{p i} \sigma^{\nu } L^{\dagger}{}_{r}^{ j}\right)\left(H^{\dagger i}\overleftrightarrow{D}^{\mu}H_k\right), \\
		& \left(\tau^I\right)^i_k W^{I}_{\mu\nu} \left(L_{p i} \sigma^{\nu } L^{\dagger}{}_{r}^{ j}\right)D_{\mu} \left(H^{\dagger k}H_j\right),\quad i\left(\tau^I\right)^i_k W^{I}_{\mu\nu} \left(L_{p i} \sigma^{\nu } L^{\dagger}{}_{r}^{ j}\right)\left(H^{\dagger k}\overleftrightarrow{D}^{\mu}H_j\right) ,\\
		& \left(\tau^I\right)^i_k \tilde{W}^{I}_{\mu\nu} \left(L_{p i} \sigma^{\nu } L^{\dagger}{}_{r}^{ j}\right)D_{\mu} \left(H^{\dagger k}H_j\right),\quad i\left(\tau^I\right)^i_k \tilde{W}^{I}_{\mu\nu} \left(L_{p i} \sigma^{\nu } L^{\dagger}{}_{r}^{ j}\right)\left(H^{\dagger k}\overleftrightarrow{D}^{\mu}H_j\right)\vspace{2ex}\\
		
		\multirow{4}*{$\mathcal{O}_{B L L^{\dagger}  HH^{\dagger}  D }^{(1\sim8)}$}
		&B_{\mu \nu} \left(L_{p i} \sigma^{\nu } L^{\dagger}{}_{r}^{ i}\right) D^{\mu }\left(H^{\dagger} H\right),\quad iB_{\mu \nu} \left(L_{p i} \sigma^{\nu } L^{\dagger}{}_{r}^{ i}\right)\left(H^{\dagger}\overleftrightarrow{D}^{\mu}H\right),\\
		&\tilde{B}_{\mu \nu} \left(L_{p i} \sigma^{\nu } L^{\dagger}{}_{r}^{ i}\right) D^{\mu }\left(H^{\dagger} H\right) ,\quad i\tilde{B}_{\mu \nu} \left(L_{p i} \sigma^{\nu } L^{\dagger}{}_{r}^{ i}\right)\left(H^{\dagger}\overleftrightarrow{D}^{\mu}H\right),\\
		&B_{\mu \nu} \left(L_{p i} \sigma^{\nu } L^{\dagger}{}_{r}^{ j}\right) D^{\mu }\left(H^{\dagger i} H_j\right),\quad iB_{\mu \nu} \left(L_{p i} \sigma^{\nu } L^{\dagger}{}_{r}^{ j}\right)\left(H^{\dagger i}\overleftrightarrow{D}^{\mu}H_j\right),\\
		&\tilde{B}_{\mu \nu} \left(L_{p i} \sigma^{\nu } L^{\dagger}{}_{r}^{ j}\right) D^{\mu }\left(H^{\dagger i} H_j\right) ,\quad i\tilde{B}_{\mu \nu} \left(L_{p i} \sigma^{\nu } L^{\dagger}{}_{r}^{ j}\right)\left(H^{\dagger i}\overleftrightarrow{D}^{\mu}H_j\right)\vspace{2ex}\\
		
		\multirow{2}*{$\mathcal{O}_{W e_{_\mathbb{C}} e_{_\mathbb{C}}^{\dagger}  HH^{\dagger} D}^{(1\sim4)}$}
		&W^{I}_{\mu \nu} \left(e_{_\mathbb{C}}{}_{p} \sigma^{\nu } e_{_\mathbb{C}}^{\dagger}{}_{r}\right) D^{\mu }\left(H^{\dagger}\tau^I H\right),\quad iW^{I}_{\mu \nu} \left(e_{_\mathbb{C}}{}_{p} \sigma^{\nu } e_{_\mathbb{C}}^{\dagger}{}_{r}\right)\left(H^{\dagger}\tau^I\overleftrightarrow{D}^{\mu}H\right),\\
		&\tilde{W}^{I}_{\mu \nu} \left(e_{_\mathbb{C}}{}_{p} \sigma^{\nu } e_{_\mathbb{C}}^{\dagger}{}_{r}\right) D^{\mu }\left(H^{\dagger}\tau^I H\right) ,\quad i\tilde{W}^{I}_{\mu \nu} \left(e_{_\mathbb{C}}{}_{p} \sigma^{\nu } e_{_\mathbb{C}}^{\dagger}{}_{r}\right)\left(H^{\dagger}\tau^I\overleftrightarrow{D}^{\mu}H\right) \vspace{2ex}\\
		
		\multirow{2}*{$\mathcal{O}_{B e_{_\mathbb{C}} e_{_\mathbb{C}}^{\dagger}  HH^{\dagger}  D}^{(1\sim4)}$}
		&B_{\mu \nu} \left(e_{_\mathbb{C}}{}_{p} \sigma^{\nu } e_{_\mathbb{C}}^{\dagger}{}_{r}\right) D^{\mu }\left(H^{\dagger}\tau^I H\right),\quad iB_{\mu \nu} \left(e_{_\mathbb{C}}{}_{p} \sigma^{\nu } e_{_\mathbb{C}}^{\dagger}{}_{r}\right)\left(H^{\dagger}\tau^I\overleftrightarrow{D}^{\mu}H\right),\\
		&\tilde{B}_{\mu \nu} \left(e_{_\mathbb{C}}{}_{p} \sigma^{\nu } e_{_\mathbb{C}}^{\dagger}{}_{r}\right) D^{\mu }\left(H^{\dagger}\tau^I H\right) ,\quad i\tilde{B}_{\mu \nu} \left(e_{_\mathbb{C}}{}_{p} \sigma^{\nu } e_{_\mathbb{C}}^{\dagger}{}_{r}\right)\left(H^{\dagger}\tau^I\overleftrightarrow{D}^{\mu}H\right),
	\end{array}\label{cl:Fpph2d1}\end{align}
while complex types with charged currents also exist:
\begin{align}\begin{array}{c|l}
		\mathcal{O}_{G  u_{_\mathbb{C}} d_{_\mathbb{C}}^{\dagger} H^2 D}^{(1,2)}
		&  \epsilon ^{ij} G^{A}_{\mu\nu } \left(u_{_\mathbb{C}}{}_{p}^{a} \sigma^{\nu }\left(\lambda ^A\right)_a^b  d_{_\mathbb{C}}^{\dagger}{}_{rb}\right) H_{i} D^{\mu } H_{j} ,\quad  \epsilon ^{ij} \tilde{G}^{A}_{\mu \nu }\left(u_{_\mathbb{C}}{}_{p}^{a} \sigma^{\nu }\left(\lambda ^A\right)_a^b d_{_\mathbb{C}}^{\dagger}{}_{rb}\right) H_{i}D^{\mu } H_{j}  \vspace{2ex}\\
		
		\mathcal{O}_{W u_{_\mathbb{C}} d_{_\mathbb{C}}^{\dagger} H^2 D}^{(1,2)}
		& \epsilon ^{jk}\left(\tau ^I\right)_k^i W^{I}_{\mu\nu } \left(u_{_\mathbb{C}}{}_{p}^{a} \sigma^{\nu } d_{_\mathbb{C}}^{\dagger}{}_{ra}\right) H_{i} D^{\mu } H_{j} ,\quad  \epsilon ^{jk} \left(\tau ^I\right)_k^i\tilde{W}^{I}_{\mu \nu } \left(u_{_\mathbb{C}}{}_{p}^{a} \sigma^{\nu } d_{_\mathbb{C}}^{\dagger}{}_{ra}\right) H_{i} D^{\mu } H_{j} \vspace{2ex}\\
		
		\mathcal{O}_{B u_{_\mathbb{C}} d_{_\mathbb{C}}^{\dagger} H^2 D }^{(1,2)}
		& \epsilon ^{ij}B_{\mu\nu } \left(u_{_\mathbb{C}}{}_{p}^{a} \sigma^{\nu } d_{_\mathbb{C}}^{\dagger}{}_{ra}\right) H_{i}D^{\mu } H_{j} ,\quad  \epsilon ^{ij}\tilde{B}_{\mu\nu } \left(u_{_\mathbb{C}}{}_{p}^{a} \sigma^{\nu } d_{_\mathbb{C}}^{\dagger}{}_{ra}\right) H_{i}D^{\mu } H_{j} .
		
	\end{array}\label{cl:Fpph2d2}\end{align}
These operators involve new Lorentz structures that were absent at lower dimensions.
The conversion to the four spinor notation for the fermion currents can be found in eq.~(\ref{eq:pgp},\ref{eq:Fierzq}).

\underline{Class $F\psi^2\phi D^2$}: There are 2 subclasses of this form. One is $F_{\rm L}\psi^2\phi D^2$, a dimension 6 class $F_{\rm L}\psi^2\phi$ with two additional derivatives, which has 2 independent Lorentz structures:
\begin{align}
	
	F_{\rm{L}1}{}{}^{\alpha }{}{}^{\beta }\psi _2{}{}^{\gamma }  \left(D \psi _3\right){}_{\alpha }{}_{\beta }{}_{\dot{\alpha }} \left(D \phi _4\right){}_{\gamma }{}{}^{\dot{\alpha }},\quad 
	F_{\rm{L}1}{}{}^{\alpha }{}{}^{\beta }\psi _2{}_{\alpha } \left(D \psi _3\right){}^{\gamma}_{\beta \dot{\alpha }} \left(D \phi _4\right)_{\gamma}^{ \dot{\alpha }} .
\end{align}
The other subclass is $F_{\rm R}\psi^2\phi D^2$, where the flip of helicity for the gauge boson is made possible by the presence of the two additional derivatives. The Lorentz structure of this subclass is unique:
\eq{ 
	\psi _1{}{}^{\alpha } \psi _2{}{}^{\beta } \left(D^2 \phi _3\right){}{}_{\dot{\alpha }}{}{}_{\dot{\beta }}{}{}_{\alpha }{}{}_{\beta }F_{\rm{R}4}{}{}^{\dot{\alpha }}{}{}^{\dot{\beta }} .
}
Converting to the $F,\tilde{F}$ basis, these two subclasses mix together. Below we present the operators of this class in the SMEFT, which are all complex types:
\begin{align}\begin{array}{c|l}
		\multirow{2}*{$\mathcal{O}_{G Q u_{_\mathbb{C}} H D^2}^{(1\sim3)} $}
		& \epsilon ^{ij}G^{A}_{\mu \lambda }\left(Q_{pa i} \sigma^{\nu }{}^{\lambda } \left(\lambda ^A\right)_b^a u_{_\mathbb{C}}{}_{r}^{b}\right) D^{\mu } D_{\nu } H_{j} ,\\
		& \epsilon ^{ij}\tilde{G}^{A}_{\mu \lambda }\left(Q_{pa i} \sigma^{\nu }{}^{\lambda } \left(\lambda ^A\right)_b^a u_{_\mathbb{C}}{}_{r}^{b}\right) D^{\mu } D_{\nu } H_{j} ,\quad \epsilon ^{ij}G^{A}_{\mu \nu } \left(Q_{pa i} \left(\lambda ^A\right)_b^a D^{\mu } u_{_\mathbb{C}}{}_{r}^{b}\right) D^{\nu } H_{j} \vspace{2ex}\\
		
		\multirow{2}*{$\mathcal{O}_{W Q u_{_\mathbb{C}} H D^2}^{(1\sim3)} $}
		& \epsilon ^{ik}\left(\tau ^I\right)_k^j W^{I}_{\mu \lambda } \left(Q_{pa i} \sigma^{\nu }{}^{\lambda } u_{_\mathbb{C}}{}_{r}^{a}\right) D^{\mu } D_{\nu } H_{j} ,\\
		& \epsilon ^{ik}\left(\tau ^I\right)_k^j \tilde{W}^{I}_{\mu \lambda } \left(Q_{pa i} \sigma^{\nu }{}^{\lambda } u_{_\mathbb{C}}{}_{r}^{a}\right) D^{\mu } D_{\nu } H_{j} ,\quad \epsilon ^{jk}\left(\tau ^I\right)_k^i W^{I}_{\mu \nu } \left(Q_{pa i} D^{\mu } u_{_\mathbb{C}}{}_{r}^{a}\right) D^{\nu } H_{j} \vspace{2ex}\\
		
		\multirow{2}*{$\mathcal{O}_{B Q u_{_\mathbb{C}} H D^2}^{(1\sim3)} $}
		& \epsilon ^{ij}B_{\mu \lambda } \left(Q_{pa i} \sigma^{\nu }{}^{\lambda } u_{_\mathbb{C}}{}_{r}^{a}\right) D^{\mu } D_{\nu } H_{j} ,\\
		& \epsilon ^{ij}\tilde{B}_{\mu \lambda } \left(Q_{pa i} \sigma^{\nu }{}^{\lambda } u_{_\mathbb{C}}{}_{r}^{a}\right) D^{\mu } D_{\nu } H_{j} ,\quad \epsilon ^{ij} B_{\mu \nu } \left(Q_{pa i} D^{\mu } u_{_\mathbb{C}}{}_{r}^{a}\right) D^{\nu } H_{j} \vspace{2ex}\\
		
		\multirow{2}*{$\mathcal{O}_{G Q d_{_\mathbb{C}} H^{\dagger}  D^2}^{(1\sim3)} $}
		& G^{A}_{\mu \lambda }\left(d_{_\mathbb{C}}{}_{p}^{a} \sigma^{\nu }{}^{\lambda } \left(\lambda ^A\right)_a^b Q_{rb i}\right) D^{\mu } D_{\nu } H^{\dagger}{}^{i} ,\\
		& \tilde{G}^{A}_{\mu \lambda }\left(d_{_\mathbb{C}}{}_{p}^{a} \sigma^{\nu }{}^{\lambda } \left(\lambda ^A\right)_a^b Q_{rb i}\right) D^{\mu } D_{\nu } H^{\dagger}{}^{i} ,\quad G^{A}_{\mu \nu }\left(d_{_\mathbb{C}}{}_{p}^{a} \left(\lambda ^A\right)_a^bD^{\mu } Q_{rb i}\right) D^{\nu } H^{\dagger}{}^{i}  \vspace{2ex}\\
		
		\multirow{2}*{$\mathcal{O}_{W Q d_{_\mathbb{C}} H^{\dagger}  D^2}^{(1\sim3)} $}
		& \left(\tau ^I\right)_j^iW^{I}_{\mu \lambda } \left(d_{_\mathbb{C}}{}_{p}^{a} \sigma^{\nu }{}^{\lambda } Q_{ra i}\right) D^{\mu } D_{\nu } H^{\dagger}{}^{j},\\
		& \left(\tau ^I\right)_j^i\tilde{W}^{I}_{\mu \lambda } \left(d_{_\mathbb{C}}{}_{p}^{a} \sigma^{\nu }{}^{\lambda } Q_{ra i}\right) D^{\mu } D_{\nu } H^{\dagger}{}^{j},\quad \left(\tau ^I\right)_j^iW^{I}_{\mu \nu } \left(d_{_\mathbb{C}}{}_{p}^{a} D^{\mu } Q_{ra i}\right) D^{\nu } H^{\dagger}{}^{j} \vspace{2ex}\\
		
		\multirow{2}*{$\mathcal{O}_{B Q d_{_\mathbb{C}} H^{\dagger}  D^2}^{(1\sim3)} $}
		& B_{\mu }{}_{\lambda } \left(d_{_\mathbb{C}}{}_{p}^{a} \sigma^{\nu }{}^{\lambda } Q_{ra i}\right) D^{\mu } D_{\nu } H^{\dagger}{}^{i},\\
		& \tilde{B}_{\mu }{}_{\lambda } \left(d_{_\mathbb{C}}{}_{p}^{a} \sigma^{\nu }{}^{\lambda } Q_{ra i}\right) D^{\mu } D_{\nu } H^{\dagger}{}^{i},\quad B_{\mu }{}_{\nu } \left(d_{_\mathbb{C}}{}_{p}^{a} D^{\mu } Q_{ra i}\right) D^{\nu } H^{\dagger}{}^{i} \vspace{2ex}\\
		
		\multirow{2}*{$\mathcal{O}_{W e_{_\mathbb{C}} L H^{\dagger}  D^2}^{(1\sim3)} $}
		& \left(\tau ^I\right)_j^iW^{I}_{\mu \lambda } \left(e_{_\mathbb{C} p} \sigma^{\nu }{}^{\lambda } L_{ri}\right)D^{\mu } D_{\nu } H^{\dagger}{}^{j} ,\\
		& \left(\tau ^I\right)_j^i\tilde{W}^{I}_{\mu \lambda } \left(e_{_\mathbb{C} p} \sigma^{\nu }{}^{\lambda } L_{ri}\right)D^{\mu } D_{\nu } H^{\dagger}{}^{j} ,\quad \left(\tau ^I\right)_j^iW^{I}_{\mu \nu } \left(e_{_\mathbb{C} p} D^{\mu } L_{ri}\right) D^{\nu } H^{\dagger}{}^{j} \vspace{2ex}\\
		
		\multirow{2}*{$\mathcal{O}_{B e_{_\mathbb{C}} L H^{\dagger}  D^2}^{(1\sim3)} $}
		& B_{\mu }{}_{\lambda } \left(e_{_\mathbb{C} p} \sigma^{\nu }{}^{\lambda } L_{ri}\right) D^{\mu } D_{\nu } H^{\dagger}{}^{i} ,\\
		& \tilde{B}_{\mu }{}_{\lambda } \left(e_{_\mathbb{C} p} \sigma^{\nu }{}^{\lambda } L_{ri}\right) D^{\mu } D_{\nu } H^{\dagger}{}^{i} ,\quad B_{\mu }{}_{\nu } \left(e_{_\mathbb{C} p} D^{\mu } L_{ri}\right)D^{\nu } H^{\dagger}{}^{i} 
	\end{array}\label{cl:Fpphd2}\end{align}
The Lorentz structures here are also new at dimension 8. To convert to four-component spinor notation, one may refer to eq.~(\ref{eq:pggp},\ref{eq:pp}).

\subsubsection{Two gauge boson involved}

\noindent\underline{CLass $F^2\psi^2\phi$}: Two subclasses are involved, with same of opposite helicities for the gauge bosons and fermions. For the subclass $F_{\rm L}^2\psi^2\phi$, we obtained 2 independent Lorentz structures:
\begin{align}
	F_{\rm{L}1}{}{}^{\alpha }{}{}^{\beta } F_{\rm{L}2}{}_{\alpha }{}{}^{\gamma }\psi _3{}_{\beta } \psi _4{}_{\gamma }\phi _5  ,\quad 
	F_{\rm{L}1}{}{}^{\alpha }{}{}^{\beta }F_{\rm{L}2}{}_{\alpha }{}_{\beta } \psi _3{}{}^{\gamma } \psi _4{}_{\gamma } \phi _5 , \label{eq:lst_FL2psi2phi}
\end{align}
while for $F_{\rm R}^2\psi^2\phi$ we have only 1 independent Lorentz structure:
\eq{
	\psi_1{}{}^{\alpha}\psi_2{}_{\alpha}\phi _3 F_{\rm{R}4}{}_{\dot\alpha\dot\beta } F_{\rm{R}5}{}^{\dot\alpha\dot\beta } .\label{eq:lst_FR2psi2phi}
}
After converting to the $F,\tilde{F}$ basis, the second in eq.~\eqref{eq:lst_FL2psi2phi} and the one in eq.~\eqref{eq:lst_FR2psi2phi} combine to the form as products of a Yukawa coupling and a gauge kinetic term, while the first in eq.~\eqref{eq:lst_FL2psi2phi} is a distinct one. The types of this class in the SMEFT can be found by adding two gauge bosons to the Yukawa terms, which are all complex:
\begin{align}\begin{array}{c|l}
		\multirow{3}*{$\mathcal{O}_{G^2 Q u_{_\mathbb{C}} H}^{(1\sim5)} $}
		&\epsilon ^{ij} d^{ABC}(G^AG^B) \left(Q_{pai} \left(\lambda ^C\right)_b^a u_{_\mathbb{C}}{}_{r}^{b}\right) H_{j} ,\quad \epsilon ^{ij}G^2 \left(Q_{pa i} u_{_\mathbb{C}}{}_{r}^{a}\right) H_{j} ,\\
		& \epsilon ^{ij}d^{ABC}(G^A\tilde{G}^B) \left(Q_{pa i}\left(\lambda ^C\right)_b^a u_{_\mathbb{C}}{}_{r}^{b}\right) H_{j} ,\quad \epsilon ^{ij}(G^A\tilde{G}^A) \left(Q_{pa i} u_{_\mathbb{C}}{}_{r}^{a}\right) H_{j},\\
		& \epsilon ^{ij} f^{ABC}G^{A}_{\mu \nu } G^{B}{}^{\mu }{}{}_{\lambda } \left(Q_{pa i} \sigma^{\nu }{}^{\lambda }\left(\lambda ^C\right)_b^a u_{_\mathbb{C}}{}_{r}^{b}\right) H_{j} \vspace{2ex}\\
		
		\multirow{2}*{$\mathcal{O}_{W^2 Q  u_{_\mathbb{C}} H}^{(1\sim3)}  $}
		&\epsilon ^{ij} W^2 \left(Q_{pa i} u_{_\mathbb{C}}{}_{r}^{a}\right)H_{j} ,\quad \epsilon ^{ij} (W^I\tilde{W}^I) \left(Q_{pa i} u_{_\mathbb{C}}{}_{r}^{a}\right)H_{j} ,\\
		&\left(\tau ^K\right)_k^i \epsilon ^{IJK} \epsilon ^{jk} W^{I}_{\mu \nu } W^{J}{}^{\mu }{}{}_{\lambda } \left(Q_{pa i} \sigma^{\nu }{}^{\lambda } u_{_\mathbb{C}}{}_{r}^{a}\right) H_{j}\vspace{2ex}\\
		
		\multirow{2}*{$\mathcal{O}_{G W Q u_{_\mathbb{C}} H}^{(1\sim3)} $} 
		&\epsilon ^{jk} \left(\tau ^I\right)_k^i (G^AW^I) \left(Q_{pa i}\left(\lambda ^A\right)_b^a u_{_\mathbb{C}}{}_{r}^{b}\right) H_{j} ,\\
		&\epsilon ^{jk} \left(\tau ^I\right)_k^i (G^A\tilde{W}^I) \left(Q_{pa i}\left(\lambda ^A\right)_b^a u_{_\mathbb{C}}{}_{r}^{b}\right) H_{j} ,\quad \epsilon ^{jk}\left(\tau ^I\right)_k^i W^{I}{}^{\mu }{}{}_{\lambda } G^{A}_{\mu \nu } \left(Q_{pa i} \sigma^{\nu }{}^{\lambda }\left(\lambda ^A\right)_b^a u_{_\mathbb{C}}{}_{r}^{b}\right) H_{j} \vspace{2ex}\\
		
		\mathcal{O}_{B^2 Q u_{_\mathbb{C}} H   }^{(1,2)}
		&\epsilon ^{ij} B^2\left(Q_{pa i} u_{_\mathbb{C}}{}_{r}^{a}\right)H_{j} ,\quad \epsilon ^{ij} (B\tilde{B})\left(Q_{pa i} u_{_\mathbb{C}}{}_{r}^{a}\right)H_{j} \vspace{2ex}\\
		
		\multirow{2}*{$\mathcal{O}_{B G Q  u_{_\mathbb{C}}  H}^{(1\sim3)} $}
		&\epsilon ^{ij}(BG^A) \left(Q_{pa i} \left(\lambda ^A\right)_b^a u_{_\mathbb{C}}{}_{r}^{b}\right) H_{j} ,\\
		&\epsilon ^{ij}(B\tilde{G}^A) \left(Q_{pa i} \left(\lambda ^A\right)_b^a u_{_\mathbb{C}}{}_{r}^{b}\right) H_{j} ,\quad  \epsilon ^{ij}B_{\mu }{}_{\nu } G^{A}{}^{\mu }{}{}_{\lambda } \left(Q_{pa i} \sigma^{\nu }{}^{\lambda } \left(\lambda ^A\right)_b^a u_{_\mathbb{C}}{}_{r}^{b}\right) H_{j} \vspace{2ex}\\
		
		\multirow{2}*{$\mathcal{O}_{B W Q  u_{_\mathbb{C}}  H}^{(1\sim3)} $}
		&\epsilon ^{jk}\left(\tau ^I\right)_k^i(BW^I) \left(Q_{pa i} u_{_\mathbb{C}}{}_{r}^{a}\right) H_{j}  ,\\
		& \epsilon ^{jk}\left(\tau ^I\right)_k^i(B\tilde{W}^I) \left(Q_{pa i} u_{_\mathbb{C}}{}_{r}^{a}\right) H_{j} ,\quad  \epsilon ^{jk}\left(\tau ^I\right)_k^i W^{I}{}^{\mu }{}{}_{\lambda } B_{\mu }{}_{\nu } \left(Q_{pa i} \sigma^{\nu }{}^{\lambda } u_{_\mathbb{C}}{}_{r}^{a}\right) H_{j} \vspace{2ex}\\
		
		\multirow{3}*{$\mathcal{O}_{G^2 Q d_{_\mathbb{C}}  H^{\dagger}  }^{(1\sim5)} $}
		&d^{ABC} (G^AG^B) \left(d_{_\mathbb{C}}{}_{p}^{a} \left(\lambda ^C\right)_a^b Q_{rb i}\right) H^{\dagger}{}^{i} ,\quad G^2 \left(d_{_\mathbb{C}}{}_{p}^{a} Q_{ra i}\right)H^{\dagger}{}^{i} ,\\
		&d^{ABC} (G^A\tilde{G}^B) \left(d_{_\mathbb{C}}{}_{p}^{a} \left(\lambda ^C\right)_a^b Q_{rb i}\right) H^{\dagger}{}^{i} ,\quad (G^A\tilde{G}^A) \left(d_{_\mathbb{C}}{}_{p}^{a} Q_{ra i}\right)H^{\dagger}{}^{i} ,\\
		& f^{ABC}G^{A}_{\mu \nu } G^{B}{}^{\mu }{}{}_{\lambda } \left(d_{_\mathbb{C}}{}_{p}^{a} \sigma^{\nu }{}^{\lambda }\left(\lambda ^C\right)_a^b Q_{rb i}\right) H^{\dagger}{}^{i} \vspace{2ex}\\
		
		\multirow{2}*{$\mathcal{O}_{W^2 Q d_{_\mathbb{C}}   H^{\dagger}}^{(1\sim3)}  $}
		&W^2\left(d_{_\mathbb{C}}{}_{p}^{a} Q_{ra i}\right)H^{\dagger}{}^{i} ,\\
		&(W^I\tilde{W}^I)\left(d_{_\mathbb{C}}{}_{p}^{a} Q_{ra i}\right)H^{\dagger}{}^{i} ,\quad  \left(\tau ^K\right)_j^i \epsilon ^{IJK}W^{I}_{\mu \nu } W^{J}{}^{\mu }{}_{\lambda } \left(d_{_\mathbb{C}}{}_{p}^{a} \sigma^{\nu }{}^{\lambda } Q_{ra i}\right) H^{\dagger}{}^{j} \vspace{2ex}\\
		
		\multirow{2}*{$\mathcal{O}_{G W Q d_{_\mathbb{C}}  H^{\dagger}}^{(1\sim3)} $}
		&\left(\tau ^I\right)_j^i (W^IG^A) \left(d_{_\mathbb{C}}{}_{p}^{a}\left(\lambda ^A\right)_a^b Q_{rb i}\right) H^{\dagger}{}^{j} ,\\
		&\left(\tau ^I\right)_j^i (W^I\tilde{G}^A) \left(d_{_\mathbb{C}}{}_{p}^{a}\left(\lambda ^A\right)_a^b Q_{rb i}\right) H^{\dagger}{}^{j} ,\quad  \left(\tau ^I\right)_j^i W^{I}{}^{\mu }{}{}_{\lambda }G^{A}_{\mu \nu } \left(d_{_\mathbb{C}}{}_{p}^{a} \sigma^{\nu }{}^{\lambda } \left(\lambda ^A\right)_a^b Q_{rb i}\right) H^{\dagger}{}^{j} \vspace{2ex}
	\end{array}\label{cl:F2pph1}\end{align}
\begin{align}\begin{array}{c|l}
		\mathcal{O}_{B^2 Q d_{_\mathbb{C}}   H^{\dagger}}^{(1,2))}
		&B^2\left(d_{_\mathbb{C}}{}_{p}^{a} Q_{ra i}\right)H^{\dagger}{}^{i} ,\quad (B\tilde{B})\left(d_{_\mathbb{C}}{}_{p}^{a} Q_{ra i}\right)H^{\dagger}{}^{i} \vspace{2ex}\\
		
		\multirow{2}*{$\mathcal{O}_{B G Q d_{_\mathbb{C}} H^{\dagger}}^{(1\sim3)} $}
		&(BG^A) \left(d_{_\mathbb{C}}{}_{p}^{a} \left(\lambda ^A\right)_a^b Q_{rb i}\right)H^{\dagger}{}^{i} ,\\
		&(B\tilde{G}^A) \left(d_{_\mathbb{C}}{}_{p}^{a} \left(\lambda ^A\right)_a^b Q_{rb i}\right)H^{\dagger}{}^{i} ,\quad  B_{\mu }{}_{\nu } G^{A}{}^{\mu }{}{}_{\lambda } \left(d_{_\mathbb{C}}{}_{p}^{a} \sigma^{\nu }{}^{\lambda } \left(\lambda ^A\right)_a^b Q_{rb i}\right) H^{\dagger}{}^{i} \vspace{2ex}\\
		
		\multirow{2}*{$\mathcal{O}_{B W Q d_{_\mathbb{C}}  H^{\dagger}}^{(1\sim3)} $}
		&\left(\tau ^I\right)_j^i (BW^I)\left(d_{_\mathbb{C}}{}_{p}^{a} Q_{ra i}\right)H^{\dagger}{}^{j} ,\\
		&\left(\tau ^I\right)_j^i (B\tilde{W}^I)\left(d_{_\mathbb{C}}{}_{p}^{a} Q_{ra i}\right)H^{\dagger}{}^{j} ,\quad  \left(\tau ^I\right)_j^iW^{I}{}^{\mu }{}{}_{\lambda } B_{\mu }{}_{\nu } \left(d_{_\mathbb{C}}{}_{p}^{a} \sigma^{\nu }{}^{\lambda } Q_{ra i}\right) H^{\dagger}{}^{j} \vspace{2ex}\\
		
		\mathcal{O}_{G^2 e_{_\mathbb{C}}  L  H^{\dagger}}^{(1,2)}
		&G^2\left(e_{_\mathbb{C} p} L_{ri}\right) H^{\dagger}{}^{i} ,\quad (G^A\tilde{G}^A)\left(e_{_\mathbb{C} p} L_{ri}\right) H^{\dagger}{}^{i} \vspace{2ex}\\
		
		\multirow{2}*{$\mathcal{O}_{W^2 e_{_\mathbb{C}}  L  H^{\dagger}}^{(1\sim3)}  $}
		& W^2\left(e_{_\mathbb{C} p} L_{ri}\right)H^{\dagger}{}^{i} ,\\
		& (W^I\tilde{W}^I)\left(e_{_\mathbb{C} p} L_{ri}\right)H^{\dagger}{}^{i} ,\quad   \left(\tau ^K\right)_j^i \epsilon ^{IJK}W^{I}_{\mu \nu } W^{J}{}^{\mu }{}{}_{\lambda } \left(e_{_\mathbb{C} p} \sigma^{\nu }{}^{\lambda } L_{ri}\right) H^{\dagger}{}^{j} \vspace{2ex}\\
		
		\mathcal{O}_{B^2 e_{_\mathbb{C}}  L  H^{\dagger} }^{(1,2)}
		&B^2 \left(e_{_\mathbb{C} p} L_{ri}\right)H^{\dagger}{}^{i} ,\quad (B\tilde{B}) \left(e_{_\mathbb{C} p} L_{ri}\right)H^{\dagger}{}^{i} \vspace{2ex}\\
		
		\multirow{2}*{$\mathcal{O}_{B W e_{_\mathbb{C}}  L  H^{\dagger}}^{(1\sim3)} $}
		&\left(\tau ^I\right)_j^i (BW^I) \left(e_{_\mathbb{C} p} L_{ri}\right) H^{\dagger}{}^{j} ,\\
		&\left(\tau ^I\right)_j^i (B\tilde{W}^I) \left(e_{_\mathbb{C} p} L_{ri}\right) H^{\dagger}{}^{j} ,\quad  \left(\tau ^I\right)_j^iW^{I}{}^{\mu }{}{}_{\lambda } B_{\mu }{}_{\nu } \left(e_{_\mathbb{C} p} \sigma^{\nu }{}^{\lambda } L_{ri}\right) H^{\dagger}{}^{j} .
		
	\end{array}\label{cl:F2pph2}\end{align}
Conversion to four-component spinor notation in this class can be found in eq.~(\ref{eq:pp},\ref{eq:pggp}).

\noindent\underline{Class $F^2\psi^2D$}: The gauge bosons can have the same or opposite helicities, leading to two subclasses $F_{\rm L}F_{\rm L/R}\psi\psi^\dagger D$, each of which contains only 1 independent Lorentz structure:
\begin{align}
	F_{\rm{L}1}{}^{\alpha\beta } F_{\rm{L}2}{}_{\alpha}{}^{\gamma } \left(D \psi _3\right)_{\beta\gamma\dot{\alpha}}\psi^{\dagger}_4{}^{\dot{\alpha }} ,\quad 
	F_{\rm{L}1}{}^{\alpha }{}^{\beta } \psi_2{}_{\alpha } \left(D \psi^{\dagger}_3\right){}_{\beta\dot\alpha\dot{\beta }} F_{\rm{R}4}{}{}^{\dot{\alpha }\dot{\beta }}.
\end{align}
Without other fields carrying hypercharges, the fermions in this class have to form a neutral current, which demands the types in the SMEFT to be all real:
\begin{align}\begin{array}{c|l}
		\multirow{3}*{$\mathcal{O}_{G^2 Q Q^{\dagger}    D}^{(1\sim5)} $}
		&i f^{ABC} G^{A}{}^{\mu }{}{}_{\nu } G^{B}{}^{\nu }{}{}_{\lambda } \left(Q_{pa i} \sigma^{\lambda }\left(\lambda ^C\right)_b^a \overleftrightarrow{D}_{\mu } Q^{\dagger}{}_{r}^{b i}\right),\quad i d^{ABC} G^{A}{}^{\mu }{}{}_{\nu } G^{B}{}^{\nu }{}{}_{\lambda } \left(Q_{pa i} \sigma^{\lambda } \left(\lambda ^C\right)_b^a \overleftrightarrow{D}_{\mu } Q^{\dagger}{}_{r}^{b i}\right)
		\\&i f^{ABC} \tilde{G}^{A}{}^{\mu }{}{}_{\nu } G^{B}{}^{\nu }{}{}_{\lambda } \left(Q_{pa i} \sigma^{\lambda }\left(\lambda ^C\right)_b^a \overleftrightarrow{D}_{\mu } Q^{\dagger}{}_{r}^{b i}\right),\quad i G^{A}{}^{\mu }{}{}_{\nu } G^{A}{}^{\nu }{}{}_{\lambda } \left(Q_{pa i} \sigma^{\lambda } \overleftrightarrow{D}_{\mu } Q^{\dagger}{}_{r}^{a i}\right),\\    
		&i f^{ABC} G^{A}{}^{\mu }{}{}_{\nu } \tilde{G}^{B}{}^{\nu }{}{}_{\lambda } \left(Q_{pa i} \sigma^{\lambda }\left(\lambda ^C\right)_b^a \overleftrightarrow{D}_{\mu } Q^{\dagger}{}_{r}^{b i}\right)\vspace{2ex}\\
		
		\multirow{2}*{$\mathcal{O}_{W^2 Q Q^{\dagger}    D}^{(1\sim4)} $}
		&i \epsilon ^{IJK} W^{I}{}^{\mu }{}{}_{\nu } W^{J}{}^{\nu }{}{}_{\lambda } \left(Q_{pa i} \sigma^{\lambda } \left(\tau ^K\right)_j^i\overleftrightarrow{D}_{\mu } Q^{\dagger}{}_{r}^{a j}\right),\quad i W^{I}{}^{\mu }{}{}_{\nu } W^{I}{}^{\nu }{}{}_{\lambda } \left(Q_{pa i} \sigma^{\lambda } \overleftrightarrow{D}_{\mu } Q^{\dagger}{}_{r}^{a i}\right) ,\\
		&i \epsilon ^{IJK} \tilde{W}^{I}{}^{\mu }{}{}_{\nu } W^{J}{}^{\nu }{}{}_{\lambda } \left(Q_{pa i} \sigma^{\lambda } \left(\tau ^K\right)_j^i\overleftrightarrow{D}_{\mu } Q^{\dagger}{}_{r}^{a j}\right),\quad i \epsilon ^{IJK} W^{I}{}^{\mu }{}{}_{\nu } \tilde{W}^{J}{}^{\nu }{}{}_{\lambda } \left(Q_{pa i} \sigma^{\lambda } \left(\tau ^K\right)_j^i\overleftrightarrow{D}_{\mu } Q^{\dagger}{}_{r}^{a j}\right)\vspace{2ex}\\
		
		\mathcal{O}_{B^2 Q Q^{\dagger}    D }
		&i B^{\mu }{}{}_{\nu } B^{\nu }{}{}_{\lambda } \left(Q_{pa i} \sigma^{\lambda } \overleftrightarrow{D}_{\mu } Q^{\dagger}{}_{r}^{a i}\right)\vspace{2ex}\\
		
		\multirow{3}*{$\mathcal{O}_{G^2 u_{_\mathbb{C}} u_{_\mathbb{C}}^{\dagger}    D}^{(1\sim5)} $}
		& i f^{ABC} G^{A}{}^{\mu }{}{}_{\nu } G^{B}{}^{\nu }{}{}_{\lambda } \left(u_{_\mathbb{C}}{}_{p}^{a} \sigma^{\lambda } \left(\lambda ^C\right)_a^b \overleftrightarrow{D}_{\mu } u_{_\mathbb{C}}^{\dagger}{}_{rb}\right),\quad i d^{ABC} G^{A}{}^{\mu }{}{}_{\nu } G^{B}{}^{\nu }{}{}_{\lambda } \left(u_{_\mathbb{C}}{}_{p}^{a} \sigma^{\lambda } \left(\lambda ^C\right)_a^b \overleftrightarrow{D}_{\mu } u_{_\mathbb{C}}^{\dagger}{}_{rb}\right),\\
		& i f^{ABC} \tilde{G}^{A}{}^{\mu }{}{}_{\nu } G^{B}{}^{\nu }{}{}_{\lambda } \left(u_{_\mathbb{C}}{}_{p}^{a} \sigma^{\lambda } \left(\lambda ^C\right)_a^b \overleftrightarrow{D}_{\mu } u_{_\mathbb{C}}^{\dagger}{}_{rb}\right),\quad i G^{A}{}^{\mu }{}{}_{\nu } G^{A}{}^{\nu }{}{}_{\lambda } \left(u_{_\mathbb{C}}{}_{p}^{a} \sigma^{\lambda } \overleftrightarrow{D}_{\mu } u_{_\mathbb{C}}^{\dagger}{}_{ra}\right),\\
		&i f^{ABC} G^{A}{}^{\mu }{}{}_{\nu } \tilde{G}^{B}{}^{\nu }{}{}_{\lambda } \left(u_{_\mathbb{C}}{}_{p}^{a} \sigma^{\lambda } \left(\lambda ^C\right)_a^b \overleftrightarrow{D}_{\mu } u_{_\mathbb{C}}^{\dagger}{}_{rb}\right)\vspace{2ex}\\
		
		\mathcal{O}_{W^2 u_{_\mathbb{C}} u_{_\mathbb{C}}^{\dagger}    D }	
		&i W^{I}{}^{\mu }{}{}_{\nu } W^{I}{}^{\nu }{}{}_{\lambda } \left(u_{_\mathbb{C}}{}_{p}^{a} \sigma^{\lambda } \overleftrightarrow{D}_{\mu } u_{_\mathbb{C}}^{\dagger}{}_{ra}\right)

	\end{array}\label{cl:F2ppdr}\end{align}

\begin{align}\begin{array}{c|l}
		
		\mathcal{O}_{B^2 u_{_\mathbb{C}} u_{_\mathbb{C}}^{\dagger}    D }
		&i B^{\mu }{}{}_{\nu } B^{\nu }{}{}_{\lambda } \left(u_{_\mathbb{C}}{}_{p}^{a} \sigma^{\lambda } \overleftrightarrow{D}_{\mu } u_{_\mathbb{C}}^{\dagger}{}_{ra}\right)\vspace{2ex}\\
		
		\multirow{3}*{$\mathcal{O}_{G^2 d_{_\mathbb{C}} d_{_\mathbb{C}}^{\dagger}    D}^{(1\sim5)} $}
		&i f^{ABC} G^{A}{}^{\mu }{}{}_{\nu } G^{B}{}^{\nu }{}{}_{\lambda } \left(d_{_\mathbb{C}}{}_{p}^{a} \sigma^{\lambda } \left(\lambda ^C\right)_a^b \overleftrightarrow{D}_{\mu } d_{_\mathbb{C}}^{\dagger}{}_{rb}\right),\quad i d^{ABC} G^{A}{}^{\mu }{}{}_{\nu } G^{B}{}^{\nu }{}{}_{\lambda } \left(d_{_\mathbb{C}}{}_{p}^{a} \sigma^{\lambda } \left(\lambda ^C\right)_a^b \overleftrightarrow{D}_{\mu } d_{_\mathbb{C}}^{\dagger}{}_{rb}\right),\\
		&i f^{ABC} \tilde{G}^{A}{}^{\mu }{}{}_{\nu } G^{B}{}^{\nu }{}{}_{\lambda } \left(d_{_\mathbb{C}}{}_{p}^{a} \sigma^{\lambda } \left(\lambda ^C\right)_a^b \overleftrightarrow{D}_{\mu } d_{_\mathbb{C}}^{\dagger}{}_{rb}\right),\quad i G^{A}{}^{\mu }{}{}_{\nu } G^{A}{}^{\nu }{}{}_{\lambda } \left(d_{_\mathbb{C}}{}_{p}^{a} \sigma^{\lambda } \overleftrightarrow{D}_{\mu } d_{_\mathbb{C}}^{\dagger}{}_{ra}\right),\\
		&i f^{ABC} G^{A}{}^{\mu }{}{}_{\nu } \tilde{G}^{B}{}^{\nu }{}{}_{\lambda } \left(d_{_\mathbb{C}}{}_{p}^{a} \sigma^{\lambda } \left(\lambda ^C\right)_a^b \overleftrightarrow{D}_{\mu } d_{_\mathbb{C}}^{\dagger}{}_{rb}\right)\vspace{2ex}\\
		
		\mathcal{O}_{W^2 d_{_\mathbb{C}} d_{_\mathbb{C}}^{\dagger}    D }
		&i W^{I}{}^{\mu }{}{}_{\nu } W^{I}{}^{\nu }{}{}_{\lambda } \left(d_{_\mathbb{C}}{}_{p}^{a} \sigma^{\lambda } \overleftrightarrow{D}_{\mu } d_{_\mathbb{C}}^{\dagger}{}_{ra}\right)\vspace{2ex}\\
		
		\mathcal{O}_{B^2 d_{_\mathbb{C}} d_{_\mathbb{C}}^{\dagger}    D }
		&i B^{\mu }{}{}_{\nu } B^{\nu }{}{}_{\lambda } \left(d_{_\mathbb{C}}{}_{p}^{a} \sigma^{\lambda } \overleftrightarrow{D}_{\mu } d_{_\mathbb{C}}^{\dagger}{}_{ra}\right)\vspace{2ex}\\
		
		\mathcal{O}_{G^2 L L^{\dagger}    D }
		&i G^{A}{}^{\mu }{}{}_{\nu } G^{A}{}^{\nu }{}{}_{\lambda } \left(L_{pi} \sigma^{\lambda } \overleftrightarrow{D}_{\mu } L^{\dagger}{}_{r}^{i}\right)\vspace{2ex}\\
		
		\multirow{2}*{$\mathcal{O}_{W^2 L L^{\dagger}    D}^{(1\sim4)} $}
		&i \epsilon ^{IJK} W^{I}{}^{\mu }{}{}_{\nu } W^{J}{}^{\nu }{}{}_{\lambda } \left(L_{pi} \sigma^{\lambda } \left(\tau ^K\right)_j^i \overleftrightarrow{D}_{\mu } L^{\dagger}{}_{r}^{j}\right),\quad i W^{I}{}^{\mu }{}{}_{\nu } W^{I}{}^{\nu }{}{}_{\lambda } \left(L_{pi} \sigma^{\lambda } \overleftrightarrow{D}_{\mu } L^{\dagger}{}_{r}^{i}\right),\\
		&i \epsilon ^{IJK} \tilde{W}^{I}{}^{\mu }{}{}_{\nu } W^{J}{}^{\nu }{}{}_{\lambda } \left(L_{pi} \sigma^{\lambda } \left(\tau ^K\right)_j^i \overleftrightarrow{D}_{\mu } L^{\dagger}{}_{r}^{j}\right),\quad i \epsilon ^{IJK} W^{I}{}^{\mu }{}{}_{\nu } \tilde{W}^{J}{}^{\nu }{}{}_{\lambda } \left(L_{pi} \sigma^{\lambda } \left(\tau ^K\right)_j^i \overleftrightarrow{D}_{\mu } L^{\dagger}{}_{r}^{j}\right)\vspace{2ex}\\
		
		\mathcal{O}_{B^2 L L^{\dagger}    D }
		&i B^{\mu }{}{}_{\nu } B^{\nu }{}{}_{\lambda } \left(L_{pi} \sigma^{\lambda } \overleftrightarrow{D}_{\mu } L^{\dagger}{}_{r}^{i}\right)\vspace{2ex}\\
		
		\mathcal{O}_{G^2 e_{_\mathbb{C}} e_{_\mathbb{C}}^{\dagger}    D }
		&i G^{A}{}^{\mu }{}{}_{\nu } G^{A}{}^{\nu }{}{}_{\lambda } \left(e_{_\mathbb{C} p} \sigma^{\lambda } \overleftrightarrow{D}_{\mu } e_{_\mathbb{C}}^{\dagger}{}_{r}\right)\vspace{2ex}\\
		
		\mathcal{O}_{W^2 e_{_\mathbb{C}} e_{_\mathbb{C}}^{\dagger}    D }
		&i W^{I}{}^{\mu }{}{}_{\nu } W^{I}{}^{\nu }{}{}_{\lambda } \left(e_{_\mathbb{C} p} \sigma^{\lambda } \overleftrightarrow{D}_{\mu } e_{_\mathbb{C}}^{\dagger}{}_{r}\right)\vspace{2ex}\\
		
		\mathcal{O}_{B^2 e_{_\mathbb{C}} e_{_\mathbb{C}}^{\dagger}    D }
		&i B^{\mu }{}{}_{\nu } B^{\nu }{}{}_{\lambda } \left(e_{_\mathbb{C} p} \sigma^{\lambda } \overleftrightarrow{D}_{\mu } e_{_\mathbb{C}}^{\dagger}{}_{r}\right)\vspace{2ex}\\
		
		\multirow{2}*{$\mathcal{O}_{G W Q Q^{\dagger}    D}^{(1\sim4)} $}
		&i G^{A\mu}{}_{\lambda}W^{I\nu\lambda} \left(Q_{pa i} \sigma_{\nu } \left(\lambda ^A\right)_b^a \left(\tau ^I\right)_j^i \overleftrightarrow{D}_{\mu}Q^{\dagger}{}_{r}^{b j}\right),\quad i G^{A\mu}{}_{\lambda}\tilde{W}^{I\nu\lambda} \left(Q_{pa i} \sigma_{\nu } \left(\lambda ^A\right)_b^a \left(\tau ^I\right)_j^i \overleftrightarrow{D}_{\mu}Q^{\dagger}{}_{r}^{b j}\right),\\
		&i G^{A\nu}{}_{\lambda}W^{I\mu\lambda} \left(Q_{pa i} \sigma_{\nu } \left(\lambda ^A\right)_b^a \left(\tau ^I\right)_j^i \overleftrightarrow{D}_{\mu}Q^{\dagger}{}_{r}^{b j}\right),\quad i G^{A\nu}{}_{\lambda}\tilde{W}^{I\mu\lambda} \left(Q_{pa i} \sigma_{\nu } \left(\lambda ^A\right)_b^a \left(\tau ^I\right)_j^i \overleftrightarrow{D}_{\mu}Q^{\dagger}{}_{r}^{b j}\right)\vspace{2ex}\\
		
		\multirow{2}*{$\mathcal{O}_{G B Q Q^{\dagger}    D}^{(1\sim4)} $}
		&i G^{A\mu}{}_{\lambda}B^{\nu\lambda} \left(Q_{pa i} \sigma_{\nu }\left(\lambda ^A\right)_b^a \overleftrightarrow{D}_{\mu} Q^{\dagger}{}_{r}^{b i}\right),\quad i G^{A\mu}{}_{\lambda}\tilde{B}^{\nu\lambda} \left(Q_{pa i} \sigma_{\nu }\left(\lambda ^A\right)_b^a \overleftrightarrow{D}_{\mu} Q^{\dagger}{}_{r}^{b i}\right),\\
		&i G^{A\nu}{}_{\lambda}B^{\mu\lambda} \left(Q_{pa i} \sigma_{\nu }\left(\lambda ^A\right)_b^a \overleftrightarrow{D}_{\mu} Q^{\dagger}{}_{r}^{b i}\right),\quad i G^{A\nu}{}_{\lambda}\tilde{B}^{\mu\lambda} \left(Q_{pa i} \sigma_{\nu }\left(\lambda ^A\right)_b^a \overleftrightarrow{D}_{\mu} Q^{\dagger}{}_{r}^{b i}\right) \vspace{2ex}\\
		
		\multirow{2}*{$\mathcal{O}_{W B Q Q^{\dagger}    D}^{(1\sim4)} $}
		&i W^{I\mu}{}_{\lambda}B^{\nu\lambda} \left(Q_{pa i} \sigma_{\nu }\left(\tau ^I\right)_j^i \overleftrightarrow{D}_{\mu} Q^{\dagger}{}_{r}^{a j}\right),\quad i W^{I\mu}{}_{\lambda}\tilde{B}^{\nu\lambda} \left(Q_{pa i} \sigma_{\nu }\left(\tau ^I\right)_j^i \overleftrightarrow{D}_{\mu} Q^{\dagger}{}_{r}^{a j}\right),\\
		&i W^{I\nu}{}_{\lambda}B^{\mu\lambda} \left(Q_{pa i} \sigma_{\nu }\left(\tau ^I\right)_j^i \overleftrightarrow{D}_{\mu} Q^{\dagger}{}_{r}^{a j}\right),\quad i W^{I\nu}{}_{\lambda}\tilde{B}^{\mu\lambda} \left(Q_{pa i} \sigma_{\nu }\left(\tau ^I\right)_j^i \overleftrightarrow{D}_{\mu} Q^{\dagger}{}_{r}^{a j}\right)\vspace{2ex}\\
		
		\multirow{2}*{$\mathcal{O}_{G B u_{_\mathbb{C}} u_{_\mathbb{C}}^{\dagger}    D}^{(1\sim4)} $}
		&i G^{A\mu}{}_{\lambda}B^{\nu\lambda} \left(u_{_\mathbb{C}}{}_{p}^{a} \sigma_{\nu } \left(\lambda ^A\right)_a^b \overleftrightarrow{D}_{\mu}u_{_\mathbb{C}}^{\dagger}{}_{rb}\right),\quad i G^{A\mu}{}_{\lambda}\tilde{B}^{\nu\lambda} \left(u_{_\mathbb{C}}{}_{p}^{a} \sigma_{\nu } \left(\lambda ^A\right)_a^b \overleftrightarrow{D}_{\mu}u_{_\mathbb{C}}^{\dagger}{}_{rb}\right),\\
		&i G^{A\nu}{}_{\lambda}B^{\mu\lambda} \left(u_{_\mathbb{C}}{}_{p}^{a} \sigma_{\nu } \left(\lambda ^A\right)_a^b \overleftrightarrow{D}_{\mu}u_{_\mathbb{C}}^{\dagger}{}_{rb}\right),\quad i G^{A\nu}{}_{\lambda}\tilde{B}^{\mu\lambda} \left(u_{_\mathbb{C}}{}_{p}^{a} \sigma_{\nu } \left(\lambda ^A\right)_a^b \overleftrightarrow{D}_{\mu}u_{_\mathbb{C}}^{\dagger}{}_{rb}\right)\vspace{2ex}\\
		
		\multirow{2}*{$\mathcal{O}_{G B d_{_\mathbb{C}} d_{_\mathbb{C}}^{\dagger}    D}^{(1\sim4)} $}
		&i G^{A\mu}{}_{\lambda}B^{\nu\lambda} \left(d_{_\mathbb{C}}{}_{p}^{a} \sigma_{\nu } \left(\lambda ^A\right)_a^b \overleftrightarrow{D}_{\mu}d_{_\mathbb{C}}^{\dagger}{}_{rb}\right),\quad i G^{A\mu}{}_{\lambda}\tilde{B}^{\nu\lambda} \left(d_{_\mathbb{C}}{}_{p}^{a} \sigma_{\nu } \left(\lambda ^A\right)_a^b \overleftrightarrow{D}_{\mu}d_{_\mathbb{C}}^{\dagger}{}_{rb}\right),\\
		&i G^{A\nu}{}_{\lambda}B^{\mu\lambda} \left(d_{_\mathbb{C}}{}_{p}^{a} \sigma_{\nu } \left(\lambda ^A\right)_a^b \overleftrightarrow{D}_{\mu}d_{_\mathbb{C}}^{\dagger}{}_{rb}\right),\quad i G^{A\nu}{}_{\lambda}\tilde{B}^{\mu\lambda} \left(d_{_\mathbb{C}}{}_{p}^{a} \sigma_{\nu } \left(\lambda ^A\right)_a^b \overleftrightarrow{D}_{\mu}d_{_\mathbb{C}}^{\dagger}{}_{rb}\right)\vspace{2ex}\\
		
		\multirow{2}*{$\mathcal{O}_{W B L L^{\dagger}    D}^{(1\sim4)} $}
		&i W^{I\mu}{}_{\lambda}B^{\nu\lambda} \left(L_{pi} \sigma_{\nu }\left(\tau ^I\right)_j^i \overleftrightarrow{D}_{\mu} L^{\dagger}{}_{r}^{j}\right),\quad i W^{I\mu}{}_{\lambda}\tilde{B}^{\nu\lambda} \left(L_{pi} \sigma_{\nu }\left(\tau ^I\right)_j^i \overleftrightarrow{D}_{\mu} L^{\dagger}{}_{r}^{j}\right),\\
		&i W^{I\nu}{}_{\lambda}B^{\mu\lambda} \left(L_{pi} \sigma_{\nu }\left(\tau ^I\right)_j^i \overleftrightarrow{D}_{\mu} L^{\dagger}{}_{r}^{j}\right),\quad i W^{I\nu}{}_{\lambda}\tilde{B}^{\mu\lambda} \left(L_{pi} \sigma_{\nu }\left(\tau ^I\right)_j^i \overleftrightarrow{D}_{\mu} L^{\dagger}{}_{r}^{j}\right)
		
	\end{array}\label{cl:F2ppdc}\end{align}
Conversion of the relevant fermion currents to 4-component notation can be found in eq.~(\ref{eq:pgp},\ref{eq:Fierzq}).

\subsection{Classes involving Four-fermions}
\label{sec:example3}

The classes of Lorentz structures with four fermions are the most populated in the dimension 8 SMEFT, thus to present in a less dense way, we separate the types in different lists by the number of quarks involved. Those with three quarks and one lepton violate both the baryon number and lepton number $\Delta B = \Delta L = \pm1$, which is the only source of these violations at dimension 8, and consequently $B-L$ is conserved for all the dimension 8 operators. 

Note that repeated fermions start to appear in this section, for which Young symmetrizers are applied to the terms to retain particular flavor symmetries, as explained in section~\ref{sec:motiv} and section~\ref{sec:prev}.

\subsubsection{Two scalars involved}
\underline{Class $\psi^4 \phi^2$}: There are two subclass in this class: $\psi^2\psi^{\dagger 2}\phi^2$ and $\psi^4\phi^2+\hc$, and the independent Lorentz structures are
\bea
\psi_1^{\alpha } \psi _2{}_{\alpha } \phi _3 \phi _4 \psi^{\dagger}_5{}_{\dot{\alpha }} \psi^{\dagger\dot{\alpha}}_6,\quad
\psi _1^{\alpha } \psi _2^{\beta } \psi _3{}_{\alpha } \psi _4{}_{\beta } \phi _5 \phi _6 ,\quad
\psi _1^{\alpha } \psi _{2\alpha } \psi _3^{\beta } \psi _{4\beta } \phi _5 \phi _6.
\eea
With the two scalars taken to be $(H^\dagger H)$, we get the same types as the four-fermion operators at dimension 6 with the additional Higgses. There are new types at dimension 8 with the two scalars taken to be the Higgses with same hypercharges $H^2$ or $H^\dagger{}^2$, whose $SU(2)_W$ indices must be symmetric to avoid the repeated field constraint. This demands at least another pair of $SU(2)_W$ doublets in the four fermions, which excludes the following types that are also Lorentz invariant gauge singlets, but with all four fermions as $SU(2)_W$ singlets:
\eq{
	d_{_\mathbb{C}} d_{_\mathbb{C}}^{\dagger}{}^2 u_{_\mathbb{C}} H^2,\quad
	d_{_\mathbb{C}}^{\dagger}  e_{_\mathbb{C}} e_{_\mathbb{C}}^{\dagger} u_{_\mathbb{C}} H^2 ,\quad
	d_{_\mathbb{C}}^{\dagger}  u_{_\mathbb{C}}{}^2 u_{_\mathbb{C}}^{\dagger}  H^2 ,\quad
	d_{_\mathbb{C}}{}^2 e_{_\mathbb{C}} u_{_\mathbb{C}} H^{\dagger}{}^2 ,\quad
	e_{_\mathbb{C}} u_{_\mathbb{C}}{}^3 H^2.
}
Operators of this class contribute to the four-fermion interactions if the Higgs fields take their vev, and operators involving two or four $L$'s are relevant to the neutrino non-standard interactions.  

\noindent 1. \underline{Operators involving only quarks:}
There are 6 real types from all combinations of the three quark currents:
\begin{align}\begin{array}{c|l}
		
		\multirow{5}*{$  \mathcal{O}_{Q{}^2 Q^{\dagger} {}^2 H H^{\dagger}}^{\left(1\sim10\right)}    $}
		
		&\mathcal{Y}\left[\tiny{\young(pr)},\tiny{\young(st)}\right] \left(Q_{pa i} Q_{rb j}\right) \left(Q^{\dagger}{}_{s}^{a j} Q^{\dagger}{}_{t}^{b k}\right)  H^{\dagger}{}^{i} H_{k}
		
		,\quad\mathcal{Y}\left[\tiny{\young(pr)},\tiny{\young(st)}\right] \left(Q_{pa i} Q_{rb j}\right) \left(Q^{\dagger}{}_{s}^{a j} Q^{\dagger}{}_{t}^{b i}\right) \left(H^{\dagger} H\right)
		
		\\&\mathcal{Y}\left[\tiny{\young(pr)},\tiny{\young(st)}\right] \left(Q_{pa i} Q_{rb j}\right) \left(Q^{\dagger}{}_{s}^{a i} Q^{\dagger}{}_{t}^{b k}\right)  H^{\dagger}{}^{j} H_{k}
		
		,\quad\mathcal{Y}\left[\tiny{\young(pr)},\tiny{\young(s,t)}\right] \left(Q_{pa i} Q_{rb j}\right) \left(Q^{\dagger}{}_{s}^{a j} Q^{\dagger}{}_{t}^{b k}\right)  H^{\dagger}{}^{i} H_{k}
		
		\\&\mathcal{Y}\left[\tiny{\young(pr)},\tiny{\young(s,t)}\right] \left(Q_{pa i} Q_{rb j}\right) \left(Q^{\dagger}{}_{s}^{a i} Q^{\dagger}{}_{t}^{b k}\right)  H^{\dagger}{}^{j} H_{k}
		
		,\quad\mathcal{Y}\left[\tiny{\young(p,r)},\tiny{\young(st)}\right] \left(Q_{pa i} Q_{rb j}\right) \left(Q^{\dagger}{}_{s}^{a j} Q^{\dagger}{}_{t}^{b k}\right)  H^{\dagger}{}^{i} H_{k}
		
		\\&\mathcal{Y}\left[\tiny{\young(p,r)},\tiny{\young(st)}\right] \left(Q_{pa i} Q_{rb j}\right) \left(Q^{\dagger}{}_{s}^{a i} Q^{\dagger}{}_{t}^{b k}\right)  H^{\dagger}{}^{j} H_{k}
		
		,\quad\mathcal{Y}\left[\tiny{\young(p,r)},\tiny{\young(s,t)}\right] \left(Q_{pa i} Q_{rb j}\right) \left(Q^{\dagger}{}_{s}^{a j} Q^{\dagger}{}_{t}^{b k}\right)  H^{\dagger}{}^{i} H_{k}
		
		\\&\mathcal{Y}\left[\tiny{\young(p,r)},\tiny{\young(s,t)}\right] \left(Q_{pa i} Q_{rb j}\right) \left(Q^{\dagger}{}_{s}^{a j} Q^{\dagger}{}_{t}^{b i}\right) \left(H^{\dagger} H\right)
		
		,\quad\mathcal{Y}\left[\tiny{\young(p,r)},\tiny{\young(s,t)}\right] \left(Q_{pa i} Q_{rb j}\right) \left(Q^{\dagger}{}_{s}^{a i} Q^{\dagger}{}_{t}^{b k}\right)  H^{\dagger}{}^{j} H_{k}
		
		\vspace{2ex}\\
		
		\multirow{2}*{$\mathcal{O}_{Q Q^{\dagger}  u_{_\mathbb{C}} u_{_\mathbb{C}}^{\dagger}  H H^{\dagger}}^{\left(1\sim4\right)}      $}
		
		& \left(Q_{pa i} u_{_\mathbb{C}}{}_{r}^{a}\right) \left(Q^{\dagger}{}_{s}^{c j} u_{_\mathbb{C}}^{\dagger}{}_{tc}\right)  H^{\dagger}{}^{i} H_{j}
		
		,\quad \left(Q_{pa i} u_{_\mathbb{C}}{}_{r}^{a}\right) \left(Q^{\dagger}{}_{s}^{c i} u_{_\mathbb{C}}^{\dagger}{}_{tc}\right) \left(H^{\dagger} H\right)
		
		\\& \left(Q_{pa i} u_{_\mathbb{C}}{}_{r}^{b}\right) \left(Q^{\dagger}{}_{s}^{a j} u_{_\mathbb{C}}^{\dagger}{}_{tb}\right)  H^{\dagger}{}^{i} H_{j}
		
		,\quad \left(Q_{pa i} u_{_\mathbb{C}}{}_{r}^{b}\right) \left(Q^{\dagger}{}_{s}^{a i} u_{_\mathbb{C}}^{\dagger}{}_{tb}\right) \left(H^{\dagger} H\right)
		
		\vspace{2ex}\\
		
		\multirow{2}*{$\mathcal{O}_{Q Q^{\dagger}  d_{_\mathbb{C}} d_{_\mathbb{C}}^{\dagger}  H H^{\dagger}}^{\left(1\sim4\right)} $}
		
		& \left(d_{_\mathbb{C}}{}_{p}^{a} Q_{ra i}\right) \left(d_{_\mathbb{C}}^{\dagger}{}_{sc} Q^{\dagger}{}_{t}^{c j}\right)  H^{\dagger}{}^{i} H_{j}
		
		,\quad \left(d_{_\mathbb{C}}{}_{p}^{a} Q_{ra i}\right) \left(d_{_\mathbb{C}}^{\dagger}{}_{sc} Q^{\dagger}{}_{t}^{c i}\right) \left(H^{\dagger} H\right)
		
		\\& \left(d_{_\mathbb{C}}{}_{p}^{a} Q_{rb i}\right) \left(d_{_\mathbb{C}}^{\dagger}{}_{sa} Q^{\dagger}{}_{t}^{b j}\right)  H^{\dagger}{}^{i} H_{j}
		
		,\quad \left(d_{_\mathbb{C}}{}_{p}^{a} Q_{rb i}\right) \left(d_{_\mathbb{C}}^{\dagger}{}_{sa} Q^{\dagger}{}_{t}^{b i}\right) \left(H^{\dagger} H\right)
		
		\vspace{2ex}\\
		
		\multirow{1}*{$\mathcal{O}_{u_{_\mathbb{C}}{}^2 u_{_\mathbb{C}}^{\dagger} {}^2   H H^{\dagger}}^{\left(1,2\right)}  $}
		
		&\mathcal{Y}\left[\tiny{\young(pr)},\tiny{\young(st)}\right] \left(u_{_\mathbb{C}}^{\dagger}{}_{sa} u_{_\mathbb{C}}^{\dagger}{}_{tb}\right) \left(u_{_\mathbb{C}}{}_{p}^{a} u_{_\mathbb{C}}{}_{r}^{b}\right) \left(H^{\dagger} H\right)
		
		,\quad\mathcal{Y}\left[\tiny{\young(p,r)},\tiny{\young(s,t)}\right] \left(u_{_\mathbb{C}}^{\dagger}{}_{sa} u_{_\mathbb{C}}^{\dagger}{}_{tb}\right) \left(u_{_\mathbb{C}}{}_{p}^{a} u_{_\mathbb{C}}{}_{r}^{b}\right) \left(H^{\dagger} H\right)
		
		\vspace{2ex}\\
		
		\multirow{1}*{$\mathcal{O}_{u_{_\mathbb{C}} u_{_\mathbb{C}}^{\dagger}  d_{_\mathbb{C}} d_{_\mathbb{C}}^{\dagger}   H H^{\dagger}}^{\left(1,2\right)}$}
		
		& \left(d_{_\mathbb{C}}^{\dagger}{}_{sb} u_{_\mathbb{C}}^{\dagger}{}_{ta}\right) \left(d_{_\mathbb{C}}{}_{p}^{a} u_{_\mathbb{C}}{}_{r}^{b}\right) \left(H^{\dagger} H\right)
		
		,\quad \left(d_{_\mathbb{C}}^{\dagger}{}_{sa} u_{_\mathbb{C}}^{\dagger}{}_{tb}\right) \left(d_{_\mathbb{C}}{}_{p}^{a} u_{_\mathbb{C}}{}_{r}^{b}\right) \left(H^{\dagger} H\right)
		
		\vspace{2ex}\\
		
		\multirow{1}*{$\mathcal{O}_{d_{_\mathbb{C}}{}^2 d_{_\mathbb{C}}^{\dagger} {}^2 H H^{\dagger}}^{\left(1,2\right)}    $}
		
		&\mathcal{Y}\left[\tiny{\young(pr)},\tiny{\young(st)}\right] \left(d_{_\mathbb{C}}^{\dagger}{}_{sa} d_{_\mathbb{C}}^{\dagger}{}_{tb}\right) \left(d_{_\mathbb{C}}{}_{p}^{a} d_{_\mathbb{C}}{}_{r}^{b}\right) \left(H^{\dagger} H\right)
		
		,\quad\mathcal{Y}\left[\tiny{\young(p,r)},\tiny{\young(s,t)}\right] \left(d_{_\mathbb{C}}^{\dagger}{}_{sa} d_{_\mathbb{C}}^{\dagger}{}_{tb}\right) \left(d_{_\mathbb{C}}{}_{p}^{a} d_{_\mathbb{C}}{}_{r}^{b}\right) \left(H^{\dagger} H\right)
		
	\end{array}\label{cl:q4h2r}\end{align}
An addition of 4 complex types exist:
\begin{align}\begin{array}{c|l}
		
		\multirow{2}*{$ \mathcal{O}_{Q{}^2 u_{_\mathbb{C}}{}^2  H{}^2}^{\left(1\sim4\right)} $}
		
		&\mathcal{Y}\left[\tiny{\young(p,r)},\tiny{\young(s,t)}\right]\epsilon ^{ik} \epsilon ^{jm} \left(Q_{pa i} u_{_\mathbb{C}}{}_{s}^{a}\right) \left(Q_{rb j} u_{_\mathbb{C}}{}_{t}^{b}\right) H_{k} H_{m}
		
		,\quad\mathcal{Y}\left[\tiny{\young(p,r)},\tiny{\young(s,t)}\right]\epsilon ^{ik} \epsilon ^{jm} \left(Q_{pa i} u_{_\mathbb{C}}{}_{s}^{b}\right) \left(Q_{rb j} u_{_\mathbb{C}}{}_{t}^{a}\right) H_{k} H_{m}
		
		\\&\mathcal{Y}\left[\tiny{\young(pr)},\tiny{\young(st)}\right]\epsilon ^{ik} \epsilon ^{jm} \left(Q_{pa i} u_{_\mathbb{C}}{}_{s}^{a}\right) \left(Q_{rb j} u_{_\mathbb{C}}{}_{t}^{b}\right) H_{k} H_{m}
		
		,\quad\mathcal{Y}\left[\tiny{\young(pr)},\tiny{\young(st)}\right]\epsilon ^{ik} \epsilon ^{jm} \left(Q_{pa i} u_{_\mathbb{C}}{}_{s}^{b}\right) \left(Q_{rb j} u_{_\mathbb{C}}{}_{t}^{a}\right) H_{k} H_{m}
		
		\vspace{2ex}\\
		
		\multirow{2}*{$ \mathcal{O}_{Q{}^2  d_{_\mathbb{C}}{}^2  H^{\dagger} {}^2}^{\left(1\sim4\right)}$}
		
		&\mathcal{Y}\left[\tiny{\young(p,r)},\tiny{\young(s,t)}\right] \left(d_{_\mathbb{C}}{}_{p}^{a} Q_{sb i}\right) \left(d_{_\mathbb{C}}{}_{r}^{b} Q_{ta j}\right) H^{\dagger}{}^{i} H^{\dagger}{}^{j}
		
		,\quad\mathcal{Y}\left[\tiny{\young(p,r)},\tiny{\young(s,t)}\right] \left(d_{_\mathbb{C}}{}_{p}^{a} Q_{sa i}\right) \left(d_{_\mathbb{C}}{}_{r}^{b} Q_{tb j}\right) H^{\dagger}{}^{i} H^{\dagger}{}^{j}
		
		\\&\mathcal{Y}\left[\tiny{\young(pr)},\tiny{\young(st)}\right] \left(d_{_\mathbb{C}}{}_{p}^{a} Q_{sb i}\right) \left(d_{_\mathbb{C}}{}_{r}^{b} Q_{ta j}\right) H^{\dagger}{}^{i} H^{\dagger}{}^{j}
		
		,\quad\mathcal{Y}\left[\tiny{\young(pr)},\tiny{\young(st)}\right] \left(d_{_\mathbb{C}}{}_{p}^{a} Q_{sa i}\right) \left(d_{_\mathbb{C}}{}_{r}^{b} Q_{tb j}\right) H^{\dagger}{}^{i} H^{\dagger}{}^{j}
		
		\vspace{2ex}\\
		
		\multirow{1}*{$\mathcal{O}_{Q Q^{\dagger}  u_{_\mathbb{C}}  d_{_\mathbb{C}}^{\dagger}   H{}^2}^{\left(1,2\right)}$}
		
		& \epsilon ^{ik} \left(d_{_\mathbb{C}}^{\dagger}{}_{sb} Q^{\dagger}{}_{t}^{a j}\right) \left(Q_{pa i} u_{_\mathbb{C}}{}_{r}^{b}\right) H_{j} H_{k}
		
		,\quad \epsilon ^{ik} \left(d_{_\mathbb{C}}^{\dagger}{}_{sb} Q^{\dagger}{}_{t}^{b j}\right) \left(Q_{pa i} u_{_\mathbb{C}}{}_{r}^{a}\right) H_{j} H_{k}
		
		\vspace{2ex}\\
		
		\multirow{4}*{$\mathcal{O}_{Q{}^2 u_{_\mathbb{C}}  d_{_\mathbb{C}}  H H^{\dagger}}^{\left(1\sim8\right)}$}
		
		&\mathcal{Y}\left[\tiny{\young(s,t)}\right]\epsilon ^{ik} \left(d_{_\mathbb{C}}{}_{p}^{a} Q_{sa j}\right) \left(Q_{rb i} u_{_\mathbb{C}}{}_{t}^{b}\right)  H^{\dagger}{}^{j} H_{k}
		
		,\quad\mathcal{Y}\left[\tiny{\young(s,t)}\right]\epsilon ^{ik} \left(d_{_\mathbb{C}}{}_{p}^{a} Q_{sb j}\right) \left(Q_{ra i} u_{_\mathbb{C}}{}_{t}^{b}\right)  H^{\dagger}{}^{j} H_{k}
		
		\\&\mathcal{Y}\left[\tiny{\young(s,t)}\right]\epsilon ^{ij} \left(d_{_\mathbb{C}}{}_{p}^{a} Q_{sa j}\right) \left(Q_{rb i} u_{_\mathbb{C}}{}_{t}^{b}\right) \left(H^{\dagger} H\right)
		
		,\quad\mathcal{Y}\left[\tiny{\young(s,t)}\right]\epsilon ^{ij} \left(d_{_\mathbb{C}}{}_{p}^{a} Q_{sb j}\right) \left(Q_{ra i} u_{_\mathbb{C}}{}_{t}^{b}\right) \left(H^{\dagger} H\right)
		
		\\&\mathcal{Y}\left[\tiny{\young(st)}\right]\epsilon ^{ik} \left(d_{_\mathbb{C}}{}_{p}^{a} Q_{sa j}\right) \left(Q_{rb i} u_{_\mathbb{C}}{}_{t}^{b}\right)  H^{\dagger}{}^{j} H_{k}
		
		,\quad\mathcal{Y}\left[\tiny{\young(st)}\right]\epsilon ^{ik} \left(d_{_\mathbb{C}}{}_{p}^{a} Q_{sb j}\right) \left(Q_{ra i} u_{_\mathbb{C}}{}_{t}^{b}\right)  H^{\dagger}{}^{j} H_{k}
		
		\\&\mathcal{Y}\left[\tiny{\young(st)}\right]\epsilon ^{ij} \left(d_{_\mathbb{C}}{}_{p}^{a} Q_{sa j}\right) \left(Q_{rb i} u_{_\mathbb{C}}{}_{t}^{b}\right) \left(H^{\dagger} H\right)
		
		,\quad\mathcal{Y}\left[\tiny{\young(st)}\right]\epsilon ^{ij} \left(d_{_\mathbb{C}}{}_{p}^{a} Q_{sb j}\right) \left(Q_{ra i} u_{_\mathbb{C}}{}_{t}^{b}\right) \left(H^{\dagger} H\right)
		
	\end{array}\label{cl:q4h2c}\end{align}

Recall the defination of Young symmetrizer $\mathcal{Y}$ in section 3.1.2, we can obtain the following relations for type $\mathcal{O}_{Q{}^2 Q^{\dagger} {}^2 H H^{\dagger}}^{\left(1\sim10\right)}$,
\eq{ \label{Y2eg}
	\mathcal{O}_{Q{}^2 Q^{\dagger} {}^2 H H^{\dagger}}^{\left(1\right)} &= \left(Q_{pa i} Q_{rb j}+Q_{ra i} Q_{pb j}\right) \left(Q^{\dagger}{}_{s}^{a j} Q^{\dagger}{}_{t}^{b k}+Q^{\dagger}{}_{t}^{a j} Q^{\dagger}{}_{s}^{b k}\right)  H^{\dagger}{}^{i} H_{k},  \\
	\mathcal{O}_{Q{}^2 Q^{\dagger} {}^2 H H^{\dagger}}^{\left(4\right)} &= \left(Q_{pa i} Q_{rb j}+Q_{ra i} Q_{pb j}\right) \left(Q^{\dagger}{}_{s}^{a j} Q^{\dagger}{}_{t}^{b k}-Q^{\dagger}{}_{t}^{a j} Q^{\dagger}{}_{s}^{b k}\right)  H^{\dagger}{}^{i} H_{k},  \\
	\mathcal{O}_{Q{}^2 Q^{\dagger} {}^2 H H^{\dagger}}^{\left(6\right)} &= \left(Q_{pa i} Q_{rb j}-Q_{ra i} Q_{pb j}\right) \left(Q^{\dagger}{}_{s}^{a j} Q^{\dagger}{}_{t}^{b k}+Q^{\dagger}{}_{t}^{a j} Q^{\dagger}{}_{s}^{b k}\right)  H^{\dagger}{}^{i} H_{k},  \\
	\mathcal{O}_{Q{}^2 Q^{\dagger} {}^2 H H^{\dagger}}^{\left(8\right)} &= \left(Q_{pa i} Q_{rb j}-Q_{ra i} Q_{pb j}\right) \left(Q^{\dagger}{}_{s}^{a j} Q^{\dagger}{}_{t}^{b k}-Q^{\dagger}{}_{t}^{a j} Q^{\dagger}{}_{s}^{b k}\right)  H^{\dagger}{}^{i} H_{k},
}
as an example of how $\mathcal{Y}$'s act on the terms.

The conversion from the two-component spinors to the four-component spinors, with extra transformation via Fierz identity, are shown by the following examples
\bea
\left(Q_{pa i} Q_{rb j}\right) \left(Q^{\dagger}{}_{s}^{a j} Q^{\dagger}{}_{t}^{b k}\right) =& \left(q_{pa i} C q_{rb j}\right) \left(\bar{q}{}_{s}^{a j} C \bar{q} {}_{t}^{b k}\right) &= \dfrac{1}{2} \left(\bar{q}{}_{s}^{a j} \gamma^\mu q_{pa i}\right) \left(\bar{q} {}_{t}^{b k} \gamma_\mu q_{rb j}\right), \nonumber \\
\left(u_{_\mathbb{C}}^{\dagger}{}_{sa} u_{_\mathbb{C}}^{\dagger}{}_{tb}\right) \left(u_{_\mathbb{C}}{}_{p}^{a} u_{_\mathbb{C}}{}_{r}^{b}\right) =& \left(u{}_{sa} C u{}_{tb}\right) \left(\bar{u}{}_{p}^{a} C \bar{u}{}_{r}^{b}\right) &= \dfrac{1}{2} \left(\bar{u}{}_{p}^{a} \gamma^\mu u{}_{sa}\right) \left(\bar{u}{}_{r}^{b} \gamma_\mu u{}_{tb}\right), \nonumber \\
\left(Q_{pa i} u_{_\mathbb{C}}{}_{r}^{a}\right) \left(Q^{\dagger}{}_{s}^{c j} u_{_\mathbb{C}}^{\dagger}{}_{tc}\right) =& \left(\bar{u}{}_{r}^{a} q_{pa i} \right) \left(\bar{q}{}_{s}^{c j} u{}_{tc}\right) &= - \dfrac{1}{2} \left(\bar{u}{}_{r}^{a} \gamma^\mu u{}_{tc}\right) \left(\bar{q}{}_{s}^{c j} \gamma_\mu q_{pa i}\right). \label{H^2}
\eea
The Hermitian conjugate of a non-Hermitian operator of this class is, for example,
\bea
\left[\epsilon ^{ik} \epsilon ^{jm} \left(Q_{pa i} u_{_\mathbb{C}}{}_{s}^{b}\right) \left(Q_{rb j} u_{_\mathbb{C}}{}_{t}^{a}\right) H_{k} H_{m}\right]^\dagger = \epsilon _{ik} \epsilon _{jm} \left(u_{_\mathbb{C}}^\dagger{}_{sb} Q^\dagger{}_{p}^{ai} \right) \left(u_{_\mathbb{C}}^\dagger{}_{ta} Q^\dagger{}_{r}^{bj} \right) H^\dagger{}^{k} H^\dagger{}^{m}.
\eea
Operators involving $d$ quark and leptons of this class can be converted similarly.

\noindent 2. \underline{Operators involving three quarks with $\Delta B=\Delta L=\pm1$:}
All the types with 3 quarks are complex, and there are 7 of them in this class:
\begin{align}\begin{array}{c|l}
		
		\multirow{4}*{$\mathcal{O}_{Q{}^3  L  H H^{\dagger}}^{\left(1\sim7\right)}   $}
		
		&\mathcal{Y}\left[\tiny{\young(rs,t)}\right]\epsilon ^{abc} \epsilon ^{im} \epsilon ^{jn} \left(Q_{ra j} Q_{tc m}\right) \left(L_{pi} Q_{sb k}\right) H^{\dagger}{}^{k} H_{n}
		
		,\quad\mathcal{Y}\left[\tiny{\young(rs,t)}\right]\epsilon ^{abc} \epsilon ^{ik} \epsilon ^{jn} \left(Q_{ra j} Q_{tc m}\right) \left(L_{pi} Q_{sb k}\right) H^{\dagger}{}^{m} H_{n}
		
		\\&\mathcal{Y}\left[\tiny{\young(rs,t)}\right]\epsilon ^{abc} \epsilon ^{ij} \epsilon ^{kn} \left(Q_{ra j} Q_{tc m}\right) \left(L_{pi} Q_{sb k}\right) H^{\dagger}{}^{m} H_{n}
		
		,\quad\mathcal{Y}\left[\tiny{\young(rst)}\right]\epsilon ^{abc} \epsilon ^{im} \epsilon ^{jn} \left(Q_{ra j} Q_{tc m}\right) \left(L_{pi} Q_{sb k}\right)H^{\dagger}{}^{k}  H_{n} 
		
		\\&\mathcal{Y}\left[\tiny{\young(rst)}\right]\epsilon ^{abc} \epsilon ^{ik} \epsilon ^{jn} \left(Q_{ra j} Q_{tc m}\right) \left(L_{pi} Q_{sb k}\right)H^{\dagger}{}^{m}  H_{n} 
		
		,\quad\mathcal{Y}\left[\tiny{\young(r,s,t)}\right]\epsilon ^{abc} \epsilon ^{im} \epsilon ^{jn} \left(Q_{ra j} Q_{tc m}\right) \left(L_{pi} Q_{sb k}\right) H^{\dagger}{}^{k} H_{n} 
		
		\\&\mathcal{Y}\left[\tiny{\young(r,s,t)}\right]\epsilon ^{abc} \epsilon ^{ik} \epsilon ^{jn} \left(Q_{ra j} Q_{tc m}\right) \left(L_{pi} Q_{sb k}\right) H^{\dagger}{}^{m} H_{n}
		
		\vspace{2ex}\\
		
		\multirow{1}*{$\mathcal{O}_{Q{}^2 u_{_\mathbb{C}}^{\dagger} e_{_\mathbb{C}}^{\dagger}   H H^{\dagger}}^{\left(1,2\right)} $}
		
		&\mathcal{Y}\left[\tiny{\young(p,r)}\right]\epsilon ^{abc} \epsilon ^{ik} \left(e_{_\mathbb{C}}^{\dagger}{}_{s} u_{_\mathbb{C}}^{\dagger}{}_{tc}\right) \left(Q_{pa i} Q_{rb j}\right)  H^{\dagger}{}^{j} H_{k}
		
		,\quad\mathcal{Y}\left[\tiny{\young(pr)}\right]\epsilon ^{abc} \epsilon ^{ik} \left(e_{_\mathbb{C}}^{\dagger}{}_{s} u_{_\mathbb{C}}^{\dagger}{}_{tc}\right) \left(Q_{pa i} Q_{rb j}\right)  H^{\dagger}{}^{j} H_{k}
		
		\vspace{2ex}\\
		
		\mathcal{O}_{Q{}^2  d_{_\mathbb{C}}^{\dagger}  e_{_\mathbb{C}}^{\dagger}  H{}^2}
		
		&\mathcal{Y}\left[\tiny{\young(p,r)}\right]\epsilon ^{abc} \epsilon ^{il} \epsilon ^{jk} \left(d_{_\mathbb{C}}^{\dagger}{}_{sc} e_{_\mathbb{C}}^{\dagger}{}_{t}\right) \left(Q_{pa i} Q_{rb j}\right) H_{k} H_{l}
		
		\vspace{2ex}\\
		
		\mathcal{O}_{Q u_{_\mathbb{C}}^{\dagger} {}^2 L  H^{\dagger} {}^2}  
		
		&\mathcal{Y}\left[\tiny{\young(s,t)}\right]\epsilon ^{abc} \left(u_{_\mathbb{C}}^{\dagger}{}_{sb} u_{_\mathbb{C}}^{\dagger}{}_{tc}\right) \left(L_{pi} Q_{ra j}\right) H^{\dagger}{}^{i} H^{\dagger}{}^{j}
		
		\vspace{2ex}\\
		
		\multirow{1}*{$\mathcal{O}_{Q u_{_\mathbb{C}}^{\dagger} d_{_\mathbb{C}}^{\dagger}  L  H H^{\dagger}}^{\left(1,2\right)} $}
		
		& \epsilon ^{abc} \epsilon ^{ij} \left(d_{_\mathbb{C}}^{\dagger}{}_{sb} u_{_\mathbb{C}}^{\dagger}{}_{tc}\right) \left(L_{pi} Q_{ra j}\right) \left(H^{\dagger} H\right)
		
		,\quad \epsilon ^{abc} \epsilon ^{ik} \left(d_{_\mathbb{C}}^{\dagger}{}_{sb} u_{_\mathbb{C}}^{\dagger}{}_{tc}\right) \left(L_{pi} Q_{ra j}\right)  H^{\dagger}{}^{j} H_{k}
		
		\vspace{2ex}\\
		
		\mathcal{O}_{Q d_{_\mathbb{C}}^{\dagger} {}^2  L  H{}^2} 
		
		& \epsilon ^{abc} \epsilon ^{il} \epsilon ^{jk} \left(d_{_\mathbb{C}}^{\dagger}{}_{sb} d_{_\mathbb{C}}^{\dagger}{}_{tc}\right) \left(L_{pi} Q_{ra j}\right) H_{k} H_{l}
		
		\vspace{2ex}\\
		
		\multirow{1}*{$\mathcal{O}_{u_{_\mathbb{C}}{}^2  d_{_\mathbb{C}} e_{_\mathbb{C}} H H^{\dagger}}^{\left(1,2\right)}   $}
		
		&\mathcal{Y}\left[\tiny{\young(st)}\right]\epsilon _{abc} \left(e_{_\mathbb{C} r} u_{_\mathbb{C}}{}_{t}^{c}\right) \left(d_{_\mathbb{C}}{}_{p}^{a} u_{_\mathbb{C}}{}_{s}^{b}\right) \left(H^{\dagger} H\right)
		
		,\quad\mathcal{Y}\left[\tiny{\young(s,t)}\right]\epsilon _{abc} \left(e_{_\mathbb{C} r} u_{_\mathbb{C}}{}_{t}^{c}\right) \left(d_{_\mathbb{C}}{}_{p}^{a} u_{_\mathbb{C}}{}_{s}^{b}\right) \left(H^{\dagger} H\right)
		
	\end{array}\label{cl:q3l1h2}\end{align}

Here are examples in the type $\mathcal{O}_{Q{}^3  L  H H^{\dagger}}^{\left(1\sim7\right)}$ about how the Young symmetrizers $\mathcal{Y}$'s act on the operators:
\eq{ \label{Y3eg}
	\mathcal{O}_{Q{}^3  L  H H^{\dagger}}^{(1)}=&\epsilon^{abc} \epsilon^{im}\epsilon^{jn} \left[\left(Q_{ra j} Q_{tc m}\right) \left(L_{pi} Q_{sb k}\right)+\left(Q_{sa j} Q_{tc m}\right) \left(L_{pi} Q_{rb k}\right)\right] H^{\dagger}{}^{k} H_{n}\\&-\epsilon^{abc} \epsilon^{im}\epsilon^{jn} \left[\left(Q_{ta j} Q_{rc m}\right) \left(L_{pi} Q_{sb k}\right)+\left(Q_{ta j} Q_{sc m}\right) \left(L_{pi} Q_{rb k}\right)\right] H^{\dagger}{}^{k} H_{n},  \\
	\mathcal{O}_{Q{}^3  L  H H^{\dagger}}^{(4)}=&\epsilon ^{abc} \epsilon ^{im} \epsilon ^{jn} \left[\left(Q_{ra j} Q_{tc m}\right) \left(L_{pi} Q_{sb k}\right)+\left(Q_{sa j} Q_{tc m}\right) \left(L_{pi} Q_{rb k}\right)\right] H^{\dagger}{}^{k}  H_{n}\\&+\epsilon ^{abc} \epsilon ^{im} \epsilon ^{jn} \left[\left(Q_{ta j} Q_{rc m}\right) \left(L_{pi} Q_{sb k}\right)+\left(Q_{ra j} Q_{sc m}\right) \left(L_{pi} Q_{tb k}\right)\right] H^{\dagger}{}^{k}  H_{n}\\&+\epsilon ^{abc} \epsilon ^{im} \epsilon ^{jn} \left[\left(Q_{ta j} Q_{sc m}\right) \left(L_{pi} Q_{rb k}\right)+\left(Q_{sa j} Q_{rc m}\right) \left(L_{pi} Q_{tb k}\right)\right] H^{\dagger}{}^{k}  H_{n} , \\
	\mathcal{O}_{Q{}^3  L  H H^{\dagger}}^{(6)}=&\epsilon ^{abc} \epsilon ^{im} \epsilon ^{jn} \left[\left(Q_{ra j} Q_{tc m}\right) \left(L_{pi} Q_{sb k}\right)-\left(Q_{sa j} Q_{tc m}\right) \left(L_{pi} Q_{rb k}\right)\right] H^{\dagger}{}^{k}  H_{n}\\&-\epsilon ^{abc} \epsilon ^{im} \epsilon ^{jn} \left[\left(Q_{ta j} Q_{rc m}\right) \left(L_{pi} Q_{sb k}\right)+\left(Q_{ra j} Q_{sc m}\right) \left(L_{pi} Q_{tb k}\right)\right] H^{\dagger}{}^{k}  H_{n}\\&+\epsilon ^{abc} \epsilon ^{im} \epsilon ^{jn} \left[\left(Q_{ta j} Q_{sc m}\right) \left(L_{pi} Q_{rb k}\right)+\left(Q_{sa j} Q_{rc m}\right) \left(L_{pi} Q_{tb k}\right)\right] H^{\dagger}{}^{k}  H_{n} .
}

\noindent 3. \underline{Operators involving two leptons and two quarks:}
There are 6 real types as combinations of the 3 quark currents and the 2 lepton currents:
\begin{align}\begin{array}{c|l}
		
		\multirow{3}*{$\mathcal{O}_{Q Q^{\dagger}  L L^{\dagger}   H H^{\dagger}}^{\left(1\sim5\right)} $}
		
		& \left(L_{pi} Q_{ra j}\right) \left(L^{\dagger}{}_{s}^{i} Q^{\dagger}{}_{t}^{a k}\right)  H^{\dagger}{}^{j} H_{k}
		
		,\quad \left(L_{pi} Q_{ra j}\right) \left(L^{\dagger}{}_{s}^{k} Q^{\dagger}{}_{t}^{a i}\right)  H^{\dagger}{}^{j} H_{k}
		
		\\& \left(L_{pi} Q_{ra j}\right) \left(L^{\dagger}{}_{s}^{j} Q^{\dagger}{}_{t}^{a k}\right)  H^{\dagger}{}^{i} H_{k}
		
		,\quad \left(L_{pi} Q_{ra j}\right) \left(L^{\dagger}{}_{s}^{j} Q^{\dagger}{}_{t}^{a i}\right)  \left(H^{\dagger} H\right)
		
		\\& \left(L_{pi} Q_{ra j}\right) \left(L^{\dagger}{}_{s}^{k} Q^{\dagger}{}_{t}^{a j}\right)  H^{\dagger}{}^{i} H_{k}
		
		\vspace{2ex}\\
		
		\multirow{1}*{$\mathcal{O}_{Q Q^{\dagger}  e_{_\mathbb{C}} e_{_\mathbb{C}}^{\dagger}   H H^{\dagger}}^{\left(1,2\right)}   $}
		
		& \left(e_{_\mathbb{C} p} Q_{ra i}\right) \left(e_{_\mathbb{C}}^{\dagger}{}_{s} Q^{\dagger}{}_{t}^{a j}\right)  H^{\dagger}{}^{i} H_{j}
		
		,\quad \left(e_{_\mathbb{C} p} Q_{ra i}\right) \left(e_{_\mathbb{C}}^{\dagger}{}_{s} Q^{\dagger}{}_{t}^{a i}\right) \left(H^{\dagger} H\right)
		
		\vspace{2ex}\\
		
		\multirow{1}*{$\mathcal{O}_{u_{_\mathbb{C}} u_{_\mathbb{C}}^{\dagger}  L L^{\dagger}  H H^{\dagger}}^{\left(1,2\right)}   $}
		
		& \left(L_{pi} u_{_\mathbb{C}}{}_{r}^{a}\right) \left(L^{\dagger}{}_{s}^{j} u_{_\mathbb{C}}^{\dagger}{}_{ta}\right)  H^{\dagger}{}^{i} H_{j}
		
		,\quad \left(L_{pi} u_{_\mathbb{C}}{}_{r}^{a}\right) \left(L^{\dagger}{}_{s}^{i} u_{_\mathbb{C}}^{\dagger}{}_{ta}\right) \left(H^{\dagger} H\right)
		
		\vspace{2ex}\\
		
		\mathcal{O}_{u_{_\mathbb{C}} u_{_\mathbb{C}}^{\dagger}  e_{_\mathbb{C}} e_{_\mathbb{C}}^{\dagger}    H H^{\dagger}}
		
		& \left(e_{_\mathbb{C}}^{\dagger}{}_{s} u_{_\mathbb{C}}^{\dagger}{}_{ta}\right) \left(e_{_\mathbb{C} p} u_{_\mathbb{C}}{}_{r}^{a}\right) \left(H^{\dagger} H\right)
		
		\vspace{2ex}\\
		
		\multirow{1}*{$\mathcal{O}_{d_{_\mathbb{C}} d_{_\mathbb{C}}^{\dagger}   L L^{\dagger}   H H^{\dagger}}^{\left(1,2\right)}  $}
		
		& \left(d_{_\mathbb{C}}{}_{p}^{a} L_{ri}\right) \left(d_{_\mathbb{C}}^{\dagger}{}_{sa} L^{\dagger}{}_{t}^{j}\right)  H^{\dagger}{}^{i} H_{j}
		
		,\quad \left(d_{_\mathbb{C}}{}_{p}^{a} L_{ri}\right) \left(d_{_\mathbb{C}}^{\dagger}{}_{sa} L^{\dagger}{}_{t}^{i}\right) \left(H^{\dagger} H\right)
		
		\vspace{2ex}\\
		
		\mathcal{O}_{d_{_\mathbb{C}} d_{_\mathbb{C}}^{\dagger}  e_{_\mathbb{C}} e_{_\mathbb{C}}^{\dagger}  H H^{\dagger}}    
		
		& \left(d_{_\mathbb{C}}^{\dagger}{}_{sa} e_{_\mathbb{C}}^{\dagger}{}_{t}\right) \left(d_{_\mathbb{C}}{}_{p}^{a} e_{_\mathbb{C} r}\right) \left(H^{\dagger} H\right)
		
	\end{array}\label{cl:q2l2h2r}\end{align}
There are also  5 complex types, in which 3 involve repeated Higgses:
\begin{align}\begin{array}{c|l}
		
		\multirow{2}*{$\mathcal{O}_{Q u_{_\mathbb{C}}  L  e_{_\mathbb{C}}  H H^{\dagger}}^{\left(1\sim4\right)} $}
		
		& \epsilon ^{ij}  \left(e_{_\mathbb{C} p} Q_{sa j}\right) \left(L_{ri} u_{_\mathbb{C}}{}_{t}^{a}\right) \left(H^{\dagger} H\right)
		
		,\quad \epsilon ^{ij}  \left(e_{_\mathbb{C} p} L_{ri}\right) \left(Q_{sa j} u_{_\mathbb{C}}{}_{t}^{a}\right) \left(H^{\dagger} H\right)
		
		\\& \epsilon ^{ik}  \left(e_{_\mathbb{C} p} Q_{sa j}\right) \left(L_{ri} u_{_\mathbb{C}}{}_{t}^{a}\right)  H^{\dagger}{}^{j} H_{k}
		
		,\quad \epsilon ^{ik}  \left(e_{_\mathbb{C} p} L_{ri}\right) \left(Q_{sa j} u_{_\mathbb{C}}{}_{t}^{a}\right)  H^{\dagger}{}^{j} H_{k}
		
		\vspace{2ex}\\
		
		\mathcal{O}_{Q u_{_\mathbb{C}} L^{\dagger}  e_{_\mathbb{C}}^{\dagger}  H{}^2}
		
		& \epsilon ^{ik}  \left(e_{_\mathbb{C}}^{\dagger}{}_{s} L^{\dagger}{}_{t}^{j}\right) \left(Q_{pa i} u_{_\mathbb{C}}{}_{r}^{a}\right) H_{j} H_{k}
		
		\vspace{2ex}\\
		
		\multirow{1}*{$\mathcal{O}_{Q  d_{_\mathbb{C}}  L e_{_\mathbb{C}}  H^{\dagger} {}^2}^{\left(1,2\right)} $}
		
		&  \left(d_{_\mathbb{C}}{}_{p}^{a} e_{_\mathbb{C} r}\right) \left(L_{si} Q_{ta j}\right) H^{\dagger}{}^{i} H^{\dagger}{}^{j}
		
		,\quad  \left(e_{_\mathbb{C} r} Q_{ta j}\right) \left(d_{_\mathbb{C}}{}_{p}^{a} L_{si}\right) H^{\dagger}{}^{i} H^{\dagger}{}^{j}
		
		\vspace{2ex}\\
		
		\multirow{1}*{$\mathcal{O}_{Q  d_{_\mathbb{C}} L^{\dagger}  e_{_\mathbb{C}}^{\dagger}  H H^{\dagger}}^{\left(1,2\right)} $}
		
		&  \left(e_{_\mathbb{C}}^{\dagger}{}_{s} L^{\dagger}{}_{t}^{j}\right) \left(d_{_\mathbb{C}}{}_{p}^{a} Q_{ra i}\right)  H^{\dagger}{}^{i} H_{j}
		
		,\quad  \left(e_{_\mathbb{C}}^{\dagger}{}_{s} L^{\dagger}{}_{t}^{i}\right) \left(d_{_\mathbb{C}}{}_{p}^{a} Q_{ra i}\right) \left(H^{\dagger} H\right)
		
		\vspace{2ex}\\
		
		\mathcal{O}_{u_{_\mathbb{C}} d_{_\mathbb{C}}^{\dagger}  L L^{\dagger}  H{}^2}
		
		& \epsilon ^{ik}  \left(d_{_\mathbb{C}}^{\dagger}{}_{sa} L^{\dagger}{}_{t}^{j}\right) \left(L_{pi} u_{_\mathbb{C}}{}_{r}^{a}\right) H_{j} H_{k}
		
	\end{array}\label{cl:q2l2h2c}\end{align}

\noindent 4. \underline{Operators involving only leptons:}
The combinations of the 2 kinds of lepton currents give 3 real types of operators in this class:
\begin{align}\begin{array}{c|l}
		
		\multirow{3}*{$  \mathcal{O}_{L{}^2 L^{\dagger} {}^2  H H^{\dagger}}^{\left(1\sim5\right)} $}
		
		&\mathcal{Y}\left[\tiny{\young(pr)},\tiny{\young(st)}\right] \left(L_{pi} L_{rj}\right) \left(L^{\dagger}{}_{s}^{j} L^{\dagger}{}_{t}^{k}\right)  H^{\dagger}{}^{i} H_{k}
		
		,\quad\mathcal{Y}\left[\tiny{\young(pr)},\tiny{\young(st)}\right] \left(L_{pi} L_{rj}\right) \left(L^{\dagger}{}_{s}^{j} L^{\dagger}{}_{t}^{i}\right) \left(H^{\dagger} H\right)
		
		\\&\mathcal{Y}\left[\tiny{\young(pr)},\tiny{\young(s,t)}\right] \left(L_{pi} L_{rj}\right) \left(L^{\dagger}{}_{s}^{j} L^{\dagger}{}_{t}^{k}\right)  H^{\dagger}{}^{i} H_{k}
		
		,\quad\mathcal{Y}\left[\tiny{\young(p,r)},\tiny{\young(st)}\right] \left(L_{pi} L_{rj}\right) \left(L^{\dagger}{}_{s}^{j} L^{\dagger}{}_{t}^{k}\right)  H^{\dagger}{}^{i} H_{k}
		
		\\&\mathcal{Y}\left[\tiny{\young(p,r)},\tiny{\young(s,t)}\right] \left(L_{pi} L_{rj}\right) \left(L^{\dagger}{}_{s}^{j} L^{\dagger}{}_{t}^{k}\right)  H^{\dagger}{}^{i} H_{k}
		
		\vspace{2ex}\\
		
		\multirow{1}*{$\mathcal{O}_{L L^{\dagger}  e_{_\mathbb{C}} e_{_\mathbb{C}}^{\dagger}   H H^{\dagger}}^{\left(1,2\right)}  $}
		
		&  \left(e_{_\mathbb{C} p} L_{ri}\right) \left(e_{_\mathbb{C}}^{\dagger}{}_{s} L^{\dagger}{}_{t}^{j}\right)  H^{\dagger}{}^{i} H_{j}
		
		,\quad  \left(e_{_\mathbb{C} p} L_{ri}\right) \left(e_{_\mathbb{C}}^{\dagger}{}_{s} L^{\dagger}{}_{t}^{i}\right) \left(H^{\dagger} H\right)
		
		\vspace{2ex}\\
		
		\mathcal{O}_{e_{_\mathbb{C}}{}^2 e_{_\mathbb{C}}^{\dagger} {}^2 H H^{\dagger}}    
		
		&\mathcal{Y}\left[\tiny{\young(pr)},\tiny{\young(st)}\right] \left(e_{_\mathbb{C} p} e_{_\mathbb{C} r}\right) \left(e_{_\mathbb{C}}^{\dagger}{}_{s} e_{_\mathbb{C}}^{\dagger}{}_{t}\right) \left( H^{\dagger} H\right)
		
	\end{array}\label{cl:l4h2r}\end{align}
There is 1 more complex type with repeated Higgses:
\begin{align}\begin{array}{c|l}
		
		\multirow{1}*{$\mathcal{O}_{L{}^2  e_{_\mathbb{C}}{}^2  H^{\dagger} {}^2}^{\left(1,2\right)} $}
		
		&\mathcal{Y}\left[\tiny{\young(p,r)},\tiny{\young(s,t)}\right] \left(e_{_\mathbb{C} p} L_{si}\right) \left(e_{_\mathbb{C} r} L_{tj}\right) H^{\dagger}{}^{i} H^{\dagger}{}^{j}
		
		,\quad\mathcal{Y}\left[\tiny{\young(pr)},\tiny{\young(st)}\right] \left(e_{_\mathbb{C} p} L_{si}\right) \left(e_{_\mathbb{C} r} L_{tj}\right) H^{\dagger}{}^{i} H^{\dagger}{}^{j}
		
	\end{array}\label{cl:l4h2c}\end{align}

\subsubsection{One derivative involved}
\underline{Class $\psi^4 \phi D$}: 
The subclass of this form must contain 3 spinors of the same helicities and 1 spinor of the opposite helicity, namely $\psi^3\psi^\dagger \phi D$. A total of 3 independent Lorentz structures exist in this subclass
\bea
\psi _1^{\alpha } \psi _2^{\beta } \left(D \psi _3\right)_{\alpha\beta\dot{\alpha } } \phi _4 \psi^{\dagger\dot{\alpha }}_5 ,\quad
\psi _1^{\alpha } \psi _2^{\beta } \psi _{3\alpha } \left(D \phi _4\right)_{\beta\dot{\alpha } } \psi^{\dagger\dot{\alpha }}_5 ,\quad
\psi _1^{\alpha } \psi _{2\alpha } \psi _3^{\beta } \left(D \phi _4\right)_{\beta\dot{\alpha }} \psi^{\dagger\dot{\alpha }}_5.
\eea
All the types in this subclass must be complex.

\noindent 1. \underline{Operators involving only quarks:}
The 6 types are all of the combinations of the 2 quark Yukawa terms and the 3 quark kinetic terms.
\begin{align}\begin{array}{c|l}
		
		\multirow{6}*{$  \mathcal{O}_{Q{}^2 Q^{\dagger}  u_{_\mathbb{C}} H D}^{\left(1\sim12\right)} $}
		
		&\mathcal{Y}\left[\tiny{\young(p,r)}\right] \epsilon ^{ik} \left(Q_{rb j} D^{\mu } u_{_\mathbb{C}}{}_{s}^{a}\right) \left(Q_{pa i} \sigma _{\mu } Q^{\dagger}{}_{t}^{b j}\right) H_{k}
		
		,\quad\mathcal{Y}\left[\tiny{\young(p,r)}\right] \epsilon ^{ij} \left(Q_{rb j} D^{\mu } u_{_\mathbb{C}}{}_{s}^{a}\right) \left(Q_{pa i} \sigma _{\mu } Q^{\dagger}{}_{t}^{b k}\right) H_{k}
		
		\\&\mathcal{Y}\left[\tiny{\young(p,r)}\right] \epsilon ^{ik} \left(Q_{pa i} u_{_\mathbb{C}}{}_{s}^{a}\right) \left(Q_{rb j} \sigma _{\mu } Q^{\dagger}{}_{t}^{b j}\right) D^{\mu } H_{k}
		
		,\quad\mathcal{Y}\left[\tiny{\young(p,r)}\right] \epsilon ^{ij} \left(Q_{pa i} u_{_\mathbb{C}}{}_{s}^{a}\right) \left(Q_{rb j} \sigma _{\mu } Q^{\dagger}{}_{t}^{b k}\right) D^{\mu } H_{k}
		
		\\&\mathcal{Y}\left[\tiny{\young(p,r)}\right] \epsilon ^{ik} \left(Q_{pa i} u_{_\mathbb{C}}{}_{s}^{b}\right) \left(Q_{rb j} \sigma _{\mu } Q^{\dagger}{}_{t}^{a j}\right) D^{\mu } H_{k}
		
		,\quad\mathcal{Y}\left[\tiny{\young(p,r)}\right] \epsilon ^{ij} \left(Q_{pa i} u_{_\mathbb{C}}{}_{s}^{b}\right) \left(Q_{rb j} \sigma _{\mu } Q^{\dagger}{}_{t}^{a k}\right) D^{\mu } H_{k}
		
		\\&\mathcal{Y}\left[\tiny{\young(pr)}\right] \epsilon ^{ik} \left(Q_{rb j} D^{\mu } u_{_\mathbb{C}}{}_{s}^{a}\right) \left(Q_{pa i} \sigma _{\mu } Q^{\dagger}{}_{t}^{b j}\right) H_{k}
		
		,\quad\mathcal{Y}\left[\tiny{\young(pr)}\right] \epsilon ^{ij} \left(Q_{rb j} D^{\mu } u_{_\mathbb{C}}{}_{s}^{a}\right) \left(Q_{pa i} \sigma _{\mu } Q^{\dagger}{}_{t}^{b k}\right) H_{k}
		
		\\&\mathcal{Y}\left[\tiny{\young(pr)}\right] \epsilon ^{ik} \left(Q_{pa i} u_{_\mathbb{C}}{}_{s}^{a}\right) \left(Q_{rb j} \sigma _{\mu } Q^{\dagger}{}_{t}^{b j}\right) D^{\mu } H_{k}
		
		,\quad\mathcal{Y}\left[\tiny{\young(pr)}\right] \epsilon ^{ij} \left(Q_{pa i} u_{_\mathbb{C}}{}_{s}^{a}\right) \left(Q_{rb j} \sigma _{\mu } Q^{\dagger}{}_{t}^{b k}\right) D^{\mu } H_{k}
		
		\\&\mathcal{Y}\left[\tiny{\young(pr)}\right] \epsilon ^{ik} \left(Q_{pa i} u_{_\mathbb{C}}{}_{s}^{b}\right) \left(Q_{rb j} \sigma _{\mu } Q^{\dagger}{}_{t}^{a j}\right) D^{\mu } H_{k}
		
		,\quad\mathcal{Y}\left[\tiny{\young(pr)}\right] \epsilon ^{ij} \left(Q_{pa i} u_{_\mathbb{C}}{}_{s}^{b}\right) \left(Q_{rb j} \sigma _{\mu } Q^{\dagger}{}_{t}^{a k}\right) D^{\mu } H_{k}
		
		\vspace{2ex}\\
		
		\multirow{6}*{$ \mathcal{O}_{Q{}^2 Q^{\dagger} d_{_\mathbb{C}}   H^{\dagger}  D}^{\left(1\sim12\right)} $}
		
		&\mathcal{Y}\left[\tiny{\young(r,s)}\right] \left(d_{_\mathbb{C}}{}_{p}^{a} Q_{rb i}\right) \left(Q_{sa j} \sigma _{\mu } Q^{\dagger}{}_{t}^{b j}\right) D^{\mu } H^{\dagger}{}^{i}
		
		,\quad\mathcal{Y}\left[\tiny{\young(r,s)}\right] \left(d_{_\mathbb{C}}{}_{p}^{a} Q_{rb i}\right) \left(Q_{sa j} \sigma _{\mu } Q^{\dagger}{}_{t}^{b i}\right) D^{\mu } H^{\dagger}{}^{j}
		
		\\&\mathcal{Y}\left[\tiny{\young(r,s)}\right] \left(d_{_\mathbb{C}}{}_{p}^{a} Q_{ra i}\right) \left(Q_{sc j} \sigma _{\mu } Q^{\dagger}{}_{t}^{c j}\right) D^{\mu } H^{\dagger}{}^{i}
		
		,\quad\mathcal{Y}\left[\tiny{\young(r,s)}\right] \left(d_{_\mathbb{C}}{}_{p}^{a} Q_{ra i}\right) \left(Q_{sc j} \sigma _{\mu } Q^{\dagger}{}_{t}^{c i}\right) D^{\mu } H^{\dagger}{}^{j}
		
		\\&\mathcal{Y}\left[\tiny{\young(r,s)}\right] \left(Q_{rb i} D^{\mu } Q_{sa j}\right) \left(d_{_\mathbb{C}}{}_{p}^{a} \sigma _{\mu } Q^{\dagger}{}_{t}^{b j}\right) H^{\dagger}{}^{i}
		
		,\quad\mathcal{Y}\left[\tiny{\young(r,s)}\right] \left(Q_{rb i} D^{\mu } Q_{sa j}\right) \left(d_{_\mathbb{C}}{}_{p}^{a} \sigma _{\mu } Q^{\dagger}{}_{t}^{b i}\right) H^{\dagger}{}^{j}
		
		\\&\mathcal{Y}\left[\tiny{\young(rs)}\right] \left(d_{_\mathbb{C}}{}_{p}^{a} Q_{rb i}\right) \left(Q_{sa j} \sigma _{\mu } Q^{\dagger}{}_{t}^{b j}\right) D^{\mu } H^{\dagger}{}^{i}
		
		,\quad\mathcal{Y}\left[\tiny{\young(rs)}\right] \left(d_{_\mathbb{C}}{}_{p}^{a} Q_{rb i}\right) \left(Q_{sa j} \sigma _{\mu } Q^{\dagger}{}_{t}^{b i}\right) D^{\mu } H^{\dagger}{}^{j}
		
		\\&\mathcal{Y}\left[\tiny{\young(rs)}\right] \left(d_{_\mathbb{C}}{}_{p}^{a} Q_{ra i}\right) \left(Q_{sc j} \sigma _{\mu } Q^{\dagger}{}_{t}^{c j}\right) D^{\mu } H^{\dagger}{}^{i}
		
		,\quad\mathcal{Y}\left[\tiny{\young(rs)}\right] \left(d_{_\mathbb{C}}{}_{p}^{a} Q_{ra i}\right) \left(Q_{sc j} \sigma _{\mu } Q^{\dagger}{}_{t}^{c i}\right) D^{\mu } H^{\dagger}{}^{j}
		
		\\&\mathcal{Y}\left[\tiny{\young(rs)}\right] \left(Q_{rb i} D^{\mu } Q_{sa j}\right) \left(d_{_\mathbb{C}}{}_{p}^{a} \sigma _{\mu } Q^{\dagger}{}_{t}^{b j}\right) H^{\dagger}{}^{i}
		
		,\quad\mathcal{Y}\left[\tiny{\young(rs)}\right] \left(Q_{rb i} D^{\mu } Q_{sa j}\right) \left(d_{_\mathbb{C}}{}_{p}^{a} \sigma _{\mu } Q^{\dagger}{}_{t}^{b i}\right) H^{\dagger}{}^{j}
		
		\vspace{2ex}\\
		
		\multirow{3}*{$  \mathcal{O}_{Q u_{_\mathbb{C}}{}^2 u_{_\mathbb{C}}^{\dagger}  H D}^{\left(1\sim6\right)} $}
		
		&\mathcal{Y}\left[\tiny{\young(r,s)}\right] \epsilon ^{ij} \left(u_{_\mathbb{C}}{}_{r}^{b} D^{\mu } u_{_\mathbb{C}}{}_{s}^{a}\right) \left(Q_{pa i} \sigma _{\mu } u_{_\mathbb{C}}^{\dagger}{}_{tb}\right) H_{j}
		
		,\quad\mathcal{Y}\left[\tiny{\young(rs)}\right] \epsilon ^{ij} \left(u_{_\mathbb{C}}{}_{r}^{b} D^{\mu } u_{_\mathbb{C}}{}_{s}^{a}\right) \left(Q_{pa i} \sigma _{\mu } u_{_\mathbb{C}}^{\dagger}{}_{tb}\right) H_{j}
		
		\\&\mathcal{Y}\left[\tiny{\young(r,s)}\right] \epsilon ^{ij} \left(u_{_\mathbb{C}}{}_{r}^{a} D^{\mu } u_{_\mathbb{C}}{}_{s}^{c}\right) \left(Q_{pa i} \sigma _{\mu } u_{_\mathbb{C}}^{\dagger}{}_{tc}\right) H_{j}
		
		,\quad\mathcal{Y}\left[\tiny{\young(rs)}\right] \epsilon ^{ij} \left(u_{_\mathbb{C}}{}_{r}^{a} D^{\mu } u_{_\mathbb{C}}{}_{s}^{c}\right) \left(Q_{pa i} \sigma _{\mu } u_{_\mathbb{C}}^{\dagger}{}_{tc}\right) H_{j}
		
		\\&\mathcal{Y}\left[\tiny{\young(r,s)}\right] \epsilon ^{ij} \left(Q_{pa i} u_{_\mathbb{C}}{}_{s}^{a}\right) \left(u_{_\mathbb{C}}{}_{r}^{b} \sigma _{\mu } u_{_\mathbb{C}}^{\dagger}{}_{tb}\right) D^{\mu } H_{j}
		
		,\quad\mathcal{Y}\left[\tiny{\young(rs)}\right] \epsilon ^{ij} \left(Q_{pa i} u_{_\mathbb{C}}{}_{s}^{a}\right) \left(u_{_\mathbb{C}}{}_{r}^{b} \sigma _{\mu } u_{_\mathbb{C}}^{\dagger}{}_{tb}\right) D^{\mu } H_{j}
		
		\vspace{2ex}\\
		
		\multirow{3}*{$\mathcal{O}_{Q u_{_\mathbb{C}} u_{_\mathbb{C}}^{\dagger}  d_{_\mathbb{C}}  H^{\dagger}  D}^{\left(1\sim6\right)} $}
		
		&   \left(d_{_\mathbb{C}}{}_{p}^{a} Q_{rb i}\right) \left(u_{_\mathbb{C}}{}_{s}^{b} \sigma^{\mu } u_{_\mathbb{C}}^{\dagger}{}_{ta}\right) D_{\mu } H^{\dagger}{}^{i}
		
		,\quad   \left(d_{_\mathbb{C}}{}_{p}^{a} Q_{ra i}\right) \left(u_{_\mathbb{C}}{}_{s}^{b} \sigma^{\mu } u_{_\mathbb{C}}^{\dagger}{}_{tb}\right) D_{\mu } H^{\dagger}{}^{i}
		
		\\&   \left(d_{_\mathbb{C}}{}_{p}^{a} u_{_\mathbb{C}}{}_{s}^{b}\right) \left(Q_{rb i} \sigma^{\mu } u_{_\mathbb{C}}^{\dagger}{}_{ta}\right) D_{\mu } H^{\dagger}{}^{i}
		
		,\quad   \left(d_{_\mathbb{C}}{}_{p}^{a} u_{_\mathbb{C}}{}_{s}^{b}\right) \left(Q_{ra i} \sigma^{\mu } u_{_\mathbb{C}}^{\dagger}{}_{tb}\right) D_{\mu } H^{\dagger}{}^{i}
		
		\\&   \left(d_{_\mathbb{C}}{}_{p}^{a} \sigma^{\mu } u_{_\mathbb{C}}^{\dagger}{}_{ta}\right) \left(Q_{rb i} D_{\mu } u_{_\mathbb{C}}{}_{s}^{b}\right) H^{\dagger}{}^{i}
		
		,\quad   \left(d_{_\mathbb{C}}{}_{p}^{a} \sigma^{\mu } u_{_\mathbb{C}}^{\dagger}{}_{tb}\right) \left(Q_{ra i} D_{\mu } u_{_\mathbb{C}}{}_{s}^{b}\right) H^{\dagger}{}^{i}
		
		\vspace{2ex}\\
		
		\multirow{3}*{$ \mathcal{O}_{Q u_{_\mathbb{C}} d_{_\mathbb{C}} d_{_\mathbb{C}}^{\dagger} H D}^{\left(1\sim6\right)} $}
		
		&  \epsilon ^{ij}  \left(d_{_\mathbb{C}}{}_{p}^{a} Q_{ra i}\right) \left(u_{_\mathbb{C}}{}_{s}^{b} \sigma^{\mu } d_{_\mathbb{C}}^{\dagger}{}_{tb}\right) D_{\mu } H_{j}
		
		,\quad  \epsilon ^{ij}  \left(d_{_\mathbb{C}}{}_{p}^{a} Q_{rb i}\right) \left(u_{_\mathbb{C}}{}_{s}^{b} \sigma^{\mu } d_{_\mathbb{C}}^{\dagger}{}_{ta}\right) D_{\mu } H_{j}
		
		\\&  \epsilon ^{ij}  \left(d_{_\mathbb{C}}{}_{p}^{a} u_{_\mathbb{C}}{}_{s}^{b}\right) \left(Q_{ra i} \sigma^{\mu } d_{_\mathbb{C}}^{\dagger}{}_{tb}\right) D_{\mu } H_{j}
		
		,\quad  \epsilon ^{ij}  \left(d_{_\mathbb{C}}{}_{p}^{a} u_{_\mathbb{C}}{}_{s}^{b}\right) \left(Q_{rb i} \sigma^{\mu } d_{_\mathbb{C}}^{\dagger}{}_{ta}\right) D_{\mu } H_{j}
		
		\\&  \epsilon ^{ij}  \left(d_{_\mathbb{C}}{}_{p}^{a} \sigma^{\mu } d_{_\mathbb{C}}^{\dagger}{}_{tb}\right) \left(Q_{ra i} D_{\mu } u_{_\mathbb{C}}{}_{s}^{b}\right) H_{j}
		
		,\quad  \epsilon ^{ij}  \left(d_{_\mathbb{C}}{}_{p}^{a} \sigma^{\mu } d_{_\mathbb{C}}^{\dagger}{}_{ta}\right) \left(Q_{rb i} D_{\mu } u_{_\mathbb{C}}{}_{s}^{b}\right) H_{j}
		
		\vspace{2ex}\\
		
		\multirow{3}*{$ \mathcal{O}_{Q d_{_\mathbb{C}}{}^2 d_{_\mathbb{C}}^{\dagger} H^{\dagger}  D}^{\left(1\sim6\right)} $}
		
		&\mathcal{Y}\left[\tiny{\young(p,r)}\right]  \left(d_{_\mathbb{C}}{}_{p}^{a} Q_{sa i}\right) \left(d_{_\mathbb{C}}{}_{r}^{b} \sigma^{\mu } d_{_\mathbb{C}}^{\dagger}{}_{tb}\right) D_{\mu } H^{\dagger}{}^{i}
		
		,\quad\mathcal{Y}\left[\tiny{\young(pr)}\right]  \left(d_{_\mathbb{C}}{}_{p}^{a} Q_{sa i}\right) \left(d_{_\mathbb{C}}{}_{r}^{b} \sigma^{\mu } d_{_\mathbb{C}}^{\dagger}{}_{tb}\right) D_{\mu } H^{\dagger}{}^{i}
		
		\\&\mathcal{Y}\left[\tiny{\young(p,r)}\right]  \left(d_{_\mathbb{C}}{}_{p}^{a} Q_{sb i}\right) \left(d_{_\mathbb{C}}{}_{r}^{b} \sigma^{\mu } d_{_\mathbb{C}}^{\dagger}{}_{ta}\right) D_{\mu } H^{\dagger}{}^{i}
		
		,\quad\mathcal{Y}\left[\tiny{\young(pr)}\right]  \left(d_{_\mathbb{C}}{}_{p}^{a} Q_{sb i}\right) \left(d_{_\mathbb{C}}{}_{r}^{b} \sigma^{\mu } d_{_\mathbb{C}}^{\dagger}{}_{ta}\right) D_{\mu } H^{\dagger}{}^{i}
		
		\\&\mathcal{Y}\left[\tiny{\young(p,r)}\right]  \left(d_{_\mathbb{C}}{}_{p}^{a} \sigma^{\mu } d_{_\mathbb{C}}^{\dagger}{}_{tb}\right) \left(d_{_\mathbb{C}}{}_{r}^{b} D_{\mu } Q_{sa i}\right) H^{\dagger}{}^{i}
		
		,\quad\mathcal{Y}\left[\tiny{\young(pr)}\right]  \left(d_{_\mathbb{C}}{}_{p}^{a} \sigma^{\mu } d_{_\mathbb{C}}^{\dagger}{}_{tb}\right) \left(d_{_\mathbb{C}}{}_{r}^{b} D_{\mu } Q_{sa i}\right) H^{\dagger}{}^{i}
		
	\end{array}\label{cl:q4hd}\end{align}

To see how $\mathcal{Y}$'s act on operators one can refer to eq.(\ref{Y2eg}) and eq.(\ref{Y3eg}). The conversion from the two-component spinors to the four-component spinors are shown by the following examples:
\bea
\left(Q_{pa i} u_{_\mathbb{C}}{}_{s}^{a}\right) \left(Q_{rb j} \sigma^{\mu } Q^{\dagger}{}_{t}^{b j}\right) &=& - \left(\bar{u}{}_{s}^{a} q_{pa i}\right) \left(\bar{q}{}_{t}^{b j} \gamma^{\mu } q_{rb j} \right), \nonumber \\
\left(Q_{pa i} u_{_\mathbb{C}}{}_{s}^{a}\right) \left(u_{_\mathbb{C}}{}_{r}^{b} \sigma^{\mu } u_{_\mathbb{C}}^{\dagger}{}_{tb}\right) &=& \left(\bar{u}{}_{s}^{a} q_{pa i}\right) \left(\bar{u}{}_{r}^{b} \gamma^{\mu } u{}_{tb}\right), \nonumber \\
\left(Q_{rbi} D_{\mu } Q_{saj}\right) \left(d_{_\mathbb{C}}{}_{p}^{a} \sigma^{\mu } Q^{\dagger}{}_{t}^{b j}\right) &=& \left(q_{rbi} C D_{\mu } q_{sa j}\right) \left(\bar{d}{}_{p}^{a} \gamma^{\mu } C \bar{q}_{t}^{b j}\right) \nonumber \\
&=& \dfrac{1}{2} \left(\bar{d}{}_{p}^{a} \gamma^{\mu } \gamma^{\nu } D_{\mu } q_{sa j}\right) \left(\bar{q}_{t}^{b j} \gamma_{\nu } q_{rbi}\right), \nonumber \\
\left(u_{_\mathbb{C}}{}_{r}^{a} D_{\mu } u_{_\mathbb{C}}{}_{s}^{c}\right) \left(Q_{pa i} \sigma^{\mu } u_{_\mathbb{C}}^{\dagger}{}_{tc}\right) &=& \left(\bar{u}{}_{r}^{a} C D_{\mu } \bar{u}{}_{s}^{c}\right) \left(q_{pa i} C \gamma^{\mu } u{}_{tc}\right) \nonumber \\
&=& \left(\bar{u}{}_{r}^{a} q_{pa i}\right) \left(D_{\mu } \bar{u}{}_{s}^{c} \gamma^{\mu } u{}_{tc}\right) - \left(\bar{u}{}_{r}^{a} \gamma^{\mu } u{}_{tc}\right) \left(D_{\mu } \bar{u}{}_{s}^{c} q_{pa i}\right). \label{HD}
\eea
The Hermitian conjugate of a non-Hermitian operator of this class is, for example,
\bea
\left[ \epsilon ^{ik}  \left(Q_{pa i} u_{_\mathbb{C}}{}_{sa}\right) \left(Q_{rb j} \sigma^{\mu } Q^{\dagger}{}_{t}^{b j}\right) D_{\mu } H_{k}\right]^\dagger = \epsilon _{ik}  \left(u_{_\mathbb{C}}^\dagger{}_{sa} Q^\dagger{}_{p}^{ai} \right) \left(Q_{tb j} \sigma^{\mu } Q^\dagger{}_{r}^{bj} \right) D_{\mu } H^\dagger{}^{k}.
\eea

\noindent 2. \underline{Operators involving three quarks with $\Delta B=\Delta L=\pm1$:}
There 7 B violating types in this class:
\begin{align}\begin{array}{c|l}
		
		\multirow{2}*{$ \mathcal{O}_{Q{}^3 e_{_\mathbb{C}}^{\dagger}  H D}^{\left(1\sim4\right)}$}
		
		&\mathcal{Y}\left[\tiny{\young(pr,s)}\right] \epsilon ^{abc} \epsilon ^{ik} \epsilon ^{jm} \left(Q_{pa i} \sigma _{\mu } e_{_\mathbb{C}}^{\dagger}{}_{t}\right) \left(Q_{rb j} D^{\mu } Q_{sc k}\right) H_{m}
		
		,\quad\mathcal{Y}\left[\tiny{\young(pr,s)}\right] \epsilon ^{abc} \epsilon ^{ij} \epsilon ^{km} \left(Q_{pa i} \sigma _{\mu } e_{_\mathbb{C}}^{\dagger}{}_{t}\right) \left(Q_{rb j} D^{\mu } Q_{sc k}\right) H_{m}
		
		\\&\mathcal{Y}\left[\tiny{\young(prs)}\right] \epsilon ^{abc} \epsilon ^{ik} \epsilon ^{jm} \left(Q_{pa i} \sigma _{\mu } e_{_\mathbb{C}}^{\dagger}{}_{t}\right) \left(Q_{rb j} D^{\mu } Q_{sc k}\right) H_{m}
		
		,\quad\mathcal{Y}\left[\tiny{\young(p,r,s)}\right] \epsilon ^{abc} \epsilon ^{ik} \epsilon ^{jm} \left(Q_{pa i} \sigma _{\mu } e_{_\mathbb{C}}^{\dagger}{}_{t}\right) \left(Q_{rb j} D^{\mu } Q_{sc k}\right) H_{m}
		
		\vspace{2ex}\\
		
		\multirow{3}*{$ \mathcal{O}_{Q{}^2 u_{_\mathbb{C}}^{\dagger} L  H^{\dagger}  D}^{\left(1\sim6\right)}$}
		
		&\mathcal{Y}\left[\tiny{\young(rs)}\right] \epsilon ^{abc} \epsilon ^{ik} \left(L_{pi} Q_{ra j}\right) \left(Q_{sb k} \sigma _{\mu } u_{_\mathbb{C}}^{\dagger}{}_{tc}\right) D^{\mu } H^{\dagger}{}^{j}
		
		,\quad\mathcal{Y}\left[\tiny{\young(r,s)}\right] \epsilon ^{abc} \epsilon ^{ik} \left(L_{pi} Q_{ra j}\right) \left(Q_{sb k} \sigma _{\mu } u_{_\mathbb{C}}^{\dagger}{}_{tc}\right) D^{\mu } H^{\dagger}{}^{j}
		
		\\&\mathcal{Y}\left[\tiny{\young(rs)}\right] \epsilon ^{abc} \epsilon ^{ij} \left(L_{pi} Q_{ra j}\right) \left(Q_{sb k} \sigma _{\mu } u_{_\mathbb{C}}^{\dagger}{}_{tc}\right) D^{\mu } H^{\dagger}{}^{k}
		
		,\quad\mathcal{Y}\left[\tiny{\young(r,s)}\right] \epsilon ^{abc} \epsilon ^{ij} \left(L_{pi} Q_{ra j}\right) \left(Q_{sb k} \sigma _{\mu } u_{_\mathbb{C}}^{\dagger}{}_{tc}\right) D^{\mu } H^{\dagger}{}^{k}
		
		\\&\mathcal{Y}\left[\tiny{\young(rs)}\right] \epsilon ^{abc} \epsilon ^{ik} \left(L_{pi} \sigma _{\mu } u_{_\mathbb{C}}^{\dagger}{}_{tc}\right) \left(Q_{ra j} D^{\mu } Q_{sb k}\right) H^{\dagger}{}^{j}
		
		,\quad\mathcal{Y}\left[\tiny{\young(r,s)}\right] \epsilon ^{abc} \epsilon ^{ik} \left(L_{pi} \sigma _{\mu } u_{_\mathbb{C}}^{\dagger}{}_{tc}\right) \left(Q_{ra j} D^{\mu } Q_{sb k}\right) H^{\dagger}{}^{j}
		
		\vspace{2ex}\\
		
		\multirow{3}*{$ \mathcal{O}_{Q{}^2 d_{_\mathbb{C}}^{\dagger}  L H D}^{\left(1\sim6\right)}$}
		
		&\mathcal{Y}\left[\tiny{\young(rs)}\right] \epsilon ^{abc} \epsilon ^{ik} \epsilon ^{jm} \left(L_{pi} Q_{ra j}\right) \left(Q_{sb k} \sigma _{\mu } d_{_\mathbb{C}}^{\dagger}{}_{tc}\right) D^{\mu } H_{m}
		
		,\quad\mathcal{Y}\left[\tiny{\young(r,s)}\right] \epsilon ^{abc} \epsilon ^{ik} \epsilon ^{jm} \left(L_{pi} Q_{ra j}\right) \left(Q_{sb k} \sigma _{\mu } d_{_\mathbb{C}}^{\dagger}{}_{tc}\right) D^{\mu } H_{m}
		
		\\&\mathcal{Y}\left[\tiny{\young(rs)}\right] \epsilon ^{abc} \epsilon ^{ij} \epsilon ^{km} \left(L_{pi} Q_{ra j}\right) \left(Q_{sb k} \sigma _{\mu } d_{_\mathbb{C}}^{\dagger}{}_{tc}\right) D^{\mu } H_{m}
		
		,\quad\mathcal{Y}\left[\tiny{\young(r,s)}\right] \epsilon ^{abc} \epsilon ^{ij} \epsilon ^{km} \left(L_{pi} Q_{ra j}\right) \left(Q_{sb k} \sigma _{\mu } d_{_\mathbb{C}}^{\dagger}{}_{tc}\right) D^{\mu } H_{m}
		
		\\&\mathcal{Y}\left[\tiny{\young(rs)}\right] \epsilon ^{abc} \epsilon ^{ik} \epsilon ^{jm} \left(L_{pi} \sigma _{\mu } d_{_\mathbb{C}}^{\dagger}{}_{tc}\right) \left(Q_{ra j} D^{\mu } Q_{sb k}\right) H_{m}
		
		,\quad\mathcal{Y}\left[\tiny{\young(r,s)}\right] \epsilon ^{abc} \epsilon ^{ik} \epsilon ^{jm} \left(L_{pi} \sigma _{\mu } d_{_\mathbb{C}}^{\dagger}{}_{tc}\right) \left(Q_{ra j} D^{\mu } Q_{sb k}\right) H_{m}
		
		\vspace{2ex}\\
		
		\multirow{2}*{$ \mathcal{O}_{Q^{\dagger}  u_{_\mathbb{C}}{}^2 e_{_\mathbb{C}} H D}^{\left(1\sim3\right)}$}
		
		&\mathcal{Y}\left[\tiny{\young(rs)}\right] \epsilon _{abc} \left(e_{_\mathbb{C} p} u_{_\mathbb{C}}{}_{r}^{a}\right) \left(u_{_\mathbb{C}}{}_{s}^{b} \sigma _{\mu } Q^{\dagger}{}_{t}^{c i}\right) D^{\mu } H_{i}
		
		,\quad\mathcal{Y}\left[\tiny{\young(rs)}\right] \epsilon _{abc} \left(u_{_\mathbb{C}}{}_{r}^{a} D^{\mu } u_{_\mathbb{C}}{}_{s}^{b}\right) \left(e_{_\mathbb{C} p} \sigma _{\mu } Q^{\dagger}{}_{t}^{c i}\right) H_{i}
		
		\\&\mathcal{Y}\left[\tiny{\young(r,s)}\right] \epsilon _{abc} \left(e_{_\mathbb{C} p} u_{_\mathbb{C}}{}_{r}^{a}\right) \left(u_{_\mathbb{C}}{}_{s}^{b} \sigma _{\mu } Q^{\dagger}{}_{t}^{c i}\right) D^{\mu } H_{i}
		
		\vspace{2ex}\\
		
		\multirow{2}*{$ \mathcal{O}_{Q^{\dagger} u_{_\mathbb{C}} d_{_\mathbb{C}} e_{_\mathbb{C}}  H^{\dagger}  D}^{\left(1\sim3\right)}$}
		
		&  \epsilon _{abc} \epsilon _{ij} \left(d_{_\mathbb{C}}{}_{p}^{a} e_{_\mathbb{C} r}\right) \left(u_{_\mathbb{C}}{}_{s}^{b} \sigma _{\mu } Q^{\dagger}{}_{t}^{c j}\right) D^{\mu } H^{\dagger}{}^{i}
		
		,\quad  \epsilon _{abc} \epsilon _{ij} \left(e_{_\mathbb{C} r} D^{\mu } u_{_\mathbb{C}}{}_{s}^{b}\right) \left(d_{_\mathbb{C}}{}_{p}^{a} \sigma _{\mu } Q^{\dagger}{}_{t}^{c j}\right) H^{\dagger}{}^{i}
		
		\\&  \epsilon _{abc} \epsilon _{ij} \left(d_{_\mathbb{C}}{}_{p}^{a} u_{_\mathbb{C}}{}_{s}^{b}\right) \left(e_{_\mathbb{C} r} \sigma _{\mu } Q^{\dagger}{}_{t}^{c j}\right) D^{\mu } H^{\dagger}{}^{i}
		
		\vspace{2ex}\\
		
		\multirow{2}*{$ \mathcal{O}_{u_{_\mathbb{C}}{}^2 d_{_\mathbb{C}} L^{\dagger}  H D}^{\left(1\sim3\right)}$}
		
		&\mathcal{Y}\left[\tiny{\young(rs)}\right] \epsilon _{abc} \left(d_{_\mathbb{C}}{}_{p}^{a} u_{_\mathbb{C}}{}_{r}^{b}\right) \left(u_{_\mathbb{C}}{}_{s}^{c} \sigma _{\mu } L^{\dagger}{}_{t}^{i}\right) D^{\mu } H_{i}
		
		,\quad\mathcal{Y}\left[\tiny{\young(rs)}\right] \epsilon _{abc} \left(u_{_\mathbb{C}}{}_{r}^{b} D^{\mu } u_{_\mathbb{C}}{}_{s}^{c}\right) \left(d_{_\mathbb{C}}{}_{p}^{a} \sigma _{\mu } L^{\dagger}{}_{t}^{i}\right) H_{i}
		
		\\&\mathcal{Y}\left[\tiny{\young(r,s)}\right] \epsilon _{abc} \left(d_{_\mathbb{C}}{}_{p}^{a} u_{_\mathbb{C}}{}_{r}^{b}\right) \left(u_{_\mathbb{C}}{}_{s}^{c} \sigma _{\mu } L^{\dagger}{}_{t}^{i}\right) D^{\mu } H_{i}
		
		\vspace{2ex}\\
		
		\multirow{2}*{$ \mathcal{O}_{u_{_\mathbb{C}} d_{_\mathbb{C}}{}^2 L^{\dagger}  H^{\dagger}  D}^{\left(1\sim3\right)}$}
		
		&\mathcal{Y}\left[\tiny{\young(pr)}\right] \epsilon _{abc} \epsilon _{ij} \left(d_{_\mathbb{C}}{}_{p}^{a} u_{_\mathbb{C}}{}_{s}^{c}\right) \left(d_{_\mathbb{C}}{}_{r}^{b} \sigma _{\mu } L^{\dagger}{}_{t}^{j}\right) D^{\mu } H^{\dagger}{}^{i}
		
		,\quad\mathcal{Y}\left[\tiny{\young(pr)}\right] \epsilon _{abc} \epsilon _{ij} \left(d_{_\mathbb{C}}{}_{p}^{a} \sigma _{\mu } L^{\dagger}{}_{t}^{j}\right) \left(d_{_\mathbb{C}}{}_{r}^{b} D^{\mu } u_{_\mathbb{C}}{}_{s}^{c}\right) H^{\dagger}{}^{i}
		
		\\&\mathcal{Y}\left[\tiny{\young(p,r)}\right] \epsilon _{abc} \epsilon _{ij} \left(d_{_\mathbb{C}}{}_{p}^{a} u_{_\mathbb{C}}{}_{s}^{c}\right) \left(d_{_\mathbb{C}}{}_{r}^{b} \sigma _{\mu } L^{\dagger}{}_{t}^{j}\right) D^{\mu } H^{\dagger}{}^{i}
		
	\end{array}\label{cl:q3l1hd}\end{align}

\noindent 3. \underline{Operators involving two leptons and two quarks:}
The combinations of 2 quark Yukawa terms and 2 lepton kinetic terms, and the combinations of 1 lepton Yukawa term and 3 quark kinetic terms, constitute 7 types here, while one more type $ {u_{_\mathbb{C}} d_{_\mathbb{C}}^{\dagger} L e_{_\mathbb{C}}  H D} $ not as such combination is present:
\begin{align}\begin{array}{c|l}
		
		\multirow{3}*{$ \mathcal{O}_{Q Q^{\dagger} L e_{_\mathbb{C}}  H^{\dagger}  D}^{\left(1\sim6\right)} $}
		
		& \left(e_{_\mathbb{C} p} \sigma _{\mu } Q^{\dagger}{}_{t}^{a j}\right) \left(L_{ri} D^{\mu } Q_{sa j}\right) H^{\dagger}{}^{i}
		
		,\quad \left(e_{_\mathbb{C} p} \sigma _{\mu } Q^{\dagger}{}_{t}^{a i}\right) \left(L_{ri} D^{\mu } Q_{sa j}\right) H^{\dagger}{}^{j}
		
		\\& \left(e_{_\mathbb{C} p} L_{ri}\right) \left(Q_{sa j} \sigma _{\mu } Q^{\dagger}{}_{t}^{a j}\right) D^{\mu } H^{\dagger}{}^{i}
		
		,\quad \left(e_{_\mathbb{C} p} L_{ri}\right) \left(Q_{sa j} \sigma _{\mu } Q^{\dagger}{}_{t}^{a i}\right) D^{\mu } H^{\dagger}{}^{j}
		
		\\& \left(e_{_\mathbb{C} p} Q_{sa j}\right) \left(L_{ri} \sigma _{\mu } Q^{\dagger}{}_{t}^{a j}\right) D^{\mu } H^{\dagger}{}^{i}
		
		,\quad \left(e_{_\mathbb{C} p} Q_{sa j}\right) \left(L_{ri} \sigma _{\mu } Q^{\dagger}{}_{t}^{a i}\right) D^{\mu } H^{\dagger}{}^{j}
		
		\vspace{2ex}\\
		
		\multirow{3}*{$ \mathcal{O}_{Q u_{_\mathbb{C}} L L^{\dagger}  H D}^{\left(1\sim6\right)} $}
		
		& \epsilon ^{ik} \left(Q_{ra j} D^{\mu } u_{_\mathbb{C}}{}_{s}^{a}\right) \left(L_{pi} \sigma _{\mu } L^{\dagger}{}_{t}^{j}\right) H_{k}
		
		,\quad \epsilon ^{ij} \left(Q_{ra j} D^{\mu } u_{_\mathbb{C}}{}_{s}^{a}\right) \left(L_{pi} \sigma _{\mu } L^{\dagger}{}_{t}^{k}\right) H_{k}
		
		\\& \epsilon ^{ik}\left(L_{pi} Q_{ra j}\right) \left(u_{_\mathbb{C}}{}_{s}^{a} \sigma _{\mu } L^{\dagger}{}_{t}^{j}\right) D^{\mu } H_{k} 
		
		,\quad \epsilon ^{ij} \left(L_{pi} Q_{ra j}\right) \left(u_{_\mathbb{C}}{}_{s}^{a} \sigma _{\mu } L^{\dagger}{}_{t}^{k}\right) D^{\mu } H_{k} 
		
		\\& \epsilon ^{ik} \left(L_{pi} u_{_\mathbb{C}}{}_{s}^{a}\right) \left(Q_{ra j} \sigma _{\mu } L^{\dagger}{}_{t}^{j}\right) D^{\mu } H_{k} 
		
		,\quad \epsilon ^{ij} \left(L_{pi} u_{_\mathbb{C}}{}_{s}^{a}\right) \left(Q_{ra j} \sigma _{\mu } L^{\dagger}{}_{t}^{k}\right) D^{\mu } H_{k} 
		
		\vspace{2ex}\\
		
		\multirow{2}*{$ \mathcal{O}_{Q u_{_\mathbb{C}} e_{_\mathbb{C}} e_{_\mathbb{C}}^{\dagger}  H D}^{\left(1\sim3\right)} $}
		
		&  \epsilon ^{ij}  \left(e_{_\mathbb{C} p} Q_{ra i}\right) \left(u_{_\mathbb{C}}{}_{s}^{a} \sigma^{\mu } e_{_\mathbb{C}}^{\dagger}{}_{t}\right) D_{\mu } H_{j}
		
		,\quad  \epsilon ^{ij}  \left(e_{_\mathbb{C} p} u_{_\mathbb{C}}{}_{s}^{a}\right) \left(Q_{ra i} \sigma^{\mu } e_{_\mathbb{C}}^{\dagger}{}_{t}\right) D_{\mu } H_{j}
		
		\\&  \epsilon ^{ij}  \left(e_{_\mathbb{C} p} \sigma^{\mu } e_{_\mathbb{C}}^{\dagger}{}_{t}\right) \left(Q_{ra i} D_{\mu } u_{_\mathbb{C}}{}_{s}^{a}\right) H_{j}
		
	\end{array}\label{cl:q2l2hd1}\end{align}

\begin{align}\begin{array}{c|l}
		
		\multirow{3}*{$ \mathcal{O}_{Q d_{_\mathbb{C}} L L^{\dagger}  H^{\dagger}  D}^{\left(1\sim6\right)} $}
		
		& \left(d_{_\mathbb{C}}{}_{p}^{a} \sigma _{\mu } L^{\dagger}{}_{t}^{j}\right) \left(L_{ri} D^{\mu } Q_{sa j}\right)  H^{\dagger}{}^{i}
		
		,\quad  \left(d_{_\mathbb{C}}{}_{p}^{a} \sigma _{\mu } L^{\dagger}{}_{t}^{i}\right) \left(L_{ri} D^{\mu } Q_{sa j}\right) H^{\dagger}{}^{j}
		
		\\&  \left(d_{_\mathbb{C}}{}_{p}^{a} L_{ri}\right) \left(Q_{sa j} \sigma _{\mu } L^{\dagger}{}_{t}^{j}\right) D^{\mu } H^{\dagger}{}^{i}
		
		,\quad  \left(d_{_\mathbb{C}}{}_{p}^{a} L_{ri}\right) \left(Q_{sa j} \sigma _{\mu } L^{\dagger}{}_{t}^{i}\right) D^{\mu } H^{\dagger}{}^{j}
		
		\\&  \left(d_{_\mathbb{C}}{}_{p}^{a} Q_{sa j}\right) \left(L_{ri} \sigma _{\mu } L^{\dagger}{}_{t}^{j}\right) D^{\mu } H^{\dagger}{}^{i}
		
		,\quad  \left(d_{_\mathbb{C}}{}_{p}^{a} Q_{sa j}\right) \left(L_{ri} \sigma _{\mu } L^{\dagger}{}_{t}^{i}\right) D^{\mu } H^{\dagger}{}^{j}
		
		\vspace{2ex}\\
		
		\multirow{2}*{$ \mathcal{O}_{Q d_{_\mathbb{C}} e_{_\mathbb{C}} e_{_\mathbb{C}}^{\dagger} H^{\dagger}  D}^{\left(1\sim3\right)} $}
		
		&   \left(d_{_\mathbb{C}}{}_{p}^{a} e_{_\mathbb{C} r}\right) \left(Q_{sa i} \sigma^{\mu } e_{_\mathbb{C}}^{\dagger}{}_{t}\right) D_{\mu } H^{\dagger}{}^{i}
		
		,\quad   \left(e_{_\mathbb{C} r} \sigma^{\mu } e_{_\mathbb{C}}^{\dagger}{}_{t}\right) \left(d_{_\mathbb{C}}{}_{p}^{a} Q_{sa i}\right) D_{\mu } H^{\dagger}{}^{i}
		
		\\&   \left(e_{_\mathbb{C} r} D_{\mu } Q_{sa i}\right) \left(d_{_\mathbb{C}}{}_{p}^{a} \sigma^{\mu } e_{_\mathbb{C}}^{\dagger}{}_{t}\right) H^{\dagger}{}^{i}
		
		\vspace{2ex}\\
		
		\multirow{2}*{$ \mathcal{O}_{u_{_\mathbb{C}} u_{_\mathbb{C}}^{\dagger} L e_{_\mathbb{C}}   H^{\dagger}  D}^{\left(1\sim3\right)} $}
		
		&   \left(e_{_\mathbb{C} p} L_{ri}\right) \left(u_{_\mathbb{C}}{}_{s}^{a} \sigma^{\mu } u_{_\mathbb{C}}^{\dagger}{}_{ta}\right) D_{\mu } H^{\dagger}{}^{i}
		
		,\quad   \left(e_{_\mathbb{C} p} u_{_\mathbb{C}}{}_{s}^{a}\right) \left(L_{ri} \sigma^{\mu } u_{_\mathbb{C}}^{\dagger}{}_{ta}\right) D_{\mu } H^{\dagger}{}^{i}
		
		\\&   \left(e_{_\mathbb{C} p} \sigma^{\mu } u_{_\mathbb{C}}^{\dagger}{}_{ta}\right) \left(L_{ri} D_{\mu } u_{_\mathbb{C}}{}_{s}^{a}\right) H^{\dagger}{}^{i}
		
		\vspace{2ex}\\
		
		\multirow{2}*{$ \mathcal{O}_{u_{_\mathbb{C}} d_{_\mathbb{C}}^{\dagger} L e_{_\mathbb{C}}  H D}^{\left(1\sim3\right)} $}
		
		&  \epsilon ^{ij}  \left(e_{_\mathbb{C} p} L_{ri}\right) \left(u_{_\mathbb{C}}{}_{s}^{a} \sigma^{\mu } d_{_\mathbb{C}}^{\dagger}{}_{ta}\right) D_{\mu } H_{j}
		
		,\quad  \epsilon ^{ij}  \left(e_{_\mathbb{C} p} u_{_\mathbb{C}}{}_{s}^{a}\right) \left(L_{ri} \sigma^{\mu } d_{_\mathbb{C}}^{\dagger}{}_{ta}\right) D_{\mu } H_{j}
		
		\\&  \epsilon ^{ij}  \left(e_{_\mathbb{C} p} \sigma^{\mu } d_{_\mathbb{C}}^{\dagger}{}_{ta}\right) \left(L_{ri} D_{\mu } u_{_\mathbb{C}}{}_{s}^{a}\right) H_{j}
		
		\vspace{2ex}\\
		
		\multirow{2}*{$ \mathcal{O}_{d_{_\mathbb{C}} d_{_\mathbb{C}}^{\dagger} L e_{_\mathbb{C}} H^{\dagger}  D}^{\left(1\sim3\right)} $}
		
		&   \left(d_{_\mathbb{C}}{}_{p}^{a} e_{_\mathbb{C} r}\right) \left(L_{si} \sigma^{\mu } d_{_\mathbb{C}}^{\dagger}{}_{ta}\right) D_{\mu } H^{\dagger}{}^{i}
		
		,\quad   \left(d_{_\mathbb{C}}{}_{p}^{a} L_{si}\right) \left(e_{_\mathbb{C} r} \sigma^{\mu } d_{_\mathbb{C}}^{\dagger}{}_{ta}\right) D_{\mu } H^{\dagger}{}^{i}
		
		\\&   \left(e_{_\mathbb{C} r} D_{\mu } L_{si}\right) \left(d_{_\mathbb{C}}{}_{p}^{a} \sigma^{\mu } d_{_\mathbb{C}}^{\dagger}{}_{ta}\right) H^{\dagger}{}^{i}
		
	\end{array}\label{cl:q2l2hd2}\end{align}

\noindent 4. \underline{Operators involving only leptons:}
The 2 following types are simply the lepton Yukawa term combined with one of the lepton kinetic terms:
\begin{align}\begin{array}{c|l}
		
		\multirow{3}*{$\mathcal{O}_{L{}^2 L^{\dagger}  e_{_\mathbb{C}}  H^{\dagger}  D}^{\left(1\sim6\right)} $}
		
		&\mathcal{Y}\left[\tiny{\young(r,s)}\right] D^{\mu } H^{\dagger}{}^{i} \left(e_{_\mathbb{C} p} L_{ri}\right) \left(L_{sj} \sigma _{\mu } L^{\dagger}{}_{t}^{j}\right)
		
		,\quad\mathcal{Y}\left[\tiny{\young(rs)}\right] D^{\mu } H^{\dagger}{}^{i} \left(e_{_\mathbb{C} p} L_{ri}\right) \left(L_{sj} \sigma _{\mu } L^{\dagger}{}_{t}^{j}\right)
		
		\\&\mathcal{Y}\left[\tiny{\young(r,s)}\right] D^{\mu } H^{\dagger}{}^{j} \left(e_{_\mathbb{C} p} L_{ri}\right) \left(L_{sj} \sigma _{\mu } L^{\dagger}{}_{t}^{i}\right)
		
		,\quad\mathcal{Y}\left[\tiny{\young(rs)}\right] D^{\mu } H^{\dagger}{}^{j} \left(e_{_\mathbb{C} p} L_{ri}\right) \left(L_{sj} \sigma _{\mu } L^{\dagger}{}_{t}^{i}\right)
		
		\\&\mathcal{Y}\left[\tiny{\young(r,s)}\right] H^{\dagger}{}^{i} \left(L_{ri} D^{\mu } L_{sj}\right) \left(e_{_\mathbb{C} p} \sigma _{\mu } L^{\dagger}{}_{t}^{j}\right)
		
		,\quad\mathcal{Y}\left[\tiny{\young(rs)}\right] H^{\dagger}{}^{i} \left(L_{ri} D^{\mu } L_{sj}\right) \left(e_{_\mathbb{C} p} \sigma _{\mu } L^{\dagger}{}_{t}^{j}\right)
		
		\vspace{2ex}\\
		
		\multirow{2}*{$ \mathcal{O}_{L e_{_\mathbb{C}}{}^2 e_{_\mathbb{C}}^{\dagger}  H^{\dagger}  D}^{\left(1\sim3\right)} $}
		
		&\mathcal{Y}\left[\tiny{\young(p,r)}\right] \left(e_{_\mathbb{C} p} L_{si}\right)  \left(e_{_\mathbb{C} r} \sigma^{\mu } e_{_\mathbb{C}}^{\dagger}{}_{t}\right) D_{\mu } H^{\dagger}{}^{i}
		
		,\quad\mathcal{Y}\left[\tiny{\young(p,r)}\right]  \left(e_{_\mathbb{C} p} \sigma^{\mu } e_{_\mathbb{C}}^{\dagger}{}_{t}\right) \left(e_{_\mathbb{C} r} D_{\mu } L_{si}\right) H^{\dagger}{}^{i}
		
		\\&\mathcal{Y}\left[\tiny{\young(pr)}\right] \left(e_{_\mathbb{C} p} L_{si}\right)  \left(e_{_\mathbb{C} r} \sigma^{\mu } e_{_\mathbb{C}}^{\dagger}{}_{t}\right) D_{\mu } H^{\dagger}{}^{i}
		
	\end{array}\label{cl:l4hd}\end{align}

\subsubsection{Two derivatives involved}
\underline{Class $\psi^4 D^2$}: 

There are two subclasses of this form: $\psi^2\psi^{\dagger 2}D^2$ and $\psi^4D^2+\hc$, and 5 independent Lorentz structures are involved:
\bea
&\psi _1^{\alpha } \psi _2^{\beta } (D \psi^{\dagger}_3)_{\alpha\dot{\alpha }\dot{\beta }} (D \psi^{\dagger}_4)_{\beta }^{\dot{\alpha }\dot{\beta }},&\quad
\psi _1^{\alpha } \psi _{2\alpha } (D \psi^{\dagger}_3)^{\beta }_{\dot{\alpha }\dot{\beta }} (D \psi^{\dagger}_4)_{\beta }^{\dot{\alpha }\dot{\beta }}, \nonumber \\
&\psi _1^{\alpha } (D \psi _2)^{\beta\gamma }_{\dot{\alpha }} \psi _3{}_{\alpha } (D \psi _4)_{\beta \gamma }^{\dot{\alpha }},&\quad
\psi _1^{\alpha } \psi _2^{\beta } (D \psi _3)_{\alpha\dot{\alpha } }^{\gamma } (D \psi _4)_{\beta \gamma }^{\dot{\alpha }},\quad
\psi _1^{\alpha } \psi _{2\alpha } (D \psi _3)^{\beta\gamma }_{\dot{\alpha }} (D \psi _4)_{\beta \gamma }^{\dot{\alpha }}.
\eea
Note that in converting to the conventional form of Lorentz structures, we avoid having parts like $\sigma^{\mu\nu} D_\mu \psi$ because they are related to $D_\nu \psi$ by the EOM redundancy. The types in this class are exactly the dimension 6 four fermion types plus two extra derivatives, which include 15 real types and 7 complex types, among which are 4 B violating types.

\noindent 1. \underline{Operators involving only quarks:}
There are 6 all-quark real types as follows
\begin{align}\begin{array}{c|l}
		
		\multirow{4}*{$\mathcal{O}_{Q{}^2 Q^{\dagger} {}^2 D^2}^{\left(1\sim8\right)} $}
		
		&\mathcal{Y}\left[\tiny{\young(p,r)},\tiny{\young(s,t)}\right] \left(D_{\mu } Q^{\dagger}{}_{s}^{a i} D_{\nu } Q^{\dagger}{}_{t}^{b j}\right) \left(Q_{pa i} \sigma ^{\mu }{}^{\nu } Q_{rb j}\right)
		,\quad\mathcal{Y}\left[\tiny{\young(p,r)},\tiny{\young(s,t)}\right] \left(D_{\mu } Q^{\dagger}{}_{s}^{a j} D_{\nu } Q^{\dagger}{}_{t}^{b i}\right) \left(Q_{pa i} \sigma ^{\mu }{}^{\nu } Q_{rb j}\right)
		
		\\&\mathcal{Y}\left[\tiny{\young(p,r)},\tiny{\young(s,t)}\right]\left(Q_{pa i} Q_{rb j}\right) \left(D_{\mu } Q^{\dagger}{}_{s}^{a i} D^{\mu } Q^{\dagger}{}_{t}^{b j}\right)
		
		,\quad\mathcal{Y}\left[\tiny{\young(p,r)},\tiny{\young(s,t)}\right]\left(Q_{pa i} Q_{rb j}\right) \left(D_{\mu } Q^{\dagger}{}_{s}^{a j} D^{\mu } Q^{\dagger}{}_{t}^{b i}\right)
		
		\\&\mathcal{Y}\left[\tiny{\young(pr)},\tiny{\young(st)}\right] \left(D_{\mu } Q^{\dagger}{}_{s}^{a i} D_{\nu } Q^{\dagger}{}_{t}^{b j}\right) \left(Q_{pa i} \sigma ^{\mu }{}^{\nu } Q_{rb j}\right)
		
		,\quad\mathcal{Y}\left[\tiny{\young(pr)},\tiny{\young(st)}\right] \left(D_{\mu } Q^{\dagger}{}_{s}^{a j} D_{\nu } Q^{\dagger}{}_{t}^{b i}\right) \left(Q_{pa i} \sigma ^{\mu }{}^{\nu } Q_{rb j}\right)
		
		\\&\mathcal{Y}\left[\tiny{\young(pr)},\tiny{\young(st)}\right]\left(Q_{pa i} Q_{rb j}\right) \left(D_{\mu } Q^{\dagger}{}_{s}^{a i} D^{\mu } Q^{\dagger}{}_{t}^{b j}\right)
		
		,\quad\mathcal{Y}\left[\tiny{\young(pr)},\tiny{\young(st)}\right]\left(Q_{pa i} Q_{rb j}\right) \left(D_{\mu } Q^{\dagger}{}_{s}^{a j} D^{\mu } Q^{\dagger}{}_{t}^{b i}\right)
		
		\vspace{2ex}\\
		
		\multirow{2}*{$\mathcal{O}_{Q Q^{\dagger}  u_{_\mathbb{C}} u_{_\mathbb{C}}^{\dagger}  D^2}^{\left(1\sim4\right)} $}
		
		& \left(Q_{pa i} u_{_\mathbb{C}}{}_{r}^{a}\right) \left(D_{\mu } Q^{\dagger}{}_{s}^{c i} D^{\mu } u_{_\mathbb{C}}^{\dagger}{}_{tc}\right)
		
		,\quad  \left(Q_{pa i} \sigma^{\mu }{}^{\nu } u_{_\mathbb{C}}{}_{r}^{a}\right) \left(D_{\mu } Q^{\dagger}{}_{s}^{c i} D_{\nu } u_{_\mathbb{C}}^{\dagger}{}_{tc}\right)
		
		\\& \left(Q_{pa i} u_{_\mathbb{C}}{}_{r}^{b}\right) \left(D_{\mu } Q^{\dagger}{}_{s}^{a i} D^{\mu } u_{_\mathbb{C}}^{\dagger}{}_{tb}\right)
		
		,\quad  \left(D_{\mu } Q^{\dagger}{}_{s}^{a i} D_{\nu } u_{_\mathbb{C}}^{\dagger}{}_{tb}\right) \left(Q_{pa i} \sigma^{\mu }{}^{\nu } u_{_\mathbb{C}}{}_{r}^{b}\right)
		
		\vspace{2ex}\\
		
		\multirow{2}*{$\mathcal{O}_{Q Q^{\dagger} d_{_\mathbb{C}} d_{_\mathbb{C}}^{\dagger}  D^2}^{\left(1\sim4\right)} $}
		
		& \left(d_{_\mathbb{C}}{}_{p}^{a} Q_{ra i}\right) \left(D_{\mu } d_{_\mathbb{C}}^{\dagger}{}_{sc} D^{\mu } Q^{\dagger}{}_{t}^{c i}\right)
		
		,\quad  \left(d_{_\mathbb{C}}{}_{p}^{a} \sigma^{\mu }{}^{\nu } Q_{ra i}\right) \left(D_{\mu } d_{_\mathbb{C}}^{\dagger}{}_{sc} D_{\nu } Q^{\dagger}{}_{t}^{c i}\right)
		
		\\& \left(d_{_\mathbb{C}}{}_{p}^{a} Q_{rb i}\right) \left(D_{\mu } d_{_\mathbb{C}}^{\dagger}{}_{sa} D^{\mu } Q^{\dagger}{}_{t}^{b i}\right)
		
		,\quad  \left(D_{\mu } d_{_\mathbb{C}}^{\dagger}{}_{sa} D_{\nu } Q^{\dagger}{}_{t}^{b i}\right) \left(d_{_\mathbb{C}}{}_{p}^{a} \sigma^{\mu }{}^{\nu } Q_{rb i}\right)
		
		\vspace{2ex}\\
		
		\multirow{2}*{$\mathcal{O}_{u_{_\mathbb{C}}{}^2 u_{_\mathbb{C}}^{\dagger} {}^2 D^2}^{\left(1\sim4\right)} $}
		
		&\mathcal{Y}\left[\tiny{\young(p,r)},\tiny{\young(s,t)}\right]i \left(D_{\mu } u_{_\mathbb{C}}^{\dagger}{}_{sa} D_{\nu } u_{_\mathbb{C}}^{\dagger}{}_{tb}\right) \left(u_{_\mathbb{C}}{}_{p}^{a} \sigma ^{\mu }{}^{\nu } u_{_\mathbb{C}}{}_{r}^{b}\right)
		
		,\quad\mathcal{Y}\left[\tiny{\young(p,r)},\tiny{\young(s,t)}\right]\left(u_{_\mathbb{C}}{}_{p}^{a} u_{_\mathbb{C}}{}_{r}^{b}\right) \left(D_{\mu } u_{_\mathbb{C}}^{\dagger}{}_{sa} D^{\mu } u_{_\mathbb{C}}^{\dagger}{}_{tb}\right)
		
		\\&\mathcal{Y}\left[\tiny{\young(pr)},\tiny{\young(st)}\right]i \left(D_{\mu } u_{_\mathbb{C}}^{\dagger}{}_{sa} D_{\nu } u_{_\mathbb{C}}^{\dagger}{}_{tb}\right) \left(u_{_\mathbb{C}}{}_{p}^{a} \sigma ^{\mu }{}^{\nu } u_{_\mathbb{C}}{}_{r}^{b}\right)
		
		,\quad\mathcal{Y}\left[\tiny{\young(pr)},\tiny{\young(st)}\right]\left(u_{_\mathbb{C}}{}_{p}^{a} u_{_\mathbb{C}}{}_{r}^{b}\right) \left(D_{\mu } u_{_\mathbb{C}}^{\dagger}{}_{sa} D^{\mu } u_{_\mathbb{C}}^{\dagger}{}_{tb}\right)
		
		\vspace{2ex}\\
		
		\multirow{2}*{$\mathcal{O}_{u_{_\mathbb{C}} u_{_\mathbb{C}}^{\dagger}  d_{_\mathbb{C}} d_{_\mathbb{C}}^{\dagger}  D^2}^{\left(1\sim4\right)} $}
		
		& \left(d_{_\mathbb{C}}{}_{p}^{a} u_{_\mathbb{C}}{}_{r}^{b}\right) \left(D_{\mu } d_{_\mathbb{C}}^{\dagger}{}_{sb} D^{\mu } u_{_\mathbb{C}}^{\dagger}{}_{ta}\right)
		
		,\quad  \left(D_{\mu } d_{_\mathbb{C}}^{\dagger}{}_{sb} D_{\nu } u_{_\mathbb{C}}^{\dagger}{}_{ta}\right) \left(d_{_\mathbb{C}}{}_{p}^{a} \sigma^{\mu }{}^{\nu } u_{_\mathbb{C}}{}_{r}^{b}\right)
		
		\\& \left(d_{_\mathbb{C}}{}_{p}^{a} u_{_\mathbb{C}}{}_{r}^{b}\right) \left(D_{\mu } d_{_\mathbb{C}}^{\dagger}{}_{sa} D^{\mu } u_{_\mathbb{C}}^{\dagger}{}_{tb}\right)
		
		,\quad  \left(D_{\mu } d_{_\mathbb{C}}^{\dagger}{}_{sa} D_{\nu } u_{_\mathbb{C}}^{\dagger}{}_{tb}\right) \left(d_{_\mathbb{C}}{}_{p}^{a} \sigma^{\mu }{}^{\nu } u_{_\mathbb{C}}{}_{r}^{b}\right)
		
		\vspace{2ex}\\
		
		\multirow{2}*{$\mathcal{O}_{d_{_\mathbb{C}}{}^2 d_{_\mathbb{C}}^{\dagger} {}^2 D^2}^{\left(1\sim4\right)} $}
		
		&\mathcal{Y}\left[\tiny{\young(p,r)},\tiny{\young(s,t)}\right]i \left(D_{\mu } d_{_\mathbb{C}}^{\dagger}{}_{sa} D_{\nu } d_{_\mathbb{C}}^{\dagger}{}_{tb}\right) \left(d_{_\mathbb{C}}{}_{p}^{a} \sigma ^{\mu }{}^{\nu } d_{_\mathbb{C}}{}_{r}^{b}\right)
		
		,\quad\mathcal{Y}\left[\tiny{\young(p,r)},\tiny{\young(s,t)}\right]\left(d_{_\mathbb{C}}{}_{p}^{a} d_{_\mathbb{C}}{}_{r}^{b}\right) \left(D_{\mu } d_{_\mathbb{C}}^{\dagger}{}_{sa} D^{\mu } d_{_\mathbb{C}}^{\dagger}{}_{tb}\right)
		
		\\&\mathcal{Y}\left[\tiny{\young(pr)},\tiny{\young(st)}\right]i \left(D_{\mu } d_{_\mathbb{C}}^{\dagger}{}_{sa} D_{\nu } d_{_\mathbb{C}}^{\dagger}{}_{tb}\right) \left(d_{_\mathbb{C}}{}_{p}^{a} \sigma ^{\mu }{}^{\nu } d_{_\mathbb{C}}{}_{r}^{b}\right)
		
		,\quad\mathcal{Y}\left[\tiny{\young(pr)},\tiny{\young(st)}\right]\left(d_{_\mathbb{C}}{}_{p}^{a} d_{_\mathbb{C}}{}_{r}^{b}\right) \left(D_{\mu } d_{_\mathbb{C}}^{\dagger}{}_{sa} D^{\mu } d_{_\mathbb{C}}^{\dagger}{}_{tb}\right)
		
	\end{array}\label{cl:q4d2r}\end{align}
\\
and 1 complex type:
\begin{align}\begin{array}{c|l}
		
		\multirow{3}*{$ \mathcal{O}_{Q{}^2 u_{_\mathbb{C}} d_{_\mathbb{C}}  D^2}^{\left(1\sim6\right)}  $}
		
		&\mathcal{Y}\left[\tiny{\young(rs)}\right]\epsilon ^{ij} \left(d_{_\mathbb{C}}{}_{p}^{a} Q_{sa j}\right) \left(D_{\mu } Q_{rb i} D^{\mu } u_{_\mathbb{C}}{}_{t}^{b}\right)
		
		,\quad\mathcal{Y}\left[\tiny{\young(rs)}\right]\epsilon ^{ij} \left(d_{_\mathbb{C}}{}_{p}^{a} Q_{sc j}\right) \left(D_{\mu } Q_{ra i} D^{\mu } u_{_\mathbb{C}}{}_{t}^{c}\right)
		
		\\&\mathcal{Y}\left[\tiny{\young(rs)}\right] \epsilon ^{ij} \left(D_{\mu } Q_{sa j} D_{\nu } u_{_\mathbb{C}}{}_{t}^{b}\right) \left(d_{_\mathbb{C}}{}_{p}^{a} \sigma ^{\mu }{}^{\nu } Q_{rb i}\right)
		
		,\quad\mathcal{Y}\left[\tiny{\young(r,s)}\right]\epsilon ^{ij} \left(d_{_\mathbb{C}}{}_{p}^{a} Q_{sa j}\right) \left(D_{\mu } Q_{rb i} D^{\mu } u_{_\mathbb{C}}{}_{t}^{b}\right)
		
		\\&\mathcal{Y}\left[\tiny{\young(r,s)}\right]\epsilon ^{ij} \left(d_{_\mathbb{C}}{}_{p}^{a} Q_{sc j}\right) \left(D_{\mu } Q_{ra i} D^{\mu } u_{_\mathbb{C}}{}_{t}^{c}\right)
		
		,\quad\mathcal{Y}\left[\tiny{\young(r,s)}\right] \epsilon ^{ij} \left(D_{\mu } Q_{sa j} D_{\nu } u_{_\mathbb{C}}{}_{t}^{b}\right) \left(d_{_\mathbb{C}}{}_{p}^{a} \sigma ^{\mu }{}^{\nu } Q_{rb i}\right)
		
	\end{array}\label{cl:q4d2c}\end{align}

To see how $\mathcal{Y}$'s act on operators one can refer to eq.(\ref{Y2eg}) and eq.(\ref{Y3eg}). The conversion from the two-component spinors to the four-component spinors follows eq.(\ref{H^2}) as
\bea
\left(D_{\mu } Q^{\dagger}{}_{s}^{a i} D_{\nu } Q^{\dagger}{}_{t}^{b j}\right) \left(Q_{pa i} \sigma^{\mu }{}^{\nu } Q_{rb j}\right) &=& \left(D_{\mu } \bar{q}{}_{s}^{a i} C D_{\nu } \bar{q}{}_{t}^{b j}\right) \left(q_{pa i} C \sigma^{\mu }{}^{\nu } q_{rb j}\right) \nonumber \\
&=& \dfrac{1}{2} \left(D_{\mu } \bar{q}{}_{s}^{a i} \gamma^\rho \sigma^{\mu }{}^{\nu } q_{rb j}\right) \left(D_{\nu } \bar{q}{}_{t}^{b j} \gamma_\rho q_{pa i}\right), \nonumber \\
\left(D_{\mu } u_{_\mathbb{C}}^{\dagger}{}_{sa} D_{\nu } u_{_\mathbb{C}}^{\dagger}{}_{tb}\right) \left(u_{_\mathbb{C}}{}_{p}^{a} \sigma^{\mu }{}^{\nu } u_{_\mathbb{C}}{}_{r}^{b}\right) &=& \left(D_{\mu } u{}_{sa} C D_{\nu } u{}_{tb}\right) \left(\bar{u}{}_{p}^{a} \sigma^{\mu }{}^{\nu } C \bar{u}{}_{r}^{b}\right) \nonumber \\
&=& \dfrac{1}{2} \left(\bar{u}{}_{p}^{a} \sigma^{\mu }{}^{\nu } \gamma^\rho D_{\nu } u{}_{tb}\right) \left(\bar{u}{}_{r}^{b} \gamma_\rho D_{\mu } u{}_{sa}\right), \nonumber \\
\left(D_{\mu } Q^{\dagger}{}_{s}^{a i} D_{\nu } u_{_\mathbb{C}}^{\dagger}{}_{tb}\right) \left(Q_{pa i} \sigma^{\mu }{}^{\nu } u_{_\mathbb{C}}{}_{r}^{b}\right) &=& \left(D_{\mu } \bar{q}{}_{s}^{a i} D_{\nu } u{}_{tb}\right) \left(\bar{u}{}_{r}^{b} \sigma^{\mu }{}^{\nu } q_{pa i}\right). \label{D^2}
\eea
The Hermitian conjugate of a non-Hermitian operator of this class is, for example,
\bea
\left[ \epsilon ^{ij} \left(D_{\mu } Q_{sa j} D_{\nu } u_{_\mathbb{C}}{}_{t}^{b}\right) \left(d_{_\mathbb{C}}{}_{p}^{a} \sigma^{\mu }{}^{\nu } Q_{rb i}\right)\right]^\dagger =  \epsilon _{ij} \left(D_{\nu } u_{_\mathbb{C}}^\dagger{}_{tb} D_{\mu } Q^\dagger{}_{s}^{aj}\right) \left(Q^\dagger{}_{r}^{bi} \bar{\sigma}^{\mu }{}^{\nu } d_{_\mathbb{C}}^\dagger{}_{pa}\right)
\eea
Operators involving leptons can be converted similarly.

\noindent 2. \underline{Operators involving one lepton and three quarks with $\Delta B=\Delta L=\pm 1$:}
The 4 B-violating types are 
\begin{align}\begin{array}{c|l}
		
		\multirow{2}*{$ \mathcal{O}_{Q{}^3 L  D^2}^{\left(1\sim4\right)}$}
		
		&\mathcal{Y}\left[\tiny{\young(rst)}\right]\epsilon ^{abc} \epsilon ^{ik} \epsilon ^{jm} \left(L_{pi} Q_{sb k}\right) \left(D_{\mu } Q_{ra j} D^{\mu } Q_{tc m}\right)
		
		,\quad\mathcal{Y}\left[\tiny{\young(rs,t)}\right]\epsilon ^{abc} \epsilon ^{ij} \epsilon ^{km} \left(L_{pi} Q_{sb k}\right) \left(D_{\mu } Q_{ra j} D^{\mu } Q_{tc m}\right)
		
		\\&\mathcal{Y}\left[\tiny{\young(rs,t)}\right] \epsilon ^{abc} \epsilon ^{ik} \epsilon ^{jm} \left(D_{\mu } Q_{sb k} D_{\nu } Q_{tc m}\right) \left(L_{pi} \sigma ^{\mu }{}^{\nu } Q_{ra j}\right)
		
		,\quad\mathcal{Y}\left[\tiny{\young(r,s,t)}\right]\epsilon ^{abc} \epsilon ^{ij} \epsilon ^{km} \left(L_{pi} Q_{sb k}\right) \left(D_{\mu } Q_{ra j} D^{\mu } Q_{tc m}\right)
		
		\vspace{2ex}\\
		
		\multirow{1}*{$\mathcal{O}_{Q{}^2 u_{_\mathbb{C}}^{\dagger} e_{_\mathbb{C}}^{\dagger}  D^2}^{\left(1,2\right)}$}
		
		&\mathcal{Y}\left[\tiny{\young(pr)}\right]\epsilon ^{abc} \epsilon ^{ij} \left(Q_{pa i} Q_{rb j}\right) \left(D_{\mu } e_{_\mathbb{C}}^{\dagger}{}_{s} D^{\mu } u_{_\mathbb{C}}^{\dagger}{}_{tc}\right)
		
		,\quad\mathcal{Y}\left[\tiny{\young(p,r)}\right]i \epsilon ^{abc} \epsilon ^{ij} \left(D_{\mu } e_{_\mathbb{C}}^{\dagger}{}_{s} D_{\nu } u_{_\mathbb{C}}^{\dagger}{}_{tc}\right) \left(Q_{pa i} \sigma^{\mu }{}^{\nu } Q_{rb j}\right)
		
		\vspace{2ex}\\
		
		\multirow{1}*{$\mathcal{O}_{Q u_{_\mathbb{C}}^{\dagger} d_{_\mathbb{C}}^{\dagger}  L  D^2}^{\left(1,2\right)}$}
		
		& \epsilon ^{abc} \epsilon ^{ij} \left(L_{pi} Q_{ra j}\right) \left(D_{\mu } d_{_\mathbb{C}}^{\dagger}{}_{sb} D^{\mu } u_{_\mathbb{C}}^{\dagger}{}_{tc}\right)
		
		,\quad  \epsilon ^{abc} \epsilon ^{ij} \left(D_{\mu } d_{_\mathbb{C}}^{\dagger}{}_{sb} D_{\nu } u_{_\mathbb{C}}^{\dagger}{}_{tc}\right) \left(L_{pi} \sigma^{\mu }{}^{\nu } Q_{ra j}\right)
		
		\vspace{2ex}\\
		
		\multirow{2}*{$ \mathcal{O}_{u_{_\mathbb{C}}{}^2 d_{_\mathbb{C}} e_{_\mathbb{C}}  D^2}^{\left(1\sim3\right)}$}
		
		&\mathcal{Y}\left[\tiny{\young(st)}\right]\epsilon _{abc} \left(d_{_\mathbb{C}}{}_{p}^{a} u_{_\mathbb{C}}{}_{s}^{b}\right) \left(D_{\mu } e_{_\mathbb{C} r} D^{\mu } u_{_\mathbb{C}}{}_{t}^{c}\right)
		
		,\quad\mathcal{Y}\left[\tiny{\young(s,t)}\right]\epsilon _{abc} \left(d_{_\mathbb{C}}{}_{p}^{a} u_{_\mathbb{C}}{}_{s}^{b}\right) \left(D_{\mu } e_{_\mathbb{C} r} D^{\mu } u_{_\mathbb{C}}{}_{t}^{c}\right)
		
		\\&\mathcal{Y}\left[\tiny{\young(s,t)}\right]\epsilon _{abc} \left(d_{_\mathbb{C}}{}_{p}^{a} e_{_\mathbb{C} r}\right) \left(D_{\mu } u_{_\mathbb{C}}{}_{s}^{b} D^{\mu } u_{_\mathbb{C}}{}_{t}^{c}\right)
		
	\end{array}\label{cl:q3l1d2}\end{align}

\noindent 3. \underline{Operators involving two leptons and two quarks:}
Combinations of 3 kinds of quark currents and 2 kinds of lepton currents provide 6 real types:
\begin{align}\begin{array}{c|l}
		
		\multirow{2}*{$\mathcal{O}_{Q Q^{\dagger} L L^{\dagger}  D^2}^{\left(1\sim4\right)} $}
		
		& \left(L_{pi} Q_{ra j}\right) \left(D_{\mu } L^{\dagger}{}_{s}^{j} D^{\mu } Q^{\dagger}{}_{t}^{a i}\right)
		
		,\quad  \left(D_{\mu } L^{\dagger}{}_{s}^{j} D_{\nu } Q^{\dagger}{}_{t}^{a i}\right) \left(L_{pi} \sigma^{\mu }{}^{\nu } Q_{ra j}\right)
		
		\\& \left(L_{pi} Q_{ra j}\right) \left(D_{\mu } L^{\dagger}{}_{s}^{i} D^{\mu } Q^{\dagger}{}_{t}^{a j}\right)
		
		,\quad  \left(D_{\mu } L^{\dagger}{}_{s}^{i} D_{\nu } Q^{\dagger}{}_{t}^{a j}\right) \left(L_{pi} \sigma^{\mu }{}^{\nu } Q_{ra j}\right)
		
		\vspace{2ex}\\
		
		\multirow{1}*{$\mathcal{O}_{Q Q^{\dagger} e_{_\mathbb{C}} e_{_\mathbb{C}}^{\dagger}  D^2}^{\left(1,2\right)} $}
		
		& \left(e_{_\mathbb{C} p} Q_{ra i}\right) \left(D_{\mu } e_{_\mathbb{C}}^{\dagger}{}_{s} D^{\mu } Q^{\dagger}{}_{t}^{a i}\right)
		
		,\quad  \left(D_{\mu } e_{_\mathbb{C}}^{\dagger}{}_{s} D_{\nu } Q^{\dagger}{}_{t}^{a i}\right) \left(e_{_\mathbb{C} p} \sigma^{\mu }{}^{\nu } Q_{ra i}\right)
		
		\vspace{2ex}\\
		
		\multirow{1}*{$\mathcal{O}_{u_{_\mathbb{C}} u_{_\mathbb{C}}^{\dagger} L L^{\dagger}    D^2}^{\left(1,2\right)} $}
		
		& \left(L_{pi} u_{_\mathbb{C}}{}_{r}^{a}\right) \left(D_{\mu } L^{\dagger}{}_{s}^{i} D^{\mu } u_{_\mathbb{C}}^{\dagger}{}_{ta}\right)
		
		,\quad  \left(D_{\mu } L^{\dagger}{}_{s}^{i} D_{\nu } u_{_\mathbb{C}}^{\dagger}{}_{ta}\right) \left(L_{pi} \sigma^{\mu }{}^{\nu } u_{_\mathbb{C}}{}_{r}^{a}\right)
		
		\vspace{2ex}\\
		
		\multirow{1}*{$\mathcal{O}_{u_{_\mathbb{C}} u_{_\mathbb{C}}^{\dagger} e_{_\mathbb{C}} e_{_\mathbb{C}}^{\dagger}  D^2}^{\left(1,2\right)} $}
		
		& \left(e_{_\mathbb{C} p} u_{_\mathbb{C}}{}_{r}^{a}\right) \left(D_{\mu } e_{_\mathbb{C}}^{\dagger}{}_{s} D^{\mu } u_{_\mathbb{C}}^{\dagger}{}_{ta}\right)
		
		,\quad  \left(D_{\mu } e_{_\mathbb{C}}^{\dagger}{}_{s} D_{\nu } u_{_\mathbb{C}}^{\dagger}{}_{ta}\right) \left(e_{_\mathbb{C} p} \sigma^{\mu }{}^{\nu } u_{_\mathbb{C}}{}_{r}^{a}\right)
		
		\vspace{2ex}\\
		
		\multirow{1}*{$\mathcal{O}_{d_{_\mathbb{C}} d_{_\mathbb{C}}^{\dagger}  L L^{\dagger}  D^2}^{\left(1,2\right)} $}
		
		& \left(d_{_\mathbb{C}}{}_{p}^{a} L_{ri}\right) \left(D_{\mu } d_{_\mathbb{C}}^{\dagger}{}_{sa} D^{\mu } L^{\dagger}{}_{t}^{i}\right)
		
		,\quad  \left(D_{\mu } d_{_\mathbb{C}}^{\dagger}{}_{sa} D_{\nu } L^{\dagger}{}_{t}^{i}\right) \left(d_{_\mathbb{C}}{}_{p}^{a} \sigma^{\mu }{}^{\nu } L_{ri}\right)
		
		\vspace{2ex}\\
		
		\multirow{1}*{$\mathcal{O}_{d_{_\mathbb{C}} d_{_\mathbb{C}}^{\dagger}  e_{_\mathbb{C}} e_{_\mathbb{C}}^{\dagger}  D^2}^{\left(1,2\right)} $}
		
		& \left(d_{_\mathbb{C}}{}_{p}^{a} e_{_\mathbb{C} r}\right) \left(D_{\mu } d_{_\mathbb{C}}^{\dagger}{}_{sa} D^{\mu } e_{_\mathbb{C}}^{\dagger}{}_{t}\right)
		
		,\quad \left(D_{\mu } d_{_\mathbb{C}}^{\dagger}{}_{sa} D_{\nu } e_{_\mathbb{C}}^{\dagger}{}_{t}\right) \left(d_{_\mathbb{C}}{}_{p}^{a} \sigma^{\mu }{}^{\nu } e_{_\mathbb{C} r}\right)
		
	\end{array}\label{cl:q2l2d2r}\end{align}
2 additional complex types are present:
\begin{align}\begin{array}{c|l}
		
		\multirow{2}*{$\mathcal{O}_{Q u_{_\mathbb{C}} L e_{_\mathbb{C}} D^2}^{\left(1\sim3\right)} $}
		
		& \epsilon ^{ij} \left(e_{_\mathbb{C} p} L_{ri}\right) \left(D_{\mu } Q_{sa j} D^{\mu } u_{_\mathbb{C}}{}_{t}^{a}\right)
		
		,\quad  \epsilon ^{ij} \left(e_{_\mathbb{C} p} \sigma^{\mu }{}^{\nu } L_{ri}\right) \left(D_{\mu } Q_{sa j} D_{\nu } u_{_\mathbb{C}}{}_{t}^{a}\right)
		
		\\& \epsilon ^{ij} \left(e_{_\mathbb{C} p} Q_{sa j}\right) \left(D_{\mu } L_{ri} D^{\mu } u_{_\mathbb{C}}{}_{t}^{a}\right)
		
		\vspace{2ex}\\
		
		\multirow{1}*{$\mathcal{O}_{Q d_{_\mathbb{C}} L^{\dagger} e_{_\mathbb{C}}^{\dagger}  D^2}^{\left(1,2\right)} $}
		
		& \left(d_{_\mathbb{C}}{}_{p}^{a} Q_{ra i}\right) \left(D_{\mu } e_{_\mathbb{C}}^{\dagger}{}_{s} D^{\mu } L^{\dagger}{}_{t}^{i}\right)
		
		,\quad  \left(D_{\mu } e_{_\mathbb{C}}^{\dagger}{}_{s} D_{\nu } L^{\dagger}{}_{t}^{i}\right) \left(d_{_\mathbb{C}}{}_{p}^{a} \sigma^{\mu }{}^{\nu } Q_{ra i}\right)
		
	\end{array}\label{cl:q2l2d2c}\end{align}

\noindent 4. \underline{Operators involving only leptons:}
2 kinds of lepton currents form 3 real types with all leptons:
\begin{align}\begin{array}{c|l}
		
		\multirow{2}*{$\mathcal{O}_{L{}^2 L^{\dagger} {}^2 D^2}^{\left(1\sim4\right)} $}
		
		&\mathcal{Y}\left[\tiny{\young(p,r)},\tiny{\young(s,t)}\right]\left(L_{pi} L_{rj}\right) \left(D_{\mu } L^{\dagger}{}_{s}^{i} D^{\mu } L^{\dagger}{}_{t}^{j}\right)
		
		,\quad\mathcal{Y}\left[\tiny{\young(p,r)},\tiny{\young(s,t)}\right] \left(D_{\mu } L^{\dagger}{}_{s}^{i} D_{\nu } L^{\dagger}{}_{t}^{j}\right) \left(L_{pi} \sigma^{\mu }{}^{\nu } L_{rj}\right)
		
		\\&\mathcal{Y}\left[\tiny{\young(pr)},\tiny{\young(st)}\right] \left(D_{\mu } L^{\dagger}{}_{s}^{i} D_{\nu } L^{\dagger}{}_{t}^{j}\right) \left(L_{pi} \sigma^{\mu }{}^{\nu } L_{rj}\right)
		
		,\quad\mathcal{Y}\left[\tiny{\young(pr)},\tiny{\young(st)}\right]\left(L_{pi} L_{rj}\right) \left(D_{\mu } L^{\dagger}{}_{s}^{i} D^{\mu } L^{\dagger}{}_{t}^{j}\right)
		
		\vspace{2ex}\\
		
		\multirow{1}*{$\mathcal{O}_{L L^{\dagger} e_{_\mathbb{C}} e_{_\mathbb{C}}^{\dagger}  D^2}^{\left(1,2\right)} $}
		
		& \left(e_{_\mathbb{C} p} L_{ri}\right) \left(D_{\mu } e_{_\mathbb{C}}^{\dagger}{}_{s} D^{\mu } L^{\dagger}{}_{t}^{i}\right)
		
		,\quad  \left(D_{\mu } e_{_\mathbb{C}}^{\dagger}{}_{s} D_{\nu } L^{\dagger}{}_{t}^{i}\right) \left(e_{_\mathbb{C} p} \sigma^{\mu }{}^{\nu } L_{ri}\right)
		
		\vspace{2ex}\\
		
		\multirow{1}*{$\mathcal{O}_{e_{_\mathbb{C}}{}^2 e_{_\mathbb{C}}^{\dagger} {}^2 D^2}^{\left(1,2\right)} $}
		
		&\mathcal{Y}\left[\tiny{\young(p,r)},\tiny{\young(s,t)}\right] \left(D_{\mu } e_{_\mathbb{C}}^{\dagger}{}_{s} D_{\nu } e_{_\mathbb{C}}^{\dagger}{}_{t}\right) \left(e_{_\mathbb{C} p} \sigma^{\mu }{}^{\nu } e_{_\mathbb{C} r}\right)
		
		,\quad\mathcal{Y}\left[\tiny{\young(pr)},\tiny{\young(st)}\right]\left(e_{_\mathbb{C} p} e_{_\mathbb{C} r}\right) \left(D_{\mu } e_{_\mathbb{C}}^{\dagger}{}_{s} D^{\mu } e_{_\mathbb{C}}^{\dagger}{}_{t}\right)
		
	\end{array}\label{cl:l4d2r}\end{align}

\subsubsection{One gauge boson involved}
\underline{Class $F \psi^4$}: 
There are two subclasses in this class: $F_{\rm L}\psi^2\psi^{\dagger 2}+\hc$ with only 1 Lorentz structure
\eq{
	F_{\rm{L}1}{}{}^{\alpha\beta } \psi _2{}_{\alpha } \psi _3{}_{\beta } \psi^{\dagger}_4{}_{\dot{\alpha }} \psi^{\dagger}_5{}{}^{\dot{\alpha }} ,
}
and $F_{\rm L}\psi^4+\hc$ with 3 independent Lorentz structures
\bea
F_{\rm{L}1}{}{}^{\alpha\beta } \psi _2{}{}^{\gamma } \psi _3{}_{\alpha } \psi _4{}_{\beta } \psi _5{}_{\gamma } ,\quad
F_{\rm{L}1}{}{}^{\alpha\beta } \psi _2{}_{\alpha } \psi _3{}{}^{\gamma } \psi _4{}_{\beta } \psi _5{}_{\gamma } ,\quad
F_{\rm{L}1}{}{}^{\alpha\beta } \psi _2{}_{\alpha } \psi _3{}_{\beta } \psi _4{}{}^{\gamma } \psi _5{}_{\gamma }.
\eea
In converting to the conventional form, the gauge boson always contracts with the $\sigma^{\mu\nu}$ (one may convert to other forms via Fierz identities, which we choose not to do), and due to the identity $\tilde{F}_{\mu\nu} \left(\sigma^{\mu\nu}\right)_\alpha{}^\beta=i F_{\mu\nu} \left(\sigma^{\mu\nu}\right)_\alpha{}^\beta$, the $F$ and $\tilde{F}$ are equivalent, hence we only use $F$ instead of $\tilde{F}$ in our operators. The types in this class are simply the dimension 6 four-fermion types with an additional gauge boson, depending on the gauge charges of the fermions: $B$ is always available as all the fermions are charged under $U(1)_Y$; $G$ is available whenever quarks are present; $W$ is available whenever $Q$ or $L$ is present.

\noindent 1. \underline{Operators involving only quarks:}
Based on the 6 real types with four quarks, $B$ and $G$ can be added to all of them, while $W$ can be added to the 3 types with $Q$. Overall, $6+6+3=15$ real types exist:
\begin{align}\begin{array}{c|l}
		
		\multirow{8}*{$\mathcal{O}_{G Q{}^2 Q^{\dagger} {}^2}^{\left(1\sim16\right)}     $}
		
		&\mathcal{Y}\left[\tiny{\young(p,r)},\tiny{\young(st)}\right] \left(\lambda ^A\right)_c^b G^{A}_{\mu \nu } \left(Q^{\dagger}{}_{s}^{c i} Q^{\dagger}{}_{t}^{a j}\right) \left(Q_{pa i} \sigma^{\mu \nu } Q_{rb j}\right)
		
		,\quad\mathcal{Y}\left[\tiny{\young(p,r)},\tiny{\young(st)}\right] \left(\lambda ^A\right)_c^b G^{A}_{\mu \nu } \left(Q^{\dagger}{}_{s}^{c j} Q^{\dagger}{}_{t}^{a i}\right) \left(Q_{pa i} \sigma^{\mu \nu } Q_{rb j}\right)
		
		\\&\mathcal{Y}\left[\tiny{\young(p,r)},\tiny{\young(st)}\right] \left(\lambda ^A\right)_b^c G^{A}_{\mu \nu } \left(Q_{ra j} Q_{pc i}\right) \left(Q^\dagger{}_{t}^{b j} \bar{\sigma}^{\mu \nu } Q^\dagger{}_{s}^{a i}\right)
		
		,\quad\mathcal{Y}\left[\tiny{\young(p,r)},\tiny{\young(st)}\right] \left(\lambda ^A\right)_b^c G^{A}_{\mu \nu } \left(Q_{ra i} Q_{pc j}\right) \left(Q^\dagger{}_{t}^{b j} \bar{\sigma}^{\mu \nu } Q^\dagger{}_{s}^{a i}\right)
		
		\\&\mathcal{Y}\left[\tiny{\young(pr)},\tiny{\young(s,t)}\right] \left(\lambda ^A\right)_c^b G^{A}_{\mu \nu } \left(Q^{\dagger}{}_{s}^{c i} Q^{\dagger}{}_{t}^{a j}\right) \left(Q_{pa i} \sigma^{\mu \nu } Q_{rb j}\right)
		
		,\quad\mathcal{Y}\left[\tiny{\young(pr)},\tiny{\young(s,t)}\right] \left(\lambda ^A\right)_c^b G^{A}_{\mu \nu } \left(Q^{\dagger}{}_{s}^{c j} Q^{\dagger}{}_{t}^{a i}\right) \left(Q_{pa i} \sigma^{\mu \nu } Q_{rb j}\right)
		
		\\&\mathcal{Y}\left[\tiny{\young(pr)},\tiny{\young(s,t)}\right] \left(\lambda ^A\right)_b^c G^{A}_{\mu \nu } \left(Q_{ra j} Q_{pc i}\right) \left(Q^\dagger{}_{t}^{b j} \bar{\sigma}^{\mu \nu } Q^\dagger{}_{s}^{a i}\right)
		
		,\quad\mathcal{Y}\left[\tiny{\young(pr)},\tiny{\young(s,t)}\right] \left(\lambda ^A\right)_b^c G^{A}_{\mu \nu } \left(Q_{ra i} Q_{pc j}\right) \left(Q^\dagger{}_{t}^{b j} \bar{\sigma}^{\mu \nu } Q^\dagger{}_{s}^{a i}\right)
		
		\\&\mathcal{Y}\left[\tiny{\young(p,r)},\tiny{\young(s,t)}\right] \left(\lambda ^A\right)_c^b G^{A}_{\mu \nu } \left(Q^{\dagger}{}_{s}^{c i} Q^{\dagger}{}_{t}^{a j}\right) \left(Q_{pa i} \sigma^{\mu \nu } Q_{rb j}\right)
		
		,\quad\mathcal{Y}\left[\tiny{\young(p,r)},\tiny{\young(s,t)}\right] \left(\lambda ^A\right)_c^b G^{A}_{\mu \nu } \left(Q^{\dagger}{}_{s}^{c j} Q^{\dagger}{}_{t}^{a i}\right) \left(Q_{pa i} \sigma^{\mu \nu } Q_{rb j}\right)
		
		\\&\mathcal{Y}\left[\tiny{\young(p,r)},\tiny{\young(s,t)}\right] \left(\lambda ^A\right)_b^c G^{A}_{\mu \nu } \left(Q_{ra j} Q_{pc i}\right) \left(Q^\dagger{}_{t}^{b j} \bar{\sigma}^{\mu \nu } Q^\dagger{}_{s}^{a i}\right)
		
		,\quad\mathcal{Y}\left[\tiny{\young(p,r)},\tiny{\young(s,t)}\right] \left(\lambda ^A\right)_b^c G^{A}_{\mu \nu } \left(Q_{ra i} Q_{pc j}\right) \left(Q^\dagger{}_{t}^{b j} \bar{\sigma}^{\mu \nu } Q^\dagger{}_{s}^{a i}\right)
		
		\\&\mathcal{Y}\left[\tiny{\young(pr)},\tiny{\young(st)}\right] \left(\lambda ^A\right)_c^b G^{A}_{\mu \nu } \left(Q^{\dagger}{}_{s}^{c i} Q^{\dagger}{}_{t}^{a j}\right) \left(Q_{pa i} \sigma^{\mu \nu } Q_{rb j}\right)
		
		,\quad\mathcal{Y}\left[\tiny{\young(pr)},\tiny{\young(st)}\right] \left(\lambda ^A\right)_c^b G^{A}_{\mu \nu } \left(Q^{\dagger}{}_{s}^{c j} Q^{\dagger}{}_{t}^{a i}\right) \left(Q_{pa i} \sigma^{\mu \nu } Q_{rb j}\right)
		
		\\&\mathcal{Y}\left[\tiny{\young(pr)},\tiny{\young(st)}\right] \left(\lambda ^A\right)_b^c G^{A}_{\mu \nu } \left(Q_{ra j} Q_{pc i}\right) \left(Q^\dagger{}_{t}^{b j} \bar{\sigma}^{\mu \nu } Q^\dagger{}_{s}^{a i}\right)
		
		,\quad\mathcal{Y}\left[\tiny{\young(pr)},\tiny{\young(st)}\right] \left(\lambda ^A\right)_b^c G^{A}_{\mu \nu } \left(Q_{ra i} Q_{pc j}\right) \left(Q^\dagger{}_{t}^{b j} \bar{\sigma}^{\mu \nu } Q^\dagger{}_{s}^{a i}\right)
		
		\vspace{2ex}\\
		
		\multirow{6}*{$\mathcal{O}_{W Q{}^2 Q^{\dagger} {}^2}^{\left(1\sim12\right)}     $}
		
		&\mathcal{Y}\left[\tiny{\young(p,r)},\tiny{\young(st)}\right] \left(\tau ^I\right)_k^j W^{I}_{\mu \nu } \left(Q^{\dagger}{}_{s}^{a i} Q^{\dagger}{}_{t}^{b k}\right) \left(Q_{pa i} \sigma^{\mu \nu } Q_{rb j}\right)
		
		,\quad\mathcal{Y}\left[\tiny{\young(p,r)},\tiny{\young(st)}\right] \left(\tau ^I\right)_j^k W^{I}_{\mu \nu } \left(Q_{rb k} Q_{pa i}\right) \left(Q^\dagger{}_{t}^{b j} \bar{\sigma}^{\mu \nu } Q^\dagger{}_{s}^{a i}\right)
		
		\\&\mathcal{Y}\left[\tiny{\young(p,r)},\tiny{\young(s,t)}\right] \left(\tau ^I\right)_k^j W^{I}_{\mu \nu } \left(Q^{\dagger}{}_{s}^{a i} Q^{\dagger}{}_{t}^{b k}\right) \left(Q_{pa i} \sigma^{\mu \nu } Q_{rb j}\right)
		
		,\quad\mathcal{Y}\left[\tiny{\young(p,r)},\tiny{\young(s,t)}\right] \left(\tau ^I\right)_k^i W^{I}_{\mu \nu } \left(Q^{\dagger}{}_{s}^{a j} Q^{\dagger}{}_{t}^{b k}\right) \left(Q_{pa i} \sigma^{\mu \nu } Q_{rb j}\right)
		
		\\&\mathcal{Y}\left[\tiny{\young(p,r)},\tiny{\young(s,t)}\right] \left(\tau ^I\right)_j^k W^{I}_{\mu \nu } \left(Q_{rb k} Q_{pa i}\right) \left(Q^\dagger{}_{t}^{b j} \bar{\sigma}^{\mu \nu } Q^\dagger{}_{s}^{a i}\right)
		
		,\quad\mathcal{Y}\left[\tiny{\young(p,r)},\tiny{\young(s,t)}\right] \left(\tau ^I\right)_i^k W^{I}_{\mu \nu } \left(Q_{rb k} Q_{pa j}\right) \left(Q^\dagger{}_{t}^{b j} \bar{\sigma}^{\mu \nu } Q^\dagger{}_{s}^{a i}\right)
		
		\\&\mathcal{Y}\left[\tiny{\young(pr)},\tiny{\young(st)}\right] \left(\tau ^I\right)_k^j W^{I}_{\mu \nu } \left(Q^{\dagger}{}_{s}^{a i} Q^{\dagger}{}_{t}^{b k}\right) \left(Q_{pa i} \sigma^{\mu \nu } Q_{rb j}\right)
		
		,\quad\mathcal{Y}\left[\tiny{\young(pr)},\tiny{\young(st)}\right] \left(\tau ^I\right)_k^i W^{I}_{\mu \nu } \left(Q^{\dagger}{}_{s}^{a j} Q^{\dagger}{}_{t}^{b k}\right) \left(Q_{pa i} \sigma^{\mu \nu } Q_{rb j}\right)
		
		\\&\mathcal{Y}\left[\tiny{\young(pr)},\tiny{\young(st)}\right] \left(\tau ^I\right)_j^k W^{I}_{\mu \nu } \left(Q_{rb k} Q_{pa i}\right) \left(Q^\dagger{}_{t}^{b j} \bar{\sigma}^{\mu \nu } Q^\dagger{}_{s}^{a i}\right)
		
		,\quad\mathcal{Y}\left[\tiny{\young(pr)},\tiny{\young(st)}\right] \left(\tau ^I\right)_i^k W^{I}_{\mu \nu } \left(Q_{rb k} Q_{pa j}\right) \left(Q^\dagger{}_{t}^{b j} \bar{\sigma}^{\mu \nu } Q^\dagger{}_{s}^{a i}\right)
		
		\\&\mathcal{Y}\left[\tiny{\young(pr)},\tiny{\young(s,t)}\right] \left(\tau ^I\right)_k^j W^{I}_{\mu \nu } \left(Q^{\dagger}{}_{s}^{a i} Q^{\dagger}{}_{t}^{b k}\right) \left(Q_{pa i} \sigma^{\mu \nu } Q_{rb j}\right)
		
		,\quad\mathcal{Y}\left[\tiny{\young(pr)},\tiny{\young(s,t)}\right] \left(\tau ^I\right)_j^k W^{I}_{\mu \nu } \left(Q_{rb k} Q_{pa i}\right) \left(Q^\dagger{}_{t}^{b j} \bar{\sigma}^{\mu \nu } Q^\dagger{}_{s}^{a i}\right)
		
		\vspace{2ex}\\
		
		\multirow{4}*{$\mathcal{O}_{B Q{}^2 Q^{\dagger} {}^2}^{\left(1\sim8\right)}     $}
		
		&\mathcal{Y}\left[\tiny{\young(p,r)},\tiny{\young(st)}\right] B_{\mu \nu } \left(Q^{\dagger}{}_{s}^{a i} Q^{\dagger}{}_{t}^{b j}\right) \left(Q_{pa i} \sigma^{\mu \nu } Q_{rb j}\right)
		
		,\quad\mathcal{Y}\left[\tiny{\young(p,r)},\tiny{\young(st)}\right] B_{\mu \nu } \left(Q^{\dagger}{}_{s}^{a j} Q^{\dagger}{}_{t}^{b i}\right) \left(Q_{pa i} \sigma^{\mu \nu } Q_{rb j}\right)
		
		\\&\mathcal{Y}\left[\tiny{\young(p,r)},\tiny{\young(st)}\right] B_{\mu \nu } \left(Q_{rbj} Q_{pai}\right) \left(Q^\dagger{}_{t}^{b j} \bar{\sigma}^{\mu \nu } Q^\dagger{}_{s}^{a i}\right)
		
		,\quad\mathcal{Y}\left[\tiny{\young(p,r)},\tiny{\young(st)}\right] B_{\mu \nu } \left(Q_{rbi} Q_{paj}\right) \left(Q^\dagger{}_{t}^{b j} \bar{\sigma}^{\mu \nu } Q^\dagger{}_{s}^{a i}\right)
		
		\\&\mathcal{Y}\left[\tiny{\young(pr)},\tiny{\young(s,t)}\right] B_{\mu \nu } \left(Q^{\dagger}{}_{s}^{a i} Q^{\dagger}{}_{t}^{b j}\right) \left(Q_{pa i} \sigma^{\mu \nu } Q_{rb j}\right)
		
		,\quad\mathcal{Y}\left[\tiny{\young(pr)},\tiny{\young(s,t)}\right] B_{\mu \nu } \left(Q^{\dagger}{}_{s}^{a j} Q^{\dagger}{}_{t}^{b i}\right) \left(Q_{pa i} \sigma^{\mu \nu } Q_{rb j}\right)
		
		\\&\mathcal{Y}\left[\tiny{\young(pr)},\tiny{\young(s,t)}\right] B_{\mu \nu } \left(Q_{rbj} Q_{pai}\right) \left(Q^\dagger{}_{t}^{b j} \bar{\sigma}^{\mu \nu } Q^\dagger{}_{s}^{a i}\right)
		
		,\quad\mathcal{Y}\left[\tiny{\young(pr)},\tiny{\young(s,t)}\right] B_{\mu \nu } \left(Q_{rbi} Q_{paj}\right) \left(Q^\dagger{}_{t}^{b j} \bar{\sigma}^{\mu \nu } Q^\dagger{}_{s}^{a i}\right)
		
		\vspace{2ex}\\
		
		\multirow{4}*{$\mathcal{O}_{G Q Q^{\dagger}  u_{_\mathbb{C}} u_{_\mathbb{C}}^{\dagger}}^{\left(1\sim8\right)}      $}
		
		& \left(\lambda ^A\right)_b^d G^{A}_{\mu \nu } \left(Q^{\dagger}{}_{s}^{a i} u_{_\mathbb{C}}^{\dagger}{}_{td}\right) \left(Q_{pa i} \sigma^{\mu \nu } u_{_\mathbb{C}}{}_{r}^{b}\right)
		
		,\quad \left(\lambda ^A\right)_d^b G^{A}_{\mu \nu } \left(u_{_\mathbb{C}}{}_{r}^{d} Q{}_{pa i}\right) \left(u_{_\mathbb{C}}^\dagger{}_{tb} \bar{\sigma}^{\mu \nu } Q^\dagger{}_{s}^{a i}\right)
		
		\\& \left(\lambda ^A\right)_c^d G^{A}_{\mu \nu } \left(Q^{\dagger}{}_{s}^{c i} u_{_\mathbb{C}}^{\dagger}{}_{td}\right) \left(Q_{pa i} \sigma^{\mu \nu } u_{_\mathbb{C}}{}_{r}^{a}\right)
		
		,\quad \left(\lambda ^A\right)_d^c G^{A}_{\mu \nu } \left(u_{_\mathbb{C}}{}_{r}^{d} Q{}_{pc i}\right) \left(u_{_\mathbb{C}}^\dagger{}_{ta} \bar{\sigma}^{\mu \nu } Q^\dagger{}_{s}^{a i}\right)
		
		\\& \left(\lambda ^A\right)_b^a G^{A}_{\mu \nu } \left(Q^{\dagger}{}_{s}^{c i} u_{_\mathbb{C}}^{\dagger}{}_{tc}\right) \left(Q_{pa i} \sigma^{\mu \nu } u_{_\mathbb{C}}{}_{r}^{b}\right)
		
		,\quad \left(\lambda ^A\right)_a^b G^{A}_{\mu \nu } \left(u_{_\mathbb{C}}{}_{r}^{c} Q{}_{pc i}\right) \left(u_{_\mathbb{C}}^\dagger{}_{tb} \bar{\sigma}^{\mu \nu } Q^\dagger{}_{s}^{a i}\right)
		
		\\& \left(\lambda ^A\right)_c^a G^{A}_{\mu \nu } \left(Q^{\dagger}{}_{s}^{c i} u_{_\mathbb{C}}^{\dagger}{}_{tb}\right) \left(Q_{pa i} \sigma^{\mu \nu } u_{_\mathbb{C}}{}_{r}^{b}\right)
		
		,\quad \left(\lambda ^A\right)_a^c G^{A}_{\mu \nu } \left(u_{_\mathbb{C}}{}_{r}^{b} Q{}_{pc i}\right) \left(u_{_\mathbb{C}}^\dagger{}_{tb} \bar{\sigma}^{\mu \nu } Q^\dagger{}_{s}^{a i}\right)
		
	\end{array}\label{cl:Fq4r1}\end{align}

\begin{align}\begin{array}{c|l}
		
		\multirow{2}*{$\mathcal{O}_{W Q Q^{\dagger}  u_{_\mathbb{C}} u_{_\mathbb{C}}^{\dagger}}^{\left(1\sim4\right)}      $}
		
		& \left(\tau ^I\right)_j^i W^{I}_{\mu \nu } \left(Q^{\dagger}{}_{s}^{c j} u_{_\mathbb{C}}^{\dagger}{}_{tc}\right) \left(Q_{pa i} \sigma^{\mu \nu } u_{_\mathbb{C}}{}_{r}^{a}\right)
		
		,\quad \left(\tau ^I\right)_i^j W^{I}_{\mu \nu } \left(u_{_\mathbb{C}}{}_{r}^{c} Q{}_{pc j}\right) \left(u_{_\mathbb{C}}^\dagger{}_{ta} \bar{\sigma}^{\mu \nu } Q^\dagger{}_{s}^{a i}\right)
		
		\\& \left(\tau ^I\right)_j^i W^{I}_{\mu \nu } \left(Q^{\dagger}{}_{s}^{a j} u_{_\mathbb{C}}^{\dagger}{}_{tb}\right) \left(Q_{pa i} \sigma^{\mu \nu } u_{_\mathbb{C}}{}_{r}^{b}\right)
		
		,\quad \left(\tau ^I\right)_i^j W^{I}_{\mu \nu } \left(u_{_\mathbb{C}}{}_{r}^{b} Q{}_{pa j}\right) \left(u_{_\mathbb{C}}^\dagger{}_{tb} \bar{\sigma}^{\mu \nu } Q^\dagger{}_{s}^{a i}\right)
		
		\vspace{2ex}\\
		
		\multirow{2}*{$\mathcal{O}_{B Q Q^{\dagger}  u_{_\mathbb{C}} u_{_\mathbb{C}}^{\dagger}}^{\left(1\sim4\right)}      $}
		
		& B_{\mu \nu } \left(Q^{\dagger}{}_{s}^{c i} u_{_\mathbb{C}}^{\dagger}{}_{tc}\right) \left(Q_{pa i} \sigma^{\mu \nu } u_{_\mathbb{C}}{}_{r}^{a}\right)
		
		,\quad B_{\mu \nu } \left(u_{_\mathbb{C}}{}_{r}^{c} Q{}_{pc i}\right) \left(u_{_\mathbb{C}}^\dagger{}_{ta} \bar{\sigma}^{\mu \nu } Q^\dagger{}_{s}^{a i}\right)
		
		\\& B_{\mu \nu } \left(Q^{\dagger}{}_{s}^{a i} u_{_\mathbb{C}}^{\dagger}{}_{tb}\right) \left(Q_{pa i} \sigma^{\mu \nu } u_{_\mathbb{C}}{}_{r}^{b}\right)
		
		,\quad B_{\mu \nu } \left(u_{_\mathbb{C}}{}_{r}^{b} Q{}_{pa i}\right) \left(u_{_\mathbb{C}}^\dagger{}_{tb} \bar{\sigma}^{\mu \nu } Q^\dagger{}_{s}^{a i}\right)
		
		\vspace{2ex}\\
		
		\multirow{4}*{$\mathcal{O}_{G Q Q^{\dagger} d_{_\mathbb{C}} d_{_\mathbb{C}}^{\dagger}}^{\left(1\sim8\right)}      $}
		
		& \left(\lambda ^A\right)_d^b G^{A}_{\mu \nu } \left(d_{_\mathbb{C}}^{\dagger}{}_{sa} Q^{\dagger}{}_{t}^{d i}\right) \left(d_{_\mathbb{C}}{}_{p}^{a} \sigma^{\mu \nu } Q_{rb i}\right)
		
		,\quad \left(\lambda ^A\right)_b^d G^{A}_{\mu \nu } \left(Q{}_{rd i} d_{_\mathbb{C}}{}_{p}^{a} \right) \left(Q^\dagger{}_{t}^{b i} \bar{\sigma}^{\mu \nu } d_{_\mathbb{C}}^\dagger{}_{sa} \right)
		
		\\& \left(\lambda ^A\right)_a^b G^{A}_{\mu \nu } \left(d_{_\mathbb{C}}^{\dagger}{}_{sc} Q^{\dagger}{}_{t}^{c i}\right) \left(d_{_\mathbb{C}}{}_{p}^{a} \sigma^{\mu \nu } Q_{rb i}\right)
		
		,\quad \left(\lambda ^A\right)_b^a G^{A}_{\mu \nu } \left(Q{}_{rc i} d_{_\mathbb{C}}{}_{p}^{c} \right) \left(Q^\dagger{}_{t}^{b i} \bar{\sigma}^{\mu \nu } d_{_\mathbb{C}}^\dagger{}_{sa} \right)
		
		\\& \left(\lambda ^A\right)_d^c G^{A}_{\mu \nu } \left(d_{_\mathbb{C}}^{\dagger}{}_{sc} Q^{\dagger}{}_{t}^{d i}\right) \left(d_{_\mathbb{C}}{}_{p}^{a} \sigma^{\mu \nu } Q_{ra i}\right)
		
		,\quad \left(\lambda ^A\right)_c^d G^{A}_{\mu \nu } \left(Q{}_{rd i} d_{_\mathbb{C}}{}_{p}^{c} \right) \left(Q^\dagger{}_{t}^{a i} \bar{\sigma}^{\mu \nu } d_{_\mathbb{C}}^\dagger{}_{sa} \right)
		
		\\& \left(\lambda ^A\right)_a^c G^{A}_{\mu \nu } \left(d_{_\mathbb{C}}^{\dagger}{}_{sc} Q^{\dagger}{}_{t}^{b i}\right) \left(d_{_\mathbb{C}}{}_{p}^{a} \sigma^{\mu \nu } Q_{rb i}\right)
		
		,\quad \left(\lambda ^A\right)_c^a G^{A}_{\mu \nu } \left(Q{}_{rb i} d_{_\mathbb{C}}{}_{p}^{c} \right) \left(Q^\dagger{}_{t}^{b i} \bar{\sigma}^{\mu \nu } d_{_\mathbb{C}}^\dagger{}_{sa} \right)
		
		\vspace{2ex}\\
		
		\multirow{2}*{$\mathcal{O}_{W Q Q^{\dagger} d_{_\mathbb{C}} d_{_\mathbb{C}}^{\dagger}}^{\left(1\sim4\right)}      $}
		
		& \left(\tau ^I\right)_j^i W^{I}_{\mu \nu } \left(d_{_\mathbb{C}}^{\dagger}{}_{sc} Q^{\dagger}{}_{t}^{c j}\right) \left(d_{_\mathbb{C}}{}_{p}^{a} \sigma^{\mu \nu } Q_{ra i}\right)
		
		,\quad \left(\tau ^I\right)_i^j W^{I}_{\mu \nu } \left(Q{}_{rc j} d_{_\mathbb{C}}{}_{p}^{c} \right) \left(Q^\dagger{}_{t}^{a i} \bar{\sigma}^{\mu \nu } d_{_\mathbb{C}}^\dagger{}_{sa} \right)
		
		\\& \left(\tau ^I\right)_j^i W^{I}_{\mu \nu } \left(d_{_\mathbb{C}}^{\dagger}{}_{sa} Q^{\dagger}{}_{t}^{b j}\right) \left(d_{_\mathbb{C}}{}_{p}^{a} \sigma^{\mu \nu } Q_{rb i}\right)
		
		,\quad \left(\tau ^I\right)_i^j W^{I}_{\mu \nu } \left(Q{}_{rb j} d_{_\mathbb{C}}{}_{p}^{a} \right) \left(Q^\dagger{}_{t}^{b i} \bar{\sigma}^{\mu \nu } d_{_\mathbb{C}}^\dagger{}_{sa} \right)
		
		\vspace{2ex}\\
		
		\multirow{2}*{$\mathcal{O}_{B Q Q^{\dagger} d_{_\mathbb{C}} d_{_\mathbb{C}}^{\dagger}}^{\left(1\sim4\right)}      $}
		
		& B_{\mu \nu } \left(d_{_\mathbb{C}}^{\dagger}{}_{sc} Q^{\dagger}{}_{t}^{c i}\right) \left(d_{_\mathbb{C}}{}_{p}^{a} \sigma^{\mu \nu } Q_{ra i}\right)
		
		,\quad B_{\mu \nu } \left(Q{}_{rc i} d_{_\mathbb{C}}{}_{p}^{c} \right) \left(Q^\dagger{}_{t}^{a i} \bar{\sigma}^{\mu \nu } d_{_\mathbb{C}}^\dagger{}_{sa} \right)
		
		\\& B_{\mu \nu } \left(d_{_\mathbb{C}}^{\dagger}{}_{sa} Q^{\dagger}{}_{t}^{b i}\right) \left(d_{_\mathbb{C}}{}_{p}^{a} \sigma^{\mu \nu } Q_{rb i}\right)
		
		,\quad B_{\mu \nu } \left(Q{}_{rb i} d_{_\mathbb{C}}{}_{p}^{a} \right) \left(Q^\dagger{}_{t}^{b i} \bar{\sigma}^{\mu \nu } d_{_\mathbb{C}}^\dagger{}_{sa} \right)
		
		\vspace{2ex}\\
		
		\multirow{4}*{$\mathcal{O}_{G u_{_\mathbb{C}}{}^2 u_{_\mathbb{C}}^{\dagger} {}^2}^{\left(1\sim8\right)}     $}
		
		&\mathcal{Y}\left[\tiny{\young(p,r)},\tiny{\young(st)}\right] \left(\lambda ^A\right)_b^d G^{A}_{\mu \nu } \left(u_{_\mathbb{C}}^{\dagger}{}_{sa} u_{_\mathbb{C}}^{\dagger}{}_{td}\right) \left(u_{_\mathbb{C}}{}_{p}^{a} \sigma^{\mu \nu } u_{_\mathbb{C}}{}_{r}^{b}\right)
		
		,\quad\mathcal{Y}\left[\tiny{\young(p,r)},\tiny{\young(st)}\right] \left(\lambda ^A\right)_d^b G^{A}_{\mu \nu } \left(u_{_\mathbb{C}}{}_{r}^{d} u_{_\mathbb{C}}{}_{p}^{a} \right) \left(u_{_\mathbb{C}}^\dagger{}_{tb} \bar{\sigma}^{\mu \nu } u_{_\mathbb{C}}^\dagger{}_{sa}\right)
		
		\\&\mathcal{Y}\left[\tiny{\young(p,r)},\tiny{\young(s,t)}\right] \left(\lambda ^A\right)_b^d G^{A}_{\mu \nu } \left(u_{_\mathbb{C}}^{\dagger}{}_{sa} u_{_\mathbb{C}}^{\dagger}{}_{td}\right)  \left(u_{_\mathbb{C}}{}_{p}^{a} \sigma^{\mu \nu } u_{_\mathbb{C}}{}_{r}^{b}\right)
		
		,\quad\mathcal{Y}\left[\tiny{\young(p,r)},\tiny{\young(s,t)}\right] \left(\lambda ^A\right)_d^b G^{A}_{\mu \nu } \left(u_{_\mathbb{C}}{}_{r}^{d} u_{_\mathbb{C}}{}_{p}^{a} \right) \left(u_{_\mathbb{C}}^\dagger{}_{tb} \bar{\sigma}^{\mu \nu } u_{_\mathbb{C}}^\dagger{}_{sa}\right)
		
		\\&\mathcal{Y}\left[\tiny{\young(pr)},\tiny{\young(st)}\right] \left(\lambda ^A\right)_b^d G^{A}_{\mu \nu } \left(u_{_\mathbb{C}}^{\dagger}{}_{sa} u_{_\mathbb{C}}^{\dagger}{}_{td}\right)  \left(u_{_\mathbb{C}}{}_{p}^{a} \sigma^{\mu \nu } u_{_\mathbb{C}}{}_{r}^{b}\right)
		
		,\quad\mathcal{Y}\left[\tiny{\young(pr)},\tiny{\young(st)}\right] \left(\lambda ^A\right)_d^b G^{A}_{\mu \nu } \left(u_{_\mathbb{C}}{}_{r}^{d} u_{_\mathbb{C}}{}_{p}^{a} \right) \left(u_{_\mathbb{C}}^\dagger{}_{tb} \bar{\sigma}^{\mu \nu } u_{_\mathbb{C}}^\dagger{}_{sa}\right)
		
		\\&\mathcal{Y}\left[\tiny{\young(pr)},\tiny{\young(s,t)}\right] \left(\lambda ^A\right)_b^d G^{A}_{\mu \nu } \left(u_{_\mathbb{C}}^{\dagger}{}_{sa} u_{_\mathbb{C}}^{\dagger}{}_{td}\right)  \left(u_{_\mathbb{C}}{}_{p}^{a} \sigma^{\mu \nu } u_{_\mathbb{C}}{}_{r}^{b}\right)
		
		,\quad\mathcal{Y}\left[\tiny{\young(pr)},\tiny{\young(s,t)}\right] \left(\lambda ^A\right)_d^b G^{A}_{\mu \nu } \left(u_{_\mathbb{C}}{}_{r}^{d} u_{_\mathbb{C}}{}_{p}^{a} \right) \left(u_{_\mathbb{C}}^\dagger{}_{tb} \bar{\sigma}^{\mu \nu } u_{_\mathbb{C}}^\dagger{}_{sa}\right)
		
		\vspace{2ex}\\
		
		\multirow{2}*{$\mathcal{O}_{B u_{_\mathbb{C}}{}^2 u_{_\mathbb{C}}^{\dagger} {}^2}^{\left(1\sim4\right)}     $}
		
		&\mathcal{Y}\left[\tiny{\young(p,r)},\tiny{\young(st)}\right] B_{\mu \nu } \left(u_{_\mathbb{C}}^{\dagger}{}_{sa} u_{_\mathbb{C}}^{\dagger}{}_{tb}\right) \left(u_{_\mathbb{C}}{}_{p}^{a} \sigma^{\mu \nu } u_{_\mathbb{C}}{}_{r}^{b}\right)
		
		,\quad\mathcal{Y}\left[\tiny{\young(p,r)},\tiny{\young(st)}\right] B_{\mu \nu } \left(u_{_\mathbb{C}}{}_{r}^{b} u_{_\mathbb{C}}{}_{p}^{a} \right) \left(u_{_\mathbb{C}}^\dagger{}_{tb} \bar{\sigma}^{\mu \nu } u_{_\mathbb{C}}^\dagger{}_{sa}\right)
		
		\\&\mathcal{Y}\left[\tiny{\young(pr)},\tiny{\young(s,t)}\right] B_{\mu \nu } \left(u_{_\mathbb{C}}^{\dagger}{}_{sa} u_{_\mathbb{C}}^{\dagger}{}_{tb}\right) \left(u_{_\mathbb{C}}{}_{p}^{a} \sigma^{\mu \nu } u_{_\mathbb{C}}{}_{r}^{b}\right)
		
		,\quad\mathcal{Y}\left[\tiny{\young(pr)},\tiny{\young(s,t)}\right] B_{\mu \nu } \left(u_{_\mathbb{C}}{}_{r}^{b} u_{_\mathbb{C}}{}_{p}^{a} \right) \left(u_{_\mathbb{C}}^\dagger{}_{tb} \bar{\sigma}^{\mu \nu } u_{_\mathbb{C}}^\dagger{}_{sa}\right)
		
		\vspace{2ex}\\
		
		\multirow{4}*{$\mathcal{O}_{G u_{_\mathbb{C}} u_{_\mathbb{C}}^{\dagger} d_{_\mathbb{C}} d_{_\mathbb{C}}^{\dagger}}^{\left(1\sim8\right)}      $}
		
		& \left(\lambda ^A\right)_b^d G^{A}_{\mu \nu } \left(d_{_\mathbb{C}}^{\dagger}{}_{sa} u_{_\mathbb{C}}^{\dagger}{}_{td}\right) \left(d_{_\mathbb{C}}{}_{p}^{a} \sigma^{\mu \nu } u_{_\mathbb{C}}{}_{r}^{b}\right)
		
		,\quad \left(\lambda ^A\right)_d^b G^{A}_{\mu \nu } \left(u_{_\mathbb{C}}{}_{r}^{d} d_{_\mathbb{C}}{}_{p}^{a} \right) \left(u_{_\mathbb{C}}^\dagger{}_{tb} \bar{\sigma}^{\mu \nu } d_{_\mathbb{C}}^\dagger{}_{sa}\right)
		
		\\& \left(\lambda ^A\right)_a^d G^{A}_{\mu \nu } \left(d_{_\mathbb{C}}^{\dagger}{}_{sb} u_{_\mathbb{C}}^{\dagger}{}_{td}\right) \left(d_{_\mathbb{C}}{}_{p}^{a} \sigma^{\mu \nu } u_{_\mathbb{C}}{}_{r}^{b}\right)
		
		,\quad \left(\lambda ^A\right)_d^a G^{A}_{\mu \nu } \left(u_{_\mathbb{C}}{}_{r}^{d} d_{_\mathbb{C}}{}_{p}^{b} \right) \left(u_{_\mathbb{C}}^\dagger{}_{tb} \bar{\sigma}^{\mu \nu } d_{_\mathbb{C}}^\dagger{}_{sa}\right)
		
		\\& \left(\lambda ^A\right)_b^c G^{A}_{\mu \nu } \left(d_{_\mathbb{C}}^{\dagger}{}_{sc} u_{_\mathbb{C}}^{\dagger}{}_{ta}\right) \left(d_{_\mathbb{C}}{}_{p}^{a} \sigma^{\mu \nu } u_{_\mathbb{C}}{}_{r}^{b}\right)
		
		,\quad \left(\lambda ^A\right)_c^b G^{A}_{\mu \nu } \left(u_{_\mathbb{C}}{}_{r}^{a} d_{_\mathbb{C}}{}_{p}^{c} \right) \left(u_{_\mathbb{C}}^\dagger{}_{tb} \bar{\sigma}^{\mu \nu } d_{_\mathbb{C}}^\dagger{}_{sa}\right)
		
		\\& \left(\lambda ^A\right)_a^c G^{A}_{\mu \nu } \left(d_{_\mathbb{C}}^{\dagger}{}_{sc} u_{_\mathbb{C}}^{\dagger}{}_{tb}\right) \left(d_{_\mathbb{C}}{}_{p}^{a} \sigma^{\mu \nu } u_{_\mathbb{C}}{}_{r}^{b}\right)
		
		,\quad \left(\lambda ^A\right)_c^a G^{A}_{\mu \nu } \left(u_{_\mathbb{C}}{}_{r}^{b} d_{_\mathbb{C}}{}_{p}^{c} \right) \left(u_{_\mathbb{C}}^\dagger{}_{tb} \bar{\sigma}^{\mu \nu } d_{_\mathbb{C}}^\dagger{}_{sa}\right)
		
		\vspace{2ex}\\
		
		\multirow{2}*{$\mathcal{O}_{B u_{_\mathbb{C}} u_{_\mathbb{C}}^{\dagger} d_{_\mathbb{C}} d_{_\mathbb{C}}^{\dagger}}^{\left(1\sim4\right)}      $}
		
		& B_{\mu \nu } \left(d_{_\mathbb{C}}^{\dagger}{}_{sb} u_{_\mathbb{C}}^{\dagger}{}_{ta}\right) \left(d_{_\mathbb{C}}{}_{p}^{a} \sigma^{\mu \nu } u_{_\mathbb{C}}{}_{r}^{b}\right)
		
		,\quad B_{\mu \nu } \left(u_{_\mathbb{C}}{}_{s}^{a} d_{_\mathbb{C}}{}_{p}^{b} \right) \left(u_{_\mathbb{C}}^\dagger{}_{tb} \bar{\sigma}^{\mu \nu } d_{_\mathbb{C}}^\dagger{}_{sa}\right)
		
		\\& B_{\mu \nu } \left(d_{_\mathbb{C}}^{\dagger}{}_{sa} u_{_\mathbb{C}}^{\dagger}{}_{tb}\right) \left(d_{_\mathbb{C}}{}_{p}^{a} \sigma^{\mu \nu } u_{_\mathbb{C}}{}_{r}^{b}\right)
		
		,\quad B_{\mu \nu } \left(u_{_\mathbb{C}}{}_{r}^{b} d_{_\mathbb{C}}{}_{p}^{a} \right) \left(u_{_\mathbb{C}}^\dagger{}_{tb} \bar{\sigma}^{\mu \nu } d_{_\mathbb{C}}^\dagger{}_{sa}\right)
		
		\vspace{2ex}\\
		
		\multirow{4}*{$\mathcal{O}_{G d_{_\mathbb{C}}{}^2 d_{_\mathbb{C}}^{\dagger} {}^2}^{\left(1\sim8\right)}     $}
		
		&\mathcal{Y}\left[\tiny{\young(p,r)},\tiny{\young(st)}\right] \left(\lambda ^A\right)_b^d G^{A}_{\mu \nu } \left(d_{_\mathbb{C}}^{\dagger}{}_{sa} d_{_\mathbb{C}}^{\dagger}{}_{td}\right)  \left(d_{_\mathbb{C}}{}_{p}^{a} \sigma^{\mu \nu } d_{_\mathbb{C}}{}_{r}^{b}\right)
		
		,\quad \mathcal{Y}\left[\tiny{\young(p,r)},\tiny{\young(st)}\right] \left(\lambda ^A\right)_d^b G^{A}_{\mu \nu } \left(d_{_\mathbb{C}}{}_{r}^{d} d_{_\mathbb{C}}{}_{p}^{a} \right)  \left(d_{_\mathbb{C}}^\dagger{}_{tb} \bar{\sigma}^{\mu \nu } d_{_\mathbb{C}}^\dagger{}_{sa}\right)
		
		\\&\mathcal{Y}\left[\tiny{\young(p,r)},\tiny{\young(s,t)}\right] \left(\lambda ^A\right)_b^d G^{A}_{\mu \nu } \left(d_{_\mathbb{C}}^{\dagger}{}_{sa} d_{_\mathbb{C}}^{\dagger}{}_{td}\right)  \left(d_{_\mathbb{C}}{}_{p}^{a} \sigma^{\mu \nu } d_{_\mathbb{C}}{}_{r}^{b}\right)
		
		,\quad \mathcal{Y}\left[\tiny{\young(p,r)},\tiny{\young(s,t)}\right] \left(\lambda ^A\right)_d^b G^{A}_{\mu \nu } \left(d_{_\mathbb{C}}{}_{r}^{d} d_{_\mathbb{C}}{}_{p}^{a} \right)  \left(d_{_\mathbb{C}}^\dagger{}_{tb} \bar{\sigma}^{\mu \nu } d_{_\mathbb{C}}^\dagger{}_{sa}\right)
		
		\\&\mathcal{Y}\left[\tiny{\young(pr)},\tiny{\young(st)}\right] \left(\lambda ^A\right)_b^d G^{A}_{\mu \nu } \left(d_{_\mathbb{C}}^{\dagger}{}_{sa} d_{_\mathbb{C}}^{\dagger}{}_{td}\right)  \left(d_{_\mathbb{C}}{}_{p}^{a} \sigma^{\mu \nu } d_{_\mathbb{C}}{}_{r}^{b}\right)
		
		,\quad \mathcal{Y}\left[\tiny{\young(pr)},\tiny{\young(st)}\right] \left(\lambda ^A\right)_d^b G^{A}_{\mu \nu } \left(d_{_\mathbb{C}}{}_{r}^{d} d_{_\mathbb{C}}{}_{p}^{a} \right)  \left(d_{_\mathbb{C}}^\dagger{}_{tb} \bar{\sigma}^{\mu \nu } d_{_\mathbb{C}}^\dagger{}_{sa}\right)
		
		\\&\mathcal{Y}\left[\tiny{\young(pr)},\tiny{\young(s,t)}\right] \left(\lambda ^A\right)_b^d G^{A}_{\mu \nu } \left(d_{_\mathbb{C}}^{\dagger}{}_{sa} d_{_\mathbb{C}}^{\dagger}{}_{td}\right)  \left(d_{_\mathbb{C}}{}_{p}^{a} \sigma^{\mu \nu } d_{_\mathbb{C}}{}_{r}^{b}\right)
		
		,\quad \mathcal{Y}\left[\tiny{\young(pr)},\tiny{\young(s,t)}\right] \left(\lambda ^A\right)_d^b G^{A}_{\mu \nu } \left(d_{_\mathbb{C}}{}_{r}^{d} d_{_\mathbb{C}}{}_{p}^{a} \right)  \left(d_{_\mathbb{C}}^\dagger{}_{tb} \bar{\sigma}^{\mu \nu } d_{_\mathbb{C}}^\dagger{}_{sa}\right)
		
		\vspace{2ex}\\
		
		\multirow{2}*{$\mathcal{O}_{B d_{_\mathbb{C}}{}^2 d_{_\mathbb{C}}^{\dagger} {}^2}^{\left(1\sim4\right)}     $}
		
		&\mathcal{Y}\left[\tiny{\young(p,r)},\tiny{\young(st)}\right] B_{\mu \nu } \left(d_{_\mathbb{C}}^{\dagger}{}_{sa} d_{_\mathbb{C}}^{\dagger}{}_{tb}\right) \left(d_{_\mathbb{C}}{}_{p}^{a} \sigma^{\mu \nu } d_{_\mathbb{C}}{}_{r}^{b}\right)
		
		,\quad\mathcal{Y}\left[\tiny{\young(p,r)},\tiny{\young(st)}\right] B_{\mu \nu } \left(d_{_\mathbb{C}}{}_{r}^{b} d_{_\mathbb{C}}{}_{p}^{a} \right)  \left(d_{_\mathbb{C}}^\dagger{}_{tb} \bar{\sigma}^{\mu \nu } d_{_\mathbb{C}}^\dagger{}_{sa}\right)
		
		\\&\mathcal{Y}\left[\tiny{\young(pr)},\tiny{\young(s,t)}\right] B_{\mu \nu } \left(d_{_\mathbb{C}}^{\dagger}{}_{sa} d_{_\mathbb{C}}^{\dagger}{}_{tb}\right) \left(d_{_\mathbb{C}}{}_{p}^{a} \sigma^{\mu \nu } d_{_\mathbb{C}}{}_{r}^{b}\right)
		
		,\quad \mathcal{Y}\left[\tiny{\young(pr)},\tiny{\young(s,t)}\right] B_{\mu \nu } \left(d_{_\mathbb{C}}{}_{r}^{b} d_{_\mathbb{C}}{}_{p}^{a} \right)  \left(d_{_\mathbb{C}}^\dagger{}_{tb} \bar{\sigma}^{\mu \nu } d_{_\mathbb{C}}^\dagger{}_{sa}\right)
		
	\end{array}\label{cl:Fq4r2}\end{align}
The only complex four-quark type with additional $G$, $W$ or $B$ constitute the 3 complex types:
\begin{align}\begin{array}{c|l}
		
		\multirow{6}*{$\mathcal{O}_{G Q{}^2 u_{_\mathbb{C}} d_{_\mathbb{C}}}^{\left(1\sim12\right)}      $}
		
		&\mathcal{Y}\left[\tiny{\young(rs)}\right] \epsilon ^{ij} \left(\lambda ^A\right)_d^b G^{A}_{\mu \nu } \left(d_{_\mathbb{C}}{}_{p}^{a} u_{_\mathbb{C}}{}_{t}^{d}\right) \left(Q_{rb i} \sigma ^{\mu \nu } Q_{sa j}\right)
		
		,\quad\mathcal{Y}\left[\tiny{\young(rs)}\right] \epsilon ^{ij} \left(\lambda ^A\right)_a^b G^{A}_{\mu \nu } \left(d_{_\mathbb{C}}{}_{p}^{a} u_{_\mathbb{C}}{}_{t}^{c}\right) \left(Q_{rb i} \sigma ^{\mu \nu } Q_{sc j}\right)
		
		\\&\mathcal{Y}\left[\tiny{\young(rs)}\right] \epsilon ^{ij} \left(\lambda ^A\right)_d^b G^{A}_{\mu \nu } \left(Q_{rb i} u_{_\mathbb{C}}{}_{t}^{d}\right) \left(d_{_\mathbb{C}}{}_{p}^{a} \sigma ^{\mu \nu } Q_{sa j}\right)
		
		,\quad\mathcal{Y}\left[\tiny{\young(rs)}\right] \epsilon ^{ij} \left(\lambda ^A\right)_a^b G^{A}_{\mu \nu } \left(Q_{rb i} u_{_\mathbb{C}}{}_{t}^{c}\right) \left(d_{_\mathbb{C}}{}_{p}^{a} \sigma ^{\mu \nu } Q_{sc j}\right)
		
		\\&\mathcal{Y}\left[\tiny{\young(rs)}\right] \epsilon ^{ij} \left(\lambda ^A\right)_d^c G^{A}_{\mu \nu } \left(Q_{ra i} u_{_\mathbb{C}}{}_{t}^{d}\right) \left(d_{_\mathbb{C}}{}_{p}^{a} \sigma ^{\mu \nu } Q_{sc j}\right)
		
		,\quad\mathcal{Y}\left[\tiny{\young(rs)}\right] \epsilon ^{ij} \left(\lambda ^A\right)_a^c G^{A}_{\mu \nu } \left(Q_{rb i} u_{_\mathbb{C}}{}_{t}^{b}\right) \left(d_{_\mathbb{C}}{}_{p}^{a} \sigma ^{\mu \nu } Q_{sc j}\right)
		
		\\&\mathcal{Y}\left[\tiny{\young(r,s)}\right] \epsilon ^{ij} \left(\lambda ^A\right)_d^b G^{A}_{\mu \nu } \left(d_{_\mathbb{C}}{}_{p}^{a} u_{_\mathbb{C}}{}_{t}^{d}\right) \left(Q_{rb i} \sigma ^{\mu \nu } Q_{sa j}\right)
		
		,\quad\mathcal{Y}\left[\tiny{\young(r,s)}\right] \epsilon ^{ij} \left(\lambda ^A\right)_a^b G^{A}_{\mu \nu } \left(d_{_\mathbb{C}}{}_{p}^{a} u_{_\mathbb{C}}{}_{t}^{c}\right) \left(Q_{rb i} \sigma ^{\mu \nu } Q_{sc j}\right)
		
		\\&\mathcal{Y}\left[\tiny{\young(r,s)}\right] \epsilon ^{ij} \left(\lambda ^A\right)_d^b G^{A}_{\mu \nu } \left(Q_{rb i} u_{_\mathbb{C}}{}_{t}^{d}\right) \left(d_{_\mathbb{C}}{}_{p}^{a} \sigma ^{\mu \nu } Q_{sa j}\right)
		
		,\quad\mathcal{Y}\left[\tiny{\young(r,s)}\right] \epsilon ^{ij} \left(\lambda ^A\right)_a^b G^{A}_{\mu \nu } \left(Q_{rb i} u_{_\mathbb{C}}{}_{t}^{c}\right) \left(d_{_\mathbb{C}}{}_{p}^{a} \sigma ^{\mu \nu } Q_{sc j}\right)
		
		\\&\mathcal{Y}\left[\tiny{\young(r,s)}\right] \epsilon ^{ij} \left(\lambda ^A\right)_d^c G^{A}_{\mu \nu } \left(Q_{ra i} u_{_\mathbb{C}}{}_{t}^{d}\right) \left(d_{_\mathbb{C}}{}_{p}^{a} \sigma ^{\mu \nu } Q_{sc j}\right)
		
		,\quad\mathcal{Y}\left[\tiny{\young(r,s)}\right] \epsilon ^{ij} \left(\lambda ^A\right)_a^c G^{A}_{\mu \nu } \left(Q_{rb i} u_{_\mathbb{C}}{}_{t}^{b}\right) \left(d_{_\mathbb{C}}{}_{p}^{a} \sigma ^{\mu \nu } Q_{sc j}\right)
		
		\vspace{2ex}\\
		
		\multirow{3}*{$\mathcal{O}_{W Q{}^2 u_{_\mathbb{C}} d_{_\mathbb{C}}}^{\left(1\sim6\right)}      $}
		
		&\mathcal{Y}\left[\tiny{\young(r,s)}\right] \epsilon ^{jk} \left(\tau ^I\right)_k^i W^{I}_{\mu \nu } \left(d_{_\mathbb{C}}{}_{p}^{a} u_{_\mathbb{C}}{}_{t}^{b}\right) \left(Q_{rb i} \sigma ^{\mu \nu } Q_{sa j}\right)
		
		,\quad\mathcal{Y}\left[\tiny{\young(r,s)}\right] \epsilon ^{jk} \left(\tau ^I\right)_k^i W^{I}_{\mu \nu } \left(Q_{rb i} u_{_\mathbb{C}}{}_{t}^{b}\right) \left(d_{_\mathbb{C}}{}_{p}^{a} \sigma ^{\mu \nu } Q_{sa j}\right)
		
		\\&\mathcal{Y}\left[\tiny{\young(r,s)}\right] \epsilon ^{jk} \left(\tau ^I\right)_k^i W^{I}_{\mu \nu } \left(Q_{ra i} u_{_\mathbb{C}}{}_{t}^{c}\right) \left(d_{_\mathbb{C}}{}_{p}^{a} \sigma ^{\mu \nu } Q_{sc j}\right)
		
		,\quad\mathcal{Y}\left[\tiny{\young(rs)}\right] \epsilon ^{jk} \left(\tau ^I\right)_k^i W^{I}_{\mu \nu } \left(d_{_\mathbb{C}}{}_{p}^{a} u_{_\mathbb{C}}{}_{t}^{b}\right) \left(Q_{rb i} \sigma ^{\mu \nu } Q_{sa j}\right)
		
		\\&\mathcal{Y}\left[\tiny{\young(rs)}\right] \epsilon ^{jk} \left(\tau ^I\right)_k^i W^{I}_{\mu \nu } \left(Q_{rb i} u_{_\mathbb{C}}{}_{t}^{b}\right) \left(d_{_\mathbb{C}}{}_{p}^{a} \sigma ^{\mu \nu } Q_{sa j}\right)
		
		,\quad\mathcal{Y}\left[\tiny{\young(rs)}\right] \epsilon ^{jk} \left(\tau ^I\right)_k^i W^{I}_{\mu \nu } \left(Q_{ra i} u_{_\mathbb{C}}{}_{t}^{c}\right) \left(d_{_\mathbb{C}}{}_{p}^{a} \sigma ^{\mu \nu } Q_{sc j}\right)
		
		\vspace{2ex}\\
		
		\multirow{3}*{$\mathcal{O}_{B Q{}^2 u_{_\mathbb{C}} d_{_\mathbb{C}}}^{\left(1\sim6\right)}      $}
		
		&\mathcal{Y}\left[\tiny{\young(rs)}\right] \epsilon ^{ij} B_{\mu \nu } \left(d_{_\mathbb{C}}{}_{p}^{a} u_{_\mathbb{C}}{}_{t}^{b}\right) \left(Q_{rb i} \sigma ^{\mu \nu } Q_{sa j}\right)
		
		,\quad\mathcal{Y}\left[\tiny{\young(rs)}\right] \epsilon ^{ij} B_{\mu \nu } \left(Q_{rb i} u_{_\mathbb{C}}{}_{t}^{b}\right) \left(d_{_\mathbb{C}}{}_{p}^{a} \sigma ^{\mu \nu } Q_{sa j}\right)
		
		\\&\mathcal{Y}\left[\tiny{\young(rs)}\right] \epsilon ^{ij} B_{\mu \nu } \left(Q_{ra i} u_{_\mathbb{C}}{}_{t}^{c}\right) \left(d_{_\mathbb{C}}{}_{p}^{a} \sigma ^{\mu \nu } Q_{sc j}\right)
		
		,\quad\mathcal{Y}\left[\tiny{\young(r,s)}\right] \epsilon ^{ij} B_{\mu \nu } \left(d_{_\mathbb{C}}{}_{p}^{a} u_{_\mathbb{C}}{}_{t}^{b}\right) \left(Q_{rb i} \sigma ^{\mu \nu } Q_{sa j}\right)
		
		\\&\mathcal{Y}\left[\tiny{\young(r,s)}\right] \epsilon ^{ij} B_{\mu \nu } \left(Q_{rb i} u_{_\mathbb{C}}{}_{t}^{b}\right) \left(d_{_\mathbb{C}}{}_{p}^{a} \sigma ^{\mu \nu } Q_{sa j}\right)
		
		,\quad\mathcal{Y}\left[\tiny{\young(r,s)}\right] \epsilon ^{ij} B_{\mu \nu } \left(Q_{ra i} u_{_\mathbb{C}}{}_{t}^{c}\right) \left(d_{_\mathbb{C}}{}_{p}^{a} \sigma ^{\mu \nu } Q_{sc j}\right)
		
	\end{array}\label{cl:Fq4c}\end{align}

To see how $\mathcal{Y}$'s act on operators one can refer to eq.(\ref{Y2eg}) and eq.(\ref{Y3eg}). The conversion from the two-component spinors to the four-component spinors are similar to eq.(\ref{D^2}). The Hermitian conjugate of a non-Hermitian operator of this class is, for example,
\begin{align}
	\left[\epsilon ^{ij} B^{\mu \nu } \left(Q_{ra i} u_{_\mathbb{C}}{}_{t}^{c}\right) \left(d_{_\mathbb{C}}{}_{p}^{a} \sigma_{\mu \nu } Q_{sc j}\right)\right]^\dagger = \epsilon _{ij} B^{\mu \nu } \left(u_{_\mathbb{C}}^\dagger{}_{tc} Q^\dagger{}_{r}^{ai}\right) \left(Q^\dagger{}_{s}^{cj} \bar{\sigma}_{\mu \nu } d_{_\mathbb{C}}^\dagger{}_{pa}\right).
\end{align}
Other operators of this class can be converted similarly.

\noindent 2. \underline{Operators involving one lepton and three quarks with $\Delta B=\Delta L=\pm1$:}
In the 4 B-violating four-fermion couplings at dimension 6, $u_{_\mathbb{C}}{}^2d_{_\mathbb{C}}e_{_\mathbb{C}}$ consists of only $SU(2)_W$ singlet, which cannot couple to $W$ in this class. Therefore we have $4+4+3=11$ types in all:
\begin{align}\begin{array}{c|l}
		
		\multirow{8}*{$\mathcal{O}_{G Q{}^3 L}^{\left(1\sim8\right)}    $}
		
		&\mathcal{Y}\left[\tiny{\young(rst)}\right] \epsilon ^{acd} \epsilon ^{ik} \epsilon ^{jm} \left(\lambda ^A\right)_d^b G^{A}_{\mu\nu} \left(Q_{sb k} Q_{tc m}\right) \left(L_{pi} \sigma^{\mu\nu} Q_{ra j}\right)
		
		\\&\mathcal{Y}\left[\tiny{\young(rst)}\right] \epsilon ^{acd} \epsilon ^{ij} \epsilon ^{km} \left(\lambda ^A\right)_d^b G^{A}_{\mu\nu} \left(L_{pi} Q_{tc m}\right) \left(Q_{ra j} \sigma^{\mu\nu} Q_{sb k}\right)
		
		\\&\mathcal{Y}\left[\tiny{\young(rs,t)}\right] \epsilon ^{acd} \epsilon ^{ik} \epsilon ^{jm} \left(\lambda ^A\right)_d^b G^{A}_{\mu\nu} \left(Q_{sb k} Q_{tc m}\right) \left(L_{pi} \sigma^{\mu\nu} Q_{ra j}\right)
		
		\\&\mathcal{Y}\left[\tiny{\young(rs,t)}\right] \epsilon ^{acd} \epsilon ^{ij} \epsilon ^{km} \left(\lambda ^A\right)_d^b G^{A}_{\mu\nu} \left(Q_{sb k} Q_{tc m}\right) \left(L_{pi} \sigma^{\mu\nu} Q_{ra j}\right)
		
		\\&\mathcal{Y}\left[\tiny{\young(rs,t)}\right] \epsilon ^{bcd} \epsilon ^{ik} \epsilon ^{jm} \left(\lambda ^A\right)_d^a G^{A}_{\mu\nu} \left(Q_{sb k} Q_{tc m}\right) \left(L_{pi} \sigma^{\mu\nu} Q_{ra j}\right)
		
		\\&\mathcal{Y}\left[\tiny{\young(rs,t)}\right] \epsilon ^{acd} \epsilon ^{ik} \epsilon ^{jm} \left(\lambda ^A\right)_d^b G^{A}_{\mu\nu} \left(L_{pi} Q_{tc m}\right) \left(Q_{ra j} \sigma^{\mu\nu} Q_{sb k}\right)
		
		\\&\mathcal{Y}\left[\tiny{\young(r,s,t)}\right] \epsilon ^{acd} \epsilon ^{ik} \epsilon ^{jm} \left(\lambda ^A\right)_d^b G^{A}_{\mu\nu} \left(Q_{sb k} Q_{tc m}\right) \left(L_{pi} \sigma^{\mu\nu} Q_{ra j}\right)
		
		\\&\mathcal{Y}\left[\tiny{\young(r,s,t)}\right] \epsilon ^{acd} \epsilon ^{ik} \epsilon ^{jm} \left(\lambda ^A\right)_d^b G^{A}_{\mu\nu} \left(L_{pi} Q_{tc m}\right) \left(Q_{ra j} \sigma^{\mu\nu} Q_{sb k}\right)
		
	\end{array}\label{cl:Fq3l11}\end{align}

\begin{align}\begin{array}{c|l}
		
		\multirow{6}*{$\mathcal{O}_{W Q{}^3 L}^{\left(1\sim6\right)}    $}
		
		&\mathcal{Y}\left[\tiny{\young(rst)}\right] \epsilon ^{abc} \epsilon ^{jn} \epsilon ^{km} \left(\tau ^I\right)_n^i W^{I}_{\mu\nu } \left(Q_{sb k} Q_{tc m}\right) \left(L_{pi} \sigma^{\mu\nu } Q_{ra j}\right)
		
		\\&\mathcal{Y}\left[\tiny{\young(rst)}\right] \epsilon ^{abc} \epsilon ^{ik} \epsilon ^{jn} \left(\tau ^I\right)_n^m W^{I}_{\mu\nu } \left(L_{pi} Q_{tc m}\right) \left(Q_{ra j} \sigma^{\mu\nu } Q_{sb k}\right)
		
		\\&\mathcal{Y}\left[\tiny{\young(rs,t)}\right] \epsilon ^{abc} \epsilon ^{jn} \epsilon ^{km} \left(\tau ^I\right)_n^i W^{I}_{\mu\nu } \left(Q_{sb k} Q_{tc m}\right) \left(L_{pi} \sigma^{\mu\nu } Q_{ra j}\right)
		
		\\&\mathcal{Y}\left[\tiny{\young(rs,t)}\right] \epsilon ^{abc} \epsilon ^{ik} \epsilon ^{jn} \left(\tau ^I\right)_n^m W^{I}_{\mu\nu } \left(Q_{sb k} Q_{tc m}\right) \left(L_{pi} \sigma^{\mu\nu } Q_{ra j}\right)
		
		\\&\mathcal{Y}\left[\tiny{\young(rs,t)}\right] \epsilon ^{abc} \epsilon ^{jn} \epsilon ^{km} \left(\tau ^I\right)_n^i W^{I}_{\mu\nu } \left(L_{pi} Q_{tc m}\right) \left(Q_{ra j} \sigma^{\mu\nu } Q_{sb k}\right)
		
		\\&\mathcal{Y}\left[\tiny{\young(r,s,t)}\right] \epsilon ^{abc} \epsilon ^{ik} \epsilon ^{jn} \left(\tau ^I\right)_n^m W^{I}_{\mu\nu } \left(Q_{sb k} Q_{tc m}\right) \left(L_{pi} \sigma^{\mu\nu } Q_{ra j}\right)
		
		\vspace{2ex}\\
		
		\multirow{2}*{$\mathcal{O}_{B Q{}^3 L}^{\left(1\sim4\right)}    $}
		
		&\mathcal{Y}\left[\tiny{\young(rs,t)}\right] \epsilon ^{abc} \epsilon ^{ik} \epsilon ^{jm} B_{\mu\nu } \left(Q_{sb k} Q_{tc m}\right) \left(L_{pi} \sigma^{\mu\nu } Q_{ra j}\right)
		
		,\quad\mathcal{Y}\left[\tiny{\young(rs,t)}\right] \epsilon ^{abc} \epsilon ^{ik} \epsilon ^{jm} B_{\mu\nu } \left(L_{pi} Q_{tc m}\right) \left(Q_{ra j} \sigma^{\mu\nu } Q_{sb k}\right)
		
		\\&\mathcal{Y}\left[\tiny{\young(rst)}\right] \epsilon ^{abc} \epsilon ^{ik} \epsilon ^{jm} B_{\mu\nu } \left(Q_{sb k} Q_{tc m}\right) \left(L_{pi} \sigma^{\mu\nu } Q_{ra j}\right)
		
		,\quad\mathcal{Y}\left[\tiny{\young(r,s,t)}\right] \epsilon ^{abc} \epsilon ^{ik} \epsilon ^{jm} B_{\mu\nu } \left(Q_{sb k} Q_{tc m}\right) \left(L_{pi} \sigma^{\mu\nu } Q_{ra j}\right)
		
		\vspace{2ex}\\
		
		\multirow{2}*{$\mathcal{O}_{G Q{}^2 u_{_\mathbb{C}}^{\dagger} e_{_\mathbb{C}}^{\dagger}}^{\left(1\sim4\right)}     $}
		
		&\mathcal{Y}\left[\tiny{\young(pr)}\right]  \epsilon ^{acd} \epsilon ^{ij} \left(\lambda ^A\right)_d^b G^{A}_{\mu \nu } \left(e_{_\mathbb{C}}^{\dagger}{}_{s} u_{_\mathbb{C}}^{\dagger}{}_{tc}\right) \left(Q_{pa i} \sigma^{\mu \nu } Q_{rb j}\right)
		
		,\quad\mathcal{Y}\left[\tiny{\young(pr)}\right] \epsilon ^{acd} \epsilon ^{ij} \left(\lambda ^A\right)_d^b G^{A}_{\mu \nu } \left(Q_{rc j} Q_{pb i}\right) \left(u_{_\mathbb{C}}^\dagger{}_{ta} \bar{\sigma}^{\mu \nu } e_{_\mathbb{C}}^\dagger{}_s\right)
		
		\\&\mathcal{Y}\left[\tiny{\young(p,r)}\right]  \epsilon ^{acd} \epsilon ^{ij} \left(\lambda ^A\right)_d^b G^{A}_{\mu \nu } \left(e_{_\mathbb{C}}^{\dagger}{}_{s} u_{_\mathbb{C}}^{\dagger}{}_{tc}\right) \left(Q_{pa i} \sigma^{\mu \nu } Q_{rb j}\right)
		
		,\quad\mathcal{Y}\left[\tiny{\young(p,r)}\right] \epsilon ^{acd} \epsilon ^{ij} \left(\lambda ^A\right)_d^b G^{A}_{\mu \nu } \left(Q_{rc j} Q_{pb i}\right) \left(u_{_\mathbb{C}}^\dagger{}_{ta} \bar{\sigma}^{\mu \nu } e_{_\mathbb{C}}^\dagger{}_s\right)
		
		\vspace{2ex}\\
		
		\mathcal{O}_{W Q{}^2 u_{_\mathbb{C}}^{\dagger} e_{_\mathbb{C}}^{\dagger}}^{\left(1,2\right)}      
		
		&\mathcal{Y}\left[\tiny{\young(pr)}\right]  \epsilon ^{abc} \epsilon ^{jk} \left(\tau ^I\right)_k^i W^{I}_{\mu \nu } \left(e_{_\mathbb{C}}^{\dagger}{}_{s} u_{_\mathbb{C}}^{\dagger}{}_{tc}\right)  \left(Q_{pa i} \sigma^{\mu \nu } Q_{rb j}\right)
		
		,\quad\mathcal{Y}\left[\tiny{\young(p,r)}\right] \epsilon ^{abc} \epsilon ^{jk} \left(\tau ^I\right)_k^i W^{I}_{\mu \nu } \left(Q_{rc j} Q_{pb i}\right) \left(u_{_\mathbb{C}}^\dagger{}_{ta} \bar{\sigma}^{\mu \nu } e_{_\mathbb{C}}^\dagger{}_s\right)
		
		\vspace{2ex}\\
		
		\mathcal{O}_{B Q{}^2 u_{_\mathbb{C}}^{\dagger} e_{_\mathbb{C}}^{\dagger}}^{\left(1,2\right)}    
		
		&\mathcal{Y}\left[\tiny{\young(p,r)}\right] \epsilon ^{abc} \epsilon ^{ij} B_{\mu \nu } \left(e_{_\mathbb{C}}^{\dagger}{}_{s} u_{_\mathbb{C}}^{\dagger}{}_{tc}\right) \left(Q_{pa i} \sigma^{\mu \nu } Q_{rb j}\right)
		
		,\quad\mathcal{Y}\left[\tiny{\young(pr)}\right] \epsilon ^{abc} \epsilon ^{ij} B_{\mu \nu } \left(Q_{rc j} Q_{pb i}\right) \left(u_{_\mathbb{C}}^\dagger{}_{ta} \bar{\sigma}^{\mu \nu } e_{_\mathbb{C}}^\dagger{}_s\right)
		
		\vspace{2ex}\\
		
		\multirow{2}*{$\mathcal{O}_{G Q u_{_\mathbb{C}}^{\dagger} d_{_\mathbb{C}}^{\dagger}  L}^{\left(1\sim4\right)}     $}
		
		& \epsilon ^{acd} \epsilon ^{ij} \left(\lambda ^A\right)_d^b G^{A}_{\mu \nu } \left(d_{_\mathbb{C}}^{\dagger}{}_{sb} u_{_\mathbb{C}}^{\dagger}{}_{tc}\right) \left(L_{pi} \sigma^{\mu \nu } Q_{ra j}\right)
		
		,\quad \epsilon ^{bcd} \epsilon ^{ij} \left(\lambda ^A\right)_d^a G^{A}_{\mu \nu } \left(d_{_\mathbb{C}}^{\dagger}{}_{sb} u_{_\mathbb{C}}^{\dagger}{}_{tc}\right) \left(L_{pi} \sigma^{\mu \nu } Q_{ra j}\right)
		
		\\& \epsilon ^{abd} \epsilon ^{ij} \left(\lambda ^A\right)_d^c G^{A}_{\mu \nu } \left(Q_{rc j} L_{pi} \right) \left(u_{_\mathbb{C}}^\dagger{}_{tb} \bar{\sigma}^{\mu \nu } d_{_\mathbb{C}}^\dagger{}_{sa} \right)
		
		,\quad \epsilon ^{acd} \epsilon ^{ij} \left(\lambda ^A\right)_d^b G^{A}_{\mu \nu } \left(Q_{rc j} L_{pi} \right) \left(u_{_\mathbb{C}}^\dagger{}_{tb} \bar{\sigma}^{\mu \nu } d_{_\mathbb{C}}^\dagger{}_{sa} \right)
		
		\vspace{2ex}\\
		
		\mathcal{O}_{W Q u_{_\mathbb{C}}^{\dagger} d_{_\mathbb{C}}^{\dagger}  L}^{\left(1,2\right)}
		
		& \epsilon ^{abc} \epsilon ^{jk} \left(\tau ^I\right)_k^i W^{I}_{\mu \nu } \left(d_{_\mathbb{C}}^{\dagger}{}_{sb} u_{_\mathbb{C}}^{\dagger}{}_{tc}\right) \left(L_{pi} \sigma^{\mu \nu } Q_{ra j}\right)
		
		,\quad \epsilon ^{abc} \epsilon ^{jk} \left(\tau ^I\right)_k^i W^{I}_{\mu \nu } \left(Q_{rc j} L_{pi} \right) \left(u_{_\mathbb{C}}^\dagger{}_{tb} \bar{\sigma}^{\mu \nu } d_{_\mathbb{C}}^\dagger{}_{sa} \right)
		
		\vspace{2ex}\\
		
		\mathcal{O}_{B Q u_{_\mathbb{C}}^{\dagger} d_{_\mathbb{C}}^{\dagger}  L}^{\left(1,2\right)} 
		
		& \epsilon ^{abc} \epsilon ^{ij} B_{\mu \nu } \left(d_{_\mathbb{C}}^{\dagger}{}_{sb} u_{_\mathbb{C}}^{\dagger}{}_{tc}\right) \left(L_{pi} \sigma^{\mu \nu } Q_{ra j}\right)
		
		,\quad \epsilon ^{abc} \epsilon ^{ij} B_{\mu \nu } \left(Q_{rc j} L_{pi} \right) \left(u_{_\mathbb{C}}^\dagger{}_{tb} \bar{\sigma}^{\mu \nu } d_{_\mathbb{C}}^\dagger{}_{sa} \right)
		
		\vspace{2ex}\\
		
		\multirow{3}*{$\mathcal{O}_{G u_{_\mathbb{C}}{}^2 d_{_\mathbb{C}} e_{_\mathbb{C}}}^{\left(1\sim6\right)}    $}
		
		&\mathcal{Y}\left[\tiny{\young(s,t)}\right] \epsilon _{acd} \left(\lambda ^A\right)_b^d G^{A}_{\mu \nu } \left(d_{_\mathbb{C}}{}_{p}^{a} u_{_\mathbb{C}}{}_{t}^{c}\right) \left(e_{_\mathbb{C} r} \sigma ^{\mu \nu } u_{_\mathbb{C}}{}_{s}^{b}\right)
		
		,\quad\mathcal{Y}\left[\tiny{\young(s,t)}\right] \epsilon _{abd} \left(\lambda ^A\right)_c^d G^{A}_{\mu \nu } \left(d_{_\mathbb{C}}{}_{p}^{a} u_{_\mathbb{C}}{}_{t}^{c}\right) \left(e_{_\mathbb{C} r} \sigma ^{\mu \nu } u_{_\mathbb{C}}{}_{s}^{b}\right)
		
		\\&\mathcal{Y}\left[\tiny{\young(s,t)}\right] \epsilon _{acd} \left(\lambda ^A\right)_b^d G^{A}_{\mu \nu } \left(e_{_\mathbb{C} r} u_{_\mathbb{C}}{}_{t}^{c}\right) \left(d_{_\mathbb{C}}{}_{p}^{a} \sigma ^{\mu \nu } u_{_\mathbb{C}}{}_{s}^{b}\right)
		
		,\quad\mathcal{Y}\left[\tiny{\young(st)}\right] \epsilon _{acd} \left(\lambda ^A\right)_b^d G^{A}_{\mu \nu } \left(d_{_\mathbb{C}}{}_{p}^{a} u_{_\mathbb{C}}{}_{t}^{c}\right) \left(e_{_\mathbb{C} r} \sigma ^{\mu \nu } u_{_\mathbb{C}}{}_{s}^{b}\right)
		
		\\&\mathcal{Y}\left[\tiny{\young(st)}\right] \epsilon _{abd} \left(\lambda ^A\right)_c^d G^{A}_{\mu \nu } \left(d_{_\mathbb{C}}{}_{p}^{a} u_{_\mathbb{C}}{}_{t}^{c}\right) \left(e_{_\mathbb{C} r} \sigma ^{\mu \nu } u_{_\mathbb{C}}{}_{s}^{b}\right)
		
		,\quad\mathcal{Y}\left[\tiny{\young(st)}\right] \epsilon _{acd} \left(\lambda ^A\right)_b^d G^{A}_{\mu \nu } \left(e_{_\mathbb{C} r} u_{_\mathbb{C}}{}_{t}^{c}\right) \left(d_{_\mathbb{C}}{}_{p}^{a} \sigma ^{\mu \nu } u_{_\mathbb{C}}{}_{s}^{b}\right)
		
		\vspace{2ex}\\
		
		\multirow{2}*{$\mathcal{O}_{B u_{_\mathbb{C}}{}^2 d_{_\mathbb{C}} e_{_\mathbb{C}}}^{\left(1\sim3\right)}    $}
		
		&\mathcal{Y}\left[\tiny{\young(st)}\right] \epsilon _{abc} B^{\mu \nu }  \left(d_{_\mathbb{C}}{}_{p}^{a} u_{_\mathbb{C}}{}_{t}^{c}\right) \left(e_{_\mathbb{C} r} \sigma_{\mu \nu } u_{_\mathbb{C}}{}_{s}^{b}\right)
		
		,\quad\mathcal{Y}\left[\tiny{\young(st)}\right] \epsilon _{abc} B^{\mu \nu }  \left(e_{_\mathbb{C} r} u_{_\mathbb{C}}{}_{t}^{c}\right) \left(d_{_\mathbb{C}}{}_{p}^{a} \sigma_{\mu \nu } u_{_\mathbb{C}}{}_{s}^{b}\right)
		
		\\&\mathcal{Y}\left[\tiny{\young(s,t)}\right] \epsilon _{abc} B^{\mu \nu }  \left(d_{_\mathbb{C}}{}_{p}^{a} u_{_\mathbb{C}}{}_{t}^{c}\right) \left(e_{_\mathbb{C} r} \sigma_{\mu \nu } u_{_\mathbb{C}}{}_{s}^{b}\right)
		
	\end{array}\label{cl:Fq3l12}\end{align}

\noindent 3. \underline{Operators involving two leptons and two quarks:}
Among the 6 real types with two leptons and two quarks, 2 involve only $SU(2)_W$ singlets. Hence we have $6+6+4=16$ real types:
\begin{align}\begin{array}{c|l}
		
		\multirow{2}*{$\mathcal{O}_{G Q Q^{\dagger} L L^{\dagger}}^{\left(1\sim4\right)}      $}
		
		& \left(\lambda ^A\right)_b^a G^{A}_{\mu \nu } \left(L^{\dagger}{}_{s}^{j} Q^{\dagger}{}_{t}^{b i}\right) \left(L_{pi} \sigma^{\mu \nu } Q_{ra j}\right)
		
		,\quad \left(\lambda ^A\right)_a^b G^{A}_{\mu \nu } \left(Q{}_{rb i} L{}_{pj} \right) \left(Q^\dagger{}_{t}^{a j} \bar{\sigma}^{\mu \nu } L^\dagger{}_{s}^{i}\right)
		
		\\& \left(\lambda ^A\right)_b^a G^{A}_{\mu \nu } \left(L^{\dagger}{}_{s}^{i} Q^{\dagger}{}_{t}^{b j}\right) \left(L_{pi} \sigma^{\mu \nu } Q_{ra j}\right)
		
		,\quad \left(\lambda ^A\right)_a^b G^{A}_{\mu \nu } \left(Q{}_{rb j} L{}_{pi} \right) \left(Q^\dagger{}_{t}^{a j} \bar{\sigma}^{\mu \nu } L^\dagger{}_{s}^{i}\right)
		
		\vspace{2ex}\\
		
		\multirow{3}*{$\mathcal{O}_{W Q Q^{\dagger} L L^{\dagger}}^{\left(1\sim6\right)}      $}
		
		& \left(\tau ^I\right)_k^j W^{I}_{\mu \nu } \left(L^{\dagger}{}_{s}^{i} Q^{\dagger}{}_{t}^{a k}\right) \left(L_{pi} \sigma^{\mu \nu } Q_{ra j}\right)
		
		,\quad \left(\tau ^I\right)_j^k W^{I}_{\mu \nu } \left(Q{}_{ra k} L{}_{pi} \right) \left(Q^\dagger{}_{t}^{a j} \bar{\sigma}^{\mu \nu } L^\dagger{}_{s}^{i}\right)
		
		\\& \left(\tau ^I\right)_k^j W^{I}_{\mu \nu } \left(L^{\dagger}{}_{s}^{k} Q^{\dagger}{}_{t}^{a i}\right) \left(L_{pi} \sigma^{\mu \nu } Q_{ra j}\right)
		
		,\quad \left(\tau ^I\right)_j^k W^{I}_{\mu \nu } \left(Q{}_{ra i} L{}_{pk} \right) \left(Q^\dagger{}_{t}^{a j} \bar{\sigma}^{\mu \nu } L^\dagger{}_{s}^{i}\right)
		
		\\& \left(\tau ^I\right)_k^i W^{I}_{\mu \nu } \left(L^{\dagger}{}_{s}^{k} Q^{\dagger}{}_{t}^{a j}\right) \left(L_{pi} \sigma^{\mu \nu } Q_{ra j}\right)
		
		,\quad \left(\tau ^I\right)_i^k W^{I}_{\mu \nu } \left(Q{}_{ra j} L{}_{pk} \right) \left(Q^\dagger{}_{t}^{a j} \bar{\sigma}^{\mu \nu } L^\dagger{}_{s}^{i}\right)
		
		\vspace{2ex}\\
		
		\multirow{2}*{$\mathcal{O}_{B Q Q^{\dagger} L L^{\dagger}}^{\left(1\sim4\right)}      $}
		
		& B_{\mu \nu } \left(L^{\dagger}{}_{s}^{j} Q^{\dagger}{}_{t}^{a i}\right) \left(L_{pi} \sigma^{\mu \nu } Q_{ra j}\right)
		
		,\quad B_{\mu \nu } \left(Q{}_{ra i} L{}_{pj} \right) \left(Q^\dagger{}_{t}^{a j} \bar{\sigma}^{\mu \nu } L^\dagger{}_{s}^{i}\right)
		
		\\& B_{\mu \nu } \left(L^{\dagger}{}_{s}^{i} Q^{\dagger}{}_{t}^{a j}\right) \left(L_{pi} \sigma^{\mu \nu } Q_{ra j}\right)
		
		,\quad B_{\mu \nu } \left(Q{}_{ra j} L{}_{pi} \right) \left(Q^\dagger{}_{t}^{a j} \bar{\sigma}^{\mu \nu } L^\dagger{}_{s}^{i}\right)
		
		\vspace{2ex}\\
		
		\mathcal{O}_{G Q Q^{\dagger} e_{_\mathbb{C}} e_{_\mathbb{C}}^{\dagger}}^{\left(1,2\right)}        
		
		& \left(\lambda ^A\right)_b^a G^{A}_{\mu \nu } \left(e_{_\mathbb{C}}^{\dagger}{}_{s} Q^{\dagger}{}_{t}^{b i}\right) \left(e_{_\mathbb{C} p} \sigma^{\mu \nu } Q_{ra i}\right)
		
		,\quad \left(\lambda ^A\right)_a^b G^{A}_{\mu \nu } \left(Q{}_{rb i} e_{_\mathbb{C}}{}_{p} \right) \left(Q^\dagger{}_{t}^{a i} \bar{\sigma}^{\mu \nu } e_{_\mathbb{C}}^\dagger{}_{s}\right)
		
		\vspace{2ex}\\
		
		\mathcal{O}_{W Q Q^{\dagger} e_{_\mathbb{C}} e_{_\mathbb{C}}^{\dagger}}^{\left(1,2\right)}        
		
		& \left(\tau ^I\right)_j^i W^{I}_{\mu \nu } \left(e_{_\mathbb{C}}^{\dagger}{}_{s} Q^{\dagger}{}_{t}^{a j}\right) \left(e_{_\mathbb{C} p} \sigma^{\mu \nu } Q_{ra i}\right)
		
		,\quad \left(\tau ^I\right)_i^j W^{I}_{\mu \nu } \left(Q{}_{ra j} e_{_\mathbb{C}}{}_{p} \right) \left(Q^\dagger{}_{t}^{a i} \bar{\sigma}^{\mu \nu } e_{_\mathbb{C}}^\dagger{}_{s}\right)
		
		\vspace{2ex}\\
		
		\mathcal{O}_{B Q Q^{\dagger} e_{_\mathbb{C}} e_{_\mathbb{C}}^{\dagger}}^{\left(1,2\right)}        
		
		& B_{\mu \nu } \left(e_{_\mathbb{C}}^{\dagger}{}_{s} Q^{\dagger}{}_{t}^{a i}\right) \left(e_{_\mathbb{C} p} \sigma^{\mu \nu } Q_{ra i}\right)
		
		,\quad B_{\mu \nu } \left(Q{}_{ra i} e_{_\mathbb{C}}{}_{p} \right) \left(Q^\dagger{}_{t}^{a i} \bar{\sigma}^{\mu \nu } e_{_\mathbb{C}}^\dagger{}_{s}\right)
		
		\vspace{2ex}\\
		
		\mathcal{O}_{G u_{_\mathbb{C}} u_{_\mathbb{C}}^{\dagger} L L^{\dagger}}^{\left(1,2\right)}        
		
		& \left(\lambda ^A\right)_a^b G^{A}_{\mu \nu } \left(L^{\dagger}{}_{s}^{i} u_{_\mathbb{C}}^{\dagger}{}_{tb}\right) \left(L_{pi} \sigma^{\mu \nu } u_{_\mathbb{C}}{}_{r}^{a}\right)
		
		,\quad \left(\lambda ^A\right)_b^a G^{A}_{\mu \nu } \left(u_{_\mathbb{C}}{}_{r}^{b} L{}_{pi} \right) \left(u_{_\mathbb{C}}^\dagger{}_{ta} \bar{\sigma}^{\mu \nu } L^\dagger{}_{s}^{i}\right)
		
		\vspace{2ex}\\
		
		\mathcal{O}_{W u_{_\mathbb{C}} u_{_\mathbb{C}}^{\dagger} L L^{\dagger}}^{\left(1,2\right)}        
		
		& \left(\tau ^I\right)_j^i W^{I}_{\mu \nu } \left(L^{\dagger}{}_{s}^{j} u_{_\mathbb{C}}^{\dagger}{}_{ta}\right) \left(L_{pi} \sigma^{\mu \nu } u_{_\mathbb{C}}{}_{r}^{a}\right)
		
		,\quad \left(\tau ^I\right)_i^j W^{I}_{\mu \nu } \left(u_{_\mathbb{C}}{}_{r}^{a} L{}_{pj} \right) \left(u_{_\mathbb{C}}^\dagger{}_{ta} \bar{\sigma}^{\mu \nu } L^\dagger{}_{s}^{i}\right)
		
		\vspace{2ex}\\
		
		\mathcal{O}_{B u_{_\mathbb{C}} u_{_\mathbb{C}}^{\dagger} L L^{\dagger}}^{\left(1,2\right)}        
		
		& B_{\mu \nu } \left(L^{\dagger}{}_{s}^{i} u_{_\mathbb{C}}^{\dagger}{}_{ta}\right) \left(L_{pi} \sigma^{\mu \nu } u_{_\mathbb{C}}{}_{r}^{a}\right)
		
		,\quad B_{\mu \nu } \left(u_{_\mathbb{C}}{}_{r}^{a} L{}_{pi} \right) \left(u_{_\mathbb{C}}^\dagger{}_{ta} \bar{\sigma}^{\mu \nu } L^\dagger{}_{s}^{i}\right)
		
		\vspace{2ex}\\
		
		\mathcal{O}_{G u_{_\mathbb{C}} u_{_\mathbb{C}}^{\dagger} e_{_\mathbb{C}} e_{_\mathbb{C}}^{\dagger}}^{\left(1,2\right)}        
		
		& \left(\lambda ^A\right)_a^b G^{A}_{\mu \nu } \left(e_{_\mathbb{C}}^{\dagger}{}_{s} u_{_\mathbb{C}}^{\dagger}{}_{tb}\right) \left(e_{_\mathbb{C} p} \sigma^{\mu \nu } u_{_\mathbb{C}}{}_{r}^{a}\right)
		
		,\quad \left(\lambda ^A\right)_b^a G^{A}_{\mu \nu } \left(u_{_\mathbb{C}}{}_{r}^{b} e_{_\mathbb{C}}{}_{p} \right) \left(u_{_\mathbb{C}}^\dagger{}_{ta} \bar{\sigma}^{\mu \nu } e_{_\mathbb{C}}^\dagger{}_s\right)
		
		\vspace{2ex}\\
		
		\mathcal{O}_{B u_{_\mathbb{C}} u_{_\mathbb{C}}^{\dagger} e_{_\mathbb{C}} e_{_\mathbb{C}}^{\dagger}}^{\left(1,2\right)}        
		
		& B_{\mu \nu } \left(e_{_\mathbb{C}}^{\dagger}{}_{s} u_{_\mathbb{C}}^{\dagger}{}_{ta}\right) \left(e_{_\mathbb{C} p} \sigma^{\mu \nu } u_{_\mathbb{C}}{}_{r}^{a}\right)
		
		,\quad B_{\mu \nu } \left(u_{_\mathbb{C}}{}_{r}^{a} e_{_\mathbb{C}}{}_{p} \right) \left(u_{_\mathbb{C}}^\dagger{}_{ta} \bar{\sigma}^{\mu \nu } e_{_\mathbb{C}}^\dagger{}_s\right)
		
		\vspace{2ex}\\
		
		\mathcal{O}_{G d_{_\mathbb{C}} d_{_\mathbb{C}}^{\dagger}  L L^{\dagger}}^{\left(1,2\right)}      
		
		& \left(\lambda ^A\right)_a^b G^{A}_{\mu \nu } \left(d_{_\mathbb{C}}^{\dagger}{}_{sb} L^{\dagger}{}_{t}^{i}\right) \left(d_{_\mathbb{C}}{}_{p}^{a} \sigma^{\mu \nu } L_{ri}\right)
		
		,\quad \left(\lambda ^A\right)_b^a G^{A}_{\mu \nu } \left(L{}_{ri} d_{_\mathbb{C}}{}_{p}^{b} \right) \left(L^\dagger{}_{t}^{i} \bar{\sigma}^{\mu \nu } d_{_\mathbb{C}}^\dagger{}_{sa} \right)
		
		\vspace{2ex}\\
		
		\mathcal{O}_{W d_{_\mathbb{C}} d_{_\mathbb{C}}^{\dagger}  L L^{\dagger}}^{\left(1,2\right)}      
		
		& \left(\tau ^I\right)_j^i W^{I}_{\mu \nu } \left(d_{_\mathbb{C}}^{\dagger}{}_{sa} L^{\dagger}{}_{t}^{j}\right) \left(d_{_\mathbb{C}}{}_{p}^{a} \sigma^{\mu \nu } L_{ri}\right)
		
		,\quad \left(\tau ^I\right)_i^j W^{I}_{\mu \nu } \left(L{}_{rj} d_{_\mathbb{C}}{}_{p}^{a} \right) \left(L^\dagger{}_{t}^{i} \bar{\sigma}^{\mu \nu } d_{_\mathbb{C}}^\dagger{}_{sa} \right)
		
		\vspace{2ex}\\
		
		\mathcal{O}_{B d_{_\mathbb{C}} d_{_\mathbb{C}}^{\dagger}  L L^{\dagger}}^{\left(1,2\right)}      
		
		& B_{\mu \nu } \left(d_{_\mathbb{C}}^{\dagger}{}_{sa} L^{\dagger}{}_{t}^{i}\right) \left(d_{_\mathbb{C}}{}_{p}^{a} \sigma^{\mu \nu } L_{ri}\right)
		
		,\quad B_{\mu \nu } \left(L{}_{ri} d_{_\mathbb{C}}{}_{p}^{a} \right) \left(L^\dagger{}_{t}^{i} \bar{\sigma}^{\mu \nu } d_{_\mathbb{C}}^\dagger{}_{sa} \right)
		
		\vspace{2ex}\\
		
		\mathcal{O}_{G d_{_\mathbb{C}} d_{_\mathbb{C}}^{\dagger}  e_{_\mathbb{C}} e_{_\mathbb{C}}^{\dagger}}^{\left(1,2\right)}      
		
		& \left(\lambda ^A\right)_a^b G^{A}_{\mu \nu } \left(d_{_\mathbb{C}}^{\dagger}{}_{sb} e_{_\mathbb{C}}^{\dagger}{}_{t}\right) \left(d_{_\mathbb{C}}{}_{p}^{a} \sigma^{\mu \nu } e_{_\mathbb{C} r}\right)
		
		,\quad \left(\lambda ^A\right)_b^a G^{A}_{\mu \nu } \left(e_{_\mathbb{C}}{}_{r} d_{_\mathbb{C}}{}_{p}^{b} \right) \left(e_{_\mathbb{C}}^\dagger{}_t \bar{\sigma}^{\mu \nu } d_{_\mathbb{C}}^\dagger{}_{sa}\right)
		
		\vspace{2ex}\\
		
		\mathcal{O}_{B d_{_\mathbb{C}} d_{_\mathbb{C}}^{\dagger}  e_{_\mathbb{C}} e_{_\mathbb{C}}^{\dagger}}^{\left(1,2\right)}      
		
		& B_{\mu \nu } \left(d_{_\mathbb{C}}^{\dagger}{}_{sa} e_{_\mathbb{C}}^{\dagger}{}_{t}\right) \left(d_{_\mathbb{C}}{}_{p}^{a} \sigma^{\mu \nu } e_{_\mathbb{C} r}\right)
		
		,\quad B_{\mu \nu } \left(e_{_\mathbb{C}}{}_{r} d_{_\mathbb{C}}{}_{p}^{a} \right) \left(e_{_\mathbb{C}}^\dagger{}_t \bar{\sigma}^{\mu \nu } d_{_\mathbb{C}}^\dagger{}_{sa}\right).
		
	\end{array}\label{cl:Fq2l2r}\end{align}
Both of the 2 complex four-fermion types with two leptons and two quarks can couple to all of the three gauge bosons, hence we have $2+2+2=6$ complex types:
\begin{align}\begin{array}{c|l}
		
		\multirow{2}*{$\mathcal{O}_{G Q u_{_\mathbb{C}} L e_{_\mathbb{C}}}^{\left(1\sim3\right)}      $}
		
		& \epsilon ^{ij} \left(\lambda ^A\right)_b^a G^{A}{}^{\mu \nu } \left(e_{_\mathbb{C} p} u_{_\mathbb{C}}{}_{t}^{b}\right) \left(L_{ri} \sigma_{\mu \nu } Q_{sa j}\right)
		
		,\quad  \epsilon ^{ij} \left(\lambda ^A\right)_b^a G^{A}{}^{\mu \nu } \left(L_{ri} u_{_\mathbb{C}}{}_{t}^{b}\right) \left(e_{_\mathbb{C} p} \sigma_{\mu \nu } Q_{sa j}\right)
		
		\\&  \epsilon ^{ij} \left(\lambda ^A\right)_b^a G^{A}{}^{\mu \nu } \left(Q_{sa j} u_{_\mathbb{C}}{}_{t}^{b}\right) \left(e_{_\mathbb{C} p} \sigma_{\mu \nu } L_{ri}\right)
		
		\vspace{2ex}\\
		
		\multirow{2}*{$\mathcal{O}_{W Q u_{_\mathbb{C}} L e_{_\mathbb{C}}}^{\left(1\sim3\right)}      $}
		
		&  \epsilon ^{jk} \left(\tau ^I\right)_k^i W^{I}{}^{\mu \nu } \left(e_{_\mathbb{C} p} u_{_\mathbb{C}}{}_{t}^{a}\right) \left(L_{ri} \sigma_{\mu \nu } Q_{sa j}\right)
		
		,\quad  \epsilon ^{jk} \left(\tau ^I\right)_k^i W^{I}{}^{\mu \nu } \left(L_{ri} u_{_\mathbb{C}}{}_{t}^{a}\right) \left(e_{_\mathbb{C} p} \sigma_{\mu \nu } Q_{sa j}\right)
		
		\\&  \epsilon ^{jk} \left(\tau ^I\right)_k^i W^{I}{}^{\mu \nu } \left(Q_{sa j} u_{_\mathbb{C}}{}_{t}^{a}\right) \left(e_{_\mathbb{C} p} \sigma_{\mu \nu } L_{ri}\right)
		
		\vspace{2ex}\\
		
		\multirow{2}*{$\mathcal{O}_{B Q u_{_\mathbb{C}} L e_{_\mathbb{C}}}^{\left(1\sim3\right)}      $}
		
		& \epsilon ^{ij} B{}^{\mu \nu } \left(e_{_\mathbb{C} p} u_{_\mathbb{C}}{}_{t}^{a}\right) \left(L_{ri} \sigma_{\mu \nu } Q_{sa j}\right)
		
		,\quad \epsilon ^{ij} B{}^{\mu \nu } \left(L_{ri} u_{_\mathbb{C}}{}_{t}^{a}\right) \left(e_{_\mathbb{C} p} \sigma_{\mu \nu } Q_{sa j}\right)
		
		\\& \epsilon ^{ij} B{}^{\mu \nu } \left(Q_{sa j} u_{_\mathbb{C}}{}_{t}^{a}\right) \left(e_{_\mathbb{C} p} \sigma_{\mu \nu } L_{ri}\right)
		
		\vspace{2ex}\\
		
		\mathcal{O}_{G Q d_{_\mathbb{C}} L^{\dagger} e_{_\mathbb{C}}^{\dagger}}^{\left(1,2\right)}      
		
		& \left(\lambda ^A\right)_a^b G^{A}_{\mu \nu } \left(e_{_\mathbb{C}}^{\dagger}{}_{s} L^{\dagger}{}_{t}^{i}\right) \left(d_{_\mathbb{C}}{}_{p}^{a} \sigma^{\mu \nu } Q_{rb i}\right)
		
		,\quad \left(\lambda ^A\right)_a^b G^{A}_{\mu \nu } \left(Q_{rb i} d_{_\mathbb{C}}{}_{p}^{a} \right) \left(L^\dagger{}_{t}^{i} \bar{\sigma}^{\mu \nu } e_{_\mathbb{C}}^\dagger{}_s\right)
		
		\vspace{2ex}\\
		
		\mathcal{O}_{W Q d_{_\mathbb{C}} L^{\dagger} e_{_\mathbb{C}}^{\dagger}}^{\left(1,2\right)}      
		
		& \left(\tau ^I\right)_j^i W^{I}_{\mu \nu } \left(e_{_\mathbb{C}}^{\dagger}{}_{s} L^{\dagger}{}_{t}^{j}\right) \left(d_{_\mathbb{C}}{}_{p}^{a} \sigma^{\mu \nu } Q_{ra i}\right)
		
		,\quad \left(\tau ^I\right)_i^j W^{I}_{\mu \nu } \left(Q_{raj} d_{_\mathbb{C}}{}_{p}^{a} \right) \left(L^\dagger{}_{t}^{i} \bar{\sigma}^{\mu \nu } e_{_\mathbb{C}}^\dagger{}_s\right)
		
		\vspace{2ex}\\
		
		\mathcal{O}_{B Q d_{_\mathbb{C}} L^{\dagger} e_{_\mathbb{C}}^{\dagger}}^{\left(1,2\right)}      
		
		& B_{\mu \nu } \left(e_{_\mathbb{C}}^{\dagger}{}_{s} L^{\dagger}{}_{t}^{i}\right) \left(d_{_\mathbb{C}}{}_{p}^{a} \sigma^{\mu \nu } Q_{ra i}\right)
		
		,\quad B_{\mu \nu } \left(Q_{ra i} d_{_\mathbb{C}}{}_{p}^{a} \right) \left(L^\dagger{}_{t}^{i} \bar{\sigma}^{\mu \nu } e_{_\mathbb{C}}^\dagger{}_s\right)
		
	\end{array}\label{cl:Fq2l2c}\end{align}

\noindent 4. \underline{Operators involving only leptons:}
There should be no $G$ coupled to the four-lepton types, hence we have $0+3+2=5$ real types as follows:
\begin{align}\begin{array}{c|l}
		
		\multirow{3}*{$\mathcal{O}_{W L{}^2 L^{\dagger} {}^2}^{\left(1\sim6\right)}     $}
		
		&\mathcal{Y}\left[\tiny{\young(p,r)},\tiny{\young(st)}\right] \left(\tau ^I\right)_k^j W^{I}_{\mu \nu } \left(L^{\dagger}{}_{s}^{i} L^{\dagger}{}_{t}^{k}\right) \left(L_{pi} \sigma^{\mu \nu } L_{rj}\right)
		
		,\quad\mathcal{Y}\left[\tiny{\young(pr)},\tiny{\young(s,t)}\right] \left(\tau ^I\right)_j^k W^{I}_{\mu \nu } \left(L{}_{rk} L{}_{pi} \right) \left(L^\dagger{}_{t}^{j} \bar{\sigma}^{\mu \nu } L^\dagger{}_{s}^{i}\right)
		
		\\&\mathcal{Y}\left[\tiny{\young(p,r)},\tiny{\young(s,t)}\right] \left(\tau ^I\right)_k^j W^{I}_{\mu \nu } \left(L^{\dagger}{}_{s}^{i} L^{\dagger}{}_{t}^{k}\right) \left(L_{pi} \sigma^{\mu \nu } L_{rj}\right)
		
		,\quad\mathcal{Y}\left[\tiny{\young(p,r)},\tiny{\young(s,t)}\right] \left(\tau ^I\right)_j^k W^{I}_{\mu \nu } \left(L{}_{rk} L{}_{pi} \right) \left(L^\dagger{}_{t}^{j} \bar{\sigma}^{\mu \nu } L^\dagger{}_{s}^{i}\right)
		
		\\&\mathcal{Y}\left[\tiny{\young(pr)},\tiny{\young(st)}\right] \left(\tau ^I\right)_k^j W^{I}_{\mu \nu } \left(L^{\dagger}{}_{s}^{i} L^{\dagger}{}_{t}^{k}\right) \left(L_{pi} \sigma^{\mu \nu } L_{rj}\right)
		
		,\quad\mathcal{Y}\left[\tiny{\young(pr)},\tiny{\young(st)}\right] \left(\tau ^I\right)_j^k W^{I}_{\mu \nu } \left(L{}_{rk} L{}_{pi} \right) \left(L^\dagger{}_{t}^{j} \bar{\sigma}^{\mu \nu } L^\dagger{}_{s}^{i}\right)
		
		\vspace{2ex}\\
		
		\multirow{2}*{$\mathcal{O}_{B L{}^2 L^{\dagger} {}^2}^{\left(1\sim4\right)}     $}
		
		&\mathcal{Y}\left[\tiny{\young(p,r)},\tiny{\young(st)}\right] B_{\mu \nu } \left(L^{\dagger}{}_{s}^{i} L^{\dagger}{}_{t}^{j}\right) \left(L_{pi} \sigma^{\mu \nu } L_{rj}\right)
		
		,\quad\mathcal{Y}\left[\tiny{\young(p,r)},\tiny{\young(st)}\right] B_{\mu \nu } \left(L{}_{rj} L{}_{pi} \right) \left(L^\dagger{}_{t}^{j} \bar{\sigma}^{\mu \nu } L^\dagger{}_{s}^{i}\right)
		
		\\&\mathcal{Y}\left[\tiny{\young(pr)},\tiny{\young(s,t)}\right] B_{\mu \nu } \left(L^{\dagger}{}_{s}^{i} L^{\dagger}{}_{t}^{j}\right) \left(L_{pi} \sigma^{\mu \nu } L_{rj}\right)
		
		,\quad\mathcal{Y}\left[\tiny{\young(pr)},\tiny{\young(s,t)}\right] B_{\mu \nu } \left(L{}_{rj} L{}_{pi} \right) \left(L^\dagger{}_{t}^{j} \bar{\sigma}^{\mu \nu } L^\dagger{}_{s}^{i}\right)
		
		\vspace{2ex}\\
		
		\mathcal{O}_{W L L^{\dagger} e_{_\mathbb{C}} e_{_\mathbb{C}}^{\dagger}}^{\left(1,2\right)}        
		
		& \left(\tau ^I\right)_j^i W^{I}_{\mu \nu } \left(e_{_\mathbb{C}}^{\dagger}{}_{s} L^{\dagger}{}_{t}^{j}\right) \left(e_{_\mathbb{C} p} \sigma^{\mu \nu } L_{ri}\right)
		
		,\quad \left(\tau ^I\right)_i^j W^{I}_{\mu \nu } \left(L{}_{rj} e_{_\mathbb{C}}{}_{p} \right) \left(L^\dagger{}_{t}^{i} \bar{\sigma}^{\mu \nu } e_{_\mathbb{C}}^\dagger{}_s\right)
		
		\vspace{2ex}\\
		
		\mathcal{O}_{B L L^{\dagger} e_{_\mathbb{C}} e_{_\mathbb{C}}^{\dagger}}^{\left(1,2\right)}        
		
		& B_{\mu \nu } \left(e_{_\mathbb{C}}^{\dagger}{}_{s} L^{\dagger}{}_{t}^{i}\right) \left(e_{_\mathbb{C} p} \sigma^{\mu \nu } L_{ri}\right)
		
		,\quad B_{\mu \nu } \left(L{}_{ri} e_{_\mathbb{C}}{}_{p} \right) \left(L^\dagger{}_{t}^{i} \bar{\sigma}^{\mu \nu } e_{_\mathbb{C}}^\dagger{}_s\right)
		
		\vspace{2ex}\\
		
		\mathcal{O}_{B e_{_\mathbb{C}}{}^2 e_{_\mathbb{C}}^{\dagger} {}^2}^{\left(1,2\right)}     
		
		&\mathcal{Y}\left[\tiny{\young(p,r)},\tiny{\young(st)}\right] B_{\mu \nu } \left(e_{_\mathbb{C}}^{\dagger}{}_{s} e_{_\mathbb{C}}^{\dagger}{}_{t}\right)  \left(e_{_\mathbb{C} p} \sigma^{\mu \nu } e_{_\mathbb{C} r}\right)
		
		,\quad\mathcal{Y}\left[\tiny{\young(p,r)},\tiny{\young(st)}\right] B_{\mu \nu } \left(e_{_\mathbb{C}}{}_{r} e_{_\mathbb{C}}{}_{p} \right)  \left(e_{_\mathbb{C}}^\dagger{}_t \bar{\sigma}^{\mu \nu } e_{_\mathbb{C}}^\dagger{}_s\right)
		
	\end{array}\label{cl:Fl4}\end{align}

\section{Conclusion}
\label{sec:conclusion}

In this paper, we provided the full result of the independent dimension 8 operator basis in the standard model effective field theory. 
Although the number of the dimension 8 operators were already counted~\cite{Lehman:2015via, Lehman:2015coa, Henning:2015daa, Henning:2015alf,Fonseca:2019yya}, and part of the list, only gauge bosons and the Higgs boson involved, was also given in Ref.~\cite{Degrande:2013kka, Hays:2018zze}, it is the first time that the two-fermion and four-fermion operators are listed in full form, that constitute over half of the complete list. 
What is more important is that the form of the operators we provide here has definite symmetry over the flavor indices, making it possible to identify independent flavor-specified operators. 
These flavor-independent operators were never obtained in the past, nor a systematic approach, for various higher dimensional operators, including the Warsaw basis in dimension 6~\cite{Grzadkowski:2010es}.

To achieve the goal, we need to overcome two main obstacles. The first is to list all the independent Lorentz structures. The methods used in literature, like the Hilbert series, are usually good for counting the number of independent Lorentz structures, but not suitable for writing down the explicit form of the operators. 
Inspired by~\cite{Henning:2019enq, Henning:2019mcv}, we introduce a $SU(N)$ transformation of the operators, which divides the space of Lorentz structures into complementary invariant subspaces, one of which consists of those with factors of total derivatives. 
The other invariant subspace, which turns out to be a single irreducible representation space, is hence a linear space of independent operators regarding the integration-by-parts. 
Group theory allows us to use the Semi-standard Young Tableau to enumerate a basis for this irreducible representation space, which is the basis of Lorentz structures we are looking for. 
It is worth mentioning that the notation of operators used in this derivation is largely inspired by the on-shell amplitudes, which is made possible by a correspondence proposed in Ref.~\cite{Shadmi:2018xan, Ma:2019gtx, Aoude:2019tzn, Durieux:2019eor, Falkowski:2019zdo}. 
This work may further imply that the on-shell language may be much closer to the essence of effective field theory than the traditional field theory language. 

The second obstacle is to get a form with definite permutation symmetries among the flavor indices. In literature, although the technique of plethysm is already widely used~\cite{Lehman:2015via, Lehman:2015coa, Henning:2015daa, Henning:2015alf,Fonseca:2019yya} to perform a systematic counting of operators with repeated fields, it is not enough for writing down the explicit form of the operators. We propose a systematic method to solve this issue. 
To obtain the basis particularly for an irreducible representation space of the permutation symmetry $\bar{S}$, which permutes fields only within the group of repeated fields, we apply the left ideal projector of the group algebra to an already-found independent basis, either for the Lorentz structure or the gauge group tensors. 
Then by use of the Clebsch-Gordan coefficients of the inner product decomposition, we combine all the symmetrized factors to get a flavor tensor with definite permutation symmetry. 
The independent flavor-specified operators are thus given by, again, the Semi-standard Young Tableau. 
This essential feature on the flavor structure makes our result more practically useful than the other papers on listing higher-dimensional operators.

After the complete list of operators is written, it is worthwhile to investigate various phenomenological applications of these operators. 
As mentioned in the introduction, if the contribution from dimension 6 operators is sub-dominant or highly constrained, the dimension 8 operators should be seriously considered, even though their Wilson coefficients are suppressed by higher inverse power of the new physics scale.   
We notice there are several new Lorentz structures that only appear at the dimension 8 level, and there are several dimension 8 operators dominant over the dimension 6 operators. These phenomenological applications deserve a closer look in the future. 

The whole procedure is implemented and automized by \textsf{Mathematica}, and our code can be easily applied to higher dimensions of SMEFT and other EFTs beyond the SM. 
In terms of efficiency, listing the dimension 8 operators only cost less than 2 min on a personal laptop.

\section*{Acknowledgements}

We are grateful to Christopher W. Murphy for correspondence on their work on the same topic, and we also thank Adam Martin and Veronica Sanz for valuable discussions. H.L.L. thanks Zhou Xu for helpful discussion on programming problems. J.H.Y. thanks Jordy de Vries for discussion which initializes this project and appreciates the hospitality of Amherst Center for Fundamental Interactions, University of Massachusetts at Amherst while working on this project. J.S. and M.L.X. thank Tom Melia for useful comments. H.L.L, Z.R. and J.H.Y. are supported by the National Science Foundation of China (NSFC) under Grants No. 11875003 and also supported by the Chinese Academy of Sciences (CAS) Hundred-Talent Program. J.S. is supported by the National Natural Science Foundation of China (NSFC) under grant No.11847612, No.11690022, No.11851302, No.11675243 and No.11761141011 and also supported by the Strategic Priority Research Program of the Chinese Academy of Sciences under grant No.XDB21010200 and No.XDB23000000. M.L.X. is supported by the National Natural Science Foundation of China (NSFC) under grant No.2019M650856 and the 2019 International Postdoctoral Exchange Fellowship Program.

\section*{Note added}

Ref.~\cite{Murphy:2020rsh} also presents a list of the dimension eight operators in the standard model effective field theory. There are two main differences between our works. First, we provide a systematic and automated method in which we obtain an independent basis directly in which the EOM and IBP redundancies are entirely absent. This would help our method apply to more complicated cases where the correctness of our result is guaranteed from the first principle. Second, in contrast to~\cite{Murphy:2020rsh}, the form of the operators we provide has definite symmetry over the flavor indices, thus the independent flavor-specified operators could be obtained easily as semi-standard Young Tableau.

\appendix

\section{Conversion between Notations}
\label{app:A}

Various people have various conventions for how operators are written, while our result is presented only in one of them. In this appendix, we provide a complete set of identities for conversions of Lorentz structures between different conventions, together with a bunch of examples, in order to make it easier for different readers to use our result. Relevant conventions are $SL(2,\mathbb{C})$ v.s. $SO(3,1)$ Lorentz indices, two-component Weyl spinor v.s. four-component Dirac spinor, various forms of four fermion couplings related by Fierz identities, and the chiral basis $F_{\rm L/R}$ v.s. Hermitian basis $F,\tilde{F}$ of the gauge bosons. 

\subsection{Identities for spinors}\label{app:a1}

\noindent\underline{1. $\sigma$ techniques}\\
This part is devoted to conversions between Lorentz structures written with all spinor indices, while every factors are in irreducible representations of $SL(2,\mathbb{C})$, and the form with the usual Lorentz indices $\mu,\nu$ etc., running over 0,1,2,3, on derivatives and the gauge bosons. The key of the conversion is at the reduction of $\sigma$ products. We adopt the following definitions: the metric is “mostly-minus” $g_{\mu\nu}={\rm diag}(+1,-1,-1,-1)$; the Levi-Civita tensors are $\epsilon^{0123}=-\epsilon_{0123}=+1$ and $\epsilon^{12}=\epsilon_{21}=+1$; the sigma matrices are defined as $\sigma^{\mu}_{\alpha\dot{\alpha}}=(\mathbbm{1}_{\alpha\dot{\alpha}},\tau^i_{\alpha\dot{\alpha}})^\mu$, $\bar{\sigma}^{\mu\dot{\alpha}\alpha}=(\mathbbm{1} ^{\dot\alpha\alpha},-\tau^{i\dot{\alpha}\alpha})^\mu$, with identity $\mathbbm{1}$ and Pauli matrices $\tau^i,\ i=1,2,3$.
The two sigma's are related by raising and lowering indices by the $\epsilon$ tensor
$$\bar\sigma^{\mu\dot\alpha\alpha}=\epsilon^{\alpha\beta}\epsilon^{\dot{\alpha}\dot{\beta}}\sigma^{\mu}_{\beta\dot{\beta}}$$
We also define
\begin{align}
	\left(\sigma^{\mu\nu}\right)_{\alpha}{}^{\beta}=&\frac{i}{2}\left(\sigma^{\mu}\bar{\sigma}^{\nu}-\sigma^{\nu}\bar{\sigma}^{\mu}\right)_{\alpha}{}^{\beta}\;,\label{eq:simunu}\\
	\left(\bar{\sigma}^{\mu\nu}\right)^{\dot{\alpha}}{}_{\dot{\beta}}=&\frac{i}{2}\left(\bar{\sigma}^{\mu}\sigma^{\nu}-\bar{\sigma}^{\nu}\sigma^{\mu}\right)^{\dot{\alpha}}{}_{\dot{\beta}}\;,\label{eq:sibarmunu}
\end{align}
which directly induce the decomposition of two $\sigma$ products:
\begin{align}
	\left(\sigma^{\mu}\bar{\sigma}^{\nu}\right)_{\alpha}{}^{\beta}=&g^{\mu\nu}\delta^{\beta}_{\alpha}-i\left(\sigma^{\mu\nu}\right)_{\alpha}{}^{\beta},\label{eq:2si}\\
	\left(\bar{\sigma}^{\mu}\sigma^{\nu}\right)^{\dot{\alpha}}{}_{\dot{\beta}}=&g^{\mu\nu}\delta^{\dot{\alpha}}_{\dot{\beta}}-i\left(\bar{\sigma}^{\mu\nu}\right)^{\dot{\alpha}}{}_{\dot{\beta}},\label{eq:2sibar}
\end{align}
For more than two $\sigma$'s multiplying as a chain, we may use the following three $\sigma$ decomposition
\begin{align}
	\left(\sigma^{\mu}\bar{\sigma}^{\nu}\sigma^{\rho}\right)_{\alpha\dot{\beta}}=&g^{\mu\nu}\sigma^{\rho}_{\alpha\dot{\beta}}-g^{\mu\rho}\sigma^{\nu}_{\alpha\dot{\beta}}+g^{\nu\rho}\sigma^{\mu}_{\alpha\dot{\beta}}+i\epsilon^{\mu\nu\rho\lambda}\sigma_{\lambda\alpha\dot{\beta}},\label{eq:3si}\\
	\left(\bar{\sigma}^{\mu}\sigma^{\nu}\bar{\sigma}^{\rho}\right)^{\dot{\alpha}\beta}=&g^{\mu\nu}\bar{\sigma}^{\nu\dot{\alpha}\beta}-g^{\mu\rho}\bar{\sigma}^{\nu\dot{\alpha}\beta}+g^{\nu\rho}\bar{\sigma}^{\mu\dot{\alpha}\beta}-i\epsilon^{\mu\nu\rho\lambda}\bar{\sigma}_{\lambda}^{\dot{\alpha}\beta},\label{eq:3sibar}
\end{align}
to recursively reduce it towards a linear combination of $\mathbbm{1},\sigma^{\mu},\bar{\sigma}^{\mu},\sigma^{\mu\nu}$, and $\bar{\sigma}^{\mu\nu}$. The Hermitian conjugates of these bilinears are given by
\eq{
	\left(\psi_1\psi_{2}\right)^{\dagger}&
	=\psi^{\dagger}_2\psi^{\dagger}_1,\\
	\left(\psi_1\sigma^{\mu}\psi_{2}^{\dagger}\right)^{\dagger}&
	=\psi_2\sigma^{\mu}\psi_1^{\dagger},\\
	\left(\psi_1\sigma^{\mu\nu}\psi_{2}\right)^{\dagger}&
	=\psi_2^{\dagger}\bar{\sigma}^{\mu\nu}\psi_1^{\dagger}\;.
}
To compute the trace of a $\sigma$'s chain, one simply reduce the chain to the above basic forms, and take the trace as follows
\eq{\label{eq:tr1}
	{\rm Tr}\;\mathbbm{1}=2,\quad {\rm Tr}\;\sigma^{\mu}=&{\rm Tr}\;\bar{\sigma}^{\mu}={\rm Tr}\;\sigma^{\mu\nu}={\rm Tr}\;\bar{\sigma}^{\mu\nu}=0\;.
}
The frequently used example of four $\sigma$ chain and trace is given as follows
\eq{
	&\sigma^{\mu}\bar{\sigma}^{\nu}\sigma^{\rho}\bar{\sigma}^{\kappa}=
	(g^{\mu\nu}g^{\rho\kappa}-g^{\mu\rho}g^{\nu\kappa}+g^{\nu\rho}g^{\mu\kappa}+i\epsilon^{\mu\nu\rho\kappa})\mathbbm{1} -i\left(g^{\mu\nu}\sigma^{\rho\kappa}-g^{\mu\rho}\sigma^{\nu\kappa}+g^{\nu\rho}\sigma^{\mu\kappa}+i\epsilon^{\mu\nu\rho\lambda}\sigma_{\lambda}{}^{\kappa}\right),\\
	&{\rm Tr}\; 
	\left(\sigma^{\mu}\bar{\sigma}^{\nu}\sigma^{\rho}\bar{\sigma}^{\kappa}\right)=2g^{\mu\nu}g^{\rho\kappa}-2g^{\mu\rho}g^{\nu\kappa}+2g^{\nu\rho}g^{\mu\kappa}+2i\epsilon^{\mu\nu\rho\kappa}, \\
	&{\rm Tr}\; \left(\bar\sigma^{\mu}{\sigma}^{\nu}\bar\sigma^{\rho}{\sigma}^{\kappa}\right)=2g^{\mu\nu}g^{\rho\kappa}-2g^{\mu\rho}g^{\nu\kappa}+2g^{\nu\rho}g^{\mu\kappa}-2i\epsilon^{\mu\nu\rho\kappa} .
}

\noindent\underline{2. Converting two-component to four-component spinor} \\

In this part, we use $\Psi,\bar{\Psi}$ to denote 4-component spinors, and $\xi,\chi$ to denote 2-component left-handed spinors, while their Hermitian conjugates $\xi^\dagger,\chi^\dagger$ are right-handed spinors. 
Generally, we may combine a left-handed Weyl spinor $\xi_{\alpha}$ and an independent right-handed Weyl spinor $\chi^{\dagger\dot{\alpha}}$ into a 4-component Dirac spinor
\begin{align}
	\Psi=\left(\begin{array}{c} \xi_{\alpha}\\\chi^{\dagger\dot{\alpha}} \end{array} \right),\quad \bar{\Psi}=\Psi^{\dagger}\gamma^0=\left(\chi^{\alpha},\;\xi^{\dagger}_{\dot{\alpha}} \right)\;.
\end{align}
We can then write down the spinor bilinears that are commonly used
\eq{\label{eq:bilinear}
	\bar{\Psi}_1\Psi_2=&\chi_1^{\alpha}\xi_{2\alpha}+\xi^{\dagger}_{1\dot{\alpha}}\chi^{\dagger\dot{\alpha}}_2\;,\\
	\bar{\Psi}_1\gamma^{\mu}\Psi_2=&\chi_1^{\alpha}\sigma^{\mu}_{\alpha\dot{\alpha}}\chi^{\dagger\dot{\alpha}}_2+\xi^{\dagger}_{1\dot{\alpha}}\bar{\sigma}^{\mu\dot{\alpha}\alpha}\xi_{2\alpha}\;,\\
	\bar{\Psi}_1\sigma^{\mu\nu}\Psi_2=&\chi_1^{\alpha}\left(\sigma^{\mu\nu}\right)_\alpha{}^\beta\xi_{2\beta}+\xi^{\dagger}_{1\dot{\alpha}}\left(\bar{\sigma}^{\mu\nu}\right)^{\dot{\alpha}}{}_{\dot{\beta}}\chi^{\dot{\beta}}_{2}\;,\\
	\Psi^{\rm{T}}_1C\Psi_2=&\xi_1^{\alpha}\xi_{2\alpha}+\chi^{\dagger}_{1\dot{\alpha}}\chi^{\dagger\dot{\alpha}}_2\;,\\
	\Psi^{\rm{T}}_1C\gamma^{\mu}\Psi_2=&\xi_1^{\alpha}\sigma^{\mu}_{\alpha\dot{\alpha}}\chi^{\dagger\dot{\alpha}}_2+\chi^{\dagger}_{1\dot{\alpha}}\bar{\sigma}^{\mu\dot{\alpha}\alpha}\xi_{2\alpha}\;,\\
	\Psi^{\rm{T}}_1C\sigma^{\mu\nu}\Psi_2=&\xi_1^{\alpha}\left(\sigma^{\mu\nu}\right)_\alpha{}^\beta\xi_{2\beta}+\chi^{\dagger}_{1\dot{\alpha}}\left(\bar{\sigma}^{\mu\nu}\right)^{\dot{\alpha}}{}_{\dot{\beta}}\chi^{\dagger\dot{\beta}}_2\;,\\
	\bar{\Psi}_1C\bar{\Psi}_2^{\rm{T}}=&\xi^{\dagger}_{1\dot{\alpha}}\xi^{\dagger\dot{\alpha}}_2+\chi^{\alpha}_1\chi_{2\alpha}\;,\\
	\bar{\Psi}_1\gamma^{\mu}C\bar{\Psi}_2^{\rm{T}}=&\chi_1^{\alpha}\sigma^{\mu}_{\alpha\dot{\alpha}}\xi^{\dagger\dot{\alpha}}_2+\xi^{\dagger}_{1\dot{\alpha}}\bar{\sigma}^{\mu\dot{\alpha}\alpha}\chi_{2\alpha}\;,\\
	\bar{\Psi}_1\sigma^{\mu\nu}C\bar{\Psi}_2^{\rm{T}}=&\xi^{\dagger}_{1\dot{\alpha}}\left(\bar{\sigma}^{\mu\nu}\right)^{\dot{\alpha}}{}_{\dot{\beta}}\xi^{\dagger\dot{\beta}}_2+\chi^{\alpha}_1\left(\sigma^{\mu\nu}\right)_\alpha{}^\beta\chi_{2\beta}\;.
}
where $C=i\gamma^0\gamma^2=\begin{pmatrix} \epsilon_{\alpha\beta}&0\\0&\epsilon^{\dot{\alpha}\dot{\beta}}\end{pmatrix}=\begin{pmatrix} -\epsilon^{\alpha\beta}&0\\0&-\epsilon_{\dot{\alpha}\dot{\beta}}\end{pmatrix}$, $\gamma^{\mu}=\begin{pmatrix}
	0&\sigma^{\mu}_{\alpha\dot{\beta}}\\\bar{\sigma}^{\mu\dot{\alpha}\beta}&0
\end{pmatrix}$ in the chiral representation, and $\sigma^{\mu\nu}=\dfrac{i}{2}[\gamma^\mu,\gamma^\nu]=\begin{pmatrix}
	\left(\sigma^{\mu\nu}\right)_\alpha{}^\beta&0\\0&\left(\bar{\sigma}^{\mu\nu}\right)^{\dot{\alpha}}{}_{\dot{\beta}}
\end{pmatrix}$. 
In the SM, the 4-component chiral fermions are related to our notations of 2-component fermions as
\begin{align}
	q_{\rm{L}}=\begin{pmatrix}Q\\0\end{pmatrix},\quad u_{\rm{R}}=\left(\begin{array}{c}0\\u_{_\mathbb{C}}^{\dagger}\end{array}\right),\quad d_{\rm R}=\left(\begin{array}{c}0\\d_{_\mathbb{C}}^{\dagger}\end{array}\right),\quad l_{\rm L}=\left(\begin{array}{c}L\\0\end{array}\right),\quad e_{\rm R}=\left(\begin{array}{c}0\\e_{_\mathbb{C}}^{\dagger}\end{array}\right).\\
	\bar{q}_{\rm{L}}=\left(0\,,\,Q^{\dagger} \right),\quad \bar{u}_{\rm{R}}=\left(u_{_\mathbb{C}}\,,\,0 \right),\quad \bar{d}_{\rm R}=\left(d_{_\mathbb{C}}\,,\,0\right),\quad \bar{l}_{\rm L}=\left(0\,,\,L^{\dagger}\right),\quad \bar{e}_{\rm R}=\left(e_{_\mathbb{C}}\,,\,0\right).
\end{align}
The conversion rules of the fermion bilinears in the SM to the 4-component notation are obtained by substituting these fields into the relations in eq.~\eqref{eq:bilinear}, 
\comments{
	Extract some simple rules:
	\begin{enumerate}
		\item $\begin{array}{cccccc}
			Q\rightarrow q,\;&u_{_\mathbb{C}}\rightarrow \bar{u},\;&d_{_\mathbb{C}}\rightarrow \bar{d},\;&L\rightarrow l,\;&e_{_\mathbb{C}}\rightarrow \bar{e},\;&\sigma^{\mu}\rightarrow\gamma^{\mu},\\
			Q^{\dagger}\rightarrow \bar{q},\;& u^{\dagger}_{_\mathbb{C}}\rightarrow u,\;&d^{\dagger}_{_\mathbb{C}}\rightarrow d,\;& L^{\dagger}\rightarrow\bar{l},\;& e^{\dagger}_{_\mathbb{C}}\rightarrow e,\;&\bar{\sigma}^{\mu}\rightarrow\gamma^{\mu}
		\end{array}$
		
		\item $
		\Psi_1\Gamma\Psi_2\rightarrow\Psi_1C\Gamma\Psi_2,\;\Psi_1\bar{\Psi_2}\rightarrow\bar{\Psi}_2\Psi_1,\;\Psi_1\gamma^{\mu}\bar{\Psi}_2\rightarrow -\bar{\Psi}_2\gamma^{\mu}\Psi_1,\;\Psi_1\sigma^{\mu\nu}\bar{\Psi_2}\rightarrow\bar{\Psi}_2\sigma^{\mu\nu}\Psi_1,\;\\
		\bar{\Psi}_1\Gamma\bar{\Psi}_2\rightarrow\bar{\Psi}_1\Gamma C\bar{\Psi}_2,
		$
	\end{enumerate}
}
such as
\begin{align}
	u_{_\mathbb{C}}\sigma^{\mu}u^{\dagger}_{_\mathbb{C}}=\bar{u}\gamma^{\mu}u,\quad e_{_\mathbb{C}}L=\bar{e}\,l,\quad u^{\dagger}_{_\mathbb{C}}d^{\dagger}_{_\mathbb{C}}=u^TCd\;.
\end{align}

\noindent\underline{3. A brief introduction to Fierz identities} \\

The following 16 bilinear forms constitute a complete basis of the $4\times4$ Hermitian matrices
\begin{align}
	\Gamma^S_1=&\mathbbm{1}\;,\\
	\Gamma^V_1\text{ through }\Gamma^V_4=&\gamma^{\mu}\;,\\
	\Gamma^T_1\text{ through }\Gamma^T_6=&\sigma^{\mu\nu}\;,\\
	\Gamma^A_1\text{ through }\Gamma^A_4=&\gamma^{\mu}\gamma_5\;,\\
	\Gamma^P_1=&\gamma_5\;.
\end{align}
Labels $S,V,T,A,P$ denote scalar, vector, tensor, axial-vector, and pseudoscalar respectively, while $\sigma^{\mu\nu}=\frac{i}{2}\left[\gamma^{\mu},\gamma^{\nu}\right],\gamma_5=i\gamma^0\gamma^1\gamma^2\gamma^3=\begin{pmatrix}
	-1&0\\0&1
\end{pmatrix}$. 
The inner product between them is defined as
\eq{\label{eq:clifford_ortho}
	{\rm Tr}\left[\Gamma^{\mathcal A}_a\Gamma^{\mathcal B}_b \right]=\delta^{\mathcal AB}g_{ab},\quad \mathcal{A},\mathcal{B}=S,V,T,A,P,\ a=1,\cdots,{\rm dim}\mathcal{A},\ b=1,\cdots,{\rm dim}\mathcal{B}\;.
} 
Regarding $g$ as the metric depending on our choice of coordinates in each subspace, and using it to raise and lower indices, the inner product induces an orthogonality relation, which allows any $4\times 4$ matrix $M$ to be expanded in this basis as $M=\sum_aM^a\Gamma_a$, with coordinates $M^a={\rm Tr}(M\Gamma^a)$.

Fierz transformations of four-fermion couplings are the linear transformations:
\begin{align}
	\sum_a(\Gamma^{\mathcal A}_a)_{ij}(\Gamma^{\mathcal{A}a})_{kl}=\sum_{\mathcal B}C_{\mathcal AB}\sum_b(\Gamma^{\mathcal B}_b)_{il}(\Gamma^{\mathcal{B}b})_{kj}\;.\label{eq:Fierz}
\end{align}
According to the orthogonality eq.~\eqref{eq:clifford_ortho}, we infer immediately $C_{\mathcal AB}=\sum_a{\rm Tr}(\Gamma^{\mathcal A}_a\Gamma^{\mathcal B}_b\Gamma^{\mathcal{A}a}\Gamma^{\mathcal{B}b})$. Calculating all the $C_{\mathcal AB}$ to get the following formula
\begin{align}
	\left(\begin{array}{c}
		\delta_{ij}\delta_{kl}\\(\gamma^{\mu})_{ij}(\gamma_{\mu})_{kl}\\\frac12(\sigma^{\mu\nu})_{ij}(\sigma_{\mu\nu})_{kl}\\(\gamma^{\mu}\gamma_5)_{ij}(\gamma_{\mu}\gamma_5)_{kl}\\(\gamma_5)_{ij}(\gamma_5)_{kl}
	\end{array} \right)=\left(\begin{array}{ccccc}
		1/4 & 1/4 & 1/4 &-1/4 & 1/4 \\
		1   &-1/2 & 0   &-1/2 &-1   \\
		3/2 & 0   &-1/2 & 0   & 3/2 \\
		-1   &-1/2 & 0   &-1/2 & 1   \\
		1/4 &-1/4 & 1/4 & 1/4 & 1/4
	\end{array} \right)\left(\begin{array}{c}
		\delta_{il}\delta_{kj}\\(\gamma^{\mu})_{il}(\gamma_{\mu})_{kj}\\\frac12(\sigma^{\mu\nu})_{il}(\sigma_{\mu\nu})_{kj}\\(\gamma^{\mu}\gamma_5)_{il}(\gamma_{\mu}\gamma_5)_{kj}\\(\gamma_5)_{il}(\gamma_5)_{kj}
	\end{array} \right).\label{eq:Fierz1}
\end{align}
Both sides of eq.~(\ref{eq:Fierz1}) contract with $\bar{\Psi}_{1i}\Psi_{2j}\bar{\Psi}_{3k}\Psi_{4l}$.
\begin{align}
	\left(\begin{array}{c}
		C_{ij}C_{kl}\\(\gamma^{\mu}C)_{ij}(C\gamma_{\mu})_{kl}\\\frac12(\sigma^{\mu\nu}C)_{ij}(C\sigma_{\mu\nu})_{kl}\\(\gamma^{\mu}\gamma_5C)_{ij}(C\gamma_{\mu}\gamma_5)_{kl}\\(\gamma_5C)_{ij}(C\gamma_5)_{kl}
	\end{array} \right)=\left(\begin{array}{ccccc}
		-1/4 & 1/4 & 1/4 & 1/4 &-1/4 \\
		-1   &-1/2 & 0   & 1/2 & 1  \\
		-3/2 & 0   &-1/2 & 0   &-3/2 \\
		1   &-1/2 & 0   & 1/2 &-1   \\
		-1/4 &-1/4 & 1/4 &-1/4 &-1/4
	\end{array} \right)\left(\begin{array}{c}
		\delta_{il}\delta_{jk}\\(\gamma^{\mu})_{il}(\gamma_{\mu})_{jk}\\\frac12(\sigma^{\mu\nu})_{il}(\sigma_{\mu\nu})_{jk}\\(\gamma^{\mu}\gamma_5)_{il}(\gamma_{\mu}\gamma_5)_{jk}\\(\gamma_5)_{il}(\gamma_5)_{jk}
	\end{array} \right).\label{eq:Fierz2}
\end{align}
Both sides of eq.~(\ref{eq:Fierz2}) contract with $\bar{\Psi}_{1i}\bar{\Psi}_{2j}\Psi_{3k}\Psi_{4l}$. 
\comments{
	Reducing them to 2-component notation, they are just
	\eq{\label{eq:2-fierz}
		& g_{\mu\nu}\sigma^{\mu}_{\alpha\dot{\alpha}}\sigma^{\nu}_{\beta\dot{\beta}}=2\epsilon_{\alpha\beta}\epsilon_{\dot{\alpha}\dot{\beta}},\\
		& \epsilon_{\alpha\beta}\delta^{\rho}_{\gamma}+\epsilon_{\beta\gamma}\delta^{\rho}_{\alpha}+\epsilon_{\gamma\alpha}\delta^{\rho}_{\beta}=0,\\
		&\epsilon_{\dot{\alpha}\dot{\beta}}\delta^{\dot{\rho}}_{\dot{\gamma}}+\epsilon_{\dot{\beta}\dot{\gamma}}\delta^{\dot{\rho}}_{\dot{\alpha}}+\epsilon_{\dot{\gamma}\dot{\alpha}}\delta^{\dot{\rho}}_{\dot{\beta}}=0.
	}
}
With these Fierz identities, some four fermion interactions can be transformed to couplings of neutral fermion currents
\begin{align}
	(\bar{d}l) (\bar{l}d) =&-\frac14(\bar{d}d)(\bar{l}l)-\frac14(\bar{d}\gamma^{\mu}d)(\bar{l}\gamma_{\mu}l)-\frac18(\bar{d}\sigma^{\mu\nu}d)(\bar{l}\sigma_{\mu\nu}l)+\frac14(\bar{d}\gamma^{\mu}\gamma_5d)(\bar{l}\gamma_{\mu}\gamma_5l)-\frac14(\bar{d}\gamma_5d)(\bar{l}\gamma_5l)\notag\\
	=&-\frac12(\bar{d}\gamma^{\mu}d)(\bar{l}\gamma_{\mu}l)\;,\\
	(\bar{l}C\bar{q})(lCq)=&-\frac14(\bar{l}l)(\bar{q}q)+\frac14(\bar{l}\gamma^{\mu}l)(\bar{q}\gamma_{\mu}q)+\frac18(\bar{l}\sigma^{\mu\nu}l)(\bar{q}\sigma_{\mu\nu}q)+\frac14(\bar{l}\gamma^{\mu}\gamma_5l)(\bar{q}\gamma_{\mu}\gamma_5q)-\frac14(\bar{l}\gamma_5l)(\bar{q}\gamma_5q)\notag\\
	=&\frac12(\bar{l}\gamma^{\mu}l)(\bar{q}\gamma_{\mu}q)\;.
\end{align}

It worths mentioning that $\Gamma_a$ may also be generators of fundamental $SU(N)$, denoted by $T_a$. Since $\{\mathbbm{1},T_a\}$ is a complete set of $N\times N$ Hermitian matrices, substituting $\Gamma^A_a=T_a,\Gamma_{I}=\mathbbm{1},{\rm Tr}(T_aT_b)=\delta_{ab}$ into eq.~(\ref{eq:Fierz}), we get the Fierz identity for $SU(N)$ group as
\begin{align}
	\sum_a (T_a)_{ij}(T_a)_{kl}=\delta_{il}\delta_{kj}-\frac{1}{N}\delta_{ij}\delta_{kl}\;.\label{eq:Fierz3}
\end{align}

\noindent\underline{4. Examples} \\

Under Fierz identities, some terms can be transformed into bilinear form which readers may be more familiar with. Here are some examples.
\begin{itemize}
	\item  Example 1, type $ \mathcal{O}_{Q Q^{\dagger} HH^{\dagger}  D^3}^{\left(1\sim4\right)} $
	\begin{align}
		i\left(Q_{pa i} \sigma_{\mu } \overleftrightarrow{D}^{\nu } Q^{\dagger}{}_{r}^{a i}\right)\left(D^{\mu}H^{\dagger}D_{\nu}H\right)=&i\left(\bar{q}^{ai}_r\gamma_{\mu}\overleftrightarrow{D}^{\nu}q_{pai}\right)\left(D^{\mu}H^{\dagger}D_{\nu}H\right)=i\left(\bar{q}_r\gamma_{\mu}\overleftrightarrow{D}^{\nu}q_p\right)\left(D^{\mu}H^{\dagger}D_{\nu}H\right)\\
		i\left(Q_{pa i} \sigma_{\mu } \overleftrightarrow{D}^{\nu } Q^{\dagger}{}_{r}^{a j}\right)D_{\nu } H_{j} D^{\mu } H^{\dagger}{}^{i}=&i\left(\bar{q}^{aj}_r\gamma_{\mu}\overleftrightarrow{D}^{\nu}q_{pai}\right) D_{\nu } H_{j} D^{\mu } H^{\dagger}{}^{i}\notag\\
		=&\frac12i\left(\bar{q}^{ai}_r\gamma_{\mu}\overleftrightarrow{D}^{\nu}q_{pai}\right) D_{\nu } H_{j} D^{\mu } H^{\dagger}{}^{j}\notag\\
		&+i\left(\left(\bar{q}^{aj}_r\gamma_{\mu}\overleftrightarrow{D}^{\nu}q_{pai}\right) D_{\nu } H_{j} D^{\mu } H^{\dagger}{}^{i}-\frac12\left(\bar{q}^{ai}_r\gamma_{\mu}\overleftrightarrow{D}^{\nu}q_{pai}\right) D_{\nu } H_{j} D^{\mu } H^{\dagger}{}^{j} \right)\notag\\
		=&\frac12i\left(\bar{q}_r\gamma_{\mu}\overleftrightarrow{D}^{\nu}q_{p}\right) \left(D^{\mu}H^{\dagger}D_{\nu}H\right)+i\left(\bar{q}_r\gamma_{\mu}\tau^I\overleftrightarrow{D}^{\nu}q_{p}\right) \left(D^{\mu}H^{\dagger}\tau^ID_{\nu}H\right)
	\end{align}
	Hence, the basis can be transformed into
	\begin{align}
		\left\{\begin{array}{c}
			i\left(Q_{pa i} \sigma_{\mu } Q^{\dagger}{}_{r}^{a i}\right)\Box \left(H^{\dagger}\overleftrightarrow{D}^{\mu}H\right)\\
			i\left(Q_{pa i} \sigma_{\mu } Q^{\dagger}{}_{r}^{a j}\right)\Box \left(H^{\dagger i}\overleftrightarrow{D}^{\mu}H_j\right) \\ 
			i\left(Q_{pa i} \sigma_{\mu } \overleftrightarrow{D}_{\nu } Q^{\dagger}{}_{r}^{a i}\right)\left(D^{\mu}H^{\dagger}D^{\nu}H\right)\\
			i\left(Q_{pa i} \sigma_{\mu } \overleftrightarrow{D}_{\nu } Q^{\dagger}{}_{r}^{a j}\right)\left(D^{\mu}H^{\dagger i}D^{\nu}H_j\right)
		\end{array}\right.
		\Longrightarrow\left\{
		\begin{array}{c}
			i\left(\bar{q}_r\gamma_{\mu} q_{p}\right)\Box \left(H^{\dagger}\overleftrightarrow{D}^{\mu}H\right)\\
			i\left(\bar{q}_r\gamma_{\mu}\tau^I q_{p}\right)\Box \left(H^{\dagger}\tau^I\overleftrightarrow{D}^{\mu}H\right)\\
			i\left(\bar{q}_r\gamma_{\mu}\overleftrightarrow{D}_{\nu}q_p\right)\left(D^{\mu}H^{\dagger}D^{\nu}H\right)\\
			i\left(\bar{q}_r\gamma_{\mu}\tau^I\overleftrightarrow{D}_{\nu}q_p\right)\left(D^{\mu}H^{\dagger i}\tau^I D^{\nu}H_j\right)
		\end{array}\right.
	\end{align}
	
	\item Example 2, type $\mathcal{O}_{Q Q^{\dagger}  u_{_\mathbb{C}} u_{_\mathbb{C}}^{\dagger}  H H^{\dagger}}^{\left(1\sim4\right)}      $
	\begin{align}
		\left(Q_{pa i} u_{_\mathbb{C}}{}_{r}^{b}\right) \left(Q^{\dagger}{}_{s}^{a i} u_{_\mathbb{C}}^{\dagger}{}_{tb}\right) \left(H^{\dagger} H\right)=&\left(\bar{u}^b_rq_{pai} \right)\left(\bar{q}^{ai}_su_{tb} \right)\left(H^{\dagger} H\right)=-\frac12\left(\bar{q}^{ai}_s\gamma^{\mu}q_{pai} \right)\left(\bar{u}^b_r\gamma_{\mu}u_{tb} \right)\left(H^{\dagger}H\right) \notag\\
		=&-\frac12\left(\bar{q}_s\gamma^{\mu}q_{p} \right)\left(\bar{u}_r\gamma_{\mu}u_{t} \right)\left(H^{\dagger}H\right)\\
		
		\left(Q_{pa i} u_{_\mathbb{C}}{}_{r}^{a}\right) \left(Q^{\dagger}{}_{s}^{c i} u_{_\mathbb{C}}^{\dagger}{}_{tc}\right) \left(H^{\dagger} H\right)=&\left(\bar{u}^a_rq_{pai} \right)\left(\bar{q}^{ci}_su_{tc} \right)\left(H^{\dagger} H\right)=-\frac12\left(\bar{q}^{ci}_s\gamma^{\mu}q_{pai} \right)\left(\bar{u}^a_r\gamma_{\mu}u_{tc} \right)\left(H^{\dagger}H\right)\notag\\
		=&-\frac12\left(\left(\bar{q}^{ci}_s\gamma^{\mu}q_{pai} \right)\left(\bar{u}^a_r\gamma_{\mu}u_{tc} \right)\left(H^{\dagger}H\right)-\frac13\left(\bar{q}^{ai}_s\gamma^{\mu}q_{pai} \right)\left(\bar{u}^c_r\gamma_{\mu}u_{tc} \right)\left(H^{\dagger}H\right) \right)\notag\\
		&-\frac16\left(\bar{q}^{ai}_s\gamma^{\mu}q_{pai} \right)\left(\bar{u}^c_r\gamma_{\mu}u_{tc} \right)\left(H^{\dagger}H\right)\notag\\
		=&-\frac12\left(\bar{q}_s\gamma^{\mu}\lambda^Aq_{p} \right)\left(\bar{u}_r\gamma_{\mu}\lambda^Au_{t} \right)\left(H^{\dagger}H\right)-\frac16\left(\bar{q}_s\gamma^{\mu}q_{p} \right)\left(\bar{u}_r\gamma_{\mu}u_{t} \right)\left(H^{\dagger}H\right)\\
		
		\left(Q_{pa i} u_{_\mathbb{C}}{}_{r}^{b}\right) \left(Q^{\dagger}{}_{s}^{a j} u_{_\mathbb{C}}^{\dagger}{}_{tb}\right)  H^{\dagger}{}^{i} H_{j}=&\left(\bar{u}^b_rq_{pai} \right)\left(\bar{q}^{aj}_su_{tb} \right)H^{\dagger i} H_j=-\frac12\left(\bar{q}^{aj}_s\gamma^{\mu}q_{pai} \right)\left(\bar{u}^b_r\gamma_{\mu}u_{tb} \right)H^{\dagger i}H_j \notag\\
		=&-\frac12\left(\left(\bar{q}^{aj}_s\gamma^{\mu}q_{pai} \right)\left(\bar{u}^b_r\gamma_{\mu}u_{tb} \right)H^{\dagger i}H_j-\frac12\left(\bar{q}^{ai}_s\gamma^{\mu}q_{pai} \right)\left(\bar{u}^b_r\gamma_{\mu}u_{tb} \right)H^{\dagger j}H_j \right)\notag\\
		&-\frac14\left(\bar{q}^{ai}_s\gamma^{\mu}q_{pai} \right)\left(\bar{u}^b_r\gamma_{\mu}u_{tb} \right)H^{\dagger j}H_j \notag\\
		=&-\frac12\left(\bar{q}_s\gamma^{\mu}\tau^Iq_{p} \right)\left(\bar{u}_r\gamma_{\mu}u_{t} \right)\left(H^{\dagger}\tau^IH\right)-\frac14\left(\bar{q}_s\gamma^{\mu}q_{p} \right)\left(\bar{u}_r\gamma_{\mu}u_{t} \right)\left(H^{\dagger}H\right)\\
		
		\left(Q_{pa i} u_{_\mathbb{C}}{}_{r}^{a}\right) \left(Q^{\dagger}{}_{s}^{c j} u_{_\mathbb{C}}^{\dagger}{}_{tc}\right)  H^{\dagger}{}^{i} H_{j}=&-\frac12\left(\bar{q}_s\gamma^{\mu}\tau^I\lambda^Aq_p\right)\left(\bar{u}_r\gamma_{\mu}\lambda^Au_t\right)\left(H^{\dagger}\tau^IH\right)-\frac16\left(\bar{q}_s\gamma^{\mu}\tau^Iq_p\right)\left(\bar{u}_r\gamma_{\mu}u_t\right)\left(H^{\dagger}\tau^IH\right)\notag\\
		&-\frac14\left(\bar{q}_s\gamma^{\mu}\lambda^Aq_p\right)\left(\bar{u}_r\gamma_{\mu}\lambda^Au_t\right)\left(H^{\dagger}H\right)-\frac{1}{12}\left(\bar{q}_s\gamma^{\mu}q_p\right)\left(\bar{u}_r\gamma_{\mu}u_t\right)\left(H^{\dagger}H\right)
	\end{align}
	Hence, the basis can be transformed into
	\begin{align}
		\left\{\begin{array}{l}
			\left(Q_{pa i} u_{_\mathbb{C}}{}_{r}^{a}\right) \left(Q^{\dagger}{}_{s}^{c j} u_{_\mathbb{C}}^{\dagger}{}_{tc}\right)  H^{\dagger}{}^{i} H_{j}\\
			\quad \left(Q_{pa i} u_{_\mathbb{C}}{}_{r}^{a}\right) \left(Q^{\dagger}{}_{s}^{c i} u_{_\mathbb{C}}^{\dagger}{}_{tc}\right) \left(H^{\dagger} H\right)\\
			\left(Q_{pa i} u_{_\mathbb{C}}{}_{r}^{b}\right) \left(Q^{\dagger}{}_{s}^{a j} u_{_\mathbb{C}}^{\dagger}{}_{tb}\right)  H^{\dagger}{}^{i} H_{j}\\
			\left(Q_{pa i} u_{_\mathbb{C}}{}_{r}^{b}\right) \left(Q^{\dagger}{}_{s}^{a i} u_{_\mathbb{C}}^{\dagger}{}_{tb}\right) \left(H^{\dagger} H\right)
		\end{array}\right.
		\Longrightarrow
		\left\{\begin{array}{l}
			\left(\bar{q}_s\gamma^{\mu}q_{p} \right)\left(\bar{u}_r\gamma_{\mu}u_{t} \right)\left(H^{\dagger}H\right)\\
			\left(\bar{q}_s\gamma^{\mu}\lambda^Aq_{p} \right)\left(\bar{u}_r\gamma_{\mu}\lambda^Au_{t} \right)\left(H^{\dagger}H\right)\\
			\left(\bar{q}_s\gamma^{\mu}\tau^Iq_{p} \right)\left(\bar{u}_r\gamma_{\mu}u_{t} \right)\left(H^{\dagger}\tau^IH\right)\\
			\left(\bar{q}_s\gamma^{\mu}\tau^I\lambda^Aq_p\right)\left(\bar{u}_r\gamma_{\mu}\lambda^Au_t\right)\left(H^{\dagger}\tau^IH\right)
		\end{array}\right.
	\end{align}
	
	\item Example 3, type $\mathcal{O}_{Q Q^{\dagger}  L L^{\dagger}   H H^{\dagger}}^{\left(1\sim5\right)} $
	\begin{align}
		\left(L_{pi} Q_{ra j}\right) \left(L^{\dagger}{}_{s}^{k} Q^{\dagger}{}_{t}^{a j}\right)  H^{\dagger}{}^{i} H_{k}=&\frac12\left(\bar{l}^k_s\gamma^{\mu}l_{pi}\right)\left(\bar{q}^{aj}_t\gamma_{\mu}q_{raj}\right)H^{\dagger i}H_k\notag\\
		=&\frac12\left(\bar{l}_s\gamma^{\mu}\tau^Il_{p}\right)\left(\bar{q}_t\gamma_{\mu}q_{r}\right)\left(H^{\dagger}\tau^IH\right)+\frac14\left(\bar{l}_s\gamma^{\mu}l_{p}\right)\left(\bar{q}_t\gamma_{\mu}q_{r}\right)\left(H^{\dagger}H\right)\\
		
		\left(L_{pi} Q_{ra j}\right) \left(L^{\dagger}{}_{s}^{j} Q^{\dagger}{}_{t}^{a i}\right)  \left(H^{\dagger} H\right)=&\frac12\left(\bar{l}_s\gamma^{\mu}\tau^Il_{p}\right)\left(\bar{q}_t\gamma_{\mu}\tau^Iq_{r}\right)\left(H^{\dagger}H\right)+\frac14\left(\bar{l}_s\gamma^{\mu}l_{p}\right)\left(\bar{q}_t\gamma_{\mu}q_{r}\right)\left(H^{\dagger}H\right)\\
		
		\left(L_{pi} Q_{ra j}\right) \left(L^{\dagger}{}_{s}^{i} Q^{\dagger}{}_{t}^{a k}\right)  H^{\dagger}{}^{j} H_{k}=&\frac12\left(\bar{l}_s\gamma^{\mu}l_{p}\right)\left(\bar{q}_t\gamma_{\mu}\tau^Iq_{r}\right)\left(H^{\dagger}\tau^IH\right)+\frac14\left(\bar{l}_s\gamma^{\mu}l_{p}\right)\left(\bar{q}_t\gamma_{\mu}q_{r}\right)\left(H^{\dagger}H\right)\\
		
		\left(L_{pi} Q_{ra j}\right) \left(L^{\dagger}{}_{s}^{j} Q^{\dagger}{}_{t}^{a k}\right)  H^{\dagger}{}^{i} H_{k}=&\frac12\left(\bar{l}^j_s\gamma^{\mu}l_{pi}\right)\left(\bar{q}^{ak}_t\gamma_{\mu}q_{raj}\right)H^{\dagger i}H_k\notag\\
		=&\frac12\left(\bar{l}_s\gamma^{\mu}\tau^Il_p\right)\left(\bar{q}^{ak}_t\gamma_{\mu}q_{ral}\right)\left(\tau^I\right)^l_mH^{\dagger m}H_k+\frac14\left(\bar{l}_s\gamma^{\mu}l_p\right)\left(\bar{q}^{ak}_t\gamma_{\mu}q_{ral}\right)H^{\dagger l}H_k\notag\\
		=&\frac12\left(\bar{l}_s\gamma^{\mu}\tau^Il_p\right)\left(\bar{q}_t\gamma_{\mu}\tau^Jq_r\right)\left(H^{\dagger}\tau^I\tau^JH\right)+\frac14\left(\bar{l}_s\gamma^{\mu}\tau^Il_p\right)\left(\bar{q}_t\gamma_{\mu}q_r\right)\left(H^{\dagger}\tau^IH\right)\notag\\
		&+\frac14\left(\bar{l}_s\gamma^{\mu}l_p\right)\left(\bar{q}_t\gamma_{\mu}\tau^Iq_r\right)\left(H^{\dagger}\tau^IH\right)+\frac18\left(\bar{l}_s\gamma^{\mu}l_p\right)\left(\bar{q}_t\gamma_{\mu}q_r\right)\left(H^{\dagger}H\right)\notag\\
		=&\frac12i\epsilon^{IJK}\left(\bar{l}_s\gamma^{\mu}\tau^Il_p\right)\left(\bar{q}_t\gamma_{\mu}\tau^Jq_r\right)\left(H^{\dagger}\tau^KH\right)+\frac14\left(\bar{l}_s\gamma^{\mu}\tau^Il_p\right)\left(\bar{q}_t\gamma_{\mu}\tau^Iq_r\right)\left(H^{\dagger}H\right)\notag\\
		+\frac14\left(\bar{l}_s\gamma^{\mu}\tau^Il_p\right)\left(\bar{q}_t\gamma_{\mu}q_r\right)&\left(H^{\dagger}\tau^IH\right)+\frac14\left(\bar{l}_s\gamma^{\mu}l_p\right)\left(\bar{q}_t\gamma_{\mu}\tau^Iq_r\right)\left(H^{\dagger}\tau^IH\right)+\frac18\left(\bar{l}_s\gamma^{\mu}l_p\right)\left(\bar{q}_t\gamma_{\mu}q_r\right)\left(H^{\dagger}H\right)\\
		
		\left(L_{pi} Q_{ra j}\right) \left(L^{\dagger}{}_{s}^{k} Q^{\dagger}{}_{t}^{a i}\right)  H^{\dagger}{}^{j} H_{k}=&\frac12i\epsilon^{IJK}\left(\bar{l}_s\gamma^{\mu}\tau^Il_p\right)\left(\bar{q}_t\gamma_{\mu}\tau^Kq_r\right)\left(H^{\dagger}\tau^JH\right)+\frac14\left(\bar{l}_s\gamma^{\mu}\tau^Il_p\right)\left(\bar{q}_t\gamma_{\mu}\tau^Iq_r\right)\left(H^{\dagger}H\right)\notag\\
		+\frac14\left(\bar{l}_s\gamma^{\mu}\tau^Il_p\right)\left(\bar{q}_t\gamma_{\mu}q_r\right)&\left(H^{\dagger}\tau^IH\right)+\frac14\left(\bar{l}_s\gamma^{\mu}l_p\right)\left(\bar{q}_t\gamma_{\mu}\tau^Iq_r\right)\left(H^{\dagger}\tau^IH\right)+\frac18\left(\bar{l}_s\gamma^{\mu}l_p\right)\left(\bar{q}_t\gamma_{\mu}q_r\right)\left(H^{\dagger}H\right)
		
	\end{align}
	Hence, the basis can be transformed into
	\begin{align}
		\left\{\begin{array}{l}
			\left(L_{pi} Q_{ra j}\right) \left(L^{\dagger}{}_{s}^{i} Q^{\dagger}{}_{t}^{a k}\right)  H^{\dagger}{}^{j} H_{k}\\
			\left(L_{pi} Q_{ra j}\right) \left(L^{\dagger}{}_{s}^{k} Q^{\dagger}{}_{t}^{a i}\right)  H^{\dagger}{}^{j} H_{k}\\
			\left(L_{pi} Q_{ra j}\right) \left(L^{\dagger}{}_{s}^{j} Q^{\dagger}{}_{t}^{a k}\right)  H^{\dagger}{}^{i} H_{k}\\
			\left(L_{pi} Q_{ra j}\right) \left(L^{\dagger}{}_{s}^{j} Q^{\dagger}{}_{t}^{a i}\right)  \left(H^{\dagger} H\right)\\
			\left(L_{pi} Q_{ra j}\right) \left(L^{\dagger}{}_{s}^{k} Q^{\dagger}{}_{t}^{a j}\right)  H^{\dagger}{}^{i} H_{k}
		\end{array}\right.
		\Longrightarrow
		\left\{\begin{array}{l}
			\left(\bar{l}_s\gamma^{\mu}l_p\right)\left(\bar{q}_t\gamma_{\mu}q_r\right)\left(H^{\dagger}H\right)\\
			\left(\bar{l}_s\gamma^{\mu}l_p\right)\left(\bar{q}_t\gamma_{\mu}\tau^Iq_r\right)\left(H^{\dagger}\tau^IH\right)\\
			\left(\bar{l}_s\gamma^{\mu}\tau^Il_p\right)\left(\bar{q}_t\gamma_{\mu}\tau^Iq_r\right)\left(H^{\dagger}H\right)\\
			\left(\bar{l}_s\gamma^{\mu}\tau^Il_p\right)\left(\bar{q}_t\gamma_{\mu}q_r\right)\left(H^{\dagger}\tau^IH\right)\\
			\epsilon^{IJK}\left(\bar{l}_s\gamma^{\mu}\tau^Il_p\right)\left(\bar{q}_t\gamma_{\mu}\tau^Jq_r\right)\left(H^{\dagger}\tau^KH\right)
		\end{array}\right.
	\end{align}
	
	\item Example 4, type $\mathcal{O}_{L{}^2 L^{\dagger} {}^2  H H^{\dagger}}^{\left(1\sim5\right)}$
	\begin{align}
		\left(L_{pi} L_{rj}\right) \left(L^{\dagger}{}_{s}^{j} L^{\dagger}{}_{t}^{i}\right) \left(H^{\dagger} H\right)=&\dfrac{1}{2} \left(\bar{l}_s\gamma^{\mu}l_r\right)\left(\bar{l}_t\gamma_{\mu}l_p\right)\left(H^{\dagger}H\right)\\
		\left(L_{pi} L_{rj}\right) \left(L^{\dagger}{}_{s}^{j} L^{\dagger}{}_{t}^{k}\right)  H^{\dagger}{}^{i} H_{k}=&\dfrac{1}{2} \left(\bar{l}_s\gamma^{\mu}l_r\right)\left(\bar{l}^k_t\gamma_{\mu}l_{pi}\right)H^{\dagger i}H_k\notag\\
		=&\frac12\left(\bar{l}_s\gamma^{\mu}l_r\right)\left(\bar{l}_t\gamma_{\mu}\tau^Il_p\right)\left(H^{\dagger}\tau^IH\right)+\frac14\left(\bar{l}_s\gamma^{\mu}l_r\right)\left(\bar{l}_t\gamma_{\mu}l_p\right)\left(H^{\dagger}H\right)
	\end{align}
	Since different $\mathcal{Y}$s indicate different flavor symmetries, operators with different $\mathcal{Y}$s should not be mixed if one do not want to confuse the flavor symmetry. $\mathcal{Y}\left[\tiny{\young(p,r)},\tiny{\young(st)}\right]$ means we will get a minus sign if we exchange $p$,$r$. The basis can be transformed into
	\begin{align}
		&\left\{\begin{array}{l}
			\mathcal{Y}\left[\tiny{\young(pr)},\tiny{\young(st)}\right] \left(L_{pi} L_{rj}\right) \left(L^{\dagger}{}_{s}^{j} L^{\dagger}{}_{t}^{k}\right)  H^{\dagger}{}^{i} H_{k}\\
			\mathcal{Y}\left[\tiny{\young(pr)},\tiny{\young(st)}\right] \left(L_{pi} L_{rj}\right) \left(L^{\dagger}{}_{s}^{j} L^{\dagger}{}_{t}^{i}\right) \left(H^{\dagger} H\right)\\
			\mathcal{Y}\left[\tiny{\young(pr)},\tiny{\young(s,t)}\right] \left(L_{pi} L_{rj}\right) \left(L^{\dagger}{}_{s}^{j} L^{\dagger}{}_{t}^{k}\right)  H^{\dagger}{}^{i} H_{k}\\
			\mathcal{Y}\left[\tiny{\young(p,r)},\tiny{\young(st)}\right] \left(L_{pi} L_{rj}\right) \left(L^{\dagger}{}_{s}^{j} L^{\dagger}{}_{t}^{k}\right)  H^{\dagger}{}^{i} H_{k}\\
			\mathcal{Y}\left[\tiny{\young(p,r)},\tiny{\young(s,t)}\right] \left(L_{pi} L_{rj}\right) \left(L^{\dagger}{}_{s}^{j} L^{\dagger}{}_{t}^{k}\right)  H^{\dagger}{}^{i} H_{k}
		\end{array}\right.\\
		\Longrightarrow&\left\{
		\begin{array}{l}
			\mathcal{Y}\left[\tiny{\young(pr)},\tiny{\young(st)}\right]\left(\bar{l}_s\gamma^{\mu}l_r\right)\left(\bar{l}_t\gamma_{\mu}l_p\right)\left(H^{\dagger}H\right)\\
			\mathcal{Y}\left[\tiny{\young(pr)},\tiny{\young(st)}\right]\left(\bar{l}_s\gamma^{\mu}l_r\right)\left(\bar{l}_t\gamma_{\mu}\tau^Il_p\right)\left(H^{\dagger}\tau^IH\right)\\
			\mathcal{Y}\left[\tiny{\young(pr)},\tiny{\young(s,t)}\right]\left(\left(\bar{l}_s\gamma^{\mu}l_r\right)\left(\bar{l}_t\gamma_{\mu}\tau^Il_p\right)\left(H^{\dagger}\tau^IH\right)+\frac12\left(\bar{l}_s\gamma^{\mu}l_r\right)\left(\bar{l}_t\gamma_{\mu}l_p\right)\left(H^{\dagger}H\right)\right)\\
			\mathcal{Y}\left[\tiny{\young(p,r)},\tiny{\young(st)}\right]\left(\left(\bar{l}_s\gamma^{\mu}l_r\right)\left(\bar{l}_t\gamma_{\mu}\tau^Il_p\right)\left(H^{\dagger}\tau^IH\right)+\frac12\left(\bar{l}_s\gamma^{\mu}l_r\right)\left(\bar{l}_t\gamma_{\mu}l_p\right)\left(H^{\dagger}H\right)\right)\\
			\mathcal{Y}\left[\tiny{\young(p,r)},\tiny{\young(s,t)}\right]\left(\left(\bar{l}_s\gamma^{\mu}l_r\right)\left(\bar{l}_t\gamma_{\mu}\tau^Il_p\right)\left(H^{\dagger}\tau^IH\right)+\frac12\left(\bar{l}_s\gamma^{\mu}l_r\right)\left(\bar{l}_t\gamma_{\mu}l_p\right)\left(H^{\dagger}H\right)\right)
		\end{array}\right.
	\end{align}
	
	\item Example 5, type $\mathcal{O}_{L{}^2 L^{\dagger} {}^2 D^2}^{\left(1\sim4\right)} $
	\begin{align}
		\left(L_{pi} L_{rj}\right) \left(D_{\mu } L^{\dagger}{}_{s}^{i} D^{\mu } L^{\dagger}{}_{t}^{j}\right) = \frac12\left(D_{\mu } \bar{l}{}_{s} \gamma_\nu l_p \right) \left(D^{\mu } \bar{l}{}_{t} \gamma^\nu l_r \right)
	\end{align}
	Operator $i \left(D_{\mu } L^{\dagger}{}_{s}^{i} D_{\nu } L^{\dagger}{}_{t}^{j}\right) \left(L_{pi} \sigma^{\mu }{}^{\nu } L_{rj}\right)$ is equivalent to $\left(L_{pi} D^{\mu } L_{rj}\right) \left(L^{\dagger}{}_{s}^{i} D_{\mu } L^{\dagger}{}_{t}^{j}\right)$ up to IBP and EOM, and
	\begin{align}
		\left(L_{pi} D^{\mu } L_{rj}\right) \left(L^{\dagger}{}_{s}^{i} D_{\mu } L^{\dagger}{}_{t}^{j}\right) = \frac12 \left( \bar{l}{}_{s} \gamma_\nu l_p \right) \left(D^{\mu } \bar{l}{}_{t} \gamma^\nu D_{\mu } l_r \right).
	\end{align}
	Hence, the basis can be transformed into
	\begin{align}
		\left\{\begin{array}{l}
			\mathcal{Y}\left[\tiny{\young(p,r)},\tiny{\young(s,t)}\right]\left(L_{pi} L_{rj}\right) \left(D_{\mu } L^{\dagger}{}_{s}^{i} D^{\mu } L^{\dagger}{}_{t}^{j}\right)\\
			\mathcal{Y}\left[\tiny{\young(p,r)},\tiny{\young(s,t)}\right] \left(D_{\mu } L^{\dagger}{}_{s}^{i} D_{\nu } L^{\dagger}{}_{t}^{j}\right) \left(L_{pi} \sigma^{\mu }{}^{\nu } L_{rj}\right)\\
			\mathcal{Y}\left[\tiny{\young(pr)},\tiny{\young(st)}\right] \left(D_{\mu } L^{\dagger}{}_{s}^{i} D_{\nu } L^{\dagger}{}_{t}^{j}\right) \left(L_{pi} \sigma^{\mu }{}^{\nu } L_{rj}\right)\\
			\mathcal{Y}\left[\tiny{\young(pr)},\tiny{\young(st)}\right]\left(L_{pi} L_{rj}\right) \left(D_{\mu } L^{\dagger}{}_{s}^{i} D^{\mu } L^{\dagger}{}_{t}^{j}\right)
		\end{array}\right.
		\Longrightarrow
		\left\{\begin{array}{l}
			\mathcal{Y}\left[\tiny{\young(p,r)},\tiny{\young(s,t)}\right]\left(D_{\mu } \bar{l}{}_{s} \gamma_\nu l_p \right) \left(D^{\mu } \bar{l}{}_{t} \gamma^\nu l_r \right)\\
			\mathcal{Y}\left[\tiny{\young(p,r)},\tiny{\young(s,t)}\right] \left( \bar{l}{}_{s} \gamma_\nu l_p \right) \left(D^{\mu } \bar{l}{}_{t} \gamma^\nu D_{\mu } l_r \right)\\
			\mathcal{Y}\left[\tiny{\young(pr)},\tiny{\young(st)}\right] \left( \bar{l}{}_{s} \gamma_\nu l_p \right) \left(D^{\mu } \bar{l}{}_{t} \gamma^\nu D_{\mu } l_r \right)\\
			\mathcal{Y}\left[\tiny{\young(pr)},\tiny{\young(st)}\right]\left(D_{\mu } \bar{l}{}_{s} \gamma_\nu l_p \right) \left(D^{\mu } \bar{l}{}_{t} \gamma^\nu l_r \right)
		\end{array}\right.
	\end{align}
	
	\item Example 6, type $\mathcal{O}_{B Q{}^3 L}^{\left(1\sim4\right)}$
	\begin{align}
		\left\{\begin{array}{l}
			\mathcal{Y}\left[\tiny{\young(rs,t)}\right]  \epsilon ^{abc}   \epsilon ^{ik} \epsilon ^{jm} B{}_{\mu }{}_{\nu } \left(L_{pi} Q_{tc m}\right) \left(Q_{ra j} \sigma^{\mu }{}^{\nu } Q_{sb k}\right)\\
			\mathcal{Y}\left[\tiny{\young(rs,t)}\right] \epsilon ^{abc}   \epsilon ^{ij} \epsilon ^{km} B{}_{\mu }{}_{\nu } \left(L_{pi} Q_{tc m}\right) \left(Q_{ra j} \sigma^{\mu }{}^{\nu } Q_{sb k}\right)\\
			\mathcal{Y}\left[\tiny{\young(rst)}\right] \epsilon ^{abc}   \epsilon ^{ik} \epsilon ^{jm} B{}_{\mu }{}_{\nu } \left(L_{pi} Q_{tc m}\right) \left(Q_{ra j} \sigma^{\mu }{}^{\nu } Q_{sb k}\right)\\
			\mathcal{Y}\left[\tiny{\young(r,s,t)}\right] \epsilon ^{abc}   \epsilon ^{ik} \epsilon ^{jm} B{}_{\mu }{}_{\nu } \left(L_{pi} Q_{tc m}\right) \left(Q_{ra j} \sigma^{\mu }{}^{\nu } Q_{sb k}\right)
		\end{array}\right.
		\Longrightarrow
		\left\{\begin{array}{l}
			\mathcal{Y}\left[\tiny{\young(rs,t)}\right]  \epsilon ^{abc}   \epsilon ^{ik} \epsilon ^{jm} B{}_{\mu }{}_{\nu }  \left(l_{pi} C q_{tc m}\right) \left(q_{ra j} C \sigma^{\mu }{}^{\nu } q_{sb k}\right)\\
			\mathcal{Y}\left[\tiny{\young(rs,t)}\right] \epsilon ^{abc}   \epsilon ^{ij} \epsilon ^{km} B{}_{\mu }{}_{\nu }  \left(l_{pi} C q_{tc m}\right) \left(q_{ra j} C \sigma^{\mu }{}^{\nu } q_{sb k}\right)\\
			\mathcal{Y}\left[\tiny{\young(rst)}\right] \epsilon ^{abc}   \epsilon ^{ik} \epsilon ^{jm} B{}_{\mu }{}_{\nu }  \left(l_{pi} C q_{tc m}\right) \left(q_{ra j} C \sigma^{\mu }{}^{\nu } q_{sb k}\right)\\
			\mathcal{Y}\left[\tiny{\young(r,s,t)}\right] \epsilon ^{abc}   \epsilon ^{ik} \epsilon ^{jm} B{}_{\mu }{}_{\nu }  \left(l_{pi} C q_{tc m}\right) \left(q_{ra j} C \sigma^{\mu }{}^{\nu } q_{sb k}\right)
		\end{array}\right.
	\end{align}
	It should be noted that $\mathcal{O}_{B Q{}^3 L}^{\left(1\sim4\right)}$ is a complex type, which means the Hermitian conjugate of independent operators of this type are still independent operators.
\end{itemize}

At last we give some examples of how $\mathcal{Y}$s act on operators, \begin{align}\begin{array}{c|l}
		
		\multirow{5}*{$  \mathcal{O}_{Q{}^2 Q^{\dagger} {}^2 H H^{\dagger}}^{\left(1\sim10\right)}    $}
		
		&\mathcal{Y}\left[\tiny{\young(pr)},\tiny{\young(st)}\right] \left(Q_{pa i} Q_{rb j}\right) \left(Q^{\dagger}{}_{s}^{a j} Q^{\dagger}{}_{t}^{b k}\right)  H^{\dagger}{}^{i} H_{k}
		
		,\quad\mathcal{Y}\left[\tiny{\young(pr)},\tiny{\young(st)}\right] \left(Q_{pa i} Q_{rb j}\right) \left(Q^{\dagger}{}_{s}^{a j} Q^{\dagger}{}_{t}^{b i}\right) \left(H^{\dagger} H\right)
		
		\\&\mathcal{Y}\left[\tiny{\young(pr)},\tiny{\young(st)}\right] \left(Q_{pa i} Q_{rb j}\right) \left(Q^{\dagger}{}_{s}^{a i} Q^{\dagger}{}_{t}^{b k}\right)  H^{\dagger}{}^{j} H_{k}
		
		,\quad\mathcal{Y}\left[\tiny{\young(pr)},\tiny{\young(s,t)}\right] \left(Q_{pa i} Q_{rb j}\right) \left(Q^{\dagger}{}_{s}^{a j} Q^{\dagger}{}_{t}^{b k}\right)  H^{\dagger}{}^{i} H_{k}
		
		\\&\mathcal{Y}\left[\tiny{\young(pr)},\tiny{\young(s,t)}\right] \left(Q_{pa i} Q_{rb j}\right) \left(Q^{\dagger}{}_{s}^{a i} Q^{\dagger}{}_{t}^{b k}\right)  H^{\dagger}{}^{j} H_{k}
		
		,\quad\mathcal{Y}\left[\tiny{\young(p,r)},\tiny{\young(st)}\right] \left(Q_{pa i} Q_{rb j}\right) \left(Q^{\dagger}{}_{s}^{a j} Q^{\dagger}{}_{t}^{b k}\right)  H^{\dagger}{}^{i} H_{k}
		
		\\&\mathcal{Y}\left[\tiny{\young(p,r)},\tiny{\young(st)}\right] \left(Q_{pa i} Q_{rb j}\right) \left(Q^{\dagger}{}_{s}^{a i} Q^{\dagger}{}_{t}^{b k}\right)  H^{\dagger}{}^{j} H_{k}
		
		,\quad\mathcal{Y}\left[\tiny{\young(p,r)},\tiny{\young(s,t)}\right] \left(Q_{pa i} Q_{rb j}\right) \left(Q^{\dagger}{}_{s}^{a j} Q^{\dagger}{}_{t}^{b k}\right)  H^{\dagger}{}^{i} H_{k}
		
		\\&\mathcal{Y}\left[\tiny{\young(p,r)},\tiny{\young(s,t)}\right] \left(Q_{pa i} Q_{rb j}\right) \left(Q^{\dagger}{}_{s}^{a j} Q^{\dagger}{}_{t}^{b i}\right) \left(H^{\dagger} H\right)
		
		,\quad\mathcal{Y}\left[\tiny{\young(p,r)},\tiny{\young(s,t)}\right] \left(Q_{pa i} Q_{rb j}\right) \left(Q^{\dagger}{}_{s}^{a i} Q^{\dagger}{}_{t}^{b k}\right)  H^{\dagger}{}^{j} H_{k}
		
\end{array}\end{align}
with
\eq{ 
	\mathcal{O}_{Q{}^2 Q^{\dagger} {}^2 H H^{\dagger}}^{\left(1\right)} &= \left(Q_{pa i} Q_{rb j}+Q_{ra i} Q_{pb j}\right) \left(Q^{\dagger}{}_{s}^{a j} Q^{\dagger}{}_{t}^{b k}+Q^{\dagger}{}_{t}^{a j} Q^{\dagger}{}_{s}^{b k}\right)  H^{\dagger}{}^{i} H_{k},  \\
	\mathcal{O}_{Q{}^2 Q^{\dagger} {}^2 H H^{\dagger}}^{\left(4\right)} &= \left(Q_{pa i} Q_{rb j}+Q_{ra i} Q_{pb j}\right) \left(Q^{\dagger}{}_{s}^{a j} Q^{\dagger}{}_{t}^{b k}-Q^{\dagger}{}_{t}^{a j} Q^{\dagger}{}_{s}^{b k}\right)  H^{\dagger}{}^{i} H_{k},  \\
	\mathcal{O}_{Q{}^2 Q^{\dagger} {}^2 H H^{\dagger}}^{\left(6\right)} &= \left(Q_{pa i} Q_{rb j}-Q_{ra i} Q_{pb j}\right) \left(Q^{\dagger}{}_{s}^{a j} Q^{\dagger}{}_{t}^{b k}+Q^{\dagger}{}_{t}^{a j} Q^{\dagger}{}_{s}^{b k}\right)  H^{\dagger}{}^{i} H_{k},  \\
	\mathcal{O}_{Q{}^2 Q^{\dagger} {}^2 H H^{\dagger}}^{\left(8\right)} &= \left(Q_{pa i} Q_{rb j}-Q_{ra i} Q_{pb j}\right) \left(Q^{\dagger}{}_{s}^{a j} Q^{\dagger}{}_{t}^{b k}-Q^{\dagger}{}_{t}^{a j} Q^{\dagger}{}_{s}^{b k}\right)  H^{\dagger}{}^{i} H_{k},
}
and
\begin{align}\begin{array}{c|l}
		
		\multirow{4}*{$\mathcal{O}_{Q{}^3  L  H H^{\dagger}}^{\left(1\sim7\right)}   $}
		
		&\mathcal{Y}\left[\tiny{\young(rs,t)}\right]\epsilon ^{abc} \epsilon ^{im} \epsilon ^{jn} \left(Q_{ra j} Q_{tc m}\right) \left(L_{pi} Q_{sb k}\right) H^{\dagger}{}^{k} H_{n}
		
		,\quad\mathcal{Y}\left[\tiny{\young(rs,t)}\right]\epsilon ^{abc} \epsilon ^{ik} \epsilon ^{jn} \left(Q_{ra j} Q_{tc m}\right) \left(L_{pi} Q_{sb k}\right) H^{\dagger}{}^{m} H_{n}
		
		\\&\mathcal{Y}\left[\tiny{\young(rs,t)}\right]\epsilon ^{abc} \epsilon ^{ij} \epsilon ^{kn} \left(Q_{ra j} Q_{tc m}\right) \left(L_{pi} Q_{sb k}\right) H^{\dagger}{}^{m} H_{n}
		
		,\quad\mathcal{Y}\left[\tiny{\young(rst)}\right]\epsilon ^{abc} \epsilon ^{im} \epsilon ^{jn} \left(Q_{ra j} Q_{tc m}\right) \left(L_{pi} Q_{sb k}\right)H^{\dagger}{}^{k}  H_{n} 
		
		\\&\mathcal{Y}\left[\tiny{\young(rst)}\right]\epsilon ^{abc} \epsilon ^{ik} \epsilon ^{jn} \left(Q_{ra j} Q_{tc m}\right) \left(L_{pi} Q_{sb k}\right)H^{\dagger}{}^{m}  H_{n} 
		
		,\quad\mathcal{Y}\left[\tiny{\young(r,s,t)}\right]\epsilon ^{abc} \epsilon ^{im} \epsilon ^{jn} \left(Q_{ra j} Q_{tc m}\right) \left(L_{pi} Q_{sb k}\right) H^{\dagger}{}^{k} H_{n} 
		
		\\&\mathcal{Y}\left[\tiny{\young(r,s,t)}\right]\epsilon ^{abc} \epsilon ^{ik} \epsilon ^{jn} \left(Q_{ra j} Q_{tc m}\right) \left(L_{pi} Q_{sb k}\right) H^{\dagger}{}^{m} H_{n}
		
\end{array}\end{align}
with
\eq{
	\mathcal{O}_{Q{}^3  L  H H^{\dagger}}^{(1)}=&\epsilon^{abc} \epsilon^{im}\epsilon^{jn} \left[\left(Q_{ra j} Q_{tc m}\right) \left(L_{pi} Q_{sb k}\right)+\left(Q_{sa j} Q_{tc m}\right) \left(L_{pi} Q_{rb k}\right)\right] H^{\dagger}{}^{k} H_{n}\\&-\epsilon^{abc} \epsilon^{im}\epsilon^{jn} \left[\left(Q_{ta j} Q_{rc m}\right) \left(L_{pi} Q_{sb k}\right)+\left(Q_{ta j} Q_{sc m}\right) \left(L_{pi} Q_{rb k}\right)\right] H^{\dagger}{}^{k} H_{n},  \\
	\mathcal{O}_{Q{}^3  L  H H^{\dagger}}^{(4)}=&\epsilon ^{abc} \epsilon ^{im} \epsilon ^{jn} \left[\left(Q_{ra j} Q_{tc m}\right) \left(L_{pi} Q_{sb k}\right)+\left(Q_{sa j} Q_{tc m}\right) \left(L_{pi} Q_{rb k}\right)\right] H^{\dagger}{}^{k}  H_{n}\\&+\epsilon ^{abc} \epsilon ^{im} \epsilon ^{jn} \left[\left(Q_{ta j} Q_{rc m}\right) \left(L_{pi} Q_{sb k}\right)+\left(Q_{ra j} Q_{sc m}\right) \left(L_{pi} Q_{tb k}\right)\right] H^{\dagger}{}^{k}  H_{n}\\&+\epsilon ^{abc} \epsilon ^{im} \epsilon ^{jn} \left[\left(Q_{ta j} Q_{sc m}\right) \left(L_{pi} Q_{rb k}\right)+\left(Q_{sa j} Q_{rc m}\right) \left(L_{pi} Q_{tb k}\right)\right] H^{\dagger}{}^{k}  H_{n} , \\
	\mathcal{O}_{Q{}^3  L  H H^{\dagger}}^{(6)}=&\epsilon ^{abc} \epsilon ^{im} \epsilon ^{jn} \left[\left(Q_{ra j} Q_{tc m}\right) \left(L_{pi} Q_{sb k}\right)-\left(Q_{sa j} Q_{tc m}\right) \left(L_{pi} Q_{rb k}\right)\right] H^{\dagger}{}^{k}  H_{n}\\&-\epsilon ^{abc} \epsilon ^{im} \epsilon ^{jn} \left[\left(Q_{ta j} Q_{rc m}\right) \left(L_{pi} Q_{sb k}\right)+\left(Q_{ra j} Q_{sc m}\right) \left(L_{pi} Q_{tb k}\right)\right] H^{\dagger}{}^{k}  H_{n}\\&+\epsilon ^{abc} \epsilon ^{im} \epsilon ^{jn} \left[\left(Q_{ta j} Q_{sc m}\right) \left(L_{pi} Q_{rb k}\right)+\left(Q_{sa j} Q_{rc m}\right) \left(L_{pi} Q_{tb k}\right)\right] H^{\dagger}{}^{k}  H_{n} .
}

\subsection{Conversion between $F_{\rm{L}/\rm{R}}$ and $F,\tilde{F}$}\label{app:a2}

From section~\ref{sec:lorentz_inv} it is clear that we strongly incline to use the chiral basis of the gauge bosons $F_{\rm{L}/\rm{R}}$, which massively simplifies our derivations. Physically, it may be due to their direct correspondence with on-shell particles with definite helicities. However, the other basis that is more commonly used, $F,\tilde{F}$, also has many privileges like its Hermiticity and definite CP. Moreover, a lot of applications are also based on the $F,\tilde{F}$ basis, like the Feynman rule calculations. In this subsection we summarize the conversion rules between the two basis. We start by writing down the definitions:
\begin{align}
	\tilde{F}^{\mu\nu}=\frac12\epsilon^{\mu\nu\rho\eta}F_{\rho\eta},\quad F_{\rm{L}/\rm{R}}=\frac12\left(F\mp i\tilde{F} \right)\;.
\end{align}
from which we can easily deduce the following useful identities
\begin{align}
	\tilde{F}_{1\mu\rho}F_2{}^{\rho\nu}
	=&-F_1{}^{\nu\rho}\tilde{F}_{2\rho\mu}-\frac12 (F_1\tilde{F}_2)\delta^{\nu}_{\mu}\;,\\
	\tilde{F}_{1\mu\rho}\tilde{F}_2{}^{\rho\nu}
	=&F_1{}^{\nu\rho}F_{2\rho\mu}+\frac12(F_1F_2)\delta^{\nu}_{\mu}\;.
\end{align}
In the following, we present various situations where we explicitly do the conversions as examples

\noindent\underline{1. Operators involving one gauge bosons} \\

When the gauge boson contracts with a two form $\mathcal{O}_{\mu\nu}$ with property $\mc{O}_{\mu\nu}^\dagger = \mc{O}_{\mu\nu}$, we have
\begin{align}
	CF_{\rm L}^{\mu\nu}\mathcal{O}_{\mu\nu}+\hc
	=&({\rm Re}\;C)\mathcal{O}_{\mu\nu}F^{\mu\nu}+({\rm Im}\;C)\mathcal{O}_{\mu\nu}\tilde{F}^{\mu\nu}.\label{Ap:1FH}
\end{align}
while for $\mc{O}_{\mu\nu}^\dagger = \mc{O}_{\nu\mu}$ we get instead
\eq{
	CF_{\rm L}^{\mu\nu}\mathcal{O}_{\mu\nu}+\hc = (\rm{Im}\;C)\mathcal{O}_{\mu\nu}F^{\mu\nu}+(\rm{Re}\;C)\mathcal{O}_{\mu\nu}\tilde{F}^{\mu\nu}.
}
In particular, when $F_{\rm L},F_{\rm R}$ contract with $\sigma_{\mu\nu}$, it is further simplified
\begin{align}
	&F_{\rm{L}}{}^{\mu\nu}\left(\sigma_{\mu\nu}\right)_{\alpha}{}^{\beta}=F^{\mu\nu}\left(\sigma_{\mu\nu}\right)_{\alpha}{}^{\beta}=-i\tilde{F}^{\mu\nu}\left(\sigma_{\mu\nu}\right)_{\alpha}{}^{\beta},\quad F_{\rm{L}}{}^{\mu\nu}\left(\bar{\sigma}_{\mu\nu}\right)^{\dot{\alpha}}{}_{\dot{\beta}}=0,\label{Ap:Fsi=0}\\
	&F_{\rm{R}}{}^{\mu\nu}\left(\sigma_{\mu\nu}\right)_{\alpha}{}^{\beta}=0, \quad F_{\rm{R}}{}^{\mu\nu}\left(\bar{\sigma}_{\mu\nu}\right)^{\dot{\alpha}}{}_{\dot{\beta}}=F^{\mu\nu}\left(\bar{\sigma}_{\mu\nu}\right)^{\dot{\alpha}}{}_{\dot{\beta}}=i\tilde{F}^{\mu\nu}\left(\bar{\sigma}_{\mu\nu}\right)^{\dot{\alpha}}{}_{\dot{\beta}}\;.\label{Ap:Fsi=Fsi}
\end{align}
Note that the basis $F,\tilde{F}$ are degenerate when contracting with $\sigma_{\mu\nu}$. In our result, for instance eq.~\eqref{cl:Fpp3}, we adopt $F$ instead of $\tilde{F}$ in the operators.

\noindent\underline{2. Operators involving two gauge boson} \\

For the $X^{\mu\nu}X_{\mu\nu}$ contractions, we have
\eq{
	F_{1\rm{L}}{}^{\mu\nu}F_{2\rm{R}\mu\nu}=0,\quad (F_{1\rm{L}}F_{2\rm{L}})=\frac12\left(F_1F_2-iF_1\tilde{F}_2 \right),\quad (F_{1\rm{R}}F_{2\rm{R}})=\frac12\left(F_1F_2+iF_1\tilde{F}_2 \right).
}
Thus in the operator they are recombined as
\begin{align}
	C(F_{1\rm{L}}F_{2\rm{L}})\mathcal{O}+\hc = ({\rm Re}\;C)(F_1F_2)\mc{O} + ({\rm Im}\;C)(F_1\tilde{F}_2)\mc{O}
\end{align}
where $\mc{O}^\dagger = \mc{O}$ is Hermitian. Contractions of the form $X_{\mu\nu}X^{\nu}_{\ \rho}$ are converted as
\begin{align}
	F_{\rm{L}\mu\rho}F_{\rm{L}}{}^{\rho\nu}=&
	\frac18\delta^{\nu}_{\mu}\left(F^2+iF\tilde{F} \right)\;,\\
	F_{\rm{R}\mu\rho}F_{\rm{R}}{}^{\rho\nu}=&\frac18\delta^{\nu}_{\mu}\left(F^2-iF\tilde{F} \right)\;,\\
	F_{\rm{L}\mu\rho}F_{\rm{R}}{}^{\rho\nu}=&
	\frac12F_{\mu\rho}F^{\rho\nu}+\frac18F^2\delta^{\nu}_{\mu}\;.
\end{align}
When $F_1$ , $F_2$ are anti-symmetric, thus $(F_1F_2)=(F_1\tilde{F}_2)=0$ (for instance they have antisymmetric group indices), we can deduce
\begin{align}
	F_{1\rm{L}\mu\rho}F_{2\rm{L}}{}^{\rho\nu}
	=&\frac14\left(2F_{1\mu\rho}F_2{}^{\rho\nu}+iF_1{}^{\nu\rho}\tilde{F}_{2\rho\mu}-iF_{1\mu\rho}\tilde{F}_2{}^{\rho\nu} \right)\;,\\
	F_{1\rm{R}\mu\rho}F_{2\rm{R}}{}^{\rho\nu}=&\frac14\left(2F_{1\mu\rho}F_2{}^{\rho\nu}-iF_1{}^{\nu\rho}\tilde{F}_{2\rho\mu}+iF_{1\mu\rho}\tilde{F}_2{}^{\rho\nu} \right)\;,\\
	F_{1\rm{L}\mu\rho}F_{2\rm{R}}{}^{\rho\nu}
	=&\frac14\left(iF_1{}^{\nu\rho}\tilde{F}_{2\rho\mu}+iF_{1\mu\rho}\tilde{F}_2{}^{\rho\nu} \right)\;.
\end{align}
For Examples to get operators in eq.~\eqref{cl:F2ppdr} we performed the following conversions:
\begin{align}
	id^{ABC} G_{\rm{L}}^{A}{}^{\mu }{}{}_{\nu } G_{\rm{R}}^{B}{}^{\nu }{}{}_{\lambda } \left(Q_{pa i} \sigma^{\lambda } \left(\lambda ^C\right)_b^a \overleftrightarrow{D}_{\mu } Q^{\dagger}{}_{r}^{b i}\right)=&\frac{i}{2} d^{ABC} G^{A}{}^{\mu }{}{}_{\nu } G^{B}{}^{\nu }{}{}_{\lambda } \left(Q_{pa i} \sigma^{\lambda } \left(\lambda ^C\right)_b^a \overleftrightarrow{D}_{\mu } Q^{\dagger}{}_{r}^{b i}\right), \\
	f^{ABC} G_{\rm{L}}^{A}{}^{\mu }{}{}_{\nu } G_{\rm{R}}^{B}{}^{\nu }{}{}_{\lambda } \left(Q_{pa i} \sigma^{\lambda }\left(\lambda ^C\right)_b^a \overleftrightarrow{D}_{\mu } Q^{\dagger}{}_{r}^{b i}\right)=&\frac{i}{4}f^{ABC}\left(G^{A\mu\nu}\tilde{G}^{B}_{\nu\lambda}+G^A_{\lambda\nu}\tilde{G}^{B\nu\mu}\right)\left(Q_{pa i} \sigma^{\lambda } \left(\lambda ^C\right)_b^a \overleftrightarrow{D}_{\mu } Q^{\dagger}{}_{r}^{b i}\right),
	
\end{align}
while for the complex type with complex Wilson coefficient $C$ we get
\eq{
	&Cf^{ABC} G_{\rm{L}}^{A}{}^{\mu }{}{}_{\nu } G_{\rm{L}}^{B}{}^{\nu }{}{}_{\lambda } \left(Q_{pa i} \sigma^{\lambda } \left(\lambda ^C\right)_b^a i\overleftrightarrow{D}_{\mu}Q^{\dagger}{}_{r}^{b i}\right)+h.c.
	=({\rm Re}\;C)f^{ABC}G^{A\mu\nu}G^{B}_{\nu\lambda}\left(Q_{pa i} \sigma^{\lambda } \left(\lambda ^C\right)_b^a i\overleftrightarrow{D}_{\mu}Q^{\dagger}{}_{r}^{b i}\right) \\
	&\hspace{2cm} +\frac12({\rm Im}\;C)f^{ABC}\left(G^{A\mu\nu}\tilde{G}^B_{\nu\lambda}-G^A_{\lambda\nu}\tilde{G}^{B\nu\mu}\right)\left(Q_{pa i} \sigma^{\lambda } \left(\lambda ^C\right)_b^a i\overleftrightarrow{D}_{\mu}Q^{\dagger}{}_{r}^{b i}\right).
}
When two $F_{\rm L}$ or $F_{\rm R}$ contract with $\sigma_{\mu\nu}$, we write the conversion rules similar with eq.~(\ref{Ap:Fsi=0}-\ref{Ap:Fsi=Fsi})
\begin{align}
	&F_{1\rm{L}}{}^{\mu\lambda}F_{2\rm{L}\lambda}{}^{\nu}\left(\sigma_{\mu\nu}\right)_{\alpha}{}^{\beta}=F_1{}^{\mu\lambda}F_{2\lambda}{}^{\nu}\left(\sigma_{\mu\nu}\right)_{\alpha}{}^{\beta}, \\ &F_{1\rm{R}}{}^{\mu\lambda}F_{2\rm{R}\lambda}{}^{\nu}\left(\bar{\sigma}_{\mu\nu}\right)^{\dot{\alpha}}{}_{\dot{\beta}}=F_1{}^{\mu\lambda}F_{2\lambda}{}^{\nu}\left(\bar{\sigma}_{\mu\nu}\right)^{\dot{\alpha}}{}_{\dot{\beta}}\;.
\end{align}

\noindent\underline{3. Operators involving more gauge bosons} \\

If all gauge bosons contract with each other, they vanish for any mixed helicity configurations
\begin{align}
	F_{1\rm{L}\mu}{}^{\nu}F_{2\rm{L}\nu}{}^{\rho}F_{3\rm{R}\rho}{}^{\mu}
	=&\;0,\quad 
	F_{1\rm{L}\mu}{}^{\nu}F_{2\rm{R}\nu}{}^{\rho}F_{3\rm{R}\rho}{}^{\mu} = 0 \;.
\end{align}
but survive when all the helicities are the same
\begin{align}
	F_{1\rm{L}\mu}{}^{\nu}F_{2\rm{L}\nu}{}^{\rho}F_{3\rm{L}\rho}{}^{\mu}
	=&\frac12\left(F_{1\mu}{}^{\nu}F_{2\nu}{}^{\rho}F_{3\rho}{}^{\mu}-iF_{1\mu}{}^{\nu}F_{2\nu}{}^{\rho}\tilde{F}_{3\rho}{}^{\mu} \right)\;,\\
	F_{1\rm{R}\mu}{}^{\nu}F_{2\rm{R}\nu}{}^{\rho}F_{3\rm{R}\rho}{}^{\mu}=&\frac12\left(F_{1\mu}{}^{\nu}F_{2\nu}{}^{\rho}F_{3\rho}{}^{\mu}+iF_{1\mu}{}^{\nu}F_{2\nu}{}^{\rho}\tilde{F}_{3\rho}{}^{\mu} \right)\;.
\end{align}
Similar feature holds for more gauge bosons contracting together. Some non-vanishing examples of 4 gauge boson contractions are 
\begin{align}
	&CB^2_{\rm R}G^2_{\rm L}+h.c.
	=\frac12({\rm Re}\;C)\left(B^2G^2+(B\tilde{B})(G\tilde{G})\right)+\frac12({\rm Im}\;C)\left(B^2(G\tilde{G})-(B\tilde{B})G^2\right)\;,\\
	&C B_{\rm{L}}{}_{\mu }{}_{\nu } B_{\rm{L}}{}^{\mu }{}{}_{\lambda } B_{\rm{L}}{}^{\lambda }{}^{\rho } B_{\rm{L}}{}^{\nu }{}{}_{\rho }+h.c.=({\rm Re}\;C)B_{\mu }{}_{\nu } B^{\mu }{}{}_{\lambda } B^{\lambda }{}^{\rho } B^{\nu }{}{}_{\rho }+({\rm Im}\;C)B_{\mu }{}_{\nu } B^{\mu }{}{}_{\lambda } B^{\lambda }{}^{\rho } \tilde{B}^{\nu }{}{}_{\rho }\;.
\end{align}

\section{Mathematical Tools}
\label{app:B}

\subsection{Convention in Permutation operation}\label{sec:permoperation}
The elements of symmetric groups $S_m$ are permutations of $m$ objects. Two most popular ways to represent the elements of the $S_m$ are Cycles notation and Matrix notation. For example, a typical element in $S_m$ that permute the first three objects and exchange the last two objects can be expressed in the following form:
\begin{eqnarray}
	\pi=(123)(45)=
	\begin{pmatrix}
		1 & 2 & 3 & 4 & 5 \\
		2 & 3 & 1 & 5 & 4
	\end{pmatrix},
	\label{eq:Rr}
\end{eqnarray} 
In the matrix notation, the numbers in the first row can be viewed as the labels or the positions of the objects and the corresponding numbers in the second row are the labels or the positions of that objects after permutation. In this sense, the permutation can also be viewed as a function that maps the numbers in the first row to the numbers in the second row, i.e. in the above example we have:
\begin{eqnarray}
	\pi(1) = 2,\ \pi(2) = 3,\ \pi(3) = 1,\ \pi(4) = 5,\ \pi(5) = 4.\nonumber \\
\end{eqnarray}
With the above point of view, which treats the group elements as a function, the group multiplication rule is inherent by the composition rule of the function such that:
\begin{eqnarray}
	\pi_i(\pi_j(k)) = (\pi_i\cdot \pi_j)(k),
\end{eqnarray}
where $\cdot$ is the ordinary group multiplication, $i$,$j$ are labels of the group  elements.
Further, we can define the group elements as an operation that permutes the order of arguments of a function such that it becomes another function of the same set of arguments with the original order:
\begin{eqnarray}
	\pi_i \circ F(p_1, p_2,\cdots, p_m) &=&
	F(p_{\pi_i(1)},p_{\pi_i(2)},\cdots , p_{\pi_i(m)}),\nonumber \\
	&\equiv & F_{\pi_i}(p_1, p_2,\cdots, p_m),
	\label{eq:defge}
\end{eqnarray}
without loss of clarity, we shorten the above notation as: $\pi_i \circ F(\{p_k\})=F(\{p_{\pi_i(k)}\})\equiv F_{\pi_i}(\{p_k\})$. More specifically, the above operation changes the $k$th argument of the function $F$ to the argument that originally seats at $\pi_i(k)$th slot in $F$, or equivalently, moves the $i$th argument to $\pi_i^{-1}(k)$ slot. 
The operation $\pi_i\circ$ is essentially a map that convert a function to another function, hence the composition rule of this map can be defined. It is easy to show that such a composition rule naturally preserves the group multiplication rule:
\begin{eqnarray}
	(\pi_i \circ \pi_j)\circ F(\{p_k\}) & \equiv &\pi_i \circ (\pi_j\circ F(\{p_k\}))\nonumber\\
	& = &\pi_i\circ F_{\pi_j}(\{p_{k}\})=\pi_i \circ F(\{p_{\pi_i(k)}\})\nonumber \\
	& = &F(\{p_{\pi_i(\pi_j(k))}\}) = F(\{p_{(\pi_i\cdot\pi_j)(k)}\})\nonumber \\
	& = &(\pi_i\cdot \pi_j)\circ F(\{p_k\}),
\end{eqnarray}
which means the correspondence between the group element $\pi_i$ and the operation $\pi_i\circ$ on functions is a homomorphism.

In section~\ref{sec:motiv}, we mention that to generate a set of bases of the Lorentz structures and the group factors transforming under a certain irrep of the symmetric group, one only needs to act onto an unsymmetrized Lorentz structure or group factor a set of group algebra elements $b^\lambda_x$ that form a basis of the same irrep in the group algebra space. Therefore, we need to generalize the concept of group elements as operations on functions to the group algebra space. 
We define a group algebra element as an operation on a function based on the definition  in Eq.~\ref{eq:defge}. 
For a generic element $r=\sum_i r^i\pi_i$ in the $\tilde{S}_m$, the corresponding operation $r\circ$ on functions is defined as:
\begin{eqnarray}
	r\circ = \sum_i r^i\pi_i\circ.
\end{eqnarray}  
This operation still changes a function to another function with the same set of arguments, while this resulting function is a linear combination of the original one with arguments permuted:
\begin{eqnarray}
	r\circ F(\{p_k\}) &=& \sum_ir^i\pi_i\circ F(\{p_k\})\nonumber \\
	&=& \sum_i r^i F(\{p_{\pi_i(k)}\}) \equiv F_r(\{p_k\}).
	\label{eq:defb}
\end{eqnarray} 
One can verify that the generalization preserves the multiplication rule in the group algebra:
\begin{eqnarray}
	\pi_i\circ F_r(\{p_k\})&=&\pi_i\circ r\circ F(\{p_k\})  \nonumber \\
	& = & F_r(\{\pi_i(k)\})\nonumber \\
	& = &  \sum_i r^j F(\{p_{(\pi_i\cdot\pi_j)(k)}\})\nonumber \\
	& = &\sum_i r^j(\pi_i\cdot\pi_j) F(\{p_{k}\})\nonumber \\
	& = &(\pi_i\cdot r)\circ  F(\{p_k\}) 
\end{eqnarray}
In this case, one can obtain a set of functions $F_x^\lambda(\{p_k\})$ by exerting $b^\lambda_x$ defined in eq.~\eqref{eq:bandD} such that:
\begin{eqnarray}
	F^\lambda_x(\{p_k\}) &\equiv& b^\lambda_x \circ F(\{p_k\})\\
	\pi_i\circ F^\lambda_x(\{p_k\}) &=& \sum_y F^\lambda_y(\{p_k\})D^\lambda_{yx}(\pi_i).
	\label{eq:defpiF}
\end{eqnarray}

As an example we show in the following how to generate the a basis of $[2,1]$ representation of $S_3$ with $T^{p_1p_2p_3p_4}=T(p_1,p_2,p_3,p_4) = \epsilon^{p_1p_2}\epsilon^{p_3p_4}$. First one can verify that the group algebra elements $b^{[2,1]}_1$ and $b^{[2,1]}_2$ below do form a basis of $[2,1]$ irrep in the group algebra space:
\begin{eqnarray}
	b^{[2,1]}_1 &=& \frac{1}{3}[e+(12)-(13)-(123)]\\
	b^{[2,1]}_2 &=& \frac{1}{3}[-(12)+(23)-(123)+(132)],
	\label{eq:defbs}
\end{eqnarray}
such that any permutation $\pi$ in $S_3$ acting on either of them will result in a linear combination of them.
This set of basis generates a matrix representation $\mc{D}^{[2,1]}(\pi)$ of $S_3$ with the two generators $(12)$ and $(123)$ given by:
\begin{eqnarray}
	\mc{D}^{[2,1]}((12))=\begin{pmatrix}
		1&-1\\
		0&-1
	\end{pmatrix},\  
	\mc{D}^{[2,1]}((123))=\begin{pmatrix}
		-1&1\\
		-1&0
	\end{pmatrix}.\nonumber \\
	\label{eq:defDs}
\end{eqnarray}
Readers can verify that the relation in eq.~\eqref{eq:bandD} does hold with the definitions in eq.~\eqref{eq:defbs} and eq.~\eqref{eq:defDs}. Under the operations $b^{[2,1]}_{1,2}\circ$, we obtain a basis from $T(p_1,p_2,p_3,p_4)$:
\begin{eqnarray}
	T^{[2,1]}_1(p_1,p_2,p_3,p_4)&=& b^{[2,1]}_{1}\circ T(p_1,p_2,p_3,p_4) \nonumber \\
	&=& \frac{1}{3}(\epsilon^{p_1p_2}\epsilon^{p_3p_4}+\epsilon^{p_2p_1}\epsilon^{p_3p_4}-\epsilon^{p_2p_4}\epsilon^{p_3p_1}-\epsilon^{p_1p_4}\epsilon^{p_3p_2})\nonumber \\
	&=& \frac{1}{3}(\epsilon^{p_1p_4}\epsilon^{p_2p_3}+\epsilon^{p_1p_3}\epsilon^{p_2p_4})\nonumber \\ \\
	T^{[2,1]}_2(p_1,p_2,p_3,p_4)&=& b^{[2,1]}_{1}\circ T(p_1,p_2,p_3,p_4) \nonumber \\
	&=& \frac{1}{3}(-\epsilon^{p_2p_1}\epsilon^{p_3p_4}+\epsilon^{p_1p_3}\epsilon^{p_2p_4}-\epsilon^{p_2p_3}\epsilon^{p_1p_4}+\epsilon^{p_3p_1}\epsilon^{p_2p_4})\nonumber \\
	&=& \frac{1}{3}(\epsilon^{p_1p_2}\epsilon^{p_3p_4}-\epsilon^{p_1p_4}\epsilon^{p_2p_3})\nonumber \\,
\end{eqnarray}
again readers can verify with the Schouten identity that they transform according to eq.~\eqref{eq:defbs} .

\subsection{Projection operator and CGCs}\label{sec:projection}
We define the projection operator in the direct product space $V=\bigotimes V_{\lambda_i} $ of the $S_m$ group:
\begin{eqnarray}
	P^j_{\lambda i} = \frac{d_\lambda}{m!}\sum_\pi \mc{D}_\lambda^{-1}(\pi)_{ji} U(\pi)
\end{eqnarray}
where $d_\lambda$ is the dimension of the $\lambda$ representation, $ m!$ is the order of the $S_m$ group, $\mc{D}_\lambda(\pi)_{ji}$ is the matrix representation of $\pi$ in irrep $\lambda$. $U(\pi)$ is the representation of $S_m$ on $V$ defined by:
\begin{eqnarray}
	U(\pi) \left(\Motimes_i \mathbf{v}_{\lambda_i}^{k_i}\right) =\sum_{j_i}\left(\Motimes_i\mathbf{v}_{\lambda_i}^{j_i}\right)  \prod_i \mc{D}_{\lambda_i}(\pi)_{j_i k_i}  
\end{eqnarray}
where $\mathbf{v}_{\lambda_i}^{j_i}$ is the $j_i$th basis vector of $\lambda_i$ irrep.

The Theorem-4.2 in ref.~\cite{tung1985group} states that: for any $\mathbf{v}\in V$, $\{P^j_{\lambda i} \mathbf{v},i=1,...,d_\lambda\}$ transform as irrep $\lambda$ if they are not null such that:
\begin{equation}
	U(\pi)\left(P^j_{\lambda i} \mathbf{v}\right)=  \sum_k \left(P^j_{\lambda k} \mathbf{v}\right) \mc{D}_{\lambda_i}(\pi)_{k i}  
\end{equation}
In practical, we chose $j=1$, and generate invariant subspaces of irrep $\lambda$ by iterating $v$ for different basis vector $\otimes \mathbf{v}_{\lambda_i}^{k_i}$ until we get the number of the linear independent subspaces equal to the number of multiplicity of irrep $\lambda$ in the inner product decomposition $\odot \lambda_i$. The CGCs can be extract from the coefficient of basis vectors of the resulting invariant subspaces of irrep $\lambda$.

\phantomsection
\addcontentsline{toc}{section}{\refname}

\bibliography{Dim8EFTref}{}
\bibliographystyle{JHEP}

\end{document}